\documentclass[modern]{aastex631}

\usepackage{pgf}

\usepackage{amsmath}

\newcommand{\Fermi}{{\textit{Fermi}}}

\newcommand{\NYpsue}{79}   
\newcommand{\NYblind}{71}  
\newcommand{\NYRQ}{\pgfmathparse{int(\NYblind-1)}\pgfmathresult} 

\newcommand{\Npsuemsps}{134}  
\newcommand{\Nblindmsps}{10}  
\newcommand{\Nyoung}{\pgfmathparse{int(\NYpsue+\NYblind)}\pgfmathresult}
\newcommand{\Nmsps}{\pgfmathparse{int(\Npsuemsps+\Nblindmsps)}\pgfmathresult}    
\newcommand{\Nall}{\pgfmathparse{int(\NYpsue+\NYblind+\Npsuemsps+\Nblindmsps)}\pgfmathresult}

\newcommand{\Ngalamsps}{427}  
\newcommand{\Npscmsps}{119}  
\newcommand{\NDCpsc}{39}  
\newcommand{\Nserendipity}{6}  
\newcommand{\NPSCtopulse}{\pgfmathparse{int(\NDCpsc-\Nserendipity)}\pgfmathresult}
\newcommand{\Ngammapscmsps}{78} 
\newcommand{\Natnf}{3359}     

\newcommand{\Nbigfile}{3436}      
\newcommand{\NbigfileHiEdot}{762} 

\newcommand{\NbigfilemspsHiEdot}{250}   
\newcommand{\NGCmsps}{254}   

\newcommand{\NmspsAll}{\pgfmathparse{int(\NGCmsps+\Ngalamsps)}\pgfmathresult}

\newcommand{\npsrfitted}{236}
\newcommand{\nwsubbandprofile}{167}
\newcommand{\npsrerestabfinal}{27}
\newcommand{\npsrerestab}{28} 

\newcommand{\nPLEC}{255} 
\newcommand{\nbfree}{116} 

\newcommand{\epeak}{E_{\mathrm{p}}}
\newcommand{\dpeak}{d_{\mathrm{p}}}
\newcommand{\xpeak}{x_{\mathrm{p}}}
\newcommand{\epiv}{E_0}
\newcommand{\edot}{\dot{E}}

\newcommand{\ergs}{\mathrm{erg\,s}^{-1}}
\newcommand{\epshi}{\epsilon_{37,\mathrm{hi}}}
\newcommand{\epslo}{\epsilon_{37,\mathrm{lo}}}

\shorttitle{$3^\mathrm{rd}$ \Fermi{}\, LAT Pulsar Catalog}
\shortauthors{Fermi-LAT Collaboration}

\begin{document}

\title{The Third \Fermi\, Large Area Telescope Catalog of Gamma-ray Pulsars}

\author[0000-0002-7833-0275]{D.~A.~Smith}
\email{David.Smith@u-bordeaux.fr}
\affiliation{Universit\'e Bordeaux, CNRS, LP2I Bordeaux, UMR 5797, F-33170 Gradignan, France}
\affiliation{Laboratoire d'Astrophysique de Bordeaux, Universit\'e de Bordeaux, CNRS, B18N, all\'ee Geoffroy Saint-Hilaire, F-33615 Pessac, France}
\author[0000-0002-6803-3605]{S.~Abdollahi}
\affiliation{IRAP, Universit\'e de Toulouse, CNRS, UPS, CNES, F-31028 Toulouse, France}
\author[0000-0002-6584-1703]{M.~Ajello}
\affiliation{Department of Physics and Astronomy, Clemson University, Kinard Lab of Physics, Clemson, SC 29634-0978, USA}
\author{M.~Bailes}
\affiliation{Centre for Astrophysics and Supercomputing, Swinburne University of Technology, PO Box 218, Hawthorn Victoria 3122, Australia}
\author[0000-0002-9785-7726]{L.~Baldini}
\affiliation{Universit\`a di Pisa and Istituto Nazionale di Fisica Nucleare, Sezione di Pisa I-56127 Pisa, Italy}
\author[0000-0002-8784-2977]{J.~Ballet}
\affiliation{Universit\'e Paris-Saclay, Universit\'e Paris Cit\'e, CEA, CNRS, AIM, F-91191 Gif-sur-Yvette Cedex, France}
\author[0000-0003-4433-1365]{M.~G.~Baring}
\affiliation{Rice University, Department of Physics and Astronomy, MS-108, P. O. Box 1892, Houston, TX 77251, USA}
\author{C.~Bassa}
\affiliation{ASTRON, The Netherlands Institute for Radio Astronomy, Postbus 2, 7990 AA, Dwingeloo, The Netherlands}
\author[0000-0002-6729-9022]{J.~Becerra~Gonzalez}
\affiliation{Instituto de Astrof\'isica de Canarias and Universidad de La Laguna, Dpto. Astrof\'isica, 38200 La Laguna, Tenerife, Spain}
\author[0000-0002-2469-7063]{R.~Bellazzini}
\affiliation{Istituto Nazionale di Fisica Nucleare, Sezione di Pisa, I-56127 Pisa, Italy}
% ...

\begin{abstract}
We present \Nall{} pulsars found in GeV data from the Large Area Telescope (LAT) on the \textit{Fermi Gamma-ray Space Telescope}.  
Another \NPSCtopulse{} millisecond pulsars (MSPs) discovered in deep radio searches of LAT  sources will likely reveal pulsations once phase-connected rotation ephemerides are achieved. 
A further dozen optical and/or X-ray binary systems co-located with LAT sources also likely harbor gamma-ray MSPs.
This catalog thus reports roughly 340 gamma-ray pulsars and candidates, 10\% of all known pulsars, compared to $\leq 11$ known before \textit{Fermi}.
Half of the gamma-ray pulsars are young. Of these, the half that are undetected in radio have a broader Galactic latitude distribution than the young radio-loud pulsars.  
The others are MSPs, with 6 undetected in radio.
Overall, $\geq \npsrfitted$ are bright enough above 50 MeV to fit the pulse profile, the energy spectrum, or both. 
For the common two-peaked profiles, the gamma-ray peak closest to the magnetic pole crossing generally has a softer spectrum.
The spectral energy distributions tend to narrow as the spindown power $\edot$ decreases to its observed minimum near $10^{33}$ erg s$^{-1}$, 
approaching the shape for synchrotron radiation from monoenergetic electrons.
We calculate gamma-ray luminosities when distances are available.
Our all-sky gamma-ray sensitivity map is useful for population syntheses.  
The electronic catalog version provides gamma-ray pulsar ephemerides, properties and fit results to guide and be compared with modeling results.
\end{abstract}


\section{Introduction}
\label{intro}


Fewer than a dozen gamma-ray pulsars were known when the \textit{Fermi Gamma-ray
Space Telescope} was launched  on  2008  June  11, and the extent and diversity
of the population and its role in Galactic dynamics were subject to debate
\citep{Thompson08}.  \Fermi{}'s primary instrument, the Large Area
Telescope \citep[LAT,][]{LATinstrument}, quickly established that the gamma-ray population is large and varied and
is the dominant GeV gamma-ray source class in the Milky Way \citep[][The First
\Fermi{} LAT Catalog of Gamma-ray Pulsars, hereafter 1PC]{1PC}. The 46 pulsars
in 1PC (6 months of data) grew to 132 in 2PC (3 years of data), the Second \Fermi{}-LAT gamma-ray pulsar catalog
\citep{2PC}. This third gamma-ray pulsar catalog (based on 12 years of data) characterizes \Nall{} confirmed gamma-ray pulsars, and tabulates \NPSCtopulse{} MSPs for which gamma-ray pulsations
have not yet been seen but likely will be once accurate rotation ephemerides are
established. We further tabulate LAT sources likely to reveal new ``spider'' MSPs, and LAT sources co-located with known pulsars, some of which may ultimately reveal gamma-ray pulsations. Roughly 340 gamma-ray pulsars and candidates are thus now known, or about 10\% of the $>$\,3400
currently known pulsars (see Table \ref{tab:tallies}).

\begin{figure}[!ht]
\centering
\includegraphics[width=0.9\textwidth]{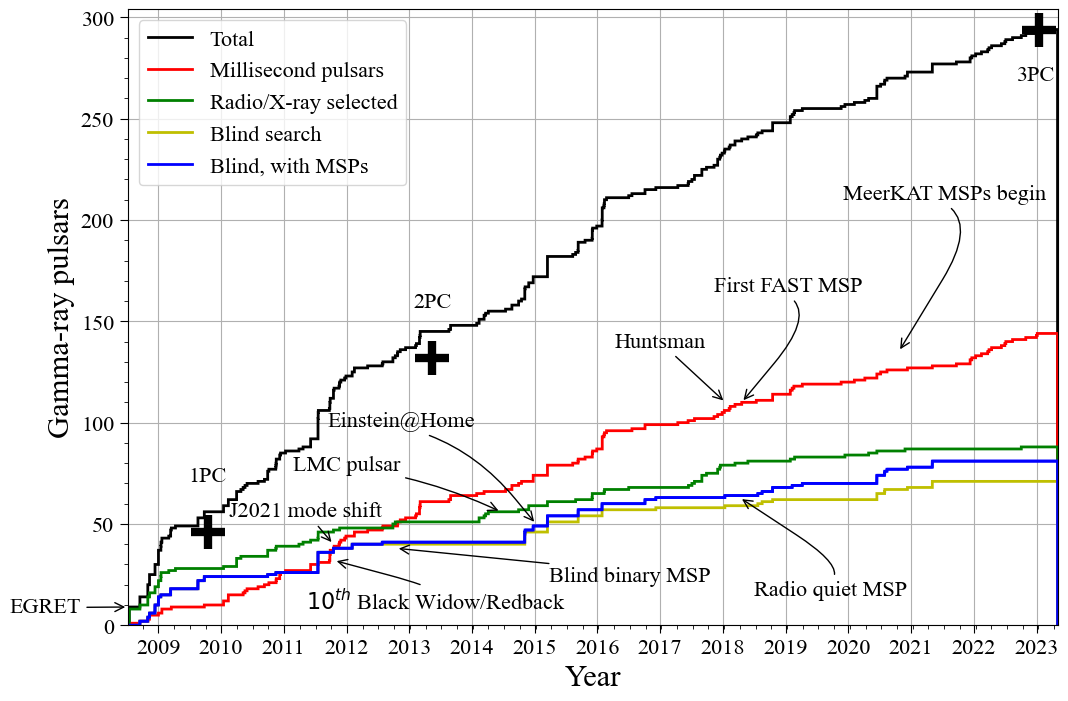}
\caption{ 
Cumulative number of known gamma-ray pulsars, beginning with the launch of \Fermi. The crosses show the numbers included in the first (1PC) and  second (2PC) catalogs of LAT pulsars and their publication dates. Some key discoveries are highlighted. 
See also Table \ref{tab:tallies}.
\label{NpsrsVsYears}}
\end{figure}

These results build on much previous work.  GeV pulsations from the Crab were glimpsed using a balloon-borne instrument at the start of the 1970s \citep[][]{CrabGammaPulseBalloon1973}, followed by the pulsed detection of Vela by the \textit{SAS-2} satellite \citep{Thompson75}. The \textit{COS-B} satellite improved the measurements \citep{COSB81}. 
In the 1990s EGRET on the \textit{Compton} Gamma-Ray Observatory (CGRO) saw six GeV pulsars 
\citep{Thompson99} while a rare pulsar that is brighter below 100 MeV than above, PSR B1509$-$58 was detected with COMPTEL on CGRO \citep{Kuiper1999, LAT_PSR_B1509-58}.
EGRET data revealed three other strong candidates: 
 PSRs J0659+1414 \citep{EGRETB0656p14} and J1048$-$5832 \citep{EGRETB1046m58},
and the millisecond pulsar PSR J0218+4232 \citep{Kuiper2000}. \textit{AGILE} discovered gamma-ray pulsations
from PSR J2021+3651 before \Fermi{}'s launch \citep{Halpern2008}.
All 11 of these pulsars were quickly confirmed using LAT data, and are noted in Figure \ref{NpsrsVsYears}.  Geminga, seen with EGRET, was undetected at radio wavelengths \citep{BnGgeminga96, Geminga1} and has turned out to be the prototype of about half of the young gamma-ray pulsars. The \Nall{} pulsars reported here are more numerous than the 271 sources, all object classes combined, in the $3^{\rm rd}$ EGRET source catalog \citep[][]{EGRET3rdCat}.

Figure \ref{NpsrsVsYears} shows that the discovery rate since launch is steady. 
Table \ref{tab:tallies} breaks the numbers down by category. 
Long-term radio observations by the ``Pulsar Timing Consortium'' \citep{TimingForFermi} enabled about half the discoveries and, importantly, also allowed for an unbiased sample
of pulsars \textit{not} seen in gamma rays \citep[][]{ThousandFold}. 
The ``Pulsar Search Consortium'' \citep{rap+12}, later joined by FAST \citep{FastSurvey,wlc+21} and the TRAPUM\footnote{\url{http://www.trapum.org/}} project on MeerKAT \citep[][]{TRAPUM_FermiSurvey}, discovered large numbers of radio millisecond pulsars (MSPs) at the positions of unidentified gamma-ray sources, leading to the subsequent detection of gamma-ray pulsations.  
Gamma-ray blind searches of unidentified sources revealed radio-quiet pulsars that make up a quarter of the current sample \citep[see e.g.][]{ClarkEatHomeI,wcp+18}. 

The discovery rate is sustained by innovations in how we detect pulsations.
Ever-improving blind search algorithms \citep{PletschClark2014,Nieder2020} allowed the first discovery of a radio-quiet MSP \citep{cpw+18}. The increasingly sophisticated use of an
optical companion's orbit reduces the parameter space searched using
Einstein@home\footnote{\url{http://einstein.phys.uwm.edu}} to discover gamma-ray
MSPs in binary systems with perturbed orbits \citep{nkc+22}. Another example is
a method that allows photon weighting (see Section \ref{obs}) even if the unpulsed gamma-ray source is undetected \citep{SearchPulsation,ThousandFold}. Underlying all analysis efforts
are the improved sensitivity and low energy reach afforded by the \texttt{Pass8} reconstruction method \citep{Pass8, improvedPass8}.  


As a result, not just the numbers but the {\it variety} of gamma-ray pulsars continues to grow. The minimum spindown power $\edot$ is now $20\times$ lower than the pre-launch expectation of $10^{34}$ erg s$^{-1}$ \citep{TimingForFermi}\footnote{$\edot = 4\pi^2 I_{0} \dot P / P^3$, for spin period $P$. We use neutron star moment of inertia $I_{0} = 10^{45}$ g cm$^2$.}. The fastest known field MSP, PSR J0952$-$0607, was found at the location of a gamma-ray source and subsequently timed with LAT data \citep{bph+17,ncb+19}. A third globular cluster, NGC 6652, was found to have gamma-ray emission dominated by a single MSP, PSR J1835$-$3259B \citep{grf+22, gammaJ1835-3259B}, with a fourth recently reported, PSR J1717+4308A in M92 \citep{gammaJ1717+4308A}. The gamma-ray flux and pulse profile of PSR J2021+4026 in the $\gamma$ Cygni supernova remnant (SNR) changed during mode transitions in 2011 and 2018 \citep[][]{gCyg, gCyg_MW}. The LAT sees more than forty `spider' MSPs and a dozen candidates (see Section \ref{not_seen}), compact binary systems where the pulsar wind ablates its companion. Spiders fall into two categories: ``black widows'' have companion masses $0.01M_\odot < M_c < 0.05 \, M_\odot$ and orbital periods $P_B < 10$ h whereas ``redbacks'' have $M_c > 0.2 \, M_\odot$ and $P_B < 1$ d.
Gamma-ray timing of 35 stable MSPs for over 12 years usefully constrains the intensity of gravitational waves from super massive black hole binaries in the hearts of distant galaxies. The upper limit should become a measurement in the coming years \citep[][]{gammaPTA}. Following the methods initially applied to PSR~J1555$-$2908 \citep[][]{nkc+22}, we may be poised to detect planets in multi-year orbits in several compact binary MSP systems.

\begin{figure}[!ht]
\centering
\includegraphics[width=1.\textwidth]{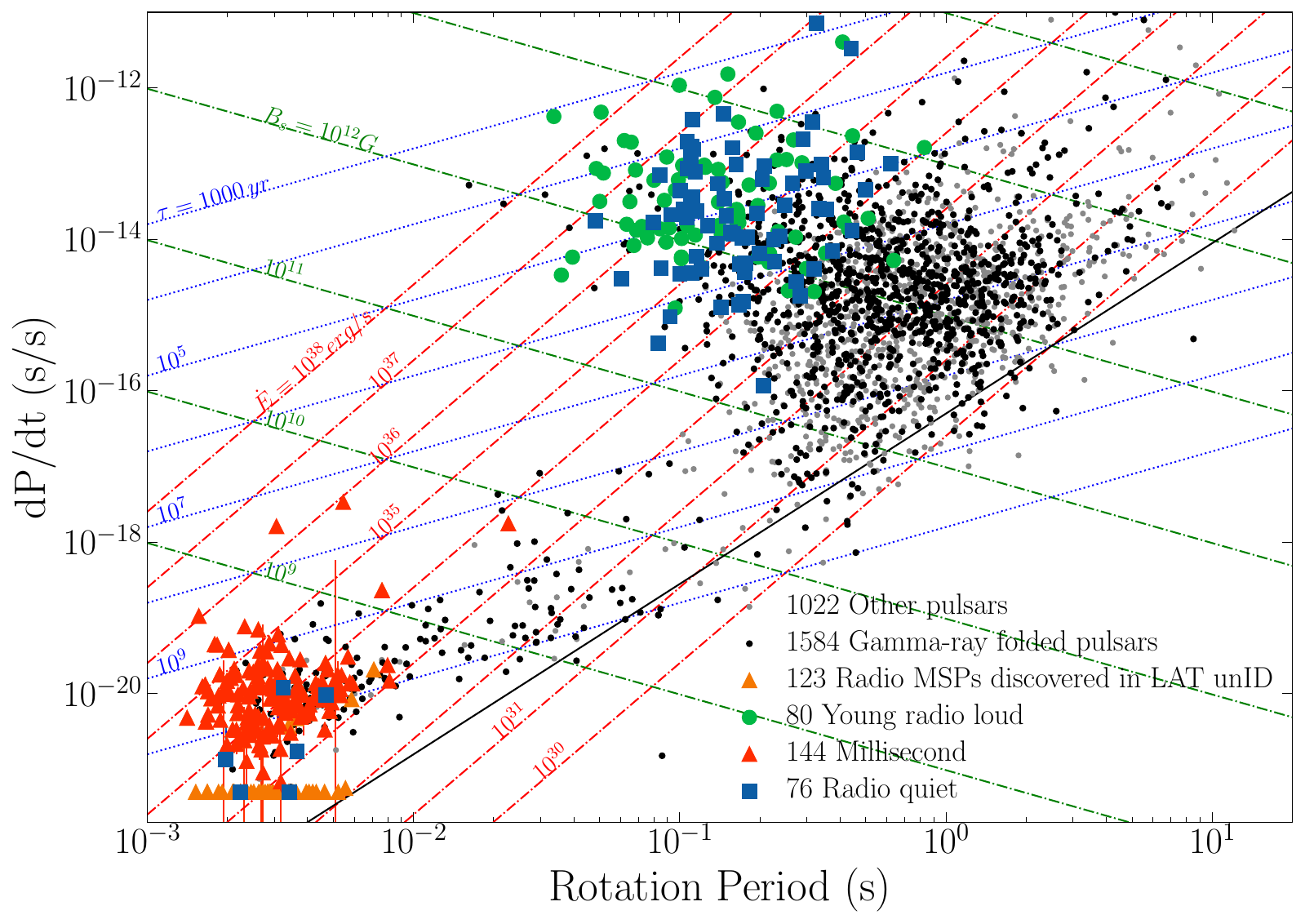}
\caption{ 
Pulsar spindown rate, $\dot P$, versus the rotation period $P$. 
Green dots indicate young, radio-loud (RL) gamma-ray pulsars and 
blue squares show `radio-quiet' (RQ) pulsars, defined as
$S_{1400}< 30\, \mu$Jy, where $S_{1400}$ is the radio flux density at 1400 MHz.
Red triangles are millisecond gamma-ray pulsars. 
Black dots indicate pulsars phase-folded in gamma rays without significant pulsations.
Phase-folding was not done for pulsars shown by gray dots. 
Orange triangles are radio MSPs discovered at the positions of previously unassociated LAT sources, hidden by red triangles when gamma pulsations were subsequently found.
The rest are listed in Table \ref{PSCnotGammaPulse}, and plotted at $\dot P = 5\times 10^{-22}$ when $\dot P$ is unavailable.
The solid black diagonal is the radio deathline of Eq. 4 of \citet{RadioDeathline}.
Shklovskii corrections to $\dot P$ have been applied only to gamma-ray MSPs
with measured proper motion (see Section \ref{doppler}).
\label{PPdotplot}}
\end{figure}

This heterogeneous population can be classified by comparing the spin period ($P$) and the period derivative ($\dot P$), shown in Figure \ref{PPdotplot}.  Throughout this paper, we call pulsars in the main population `young' to distinguish them from the much older MSPs, thought to be spun up to rapid periods via accretion from a companion \citep{Alpar1982}, although e.g. the accretion-induced collapse of white dwarfs might also create MSPs \citep{AccretionInducedMSPbirth}.

\begin{figure}[htbp]
\centering
\includegraphics[width=0.9\textwidth]{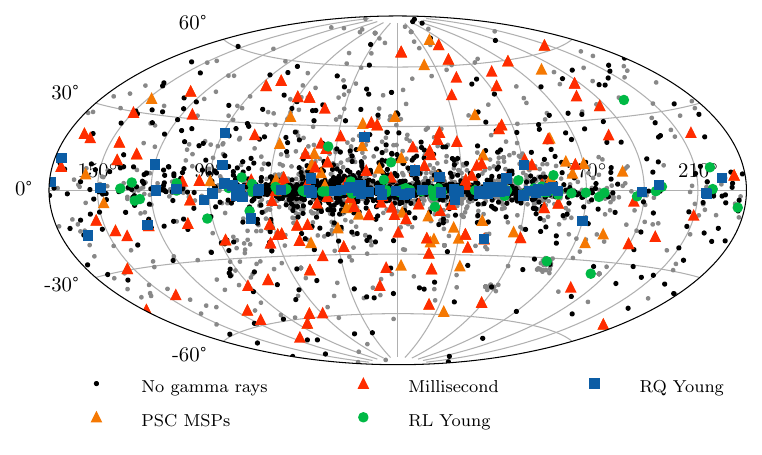}
\caption{Pulsar sky map in Galactic coordinates (Hammer projection). Symbols as in Figure \ref{PPdotplot}. 
\label{Aitoff}}
\end{figure}

All known gamma-ray pulsars are rotation-powered (RPPs): LAT has not yet detected accretion-powered pulsars nor the magnetars that populate the upper-right portion of the $P\dot P$ plane, for which the dominant energy source is magnetic field decay \citep{FermiMagnetars}.  An interesting exception is a LAT detection of a few photons for a few minutes from an extragalactic magnetar giant flare \citep{LATmagnetarFlare}.
The locations of all \Nall{} gamma-ray pulsars on the sky are shown in Figure \ref{Aitoff}. The $P\dot{P}$ diagram shows diagonal lines of constant $\edot$, $\tau_c$, and $B_{\rm S}$ derived from the timing information as follows.  For an orthogonal rotator, the magnetic field on the neutron star surface at the magnetic equator (the rotation pole) is $B_{\rm S} = (1.5 I_{0} c^{3} P \dot{P})^{1/2}/(2\pi R_{\rm NS}^{3}) \simeq 3.2\times 10^{19} {\rm G} \sqrt{P \dot P}$. The ``characteristic age'' $\tau_c = P/(2\dot{P})$ assumes that magnetic dipole braking is the only energy-loss mechanism, that the magnetic moment and inclination do not change, and that the initial spin period was much less than the current period.  $\tau_c$ thus approximates true age well for some young pulsars, and poorly for MSPs. We set the neutron star radius to $R_{\rm NS} = 10$ km, and $c$ is the speed of light in a vacuum.

The 4$^\mathrm{th}$ \Fermi-LAT source catalog \citep{4FGL}, and specifically Data Release 3 \citep[DR3,][hereafter 4FGL]{4FGL-DR3}\footnote{\url{https://fermi.gsfc.nasa.gov/ssc/data/access/lat/12yr_catalog}} characterizes 6658 point and extended sources using 12 years of LAT data. Half of the sources are various blazar classes of active galactic nuclei, but a third remain unassociated with objects known at other wavelengths. 
Radio and gamma-ray pulsation searches at the positions of unidentified sources have yielded fully half of the gamma-ray pulsars. The discoveries continue, as detailed in Sections \ref{pulsedisc} and \ref{not_seen}. The 4FGL spectral, flux, and variability measurements are used throughout this work for the pulsar searches and characterization.

We provide the pulsar catalog as in FITS\footnote{\url{http://fits.gsfc.nasa.gov/}} and spreadsheet file formats, along with other information, 
available in the journal electronic supplement as well as on the \Fermi\, Science Support Center (FSSC) servers\footnote{\url{http://fermi.gsfc.nasa.gov/ssc/data/access/lat/3rd_PSR_catalog/} (Password protected into August 2023.) }.  Appendix \ref{online} describes the online material.

\section{Observations}
\label{obs}

\begin{deluxetable}{lrc}
\tablecaption{Pulsar varieties\label{tab:tallies}}
\tablewidth{0pt}
\tabletypesize{\scriptsize}
\tablehead{
\colhead{Category}      & 
\colhead{Count}   & 
\colhead{Sub-count}}
\startdata
Known rotation-powered pulsars (RPPs)\tablenotemark{a} &  \Nbigfile &   \\
\hspace{1cm}with measured $\dot E > 3\times 10^{33}$ erg s$^{-1}$       & & \NbigfileHiEdot  \\
 \\
Millisecond pulsars (MSPs, $P < 30 $ ms) & \NmspsAll &   \\
\hspace{1cm}with measured $\dot E > 3\times 10^{33}$ erg s$^{-1}$  & &\NbigfilemspsHiEdot  \\
\hspace{1cm}Field MSPs\tablenotemark{b}  & & \Ngalamsps   \\
\hspace{1cm}MSPs in globular clusters\tablenotemark{c}   & & \NGCmsps \\
 \\
\textbf{Gamma-ray pulsars in this catalog}\tablenotemark{d}  &\textbf{\Nall} &  \\
\hspace{1cm}Spectral fits (with free $b$ parameter) \tablenotemark{f}&  & \nPLEC \, (\nbfree)    \\
\hspace{1cm}Profile fits in $\geq \, 1, \, 2, \, 6$ energy bands &  & \npsrfitted , \, \nwsubbandprofile , \, {\pgfmathparse{int(\npsrerestab)}\pgfmathresult} \\
\\
Young gamma-ray pulsars &  \Nyoung &  \\
\hspace{1cm}Radio-quiet\tablenotemark{e}  &  &  \NYRQ  \\
Gamma-ray MSPs   & \Nmsps & \\
\hspace{1cm}Isolated, Binary  &  & 32, 112 \\
\hspace{1cm}Discovered in LAT blind searches & & \Nblindmsps \\
\hspace{1cm}Radio-quiet  &  &  6 \\
\hspace{1cm}Black Widows, Redbacks: & &  32, 13   \\
 \\
Radio MSPs discovered in LAT sources & \Npscmsps &  \\
\hspace{1cm}with gamma-ray pulsations &  & \Ngammapscmsps  \\
\hspace{1cm}waiting for ephemeris phase-connection\tablenotemark{d}  & & \NPSCtopulse \\
\enddata

\tablenotetext{a}{Includes the \Natnf\, pulsars, which are all RPPs, in \textit{psrcat}, the ATNF Pulsar Catalog \citep[v1.69, ][]{ATNFcatalog}, 
and as-yet unpublished discoveries.}
\tablenotetext{b}{\url{http://astro.phys.wvu.edu/GalacticMSPs}}
\tablenotetext{c}{\url{http://www.naic.edu/~pfreire/GCpsr}}
\tablenotetext{d}{Table \ref{PSCnotGammaPulse} lists \NDCpsc{} MSPs discovered in radio searches of bright 4FGL sources with pulsar-like spectra. At least 6 were serendipitous. The rest will likely show pulsations once radio timing allows gamma-ray phase-folding. Table \ref{ColocatedNotPulse} lists additional pulsars co-located with 4FGL sources, some of which may reveal pulsations in the future. Table \ref{spiders} lists 13 ``spider'' MSP candidates co-located with LAT sources. The number of detected gamma-ray pulsars thus likely exceeds 340, including unpulsed detections.}
\tablenotetext{e}{$S_{1400}< 30\, \mu$Jy, where $S_{1400}$ is the radio flux density at 1400 MHz.}
\tablenotetext{f}{Sections \ref{profiles} and \ref{spectralSection} describe the pulse profile fits and energy spectral fits, respectively.}


\end{deluxetable}

\citet{LATinstrument} describe the \Fermi{} LAT, and \citet{OnOrbit}, \citet{P7Paper}, and \citet{LAT10year} report on-orbit performance.  \Fermi{} carries another instrument, the Gamma-ray Burst Monitor \citep[GBM;][]{GBMinstrument}, which was not used to prepare this catalog.

The LAT is a pair-production
telescope composed of a $4\times4$ grid of towers. Each tower consists of a stack of tungsten foil converters interleaved with silicon-strip particle tracking detectors, mated with a hodoscopic cesium-iodide calorimeter. A segmented plastic scintillator anti-coincidence detector covers the grid to help discriminate charged particle backgrounds from gamma-ray photons. 
The LAT field of view is $\sim$2.4 sr. 
For most of the  \Fermi{}  mission, the primary operational mode has been a sky survey where the satellite rocks between a pointing above the orbital plane and one below the plane after each orbit.
In this mode, the entire sky is imaged every two orbits ($\sim$3 hours) and any given point on the sky is observed $\sim1/6^{\rm th}$ of the time. For one year beginning 2013 December, the normal survey mode was changed to favor exposure to the Galactic center region\footnote{\url{https://fermi.gsfc.nasa.gov/ssc/observations/types/exposure/}}. Survey mode was again modified after a solar panel rotation drive stopped moving on 2018 March 16, detailed in Section 2.1 of \citet{LAT10year}.

The LAT is sensitive to gamma rays with energies $E$ from 20\,MeV to over 300\,GeV, with an on-axis effective area of $\sim$8000\,cm$^{2}$ above 1 GeV. Multiple Coulomb scattering of the electron-positron pairs created by converted gamma rays degrades the per-photon angular resolution, with the average 68\% containment radius varying as $E^{-0.8}$ from 5$^\circ$ at 100\,MeV to $0\fdg1$ at 10\,GeV\footnote{\url{https://www.slac.stanford.edu/exp/glast/groups/canda/lat_Performance.htm}}.

 This energy-dependent point spread function (PSF), the backgrounds from the complex diffuse emission and nearby sources, and the source spectrum are encapsulated in the \textit{photon weights} \citep[$w_i$,][]{KerrWeighted}, which give the probability that a photon originates from a particular source.  Weights are used to optimize pulsed signal significance while minimizing event selection trials penalties. This powerful tool has been extended (``\textit{simple weights''} and ``\textit{model weights''}) to sky locations with no point source \citep{SearchPulsation}.  Weighting is used for all of the discovery techniques (Section \ref{pulsedisc}), and to characterize the pulse profiles (Section \ref{profiles}).

Events recorded by the LAT have timestamps derived from GPS clocks integrated into the satellite's Guidance, Navigation, and Control (GNC) subsystem, accurate to $\approx$\,300\,ns relative to UTC \citep{OnOrbit,LAT10year}.  GNC provides the instantaneous spacecraft position with $\approx$\,60\,m accuracy. Generally, we compute pulsar rotational phases $\phi_i$ using \textsc{Tempo2} \citep{Hobbs2006} with the \texttt{fermi} plug-in \citep{Ray2011}, or with PINT \citep{Luo21}.  These use the recorded times and spacecraft positions combined with a pulsar rotational ephemeris (specified in a \textsc{Tempo2} parameter, or `par', file). The timing chain from the instrument clocks through the barycentering and epoch folding software is accurate to better than a $\mu$s \citep{TimingForFermi}.  

We use different data selections and processing for the various analyses presented here, though all cases make use of `Source' class events reconstructed using \texttt{Pass8} revision 3 \citep{Pass8, improvedPass8}.  
The pulsation searches and pulsar timing described in Section \ref{pulsedisc} generally make use of all available data at the time of analysis and employ a variety of data processing methods.  The effects of this heterogeneity on e.g. pulsar discovery efficiency or pulse profile inference are minor, and we do not attempt to provide further details.

For the pulse profiles (see Section \ref{profiles}), the data span $11.8$ years, from MJD 54682 (2008 August 4, when LAT data taking began) to MJD 59000 (2020 May 31).  We exclude gamma rays collected when the LAT was not in nominal science operations mode or when the spacecraft rocking angle exceeded $52^{\circ}$, and we accept photons with reconstructed energies from $0.05$ to 100 GeV, within 15$^{\circ}$ of the pulsar positions.  We used Model Weights, requiring weights $w_i>0.001$. 

The spectral measurements (see Section \ref{spectralSection}) use the same data set as 4FGL-DR3.  In brief, the data span 12 years, to MJD 59063 (2020 August 2) and are selected with energies 50\,MeV--1\,TeV.  The catalog analysis further uses a heterogeneous zenith angle cut ranging from $80^{\circ}$ for energies below 100 MeV to $105^{\circ}$ above 1 GeV.

\section{Discovery and Timing}
\label{pulsedisc}

Inclusion in the main catalog requires a statistically
significant pulsation detection in the \Fermi{} LAT gamma-ray data. The following
subsections describe the paths to detection, as well as a brief description of gamma-ray timing that often strengthens the initial signal.

In all cases, detecting and characterizing pulsations requires a rotation ephemeris (or a ``timing model'') to convert photon arrival times $t_i$ to neutron star rotational phases $\phi_i$. 
This ``folding'', which ``stacks'' photons at the same rotational phase, allows a pulse to rise above the background since $\ll$1 photons are collected per pulse even from the brightest pulsars \citep{Kerr22}. We detect pulsations with the H-test \citep{DeJager1989, deJager2010}, a statistical test for discarding the null hypothesis that a set of photon phases is uniformly distributed. For $N_{\gamma}$ gamma rays, the $m=20$ harmonic weighted version of the H-test statistic \citep{KerrWeighted} is 
\begin{equation}
H \equiv \max(Z^2_{mw} - 4\times(m-1), 1 \leq m \leq 20),
\end{equation}
with 
\begin{equation}
Z^2_{w}\equiv \frac{2}{\sum_i w_i^2}\sum_{k=1}^{m=20}(\alpha_{kw}^2+\beta_{kw}^2),
\label{Eq:Ztest}
\end{equation}
and $\alpha_{kw}$ and $\beta_{kw}$ the empirical trigonometric coefficients 
$\alpha_{kw}\equiv \sum_{i=1}^{N_{\gamma}}
w_i\sin(2\pi k\phi_i)$ and $\beta_{kw}\equiv \sum_{i=1}^{N_{\gamma}}
w_i\cos(2\pi k\phi_i)$. The $w$ subscripts indicate that this is a photon-weighted version of the test, and the $w_i$ are the photon weights, evaluated via spectral analysis \citep[see also 2PC,\,][]{2PC}.  For $m=20$, which we adopt universally, the cumulative distribution function for $H$ in the asymptotic limit is $P(H \geq x)=\exp(-0.398405 x)$ \citep{KerrWeighted}, and 3$\sigma$, 4$\sigma$, and 5$\sigma$ thresholds correspond to $H=14$, $H=24$, and $H=36$. 
The H-test is unbinned, well-suited to the extremely sparse gamma-ray pulsar data:  the LAT often detects only one photon in tens of thousands (millions, for MSPs) of pulsar rotations. \citet{SearchPulsation} gives corrections for small photon counts. 
Pulsars with narrow, sharp peaks are easier to detect than pulsars with broad peaks \citep[see Figure 5 of][]{SixWeak}. All pulsars in our sample are detected with $m=8$. Using $m=20$ incurs little computational cost and insures sensitivity to putative exotic profile shapes. \citet{KerrWeighted} also showed that large $m$ does not cause false positive detections.

\begin{figure}[tbp]
\centering
\includegraphics[width=0.8\textwidth]{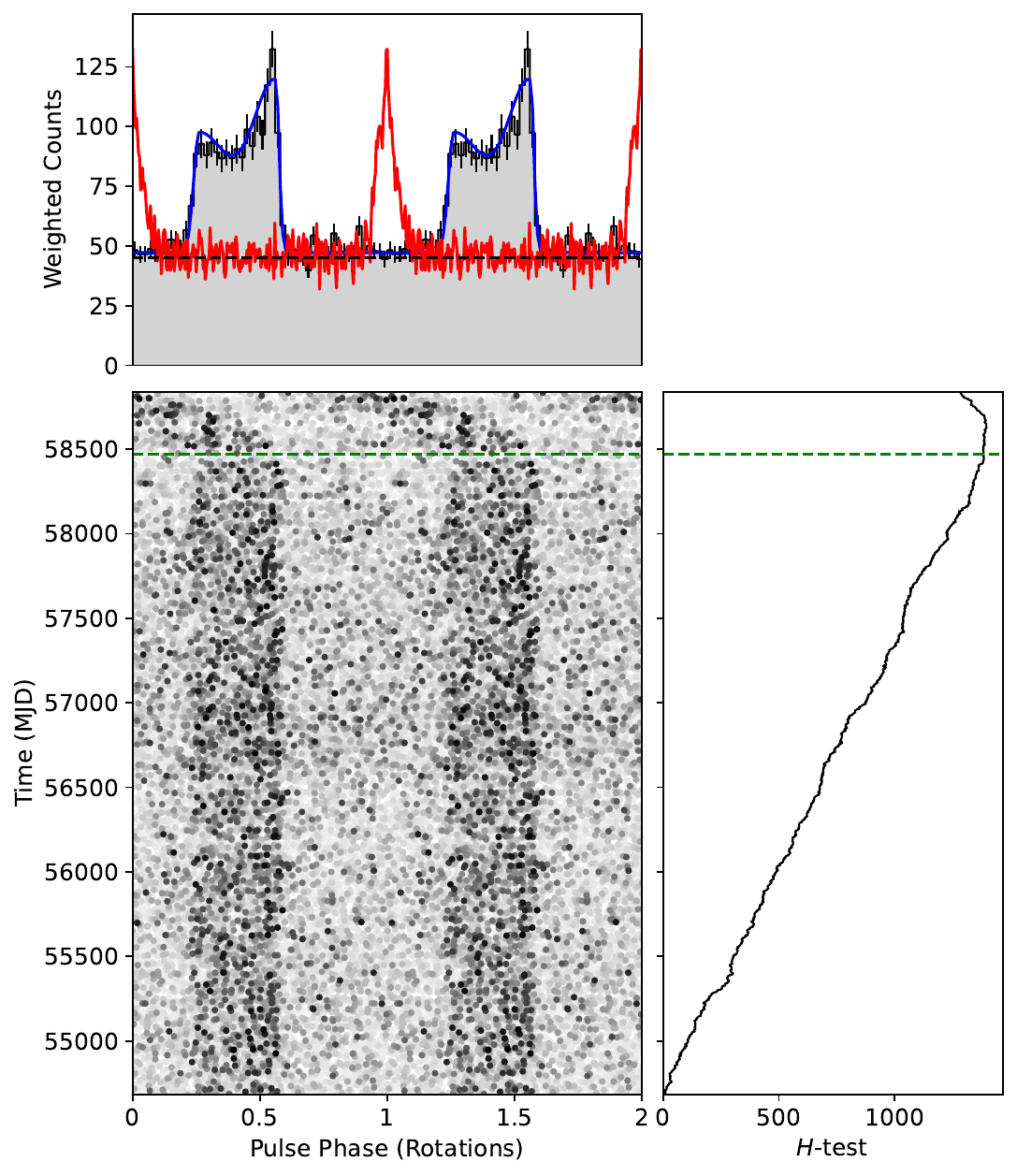}
\caption{Top: gamma-ray phase histogram for PSR J1648-4611 discovered by \citet{kbm+03}, overlaid with the $1.4$ GHz profile (red) obtained during Parkes radio telescope timing \citep[][]{psj+19}. The blue curve shows a fit to the histogram, and the horizontal dashed line is an estimate of the background level (see Section \ref{profiles:lcs}).
Bottom-left: Phaseogram over the course of the mission. Dots indicate photons, with the grayscale set according to the photon weight. Right: H-test significance accumulated over the course of the mission. 
The green horizontal dashed line shows when the ephemeris validity finishes (see Section \ref{radioselected}). 
 \label{SampleLC}}
\end{figure}

Figure \ref{SampleLC} highlights some aspects of gamma-ray phase folding. 
Figure \ref{profiles:example_lightcurve_0030} and Appendix \ref{App-Samples} show other example profiles.
The top-most frame shows a weighted phase histogram, duplicated over a second rotation. 
Section \ref{profiles} describes the profile fit overlaid in blue. The phase-aligned $1.4$ GHz 
radio pulse overlaid in red comes from the radio timing observations used to create the rotation ephemeris, in this specific case by \citet{psj+19}. The horizontal dashed line shows the gamma-ray background level, estimated from the photon weights as $\frac{1}{\mathrm{N_{bin}}}\left(\sum_i w_i - \sum_i w_i^2\right)$, the sum of the expected contribution to the weights of background photons not associated with the pulsar. A phase histogram baseline exceeding the background level may indicate the presence of unpulsed magnetospheric emission. Section \ref{profiles:lcs} gives details.

The next frame below it shows the phase drifting after the last radio time of arrival used to model the neutron star rotation,
indicated by the green horizontal dashed line. The start of this timing model's ``validity'' range, 
before \Fermi's launch, is not shown. Pulsars with irregular spindown, as for this young high-$\dot E$ pulsar, require extra parameters to model the rotation, and accuracy of the extrapolation past the validity range rapidly degrades.  Stable pulsars can be modeled with few parameters, often accurately predicting the neutron star rotation for years before and/or after validity. 
The right-hand frame shows the weighted H-test increasing as data accumulated over the years, 
a nearly straight line for most pulsars. Changes in slope can result from phase drifts, as in this case, 
or from increased background due to e.g. a nearby flaring blazar (see Section \ref{variability}), or
from changes in the LAT's exposure to the pulsar's sky position.
Exposure per unit time increased for this pulsar during 2014 (mid-year was MJD 56810) when LAT pointed more frequently towards the Galactic center, visible in the Time versus Phase plot, without however affecting the H-test growth.
\citet{ThousandFold} show simulations for which Poisson fluctuations in the very low rate of photon arrivals for pulsars near detection threshold significantly perturb the H-test time evolution.
Slope variations due to pulsar flux changes nearly never happen: rare exceptions are the young pulsar PSR~J2021+4026 \citep[][]{gCyg, gCyg_MW} and transitional MSPs like PSR~J1023$+$0038 \citep[see][and Appendix A]{J1023transition} and J1227$-$4853 \citep{rrb+15, jrr+15}, which lack detectable gamma-ray pulsation during their accretion states.

If different analyses were tried, the number of trials must be accounted for in the pulsation significance calculation, requiring a higher H-test value to claim a detection.
Before the advent of photon weighting, we sometimes varied the minimum energy and/or the
angular extent of our dataset, to explore the pulsar's spectral hardness and local background level.
Using a LAT source's measured spectrum to calculate weights allows one single trial
for this parameter space. 

When the source spectrum is too faint to measure, we use ``simple weights'' with a parameter $\mu_w$ of unknown optimal value, and thus, possible additional trials. The $\mu_w$ parameter is the logarithm of the energy at which the log-Gaussian weighting function peaks.
Since \citet{ThousandFold} we have been using six trials for simple weight gamma-ray pulsation searches using radio or X-ray ephemerides, and an H-test $>25$ ($p=4.7\times10^{-5}$, or $>4.1\sigma$) detection threshold. We fold the entire data set 3 times, using $\mu_w = (3.2, 3.6, 4.0)$, and 3 more times restricting the data to the ephemeris validity period. Although these 6 trials are not independent, a conservative estimate for the chance probability for a false positive detection with a H-test$>25$ threshold is $6\times4.7\times10^{-5}$, or 0.3 for a sample of 1000 pulsars. In 2PC we required H-test $>36$ ($>5\sigma$). \citet{ThousandFold} discovered 16 new gamma-ray pulsars, in part because of this refined, relaxed threshold. Several more pulsars in this catalog were later found in the same way. 

This prescription may miss some pulsars if the ephemeris extrapolates well enough that the accumulation of  additional data beyond the validity range yields $H$$>$25, but poorly enough that the full data yield $H$$<$25. Such cases also arise from positive fluctuations, so we avoid further trials factors by limiting our search to the two combinations just described.

As further explained in Section \ref{blindsearch}, when the ephemeris is established independently of the LAT, using radio or X-ray observations (Section \ref{radioselected}), the pulsar can be up to 20 times fainter in gamma rays than if an unidentified LAT point source (Section \ref{targets}) guides a ``blind'' gamma-ray search.
Once pulsations are found during some epoch of LAT data, gamma-ray timing can extend and improve the ephemeris (Section \ref{LATtiming}).

Using the above detection criteria, we report \Nall{} gamma-ray pulsars in the breakdown tabulated in
Table \ref{tab:tallies} with their distribution on the sky shown in Figure \ref{Aitoff}. The \Nyoung{} young pulsars are named in Table \ref{tbl-charPSR}, and the \Nmsps\ MSPs are in Table \ref{tbl-charMSP}.
The new pulsars in 3PC were mostly more ``difficult'' to discover than those reported in the earlier catalogs. For pulsars found with just a few foldings using a radio ephemeris, discussed in Section \ref{radioselected}, \textit{difficulty} stems from their extreme faintness ($<1$ photon per month): an analysis using many trials (for example, from exploring different data intervals) yields significance too low to distinguish from statistical fluctuations. Furthermore, we continue to discover radio MSPs at the positions of unidentified FGL sources, described in Section \ref{psc}. These are \textit{difficult} both in radio, because of eclipses, scintillation, and intrinsic faintness, and in gamma rays if their signal-to-noise ratio is low and the ephemeris validity is brief, that is, if the radio observations cover only a few years of the LAT mission.
The blind searches, discussed in Section \ref{blindsearch}, are now finding MSPs, including those with short, varying orbital periods, feared as perhaps impossible before launch \citep{RansomPreLaunchBlindSearch}. These searches require many orders of magnitude more computing power than for young pulsars.  The catalog includes 44 ``spider'' systems, and in Section \ref{xraypsrs} we tabulate several candidates co-located with gamma-ray sources, for which pulsations will likely be seen in the coming years.
Timing models for ``noisy'' pulsars require many parameters which presents \textit{difficulty} in maintaining a coherent timing solution, e.g. with frequent radio monitoring.    Our methodical pursuit of ever more \textit{difficult} gamma-ray pulsars in the LAT's unbiased all-sky data set means that our notion of gamma-ray pulsars is greatly enriched compared to 1PC.

\clearpage
\startlongtable
\begin{deluxetable}{llrrrrrcr}
\tabletypesize{\scriptsize}
\tablewidth{0pt}
\tablecaption{Some parameters of young LAT-detected pulsars
\label{tbl-charPSR}}

\tablehead{
\colhead{PSR} &   \colhead{Codes} &    \colhead{$l$} & \colhead{$b$} & \colhead{$P$} & \colhead{$\dot P$}  & \colhead{$10^{-33}\dot E$}& \colhead{$S_{1400}$}  \\
&      &  \colhead{($^\circ$)} & \colhead{($^\circ$)} & \colhead{(ms)} & \colhead{($10^{-15}$)}  & \colhead{(erg s$^{-1}$)} & \colhead{(mJy)} 
}

\startdata
 J0002+6216 & GUr  & 117.33 & -0.07 &  115.4 & 6.0 &    153. &  0.02 \\ 
 J0007+7303 & Gq  & 119.66 & 10.46 &  315.9 & 355.92 &    445. & $<$0.005 \\ 
 J0106+4855 & GUr  & 125.47 & -13.87 &  83.2 & 0.4 &   29.4 &  0.01 \\ 
 J0117+5914 & Rr  & 126.28 & -3.46 &  101.4 & 5.8 &    221. &  0.30 \\ 
 J0139+5814 & Rr  & 129.22 & -4.04 &  272.5 & 10.7 &   20.9 &  4.60 \\ 
 J0205+6449 & Xrx  & 130.72 & 3.08 &  65.7 & 192.1 &  26688. &  0.05 \\ 
 J0248+6021 & Rr  & 136.90 & 0.70 &  217.1 & 55.2 &    212. & 13.70 \\ 
 J0357+3205 & GUq  & 162.76 & -16.01 &  444.1 & 13.10 &   5.90 & $<$0.004 \\ 
 J0359+5414 & GUq  & 148.23 & 0.88 &  79.4 & 16.7 &   1317. &           \\ 
 J0514$-$4408 & Rr  & 249.51 & -35.36 &  320.3 & 2.0 &   2.45 &  0.71 \\ 
 J0534+2200 & RErx  & 184.56 & -5.78 &  33.7 & 420.2 & 435319. & 14.00 \\ 
 J0540$-$6919 & Xrx  & 279.72 & -31.52 &  50.6 & 478.9 & 145529. &  0.10 \\ 
 J0554+3107 & GUq  & 179.06 & 2.70 &  465.0 & 142.60 &   56.0 & $<$0.066 \\ 
 J0622+3749 & GUq  & 175.88 & 10.96 &  333.2 & 25.42 &   27.1 & $<$0.012 \\ 
 J0631+0646 & GUr  & 204.68 & -1.24 &  111.0 & 3.6 &    104. & 0.025$^a$ \\ 
 J0631+1036 & Rr  & 201.22 & 0.45 &  287.8 & 102.7 &    170. &  1.11 \\ 
 J0633+0632 & GUq  & 205.09 & -0.93 &  297.4 & 79.57 &    119. & $<$0.003 \\ 
 J0633+1746 & XExq  & 195.13 & 4.27 &  237.1 & 10.97 &   32.5 & $<$0.507 \\ 
 J0659+1414 & Rrx  & 201.11 & 8.26 &  384.9 & 55.0 &   38.0 &  2.70 \\ 
 J0729$-$1448 & Rr  & 230.39 & 1.42 &  251.7 & 113.4 &    280. &  0.83 \\ 
 J0729$-$1836 & Rr  & 233.76 & -0.34 &  510.2 & 18.9 &   5.63 &  1.90 \\ 
 J0734$-$1559 & GUq  & 232.06 & 2.02 &  155.1 & 12.51 &    132. & $<$0.005 \\ 
 J0742$-$2822 & Rr  & 243.77 & -2.44 &  166.8 & 16.7 &    141. & 26.00 \\ 
 J0744$-$2525 & GUq  & 241.35 & -0.73 &  92.0 & 1.0 &   48.3 &           \\ 
 J0802$-$5613 & GUq  & 269.98 & -13.19 &  274.1 & 2.8 &   5.30 &           \\ 
 J0834$-$4159 & Rr  & 260.89 & -1.04 &  121.1 & 4.3 &   95.1 &  0.28 \\ 
 J0835$-$4510 & RErx  & 263.55 & -2.79 &  89.4 & 122.3 &   6763. & 1050.00 \\ 
 J0908$-$4913 & Rr  & 270.27 & -1.02 &  106.8 & 15.1 &    490. & 20.00 \\ 
 J0922+0638 & Rr  & 225.42 & 36.39 &  430.6 & 13.7 &   6.77 & 10.00 \\ 
 J0940$-$5428 & Rr  & 277.51 & -1.29 &  87.6 & 32.8 &   1928. &  0.66 \\ 
 J1016$-$5857 & Rr  & 284.08 & -1.88 &  107.4 & 80.4 &   2563. &  0.90 \\ 
 J1019$-$5749 & Rr  & 283.84 & -0.68 &  162.5 & 20.1 &    184. &  3.80 \\ 
 J1023$-$5746 & GUq  & 284.17 & -0.41 &  111.5 & 379.89 &  10820. & $<$0.030 \\ 
 J1028$-$5819 & Rr  & 285.06 & -0.50 &  91.4 & 14.2 &    734. &  0.24 \\ 
 J1044$-$5737 & GUq  & 286.57 & 1.16 &  139.0 & 54.57 &    801. & $<$0.020 \\ 
 J1048$-$5832 & Rr  & 287.43 & 0.58 &  123.7 & 95.5 &   1992. &  9.10 \\ 
 J1055$-$6028 & Rr  & 289.13 & -0.74 &  99.7 & 29.5 &   1176. &  0.95 \\ 
 J1057$-$5226 & RErx  & 285.98 & 6.65 &  197.1 & 5.8 &   30.1 &  4.40 \\ 
 J1057$-$5851 & GUq  & 288.61 & 0.80 &  620.4 & 100.6 &   16.6 &           \\ 
 J1105$-$6037 & GUq  & 290.24 & -0.40 &  194.9 & 21.8 &    116. &           \\ 
 J1105$-$6107 & Rr  & 290.49 & -0.85 &  63.2 & 15.8 &   2475. &  1.20 \\ 
 J1111$-$6039 & GUq  & 291.02 & -0.11 &  106.7 & 195.2 &   6346. &           \\ 
 J1112$-$6103 & Rr  & 291.22 & -0.46 &  65.0 & 31.5 &   4537. &  2.30 \\ 
 J1119$-$6127 & Rrx  & 292.15 & -0.54 &  409.1 & 4042.4 &   2330. &  1.09 \\ 
 J1124$-$5916 & Rrx  & 292.04 & 1.75 &  135.5 & 751.5 &  11914. &  0.08 \\ 
 J1135$-$6055 & GUq  & 293.79 & 0.58 &  114.5 & 78.23 &   2057. & $<$0.030 \\ 
 J1139$-$6247 & GUq  & 294.79 & -1.06 &  120.4 & 4.1 &   91.6 &           \\ 
 J1151$-$6108 & Rr  & 295.81 & 0.91 &  101.6 & 10.3 &    386. &  0.06 \\ 
 J1203$-$6242 & GUq  & 297.52 & -0.34 &  100.6 & 44.1 &   1709. &           \\ 
 J1208$-$6238 & GUq  & 297.99 & -0.18 &  440.7 & 3309.57 &   1526. & $<$0.017 \\ 
 J1224$-$6407 & Rr  & 299.98 & -1.41 &  216.5 & 5.0 &   19.3 &  8.90 \\ 
 J1231$-$5113 & GUq  & 299.76 & 11.52 &  206.4 & 0.1 &  0.525 &           \\ 
 J1231$-$6511 & GUq  & 300.87 & -2.40 &  247.4 & 28.4 &   74.0 &           \\ 
 J1253$-$5820 & Rr  & 303.20 & 4.53 &  255.5 & 2.1 &   4.98 &  4.10 \\ 
 J1341$-$6220 & Rr  & 308.73 & -0.03 &  193.4 & 253.0 &   1379. &  2.70 \\ 
 J1350$-$6225 & GUq  & 309.73 & -0.34 &  138.2 & 8.9 &    132. &           \\ 
 J1357$-$6429 & Rrx  & 309.92 & -2.51 &  166.2 & 354.4 &   3047. &  0.52 \\ 
 J1358$-$6025 & GUq  & 311.11 & 1.37 &  60.5 & 3.0 &    536. &           \\ 
 J1410$-$6132 & Rr  & 312.19 & -0.09 &  50.1 & 31.8 &  10000. &  1.90 \\ 
 J1413$-$6205 & GUq  & 312.37 & -0.74 &  109.7 & 27.39 &    818. & $<$0.024 \\ 
 J1418$-$6058 & GUq  & 313.32 & 0.13 &  110.6 & 171.00 &   4992. & $<$0.029 \\ 
 J1420$-$6048 & Rrx  & 313.54 & 0.23 &  68.2 & 82.4 &  10250. &  1.19 \\ 
 J1422$-$6138 & GUq  & 313.52 & -0.66 &  341.0 & 96.79 &   96.4 & $<$0.060 \\ 
 J1429$-$5911 & GUq  & 315.26 & 1.30 &  115.8 & 23.88 &    606. & $<$0.021 \\ 
 J1447$-$5757 & GUq  & 317.85 & 1.51 &  158.7 & 11.7 &    115. &           \\ 
 J1459$-$6053 & GUqx  & 317.89 & -1.79 &  103.2 & 25.26 &    908. & $<$0.037 \\ 
 J1509$-$5850 & Rr  & 319.97 & -0.62 &  88.9 & 9.2 &    516. &  0.21 \\ 
 J1513$-$5908 & XErx  & 320.32 & -1.16 &  151.6 & 1526.2 &  17290. &  1.43 \\ 
 J1522$-$5735 & GUq  & 322.05 & -0.41 &  204.3 & 62.46 &    289. & $<$0.034 \\ 
 J1528$-$5838 & GUq  & 322.17 & -1.75 &  355.7 & 24.8 &   21.7 &           \\ 
 J1531$-$5610 & Rr  & 323.90 & 0.03 &  84.2 & 13.8 &    911. &  0.87 \\ 
 J1614$-$5048 & Rr  & 332.21 & 0.17 &  231.9 & 492.0 &   1557. &  4.10 \\ 
 J1615$-$5137 & GUq  & 331.76 & -0.54 &  179.3 & 10.6 &   72.8 &           \\ 
 J1620$-$4927 & GUq  & 333.89 & 0.41 &  171.9 & 10.49 &   81.5 & $<$0.040 \\ 
 J1623$-$5005 & GUq  & 333.72 & -0.31 &  85.1 & 4.2 &    266. &           \\ 
 J1624$-$4041 & GUq  & 340.56 & 6.15 &  167.9 & 4.7 &   39.4 &           \\ 
 J1641$-$5317 & GUq  & 333.29 & -4.56 &  175.1 & 3.7 &   27.2 &           \\ 
 J1646$-$4346 & Rr  & 341.11 & 0.98 &  231.7 & 111.8 &    355. &  1.25 \\ 
 J1648$-$4611 & Rr  & 339.44 & -0.79 &  165.0 & 23.7 &    208. &  0.61 \\ 
 J1650$-$4601 & GUq  & 339.78 & -0.95 &  127.1 & 15.1 &    291. &           \\ 
 J1702$-$4128 & Rr  & 344.74 & 0.12 &  182.2 & 52.3 &    341. &  1.17 \\ 
 J1705$-$1906 & Rr  & 3.19 & 13.03 &  299.0 & 4.1 &   6.11 &  5.66 \\ 
 J1709$-$4429 & RErx  & 343.10 & -2.69 &  102.5 & 94.8 &   3475. & 12.10 \\ 
 J1714$-$3830 & GUq  & 348.44 & 0.14 &  84.1 & 70.3 &   4657. &           \\ 
 J1718$-$3825 & Rr  & 348.95 & -0.43 &  74.7 & 13.2 &   1248. &  1.70 \\ 
 J1730$-$3350 & Rr  & 354.13 & 0.09 &  139.5 & 84.1 &   1222. &  4.30 \\ 
 J1731$-$4744 & Rr  & 342.56 & -7.67 &  829.9 & 163.5 &   11.3 & 27.00 \\ 
 J1732$-$3131 & GUr  & 356.31 & 1.01 &  196.5 & 28.0 &    145. &  0.05 \\ 
 J1736$-$3422 & GUq  & 354.33 & -1.18 &  346.9 & 65.5 &   62.0 &           \\ 
 J1739$-$3023 & Rr  & 358.09 & 0.34 &  114.4 & 11.4 &    300. &  1.01 \\ 
 J1740+1000 & Rr  & 34.01 & 20.27 &  154.1 & 21.3 &    230. &  2.70 \\ 
 J1741$-$2054 & GUrx  & 6.42 & 4.91 &  413.7 & 17.0 &   9.48 &  0.16 \\ 
 J1742$-$3321 & GUq  & 355.85 & -1.69 &  143.3 & 1.3 &   17.0 &           \\ 
 J1746$-$3239 & GUq  & 356.96 & -2.17 &  199.5 &  6.56 &   32.6 & $<$0.034 \\ 
 J1747$-$2958 & Rrx  & 359.31 & -0.84 &  98.8 & 61.3 &   2506. &  0.25 \\ 
 J1748$-$2815 & GUq  & 0.91 & -0.19 &  100.2 & 3.5 &    138. &           \\ 
 J1757$-$2421 & Rr  & 5.28 & 0.06 &  234.1 & 12.7 &   39.2 &  7.20 \\ 
 J1801$-$2451 & Rr  & 5.27 & -0.87 &  125.0 & 89.5 &   1810. &  1.46 \\ 
 J1803$-$2149 & GUq  & 8.14 & 0.19 &  106.3 & 19.50 &    640. & $<$0.024 \\ 
 J1809$-$2332 & Gq  & 7.39 & -2.00 &  146.8 & 34.39 &    429. & $<$0.025 \\ 
 J1813$-$1246 & GUq  & 17.24 & 2.44 &  48.1 & 17.56 &   6238. & $<$0.017 \\ 
 J1816$-$0755 & Rr  & 21.87 & 4.09 &  217.6 & 6.5 &   24.8 &  0.17 \\ 
 J1817$-$1742 & GUq  & 13.34 & -0.70 &  149.7 & 20.6 &    241. &           \\ 
 J1826$-$1256 & Gxq  & 18.56 & -0.38 &  110.2 & 120.95 &   3564. & $<$0.013 \\ 
 J1827$-$1446 & GUq  & 17.08 & -1.50 &  499.2 & 45.3 &   14.4 &           \\ 
 J1828$-$1101 & Rr  & 20.49 & 0.04 &  72.1 & 14.8 &   1562. &  2.30 \\ 
 J1831$-$0952 & Rr  & 21.90 & -0.13 &  67.3 & 8.3 &   1078. &  0.35 \\ 
 J1833$-$1034 & Rr  & 21.50 & -0.89 &  61.9 & 202.0 &  33623. &  0.07 \\ 
 J1835$-$1106 & Rr  & 21.22 & -1.51 &  165.9 & 20.6 &    178. &  2.50 \\ 
 J1836+5925 & Gq  & 88.88 & 25.00 &  173.3 &  1.50 &   11.4 & $<$0.004 \\ 
 J1837$-$0604 & Rr  & 25.96 & 0.27 &  96.3 & 44.9 &   1982. &  0.75 \\ 
 J1838$-$0537 & GUq  & 26.51 & 0.21 &  145.8 & 450.71 &   5746. & $<$0.017 \\ 
 J1841$-$0524 & Rr  & 27.02 & -0.33 &  445.8 & 233.2 &    103. &  0.20 \\ 
 J1844$-$0346 & GUq  & 28.79 & -0.19 &  112.9 & 154.7 &   4249. &           \\ 
 J1846+0919 & GUq  & 40.69 & 5.34 &  225.6 &  9.93 &   34.2 & $<$0.005 \\ 
 (\textit{J1846$-$0258})$^b$ & Xxq  & 29.71 & -0.24 &  326.6 & 7107.1 &   8055. &           \\ 
 J1853$-$0004 & Rr  & 33.09 & -0.47 &  101.4 & 5.6 &    210. &  0.70 \\ 
 J1856+0113 & Rr  & 34.56 & -0.50 &  267.5 & 205.9 &    424. &  0.19 \\ 
 J1857+0143 & Rr  & 35.17 & -0.57 &  139.8 & 31.0 &    448. &  0.74 \\ 
 J1906+0722 & GUq  & 41.22 & 0.03 &  111.5 & 35.86 &   1020. & $<$0.021 \\ 
 J1907+0602 & Gr  & 40.18 & -0.89 &  106.6 & 86.5 &   2814. &  0.00 \\ 
 J1913+0904 & Rr  & 43.50 & -0.68 &  163.3 & 17.6 &    159. &  0.40 \\ 
 J1913+1011 & Rr  & 44.48 & -0.17 &  35.9 & 3.4 &   2882. &  0.90 \\ 
 J1925+1720 & Rr  & 52.18 & 0.59 &  75.7 & 10.5 &    954. &  0.07 \\ 
 J1928+1746 & Rr  & 52.93 & 0.11 &  68.7 & 13.2 &   1602. &  0.28 \\ 
 J1932+1916 & GUq  & 54.66 & 0.08 &  208.2 & 93.17 &    407. & $<$0.075 \\ 
 J1932+2220 & Rr  & 57.36 & 1.55 &  144.5 & 57.0 &    746. &  1.20 \\ 
 J1935+2025 & Rr  & 56.05 & -0.05 &  80.0 & 60.4 &   4638. &  0.50 \\ 
 J1952+3252 & RErx  & 68.77 & 2.82 &  39.5 & 5.8 &   3723. &  1.00 \\ 
 J1954+2836 & GUq  & 65.24 & 0.38 &  92.7 & 21.16 &   1048. & $<$0.005 \\ 
 J1954+3852 & Rr  & 74.04 & 5.70 &  352.9 & 6.6 &   5.93 &  1.07 \\ 
 J1957+5033 & GUq  & 84.58 & 11.01 &  374.8 &  7.08 &   5.31 & $<$0.010 \\ 
 J1958+2846 & GUq  & 65.88 & -0.35 &  290.4 & 211.89 &    341. & $<$0.006 \\ 
 J2006+3102 & Rr  & 68.67 & -0.53 &  163.7 & 24.9 &    223. &  0.27 \\ 
 J2017+3625 & GUq  & 74.51 & 0.39 &  166.7 & 1.4 &   11.6 &           \\ 
 J2021+3651 & Rr  & 75.22 & 0.11 &  103.7 & 94.8 &   3351. &  0.10 \\ 
 J2021+4026 & Gq  & 78.23 & 2.09 &  265.3 & 55.60 &    117. & $<$0.020 \\ 
 J2022+3842 & Xrx  & 76.89 & 0.96 &  48.6 & 86.2 &  29707. &  0.11 \\ 
 J2028+3332 & GUq  & 73.36 & -3.01 &  176.7 &  4.86 &   34.8 & $<$0.005 \\ 
 J2030+3641 & RUr  & 76.12 & -1.44 &  200.1 & 6.5 &   32.0 &  0.15 \\ 
 J2030+4415 & GUq  & 82.34 & 2.89 &  227.1 &  5.05 &   17.0 & $<$0.008 \\ 
 J2032+4127 & GUbr  & 80.22 & 1.03 &  143.2 & 11.6 &    156. &  0.23 \\ 
 J2043+2740 & Rr  & 70.61 & -9.15 &  96.1 & 1.2 &   55.0 &  0.34 \\ 
 J2055+2539 & GUq  & 70.69 & -12.52 &  319.6 &  4.10 &   4.96 & $<$0.007 \\ 
 J2111+4606 & GUq  & 88.31 & -1.45 &  157.8 & 162.63 &   1632. & $<$0.013 \\ 
 J2139+4716 & GUq  & 92.63 & -4.02 &  282.8 &  1.78 &   3.11 & $<$0.014 \\ 
 J2208+4056 & Rr  & 92.57 & -12.11 &  637.0 & 5.3 &  0.807 &  0.04 \\ 
 J2229+6114 & Rrx  & 106.65 & 2.95 &  51.7 & 75.3 &  21557. &  0.25 \\ 
 J2238+5903 & GUq  & 106.56 & 0.48 &  162.7 & 96.85 &    887. & $<$0.011 \\ 
 J2240+5832 & Rr  & 106.57 & -0.11 &  139.9 & 15.3 &    219. &  2.70 \\ 

\enddata

\tablecomments{
Column 2 gives the discovery and detection codes:
G=discovered in Fermi-LAT gamma-ray data,
R=discovered in the radio and/or gamma-ray pulsations detected using the radio ephemeris,
X=discovered in the X-ray and/or gamma-ray pulsations detected using the X-ray ephemeris,
E=pulsar was detected in gamma rays by EGRET/COMPTEL, 
P=discovered by the Pulsar Search Consortium, U=discovered using a Fermi-LAT seed position, r=pulsations detected in the radio band,
x=pulsations detected in the X-ray band, b=binary, q=no radio detection.
In \citet{pga+13} and \citet{LATtiming}, PSR J1522$-$5735 erroneously has half the spin period shown here and 4$\times$ higher $\dot E$.  Our larger gamma-ray photon sample made clear that the fundamental spin frequency is half of the discovery value, reflected in the ephemeris we distribute.
Columns 3 and 4 give Galactic coordinates for each pulsar. 
Columns 5 and 6 list the period ($P$) and its first derivative ($\dot{P}$), and  
Column 7 gives the spindown luminosity $\dot E$. 
The Shklovskii correction to $\dot{P}$ and $\dot E$ is negligible for these young pulsars (see Section \ref{doppler}).
Column 8 gives the radio flux density (or upper limit) at 1400 MHz ($S_{1400}$, see Section \ref{S1400}). 
PSR~J1509$-$5850 should not be confused with PSR~B1509$-$58 (= J1513$-$5908) studied with the \textit{Compton Gamma-Ray Observatory}.
(a) Jason Wu, private communication. 
(b) \citet{Kuiper_J1846-0258} and Section \ref{J1846-0258section} describe the detection of pulsed $<$100\,MeV gamma rays from PSR~J1846$-$0258.
}


\end{deluxetable}

\clearpage
\startlongtable
\tabletypesize{\scriptsize}
\begin{deluxetable}{llrrrrrcc}
\tablewidth{0pt}
\tablecaption{Some parameters of the LAT-detected millisecond pulsars
\label{tbl-charMSP}}

\tablehead{
\colhead{PSR} &   \colhead{Codes} &    \colhead{$l$} & \colhead{$b$} & \colhead{$P$} & \colhead{$\dot P_\mathrm{obs}$}  & \colhead{$10^{-33} \dot{E}_\mathrm{obs}$}& \colhead{$S_{1400}$}  \\
&  \colhead{}  &  \colhead{($^\circ$)} &  \colhead{($^\circ$)} & \colhead{(ms)} & \colhead{($10^{-20}$)}  & \colhead{(erg s$^{-1}$)} & \colhead{(mJy)} 
}

\startdata
 J0023+0923   & RUPbwr & 111.38 & -52.85 &  3.05 & 1.08 &   15.9 &  0.73 \\ 
 J0030+0451   & Rrx & 113.14 & -57.61 &  4.87 & 1.02 &   3.49 &  1.09 \\ 
 J0034$-$0534   & Rbr & 111.49 & -68.07 &  1.88 & 0.50 &   29.7 &  0.21 \\ 
 J0101$-$6422   & RUPbr & 301.19 & -52.72 &  2.57 & 0.48 &   11.8 &  0.16 \\ 
 J0102+4839   & RUPbr & 124.87 & -14.17 &  2.96 & 1.17 &   17.3 &  0.06 \\ 
 J0154+1833   & Rr & 143.18 & -41.81 &  2.36 & 0.28 &   8.47 &  0.11 \\ 
 J0218+4232   & Rbrx & 139.51 & -17.53 &  2.32 & 7.74 &    243. &  0.90 \\ 
 J0248+4230   & RUPr & 144.88 & -15.32 &  2.60 & 1.68 &   37.8 &  0.06 \\ 
 J0251+2606   & RUPbwr & 153.88 & -29.49 &  2.54 & 0.76 &   18.2 &  0.02 \\ 
 J0307+7443   & RUPbr & 131.70 & 14.22 &  3.16 & 1.71 &   21.7 &  0.04 \\ 
 J0312$-$0921   & RUPbrw & 191.51 & -52.38 &  3.70 & 1.97 &   15.3 &          \\ 
 J0318+0253   & RUr & 178.46 & -43.60 &  5.19 & 1.76 &   4.98 &  0.01 \\ 
 J0340+4130   & RUPr & 153.78 & -11.02 &  3.30 & 0.59 &   7.74 &  0.31 \\ 
 J0418+6635   & RUPr & 141.52 & 11.54 &  2.91 & 1.37 &   21.9 &  1.84 \\ 
 J0437$-$4715   & Rbrx & 253.39 & -41.96 &  5.76 & 5.73 &   11.8 &  150.20 \\ 
 J0533+6759   & RUPr & 144.78 & 18.18 &  4.39 & 1.26 &   5.90 &  0.10 \\ 
 J0605+3757   & RUPbr & 174.19 & 8.02 &  2.73 & 0.47 &   9.17 &  0.14 \\ 
 J0610$-$2100   & Rbwr & 227.75 & -18.18 &  3.86 & 1.23 &   8.45 &  0.65 \\ 
 J0613$-$0200   & Rbr & 210.41 & -9.30 &  3.06 & 0.96 &   13.2 &  2.25 \\ 
 J0614$-$3329   & RUPbrx & 240.50 & -21.83 &  3.15 & 1.78 &   22.0 &  0.68 \\ 
 J0621+2514   & RUPbr & 187.12 & 5.07 &  2.72 & 2.49 &   48.7 &  0.08 \\ 
 J0636+5128   & Rbwrx & 163.91 & 18.64 &  2.86 & 0.34 &   5.76 &  1.00 \\ 
 J0653+4706   & RUPbr & 169.26 & 19.78 &  4.75 & 2.08 &   7.61 &  0.10 \\ 
 J0737$-$3039A   & Rbr & 245.24 & -4.50 &  22.70 & 176.00 &   5.94 &  3.06 \\  
 J0740+6620   & Rbrx & 149.73 & 29.60 &  2.88 & 1.22 &   20.0 &  1.10 \\ 
 J0751+1807   & Rbrx & 202.73 & 21.09 &  3.48 & 0.78 &   7.30 &  1.35 \\ 
 J0931$-$1902   & Rr & 251.00 & 23.05 &  4.64 & 0.36 &   1.43 &  0.52 \\ 
 J0952$-$0607   & RUPbwr & 243.65 & 35.38 &  1.41 & 0.48 &   66.5 &  0.02 \\ 
 (\textit{J0955$-$3947})$^1$   & RUbrk & 269.93 & 11.54 &  2.02 & 3.73 &    178. &          \\ 
 J0955$-$6150   & RUPbr & 283.68 & -5.74 &  1.99 & 1.43 &   70.4 &  0.64 \\ 
 J1012$-$4235   & RUr & 274.22 & 11.22 &  3.10 & 0.66 &   8.68 &  0.26 \\ 
 (\textit{J1023+0038})$^2$   & Rbrk & 243.49 & 45.78 &  1.69 & 0.68 &   55.9 &  15.01 \\ 
 J1024$-$0719   & Rrx & 251.70 & 40.52 &  5.16 & 1.86 &   5.33 &  1.50 \\ 
 J1035$-$6720   & GUr & 290.37 & -7.84 &  2.87 & 4.65 &   77.4 &  0.06 \\ 
 J1036$-$8317   & RUPbr & 298.94 & -21.50 &  3.41 & 3.06 &   30.5 &  0.45 \\ 
 J1048+2339   & RUPbrk & 213.17 & 62.14 &  4.67 & 3.01 &   11.7 &  0.17 \\ 
 J1124$-$3653   & RUPbwr & 284.09 & 22.76 &  2.41 & 0.58 &   17.0 &  0.04 \\ 
 J1125$-$5825   & Rbr & 291.89 & 2.60 &  3.10 & 6.09 &   80.6 &  1.00 \\ 
 J1125$-$6014   & Rbr & 292.50 & 0.89 &  2.63 & 0.37 &   8.12 &  1.32 \\ 
 J1137+7528   & RUPbr & 129.00 & 40.77 &  2.51 & 0.32 &   7.96 &  0.10 \\ 
 J1142+0119   & RUPbr & 267.54 & 59.40 &  5.08 & 1.50 &   4.52 &  0.05 \\ 
 J1207$-$5050   & RUPr & 295.86 & 11.42 &  4.84 & 0.61 &   2.12 &  0.39 \\ 
 J1221$-$0633   & Rbr & 289.68 & 55.53 &  1.93 & 1.09 &   59.2 &  0.09 \\ 
 J1227$-$4853   & RUPbkr & 298.97 & 13.80 &  1.69 & 1.33 &    109. &  1.56 \\ 
 J1231$-$1411   & RUPbrx & 295.53 & 48.39 &  3.68 & 2.12 &   17.9 &  0.29 \\ 
 (\textit{J1259$-$8148})$^1$   & RUbrw & 303.24 & -18.95 &  2.09 & 0.33 &   14.5 &          \\ 
 J1301+0833   & RUPbwr & 310.81 & 71.28 &  1.84 & 1.05 &   66.5 &          \\ 
 J1302$-$3258   & RUPbr & 305.59 & 29.84 &  3.77 & 0.66 &   4.83 &  0.17 \\ 
 (\textit{J1306$-$6043})$^3$   & Rbr & 304.75 & 2.09 &  5.67 & 3.04 &   6.58 &          \\ 
 J1311$-$3430   & GUPbwr & 307.68 & 28.18 &  2.56 & 2.09 &   49.0 &  0.11 \\ 
 J1312+0051   & RUPbr & 314.84 & 63.23 &  4.23 & 1.71 &   9.15 &  0.19 \\ 
 J1327$-$0755   & Rbr & 318.38 & 53.85 &  2.68 & 0.18 &   3.72 &  0.19 \\ 
 J1335$-$5656   & GUq & 308.89 & 5.43 &  3.24 & 1.21 &   14.0 &          \\ 
 J1400$-$1431   & Rbr & 326.99 & 45.09 &  3.08 & 0.72 &   9.74 &  0.17 \\ 
 (\textit{J1402+1306})$^4$   & RUPbr & 356.42 & 68.22 &  5.89 & 1.35 &   2.60 &          \\ 
 J1431$-$4715   & Rbkr & 320.05 & 12.25 &  2.01 & 1.41 &   68.4 &  0.67 \\ 
 J1446$-$4701   & Rbwr & 322.50 & 11.43 &  2.19 & 0.98 &   36.6 &  0.46 \\ 
 J1455$-$3330   & Rbr & 330.72 & 22.56 &  7.99 & 2.43 &   1.88 &  0.73 \\ 
 J1513$-$2550   & RUPbwr & 338.82 & 26.96 &  2.12 & 2.15 &   89.1 &  0.31 \\ 
 J1514$-$4946   & RUPbr & 325.25 & 6.81 &  3.59 & 1.87 &   15.9 &  0.25 \\ 
 (\textit{J1526$-$2744})$^5$   & RUb & 340.23 & 23.67 &  2.49 & 0.35 &   9.05 &          \\ 
 J1536$-$4948   & RUPbr & 328.20 & 4.79 &  3.08 & 2.12 &   28.6 &  0.09 \\ 
 J1543$-$5149   & Rbr & 327.92 & 2.48 &  2.06 & 1.62 &   73.3 &  0.82 \\ 
 J1544+4937   & RUPbwr & 79.17 & 50.17 &  2.16 & 0.31 &   11.0 &  2.13 \\ 
 J1552+5437   & Rr & 85.59 & 47.21 &  2.43 & 0.28 &   7.79 &  0.04 \\ 
 J1555$-$2908   & RUPbrw & 344.48 & 18.50 &  1.79 & 4.45 &    307. &  0.20 \\ 
 J1600$-$3053   & Rbr & 344.09 & 16.45 &  3.60 & 0.95 &   8.05 &  2.44 \\ 
 J1614$-$2230   & Rbrx & 352.64 & 20.19 &  3.15 & 0.96 &   12.1 &  1.14 \\ 
 J1622$-$0315   & RUPbkr & 10.71 & 30.68 &  3.85 & 1.14 &   7.93 &          \\ 
 (\textit{J1623$-$6936})$^5$   & RUbr & 319.61 & -13.96 &  2.41 & 0.91 &   25.5 &          \\ 
 J1625$-$0021   & RUPbr & 13.89 & 31.83 &  2.83 & 2.13 &   37.0 &  0.19 \\ 
 (\textit{J1627+3219})$^6$   & RUb & 52.97 & 43.21 &  2.18 & 0.55 &   20.8 &          \\ 
 J1628$-$3205   & RUPbkr & 347.43 & 11.48 &  3.21 & 1.48 &   14.2 &          \\ 
 J1630+3734   & RUPbr & 60.24 & 43.21 &  3.32 & 1.07 &   11.6 &  0.02 \\ 
 J1640+2224   & Rbr & 41.05 & 38.27 &  3.16 & 0.28 &   3.51 &  0.46 \\ 
 J1641+8049   & Rbwr & 113.84 & 31.76 &  2.01 & 0.98 &   46.8 &  0.15 \\ 
 J1649$-$3012   & GUq & 351.96 & 9.21 &  3.42 & 1.32 &   13.0 &          \\ 
 J1653$-$0158   & GUbwq & 16.61 & 24.94 &  1.97 & 0.24 &   12.4 &   $<$0.008 \\ 
 J1658$-$5324   & RUPr & 334.87 & -6.63 &  2.44 & 1.10 &   30.4 &  0.43 \\ 
 J1713+0747   & Rbr & 28.75 & 25.22 &  4.57 & 0.85 &   3.53 &  8.30 \\ 
 J1730$-$2304   & Rr & 3.14 & 6.02 &  8.12 & 2.02 &   1.49 &  4.00 \\ 
 J1732$-$5049   & Rbr & 340.03 & -9.45 &  5.31 & 1.42 &   3.74 &  2.11 \\ 
 J1741+1351   & Rbr & 37.89 & 21.64 &  3.75 & 3.02 &   22.7 &  0.29 \\ 
 J1744$-$1134   & Rr & 14.79 & 9.18 &  4.07 & 0.89 &   5.21 &  2.60 \\ 
 J1744$-$7619   & GUq & 317.11 & -22.46 &  4.69 & 0.97 &   3.71 &   $<$0.023 \\ 
 J1745+1017   & RUPbwr & 34.87 & 19.25 &  2.65 & 0.25 &   6.45 &  0.51 \\ 
 J1747$-$4036   & RUPr & 350.21 & -6.41 &  1.65 & 1.33 &    116. &  1.51 \\ 
 (\textit{J1757$-$6032})$^5$   & RUbr & 332.98 & -17.18 &  2.91 & 0.30 &   4.77 &          \\ 
 (\textit{J1803$-$6707})$^5$   & RUbk & 326.85 & -20.34 &  2.13 & 1.85 &   75.0 &          \\ 
 J1805+0615   & RUPbwr & 33.35 & 13.01 &  2.13 & 2.28 &   93.2 &  0.36 \\ 
 J1810+1744   & RUPbwr & 44.64 & 16.81 &  1.66 & 0.46 &   38.4 &  0.28 \\ 
 J1811$-$2405   & Rbr & 7.07 & -2.56 &  2.66 & 1.34 &   28.0 &  1.33 \\ 
 J1816+4510   & RUbkr & 72.83 & 24.74 &  3.19 & 4.31 &   52.2 &  0.04 \\ 
 J1823$-$3021A   & Rr & 2.79 & -7.91 &  5.44 & 337.62 &    827. &  0.72 \\ 
 J1824+1014   & RUPbr & 39.08 & 10.65 &  4.07 & 0.55 &   3.21 &  0.01 \\ 
 (\textit{J1824$-$0621})$^7$   & Rrb & 24.16 & 3.10 &  3.23 & 0.91 &   10.7 &          \\ 
 J1824$-$2452A   & Rrx & 7.80 & -5.58 &  3.05 & 161.89 &   2243. &  2.30 \\ 
 J1827$-$0849   & GUq & 22.36 & 1.22 &  2.24 & 1.10 &   38.4 &   $<$0.800 \\ 
 J1832$-$0836   & Rr & 23.11 & 0.26 &  2.72 & 0.83 &   16.2 &  0.92 \\ 
 J1833$-$3840   & RUPbrw & 356.01 & -13.26 &  1.87 & 1.77 &    107. &          \\ 
 (\textit{J1835$-$3259B})$^8$   & Rr & 1.53 & -11.38 &  1.83 & 4.34 &    279. &          \\ 
 J1843$-$1113   & Rr & 22.05 & -3.40 &  1.85 & 0.96 &   60.0 &  0.10 \\ 
 (\textit{J1852$-$1310})$^2$   & RUPr & 21.27 & -6.15 &  4.31 & 1.02 &   5.01 &          \\ 
 J1855$-$1436   & RUPbr & 20.36 & -7.57 &  3.59 & 1.09 &   9.29 &  0.05 \\ 
 (\textit{J1857+0943})$^2$   & Rbr & 42.29 & 3.06 &  5.36 & 1.78 &   4.57 &  5.00 \\ 
 J1858$-$2216   & RUPbr & 13.58 & -11.39 &  2.38 & 0.39 &   11.3 &  0.06 \\ 
 (\textit{J1858$-$5422})$^5$   & RUbr & 342.07 & -22.69 &  2.36 & 0.41 &   12.4 &          \\ 
 J1901$-$0125   & GUr & 32.82 & -2.90 &  2.79 & 3.58 &   64.8 &  2.50 \\ 
 J1902$-$5105   & RUPbr & 345.65 & -22.38 &  1.74 & 0.90 &   68.7 &  1.01 \\ 
 J1903$-$7051   & RUPbr & 324.39 & -26.51 &  3.60 & 1.04 &   8.80 &  0.96 \\ 
 J1908+2105   & RUPbwr & 53.69 & 5.78 &  2.56 & 1.38 &   32.4 &  0.04 \\ 
 J1909$-$3744   & Rbr & 359.73 & -19.60 &  2.95 & 1.40 &   21.6 &  1.80 \\ 
 J1921+0137   & RUPbr & 37.83 & -5.94 &  2.50 & 1.88 &   47.6 &  0.10 \\ 
 J1921+1929   & Rbr & 53.62 & 2.45 &  2.65 & 3.82 &   81.4 &  0.20 \\ 
 J1939+2134   & Rrx & 57.51 & -0.29 &  1.56 & 10.51 &   1097. &  13.90 \\ 
 J1946+3417   & Rbr & 69.29 & 4.71 &  3.17 & 0.32 &   3.90 &  0.90 \\ 
 J1946$-$5403   & RUPbwr & 343.88 & -29.58 &  2.71 & 0.27 &   5.33 &  0.35 \\ 
 J1959+2048   & Rbwrx & 59.20 & -4.70 &  1.61 & 1.68 &    159. &  0.29 \\ 
 J2006+0148   & RUPbr & 43.40 & -15.76 &  2.16 & 0.33 &   12.8 &  0.21 \\ 
 J2017+0603   & RUPbr & 48.62 & -16.03 &  2.90 & 0.83 &   13.0 &  0.18 \\ 
 J2017$-$1614   & RUPbwr & 27.31 & -26.22 &  2.31 & 0.24 &   7.67 &  0.10 \\ 
 (\textit{J2029$-$4239})$^1$   & RUr & 358.20 & -35.51 &  5.31 & 0.94 &   2.48 &          \\ 
 J2034+3632   & GUq & 76.60 & -2.34 &  3.65 & 0.17 &   1.40 &          \\ 
 J2039$-$3616   & Rbr & 6.33 & -36.52 &  3.27 & 0.84 &   9.47 &  0.50 \\ 
 J2039$-$5617   & GUbkr & 341.27 & -37.15 &  2.65 & 1.42 &   30.0 &  0.58 \\ 
 J2042+0246   & RUbr & 48.99 & -23.02 &  4.53 & 1.41 &   5.98 &  0.06 \\ 
 J2043+1711   & RUPbr & 61.92 & -15.31 &  2.38 & 0.57 &   15.4 &  0.12 \\ 
 J2047+1053   & RUPbwr & 57.05 & -19.68 &  4.29 & 2.10 &   10.4 &          \\ 
 J2051$-$0827   & Rbwr & 39.19 & -30.41 &  4.51 & 1.27 &   5.49 &  2.80 \\ 
 J2052+1219   & RUPbwr & 59.14 & -19.99 &  1.99 & 0.67 &   33.9 &  0.44 \\ 
 J2115+5448   & RUPbwr & 95.04 & 4.11 &  2.61 & 7.49 &    167. &  0.46 \\ 
 (\textit{J2116+1345})$^9$   & RUb & 64.15 & -23.82 &  2.22 & 0.26 &   9.57 &          \\ 
 J2124$-$3358   & Rrx & 10.92 & -45.44 &  4.93 & 2.06 &   6.77 &  4.50 \\ 
 J2129$-$0429   & RUPbkr & 48.91 & -36.94 &  7.61 & 23.70 &   29.3 &  0.00 \\ 
 J2205+6012   & Rbr & 103.69 & 3.70 &  2.41 & 1.98 &   55.4 &  0.49 \\ 
 J2214+3000   & RUPbwrx & 86.86 & -21.67 &  3.12 & 1.50 &   19.2 &  0.53 \\ 
 J2215+5135   & RUPbkr & 99.87 & -4.16 &  2.61 & 2.34 &   62.7 &  0.16 \\ 
 J2234+0944   & RUPbwr & 76.28 & -40.44 &  3.63 & 1.96 &   16.6 &  1.90 \\ 
 J2241$-$5236   & RUPbwrx & 337.46 & -54.93 &  2.19 & 0.87 &   26.0 &  1.83 \\ 
 J2256$-$1024   & Rbwr & 59.23 & -58.29 &  2.29 & 1.14 &   37.1 &  0.73 \\ 
 J2302+4442   & RUPbr & 103.40 & -14.00 &  5.19 & 1.33 &   3.91 &  1.40 \\ 
 J2310$-$0555   & RUPbr & 69.70 & -57.91 &  2.61 & 0.50 &   11.0 &  0.07 \\ 
 J2317+1439   & Rbr & 91.36 & -42.36 &  3.45 & 0.24 &   2.35 &  0.60 \\ 
 J2339$-$0533   & RUPbkr & 81.35 & -62.48 &  2.88 & 1.41 &   23.2 &          \\ 

\enddata

\tablecomments{ 
In Column 1, pulsars with names in italics, in parentheses, revealed gamma-ray pulsations after the initial sample was defined,
and are not analyzed in this catalog. Superscript numbers denote the following references: 
(1) TRAPUM Collaboration (in prep.)
(2) Appendix A
(3) \citet{pbs+23}
(4) \citet{ThankfulThesis}
(5) \citet{TRAPUM_FermiSurvey}
(6) Wang et al. (in prep.)
(7) \citet{mzl+23}
(8) \citet{grf+22, gammaJ1835-3259B}
(9) \citet{lewis2023_4AO_MSPs}.
Column 2 gives the discovery and detection codes, as in Table \ref{tbl-charPSR},
with in addition w=black widow and k=redback.
Columns 3 and 4 give the Galactic coordinates, with the rotation period $P$ in column 5. 
The first period time derivative $ \dot{P}_\mathrm{obs}$ and the spindown luminosity $\edot_\mathrm{obs}$ in Columns 6 and 7 are \textit{uncorrected} for the Shklovskii effect. Corrected values are in Table \ref{tbl-doppler}, Section \ref{doppler}. 
Column 9 gives the radio flux density (or upper limit) at 1400 MHz (see Section \ref{S1400}).
Some MSPs without an $S_{1400}$ flux measurement suffer intense scintillation, making few-epoch measurements unreliable.
}



%
\end{deluxetable}

\subsection{Using Known Rotation Ephemerides}
\label{radioselected}

The most straightforward way to search for gamma-ray pulsations is to fold the LAT data using an existing rotation ephemeris that is valid for the full 12 or more years on orbit.
To that end, radio and X-ray astronomers shared $>$\,1400 rotation ephemerides with the LAT team,  $\sim$40\% of the over $3400$ known rotation-powered pulsars \citep[mostly from the ATNF Pulsar Catalog v1.69\footnote{\url{http://www.atnf.csiro.au/research/pulsar/psrcat}},][as well as some unpublished pulsars, see Table \ref{tab:tallies}]{ATNFcatalog}. 
Table \ref{tab:radio-search} includes radio telescopes that contributed timing models used in the LAT pulsar searches.
These timing models yielded the discovery of $\sim$110 gamma-ray pulsars. 
A special effort was made to search for LAT pulsations from the energetic ($\dot E > 10^{34}$ erg s$^{-1}$) subset of the pulsar population, representing about 10\% of all pulsars.
We have folded $>$90\% of the high-$\dot E$ pulsars (see Figure \ref{EdotFractions}), 
revealing gamma-ray pulsations in about two thirds of them. Importantly, the ephemerides 
provided by the radio community sample the entire $P-\dot P$ plane. The $>$\,1200 pulsars that
were phase-folded without seeing gamma-ray pulsations are shown as black dots in 
Figure \ref{PPdotplot}. Folding so many pulsars gives a largely unbiased view of known pulsars which may emit gamma rays not visible from earth, discussed by \citet{JohnstonSmith2020}. 

The availability of so many ephemerides is possible because of the long-term timing campaigns of several radio telescopes, summarized in \citet{ThousandFold}. Astronomers originally organized support for the LAT as a ``Pulsar Timing Consortium'' \citep{TimingForFermi} and have continued as the \Fermi{} mission proceeds. We aim for $P/50$ ephemeris accuracy ($0.02$ in phase), which we have generally achieved: possible gamma-ray pulse width smearing at this level is unlikely to impede
pulsation discovery, given that the narrowest known pulse is $P/33$  ($0.03$ in phase) for PSR J1959+2048 \citep{GuillemotBlackWidow_2012}.
Gamma-ray pulse widths and background rates are such that phase histograms generally require far fewer than 100 bins (Figure \ref{SampleLC} uses 50 bins). 
Of the $\sim$110 gamma-ray pulsars found by phase-folding using radio ephemerides,
20 are too faint for the 4FGL-DR3 catalog: they rise above the background only when the photons accumulate in a restricted phase range, and thus probe a population beyond that accessible by the source catalog methods.  We do not include additional spectral analysis of these sources, so they are absent from the analysis of Section \ref{spectralSection}.

Fully describing the rotation of young pulsars, which have rotational irregularities known as timing noise, over a timespan of many years often requires many noise-smoothing parameters, which may be covariant with parameters that have a physical interpretation. That is to say, ``whitened'' solutions used for our analysis and distributed with this paper accurately
calculate rotational phase, but should be used with care when studying braking indices, proper motions,
or other pulsar properties. 

For a few pulsars, X-ray timing was either combined with radio timing to construct an ephemeris \citep[e.g. PSR J2022+3842, ][]{Limyansky_thesis}, or was used exclusively \citep[e.g. PSR J0540$-$6919 in the Large Magellanic Cloud,][]{B0540_braking_index}. Section \ref{xraypsrs} further addresses searches for GeV pulsations in X-ray bright pulsars. 

Finally, the timing models enable phase alignment between observations at different wavelengths from different observatories.  The absolute phase reference is given by the  \textsc{Tempo2} parameters
for the arrival time (TZRMJD) and location (TZRSITE) of a reference pulse of a particular frequency (TZRFREQ).  The dispersion measure (DM) determines the
frequency-dependent delays in the interstellar medium and is necessary to align radio and high-energy data.  
Carefully-aligned pulse profiles provide information about the relative geometry
of the different emission regions. 

\subsection{Pulsar Search Targets}
\label{targets}

About half of the gamma-ray pulsars were discovered in searches around LAT sources with pulsar-like properties. In addition, many LAT sources are co-located with previously known radio and/or X-ray pulsars (see Table \ref{ColocatedNotPulse}). Section \ref{not_seen} discusses the prospects for finding gamma-ray pulsations from the latter, as well as from ``spider''-like binary systems found in LAT unidentified sources, likely to yield still more MSP discoveries. 
This section describes how we select and improve candidate targets for the deep radio searches and the gamma-ray blind searches described in Sections \ref{psc} and \ref{blindsearch}.

\subsubsection{Pulsar-like LAT Sources} 
\label{4fgl}

The LAT source catalog lists ``associations'' and ``identifications'' for
two-thirds of the sources. ``Association'' means that the probability  that an
object known at some other wavelength is responsible for the LAT emission is
estimated to be $>$\,80\%, where the prior for the Bayesian probability is the sky
distribution for each population class \citep[][]{4FGL}. ``Identification'' requires either
detection of gamma-ray pulsations (in the case of pulsars) or an additional
match with observations at another wavelength. Examples are simultaneous blazar
flares or supernova remnant morphology. Pulsars are the largest class of
identified sources, assigned `PSR' or `MSP' classes in the 4FGL catalog (capital letters indicate identification). A spatial match between a known
pulsar and a catalog source without gamma-ray pulsations yields association
classes `psr' and `msp' (lower case indicates association without
confirmed identification).  The 4FGL catalog table has a second association
column for low probability ($<$\,20\%) associations. 
Table \ref{ColocatedNotPulse} is complementary to the association process, listing 
simple co-locations without regard for non-pulsar populations.

Additional observables are used to rank candidates in the long list of unidentified sources. 
Pulsar spectral energy distributions (SEDs), $\nu F_\nu = E^2\frac{dN}{dE}$,
have sharp cut-offs in the GeV range (Section \ref{spectra:shapes}) that distinguish them from other categories of gamma-ray sources. 
The 4FGL catalog includes a variability index for each
source, which can help distinguish between pulsars, whose fluxes are stable (see Section \ref{variability}), and the $10\times$ more common blazars, which can exhibit gamma-ray flares.  
Early successes in selecting pulsar candidates from among
the unidentified LAT catalog sources exploited SED curvature versus flux
variability correlations \citep{UNASSOC_SOURCE_PAPER}. Subsequent works used
machine learning methods to refine the selections \citep{MirabalMining,KJLee, PabloRanking,wcp+18,Luo2020, Finke2021}, 
or visual inspection and ranking of the LAT source spectra
\citep{ckr+15}. Regardless of the ranking scheme, these lists of pulsar-like
unassociated LAT sources have provided a large number of candidate pulsar
positions that have been targeted by radio, X-ray, and gamma-ray searches. As
one recent example, PSR J1653$-$0158 \citep{nck+20} was recently found in the
brightest remaining pulsar candidate of \citet{PabloRanking}.

In 4FGL-DR3, about 80\% of the unidentified sources have error ellipses 
with semi-major axes between $0.05$ and $0.15$ degrees. Improved localization can be
valuable for the high-frequency radio observations needed to pierce the thick
electron column density (i.e. high DM values) in, for example, the direction of
the Galactic center. Improved localization also reduces the computing cost of
blind period searches, and it can allow a match with optical or X-ray sources
necessary to motivate time-consuming observations and analyses (see Section \ref{sec:opticalx-ray}). Radio telescope arrays can coherently combine data from multiple antennas to form tied array beams which are typically much smaller than the primary beam.  These also
benefit from improved localization, since computational resources generally
limit the number of synthesized beams that can be formed and searched.

\begin{figure}[!ht]
\centering
\includegraphics[width=0.9\textwidth]{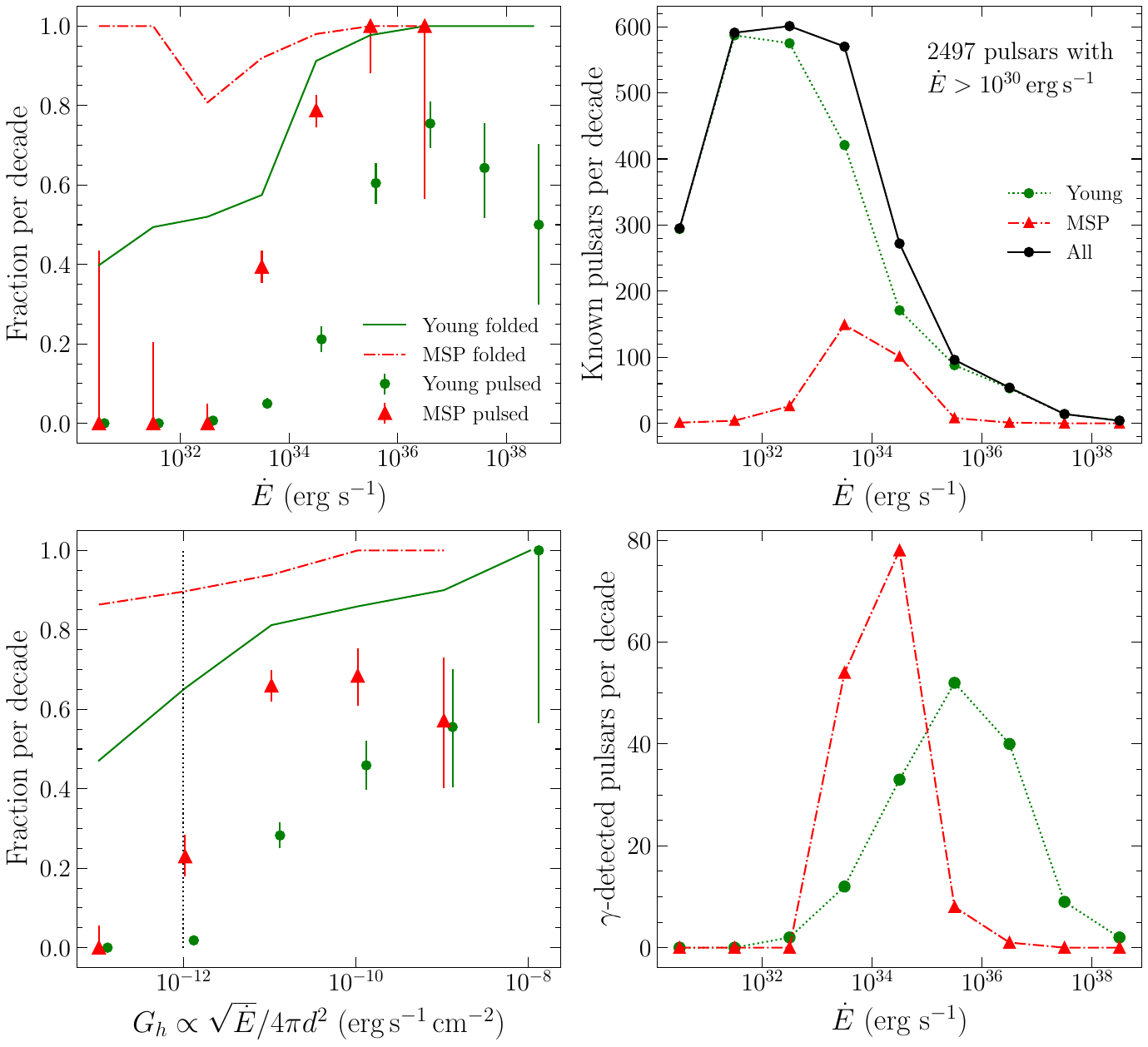}
\caption{ 
Left-hand plots: 
the lines show the fractions of known field pulsars that have been gamma-folded, versus spindown power $\edot$ (top) and, when a distance estimate exists, the heuristic gamma-ray flux $G_h$ defined in Section \ref{xraypsrs} (bottom). The points show the fractions of folded pulsars for which pulsations have been detected. 
The vertical dotted line approximates the minimum detectable integral energy flux $G_{100}$.
Right-hand plots: Numbers of known and gamma-ray detected field pulsars versus $\dot E$. (A quarter of ATNF \textit{psrcat} pulsars have unmeasured $\dot E$.)
An earlier version of this figure appears in \citet{LaffonNewPSRs}.
\label{EdotFractions}}
\end{figure}

\subsubsection{Optical, X-ray, and Radio Studies to Refine Targets} 
\label{sec:opticalx-ray}


Fewer than 1\% of known neutron stars pulse in the optical band, so optical observations of gamma-ray sources might seem unproductive. However, relative to pulsars discovered in radio surveys, the LAT pulsar sample includes many more members of the subclass of interacting binaries---black widows, redbacks, transitional MSPs, and related systems. The companions in these systems can have relatively bright optical magnitudes and (importantly) often exhibit modulation at the orbital period in brightness, color and radial velocity.  Consequently, searches for periodic optical sources within LAT source error regions have identified many candidate gamma-ray pulsars. Some of these have been followed up with radio detections, while others have enabled direct gamma-ray periodicity detections by using the precise position and orbital parameters to reduce the vast parameter space to be searched, as detailed in Section \ref{blindsearch}. Others remain strong candidates that are likely powering the gamma-ray sources, but have yet to be confirmed as pulsars (see Section \ref{not_seen}). 
Some examples of the power of optical studies of LAT source regions include:
\begin{itemize}
\item Optical and X-ray studies strongly indicated that 1FGL J2339.7$-$0531 hosts a black-widow MSP \citep{Romani_2011,Kong_2012}. After LAT blind searches found no pulsar, \citet{rpp+20} discovered radio pulsations.  PSR J2339$-$0533 turned out to be a redback with a companion mass outside the parameter space covered by the LAT searches. The resulting ephemeris yielded gamma-ray pulsations as well \citep{LATJ2339}.
\item Optical studies by \citet{Romani2012}
led to the first discovery of a millisecond pulsar, PSR J1311$-$3430, in a blind search of LAT data \citep{Pletsch_J1311-3430}. Detection of radio pulsations followed shortly \citep{RayJ1311}. Two further binary MSPs have been discovered in this way: the black-widow PSR~J1653$-$0158 \citep{Nieder2020} from an optical candidate discovered by \citet{Romani2014+J1653} and \citet{Kong2014+J1653}; and the redback PSR~J2039$-$5617 \citep{cnv+21} from an optical candidate discovered by \citet{Romani2015+J2039} and \citet{Salvetti2015+J2039}. 

\item  Four candidate gamma-ray redbacks identified in optical and X-ray searches were recently identified as radio MSPs by TRAPUM \citep[][]{ProbablePSRJ0838,ProbablePSRJ0955,ProbableJ2333, ProbableJ1910}, listed in Table \ref{PSCnotGammaPulse}. 

\item Optical studies of an unidentified LAT source by \citet{Strader_J1417} revealed an accretion disk in a binary system with a $5.4$ d period. \citet{J1417-4402} subsequently discovered PSR J1417-4402 using the Parkes radio telescope. First thought to be a redback, its orbit is wide and the wind interactions differ from other spiders. \citet{Huntsman} argued that it is a possible progenitor of normal field MSP binaries and dubbed it the ``huntsman'' as the archetype of an emerging class. Unfortunately, it is difficult to time and a phase-connected ephemeris which would yield gamma-ray pulsations is not anticipated soon.

\end{itemize}

Neutron stars can also power X-ray emission either directly from their surfaces or magnetospheres, or via shock emission when the pulsar wind interacts with a companion or the interstellar medium  \citep[][]{Gentile2014, Marelli2015,CotiZelati2020}. The non-thermal shock spectra, sometimes modulated at the orbital period, provide strong signatures. Consequently, X-ray surveys also provide valuable information as demonstrated by varied studies of the error ellipses of unidentified LAT sources. For example, the \textit{Swift} XRT systematically studied a large number of LAT unassociated source regions\footnote{\url{https://www.swift.psu.edu/unassociated/}}. This dataset has been mined using machine learning techniques to identify promising pulsar candidates for radio follow up \citep{Kaur2019,Kaur2021}.

Pulsars are steep-spectrum radio point sources \citep[][]{Jankowski2018}. This was used to identify the source (4C~21.53) that was later found to be the first millisecond
pulsar (PSR B1937+21; \citealt{1982Natur.300..615B}). The availability of radio surveys covering nearly the whole sky was recently harnessed to look for pulsar candidates
associated with LAT source regions
\citep{FrailUnIds,FrailUnIds2,Bruzewski2021,rnc+22}. These candidate lists have high
precision positions from radio interferometric imaging and well-measured fluxes,
enabling both gamma-ray blind searches with reduced parameter spaces
and radio pulsation searches that discovered several new MSPs. 

\subsection{Targeted Radio Searches} 
\label{psc}

Shortly after LAT operations began, the \Fermi{} Pulsar Search Consortium
(PSC) was organized, bringing together the LAT team and many international
collaborators with pulsar searching expertise and access to the world's largest radio telescopes \citep{rap+12}. The aim was to coordinate radio pulsation searches of LAT sources, and to see which pulsars found in gamma-ray blind searches are truly radio quiet. The PSC has been exceptionally successful at efficiently discovering millisecond pulsars in searches of LAT sources. Beginning in 2010, the TRAPUM collaboration formed to exploit MeerKAT \citep{sk16}. It is not formally part of the PSC but includes searches of LAT sources. The FAST radio telescope in China also joined the searches under a separate arrangement. Radio astronomers are provided with early access to LAT source lists as well as the parameters of new pulsars discovered in gamma-ray blind searches (see Section \ref{blindsearch}). 

LAT gamma-ray source localizations are much better than was possible with EGRET,
making it often possible to search an entire 95\% confidence error region with a
single radio telescope pointing. Of course, the primary beam of a radio
telescope depends on the observation frequency and the dish diameter. Thus radio
astronomers could choose telescope/frequency combinations optimized for the error
regions being searched. Initially, the telescopes were all single-dish (Green
Bank Telescope (GBT), Parkes, Effelsberg, Nan\c{c}ay Radio Telescope (NRT),
Arecibo Telescope, and the Lovell Telescope at Jodrell Bank), but array
telescopes were soon added, the first being the Giant Metrewave Radio Telescope
(GMRT), followed by LOFAR and MeerKAT (see Table \ref{tab:radio-search}). 

\begin{deluxetable}{lrrrrr}
\tabletypesize{\scriptsize}
\tablewidth{0pt}
\tablecaption{Radio telescopes searching for new pulsars in unidentified LAT sources, and timing discoveries. 
\label{tab:radio-search}}
\tablehead{
\colhead{Telescope} & \colhead{Frequencies}  & Beam HWHM & \colhead{Decl.\ range\tablenotemark{a}}	  &  $N_\mathrm{disc}/N_\mathrm{3PC}$\tablenotemark{b} & \colhead{Refs.\tablenotemark{c}}  \\
 & \colhead{(MHz)} & \colhead{(arcmin)} }
\startdata
Pulsar Searching \\
\hline
Parkes       & 1400           & 7            & $-90^\circ < \delta < 33^\circ$ & 18/13 & [1] \\
Jodrell Bank & 1400           & 5             & $-37^\circ < \delta < 90^\circ$  & 0/0 & [2] \\
Nan\c{c}ay   & 1400           & $4 \times 22$  & $-39^\circ < \delta < 90^\circ$  & 3/3 & [3] \\
Green Bank   & 350, 820, 2000 & 18.5, 7.9, 3.1 & $-46^\circ < \delta < 90^\circ$  & 46/35 & [4] \\
Effelsberg   & 1400           & 5             & $-31^\circ < \delta < 90^\circ$  & 1/1 & [5] \\
GMRT         & 325, 610       & 85, 44               & $-55^\circ < \delta < 90^\circ$  & 9/5  & [6] \\
Arecibo      & 327, 1400      & 14, $3.3\times 3.8$ & $-1.3^\circ < \delta < 38^\circ $  &  16/9 & [7] \\
Molonglo & 840       & $0.8 \times 84$   & $-90^\circ < \delta < 18^\circ$  & 0/0 & [8] \\
LOFAR\tablenotemark{d}        & 140            & 10              & $-7^\circ < \delta < 90^\circ$  & 3/3  & [9] \\
MeerKAT\tablenotemark{e}      & 544--1088, 856--1712           & 0.25--0.5,  0.15--0.3            & $-90^\circ < \delta < 44^\circ$  & 21/2  & [10] \\
FAST         & 1400           & 3              & $-15^\circ < \delta < 65^\circ$  &  3/2 & [11]\\
\hline
\enddata
\tablenotetext{a}{The declination ranges for long-duration deep searches are typically $20^\circ$ narrower than the maximum pointing ranges listed here.}
\tablenotetext{b}{Total numbers of pulsars discovered in LAT-targeted searches and the subset confirmed as gamma-ray pulsars and included in this catalog. Table \ref{PSCnotGammaPulse}  lists discoveries yet to reveal gamma-ray pulsations.}
\tablenotetext{c}{[1] \citet{ParkesFermiTiming}, \citet{J1417-4402}, \citet{ckr+15}, \citet{Kerr5MSPs}, \citet{kjr+11}; [2] \citet{Jodrell}; [3] \citet{gfc+12}, \citet{Cognard2011}; [4] \citet{rrc+11}, \citet{trr+21}, \citet{rpp+20}, \citet{bbc+22}, \citet{TuckThesis}; [5] \citet{BarrJ1745};
[6] \citet{brj+21}, \citet{rrb+15}, \citet{Bhattacharyya}; [7] \citet{Cromartie_6FermiMSPs}; 
[8] UTMOST, \citet{UTMOST_I,UTMOST_II} ; [9] \citet{LOFARmspSearching}; [10] \citet{sk16}; [11]  \citet{wlc+21}  }
\tablenotetext{d}{The value reported here is typical of synthesized beams using the ``Superterp'' stations \citep{scb+19}.}
\tablenotetext{e}{Beam widths for MeerKAT are those of a coherent tied-array beam (TAB), and are position dependent, hence we quote a range of values corresponding to elevations above 30\degr. Several hundred TABs can be recorded simultaneously, so LAT source regions with semi-major axes up to $\sim7\arcmin$ can be covered in single pointings. }
\end{deluxetable}

Array-based telescopes offer increasingly powerful and flexible methods for pulsar searches.  Incoherently combining the beams allows for very efficient but less sensitive searches of large error boxes. Some modern arrays have backends powerful enough to search hundreds of coherent
(tied-array) beams arranged to cover a large primary beam, enabling extremely
efficient full sensitivity searches of large error boxes \citep[][]{TRAPUM_FermiSurvey}.  Gated imaging
\citep{rb13} or tied-array beamforming \citep{sk16,scb+19} can localize newly-discovered pulsars extremely well, without the need for long-term timing campaigns.  

A key feature of radio pulsar searches, compared to LAT blind searches, is that
it is computationally feasible to maintain sensitivity to short-period binary
pulsars even when none of the system parameters are known (as described in
 Section \ref{blindsearch}, binary MSPs can be found in gamma-ray blind searches when approximate
orbital parameters are known from optical studies). Because the observation
durations are generally short compared to the orbital period, this can be done
through Fourier search techniques that assume a constant acceleration
\citep{RansomLongOrbit2002}, or a constant jerk \citep{JerkMethod}. Recently, an
analysis of PSC search data using the jerk search method resulted in the
discovery of 11 new MSPs, at least one of which would likely have been missed in
a standard acceleration search \citep{trr+21}.


At least 110 new pulsars have been
discovered in the radio searches targeting LAT unassociated sources. All except
8 are MSPs -- MSPs dominate the high Galactic latitudes where radio surveys have been less thorough. At present, only one of the slow pulsars has been confirmed as a
gamma-ray pulsar (PSR J2030+3641; \citealt{Camilo2012}). Given the many
observations, some chance coincidences are expected, especially for telescopes
with wide fields of view \citep[particularly arrays of smaller telescopes combined incoherently, like the GMRT, e.g.][]{Bhattacharyya2022+GMRT}. Table \ref{PSCnotGammaPulse} lists \NDCpsc{} MSPs found
in the targeted radio searches for which gamma-ray pulsations have not yet been
detected. An X indicates \Nserendipity{} where it has been concluded that the MSP is not associated with
the target source. Most of the others were discovered after 2016,
so the lack of gamma-ray pulsations from these sources is likely due to the lag
in obtaining a rotation ephemeris with which to fold LAT data.

Developing such an ephemeris is an involved process, as the timing model
must extrapolate well over the long span of LAT data required for the accumulated signal to become significant.  Especially if the position is not
well constrained by radio imaging or an X-ray or optical counterpart, then the radio campaign requires a dense set of observations to measure the spin period well and longer intervals (0.5--1.0\,yr) to measure the position and $\dot{P}$.  Pulsars with orbital period variability may require years of monitoring.  Such follow-up campaigns can require multiple proposal cycles.

In most cases, after getting a good timing solution, LAT pulsations
have been detected from the MSPs. If pulsations are detected in a fraction of
the LAT data, the solution can often be extended to cover the full mission using
LAT timing methods (see Section \ref{LATtiming}). We expect that a large
fraction of the recent MSPs in Table \ref{PSCnotGammaPulse} will be identified
as gamma-ray pulsars in the future. Some of the pulsars in Table \ref{ColocatedNotPulse} may also someday reveal gamma-ray pulsations.


\subsection{Blind Periodicity Searches}
\label{blindsearch}

In addition to the detection of gamma-ray pulsations from known radio or X-ray pulsars via folding, the LAT data also allow for the discovery of entirely new pulsars through direct searches for their gamma-ray pulsations, without prior knowledge of their spin periods or spin-down rates.  These searches complement and contrast the efforts described above in Sections~\ref{radioselected} and \ref{psc}, providing access to a population of otherwise undetectable pulsars. However, these searches are less sensitive than following-up a known radio or X-ray pulsar, and are extremely computationally demanding: searching for isolated young pulsars/MSPs in a single \textit{Fermi}-LAT source typically takes a few days/weeks, respectively, using tens of thousands of CPU cores, while a search for a single binary MSP can run for months on thousands of GPUs.

Of the pulsars in this catalog, 81 were first discovered in the LAT data, compared to 36 in 2PC. The majority are young pulsars, located near the Galactic plane. Notably, this includes 10 LAT-discovered millisecond pulsars of which three are in binary systems, whereas 2PC contained just one such object. That the discovery rate from direct gamma-ray pulsation searches has not fallen off is perhaps surprising when one considers how the computational cost and sensitivity of these searches increases with the data duration.

The computational cost depends on the dimensionality and volume of the parameter space which must be searched, and the required density of search points covering this volume. At minimum, these searches are 2-dimensional, covering the range of spin frequencies (or periods) and spin-down rates spanned by the known pulsar population. A sub-arcsecond precision position---much better than can be provided by the LAT---is required to correct for the Earth's orbital R\"{o}mer delay.  Unless one is available from multi-wavelength observations, an additional 2-dimensional grid of sky locations covering the localization region must be searched. The required positional precision in each dimension is proportional to the spin frequency, $f$.  This would make searches for MSPs thousands of times more costly than searching for young pulsars, were this not partially mitigated by the far smaller range of spin-down rate $\dot{f}$ over which MSPs are found.

When searching for pulsars in circular binaries, at least three further parameters are required (the orbital period, phase and radius), with two additional parameters required for eccentric systems. The search grid density in all binary parameters depends on $f$, again making rapidly spinning millisecond pulsars substantially more costly to find. The binary search parameter space is in fact so large that entirely uninformed searches are as yet infeasible, with searches only becoming possible when a candidate binary system is identified and its orbital parameters constrained via multi-wavelength observations.

The required search-grid density in most dimensions depends on the duration $T_{\rm obs}$ of the data over which the signal is integrated \citep{PletschClark2014}. In $f$, and $\dot{f}$, the number of grid points scales linearly and quadratically, respectively, with $T_{\rm obs}$. For $T_{\rm obs} < 1$\,yr, the number of sky positions which must be searched scales quadratically with $T_{\rm obs}$, but this saturates at longer timescales. For binary pulsar searches, the required density of orbital period search points also scales linearly with $T_{\rm obs}$ \citep{Nieder2020}. The result is that the computing cost scales with at least $T_{\rm obs}^3$ for isolated pulsars and $T_{\rm obs}^4$ for binaries. This rapid scaling leads to fully-coherent searches of data lasting more than a few months being entirely infeasible.

A compromise is achieved using ``semi-coherent'' search methods, where the signal powers from many shorter windows (of duration $T_{\rm coh}$, typically a few weeks long) are combined incoherently \citep{Atwood2006,pga+12}.
For a semi-coherent search, the numbers of grid points in the frequency and spin-down dimensions are each reduced by a factor of $T_{\rm obs} / T_{\rm coh}$, while the number of sky positions that must be searched is reduced by a factor of $(1\,{\rm yr})^2 / T_{\rm coh}^2$ (assuming $T_{\rm coh} < 1 \,{\rm yr} < T_{\rm obs}$). This semi-coherent search is performed as an initial scan over the entire parameter space, with more sensitive fully-coherent stages being used to ``follow-up'' a smaller number of semi-coherent candidates to increase their significance.

The compromise is that the sensitivity of a semi-coherent search is far worse than that of a fully-coherent search (e.g. for pulsations from a known pulsar). When folding with a known ephemeris, the signal-to-noise ratio is proportional to $T_{\rm obs}^{1/2}$. In a semi-coherent search, this scaling becomes $(T_{\rm coh} T_{\rm obs})^{1/4}$. Combining this with the computing cost scaling, we see that the recovered signal-to-noise ratio (a proxy for sensitivity) increases very slowly as the computing cost, $C(T_{\rm obs},T_{\rm coh})$, increases. For fixed $T_{\rm obs} > 1\,{\rm yr}$, ${\rm S/N} \propto C(T_{\rm coh})^{a}$, where $a=1/8$ if positional information is available, or $a=1/16$ otherwise. For fixed $T_{\rm coh} < 1\,{\rm yr}$, ${\rm S/N} \propto C(T_{\rm obs})^{b}$ where $b=1/4$ for isolated pulsars, and $b=1/8$ for binary pulsars.

Two additional effects further limit this sensitivity. First, the vast parameter space which must be searched over introduces a large ``trials'' factor, meaning that pulsed signals need to be detected with far higher test statistic values in order to be statistically significant. Where pulsars can be found in single-trial ephemeris folding searches at $H=25$, the associated trials factor of a semi-coherent search raises this significance threshold. To be statistically significant, these searches require the coherent power at the fundamental frequency to be $P_1 \gtrsim 120$. For typical gamma-ray pulse profile shapes, this corresponds to $H \gtrsim 200$. $T_{\rm coh}$ can be chosen such that signals above this level will be detectable in the initial semi-coherent stage.

Second, the accumulation of a pulsed signal can be disrupted by more complex timing behavior such as timing noise, glitching, or orbital period variations in the case of redback binary systems. However, statistical noise still accumulates over the full data duration. For timing behavior that results in a signal being ``well behaved'' only over a timescale $\tau$, this means that the signal-to-noise is further reduced by a factor of $\sqrt{\tau/T_{\rm obs}}$. This latter effect does not occur in isolated millisecond pulsars, whose spin-down rates are extremely stable over time, whereas energetic young pulsars, which tend to exhibit the most timing noise and glitches, will be the worst affected. \citet{ClarkEatHomeI} predicted that the detection rates of isolated MSPs would likely increase relative to those of young pulsars as a result.

In summary, as the LAT data duration increases, these searches become more expensive, and the minimum detectable flux density decreases more slowly than that of an ephemeris-folding search. The ratio between these sensitivities depends on the background photon flux, and hence on the location of the targeted source in the sky, but varies between a factor of $\sim3$--$4$ for the ongoing \textit{Einstein@Home} pulsar surveys (with $T_{\rm obs}=10\,{\rm yr}$, $T_{\rm coh}=2^{22}\,{\rm s}$). The difference in sensitivities between these two methods is emphasized by the difference in the gamma-ray fluxes of the radio-loud and radio-quiet populations shown in Figure~\ref{G100_S1400}.

To tackle these growing challenges, search efforts require increasing computing resources and methodological developments to increase efficiency and sensitivity. Fortunately, the available computing power does increase over time as technology progresses, and as search efforts have shifted to larger computing systems. The first successful searches in the early \Fermi{} mission either searched the center of the localization region for each source, or targeted the location of a plausible X-ray counterpart \citep{BSP, Saz_Parkinson_2010}. After the first two years of the \Fermi{} mission, these discoveries became less frequent. Newer searches \citep{pga+12,Pletsch_2012b} mitigated dependence on event selection criteria and source localization by weighting events and shifting to large computing clusters to enable searching over a grid of positions. These searches were later migrated to the \textit{Einstein@Home} distributed volunteer computing system, allowing for longer $T_{\rm coh}$ and hence more sensitivity \citep{pga+13,ClarkJ1906+0722,Clark_J1208,ClarkEatHomeI,cpw+18}. 

Additional sensitivity can be gained not just through computing power, but also through careful study. While semi-coherent search methods were developed prior to \Fermi's launch \citep{Atwood2006}, improvements to their sensitivity and efficiency have been made throughout the LAT mission \citep{PletschClark2014,Nieder2020}. Improvements to the LAT event reconstruction and background rejection increase the recoverable signal strengths \citep{Pass8,improvedPass8}, as do higher-level processing developments such as photon weighting \citep{KerrWeighted,SearchPulsation}. 
An unbinned likelihood method improved LAT's sensitivity at all time scales, useful for a broad range of astrophysical targets in addition to pulsars \citep[][]{KerrWeightedUnbinnedAna}.

Extensive multi-wavelength searches for possible counterparts at which to target pulsation searches continue to lead to new discoveries. To give some recent examples: targeted searches of X-ray sources discovered energetic gamma-ray pulsars within two supernova remnants (PSRs~J1111$-$6039 and J1714$-$3830), and searches targeting steep-spectrum radio sources associated with LAT sources discovered two isolated MSPs \citep{FrailUnIds2}.

Searches for periodic optical and X-ray counterparts to pulsar-like LAT sources have identified several promising black-widow and redback candidates, detailed in Section~\ref{not_seen}. The precise orbital constraints from long-term optical observations are key inputs for binary searches, although they remain the most computationally demanding projects. Nevertheless, the development of efficient search grid designs \citep{Nieder2020} and a GPU-accelerated\footnote{Graphic Processor Unit} search algorithm deployed on \textit{Einstein@Home} has resulted in two binary MSP discoveries \citep{nck+20, cnv+21},  in addition to PSR~J1311$-$3430 \citep{pga+12}, which was included in 2PC.

The complementary nature of the blind-search and ephemeris-folding populations is highlighted by the number of radio detections of blind-search pulsars. In 2PC, there were radio detections of just four pulsars discovered in gamma-ray searches. This fraction remains small in this catalog, only 11 radio detections for 81 gamma-ray blind-search pulsars, despite deep radio follow-ups.  Of those detected, most have radio flux densities lying below typical radio survey detection thresholds, which are above the 30 $\mu$Jy limit we use to define ``radio quiet'', shown in Figures \ref{G100_S1400} and \ref{S1400Dist}. This is perhaps unsurprising: most pulsars discovered in gamma-ray searches are young pulsars located in the Galactic plane, which has been extensively surveyed by radio telescopes, and therefore we would expect most radio-loud pulsars in this region to be known. Most gamma-ray pulsars found off the Galactic plane are millisecond pulsars, of which most are found in binary systems, and therefore inaccessible to gamma-ray pulsation searches unless orbital constraints from an optical counterpart can be derived. A notable addition since 2PC are the 6 MSPs that remain undetected in radio observations, including the black-widow PSR~J1653-0158 \citep{nck+20}. 
In such spider binaries, the dense ionized wind can obscure the pulsed radio signal for much, or at times all, of the orbit, hindering radio detection.

\subsection{Timing Pulsars using LAT data}
\label{LATtiming}
The timing validity range of a recently discovered radio or X-ray pulsar will likely not cover the entire LAT data set, limiting the statistics available for profile characterization, and limiting the discovery potential for glitches, profile changes, spin-ups \citep{B0540_spindown_change} or other rare behavior. 
Often, pulsar timing using LAT data allows an accurate phase-connected ephemeris covering the entire mission duration.

LAT pulsar timing has significantly improved, with new techniques to handle timing noise and faint pulsars developed by \citet{Ray2011} and \citet{LATtiming}. They use pulse time of arrival (ToA) estimation and time-domain models of timing noise. For faint pulsars, ToA estimates for short data segments are not possible, and we adopted the unbinned methods used in \citet{gammaPTA}. They optimize the timing model by maximizing the likelihood, using PINT \citep{Luo21} to compute the rotational phase for each photon. Parameter uncertainties are derived from the likelihood curvature near the optimum. Particularly useful for MSPs, we have been able to estimate some precise proper motions directly from LAT data, necessary for the Doppler corrections presented in Section \ref{doppler}.

When both radio and LAT timing of comparable quality are available, we use the radio ephemeris. LAT timing was used for 142 of the pulsars in this catalog. Two-thirds of those were made with the ToA method. The fraction of timing solutions using the unbinned method is increasing, as analysts adopt the tool, as more difficult pulsars are added to the sample, and as the datasets lengthen.
The ephemerides we used to fold the \Nall{} pulsars are included in the supplementary on-line data, at \url{https://fermi.gsfc.nasa.gov/ssc/data/access/lat/3rd_PSR_catalog/} (password protected into August 2023), and at \url{https://fermi.gsfc.nasa.gov/ssc/data/access/lat/ephems/}. See Appendix \ref{online}.

\clearpage

\startlongtable
\begin{singlespace}
\begin{deluxetable}{llrrrrlll}
\centerwidetable
\tabletypesize{\scriptsize}
\tablecaption{Galactic pulsars colocated with LAT sources, i.e.\ within $f = \Delta \theta/r_{95} < 1.2$ of a 4FGL-DR3 source, and having efficiency $\epsilon = L_\gamma/ \dot E < 10$, assuming the pulsar distance (73 matches were over-luminous).
For association classes `psr' or `msp' we list the pulsar regardless of the efficiency. 
These pulsars show no gamma-ray pulsations, although many have not been gamma-ray folded, indicated by no $\dot E$ value or by * in the PSR column. 
$ \Delta \theta$ is the angular separation between the pulsar and the source, and $r_{95}$ is the 95\% confidence level semi-major axis of the LAT error ellipse. $E_{\rm peak}$ is the peak of the 4FGL spectral energy distribution when the PLEC fit was reported in 4FGL-DR3. 
The last column lists the 4FGL-DR3 association for the LAT source, `sic' means that it matches the pulsar name.
 V indicates \texttt{Variability\_Index} $>24.7$ . Pulsars with a `g' suffix come from the FAST Galactic plane pulsar survey (GPPS) \citep[][]{hww+21}.
\label{ColocatedNotPulse}}
\tablehead{
\colhead{PSR} & \colhead{4FGL }  & \colhead{$P$ }  & \colhead{$10^{-33}\dot E$} & \colhead{Dist} & \colhead{$\epsilon =$}	& \colhead{$E_{\rm peak}$} & \colhead{$f =$} &  \colhead{Association} \\
          &    & \colhead{(ms)} & \colhead{(erg s$^{-1}$)} & \colhead{(kpc)}    &    \colhead{$L_\gamma/\dot E$}  &    \colhead{(GeV)}     &    \colhead{$ \Delta \theta/r_{95}$  }    &  
}
\startdata
     J0301+35 * &  J0300.4+3450 &   147.06 &   \nodata  & 3.1 &   & 0.12 &  1.19 &     \\
   J0921-5202  & J0924.1-5202c &     9.68 &      0.734 & 0.4 &  0.51 &   &  0.66 &     \\
   J1054-5943 * &  J1054.0-5938 &   346.91 &       3.85 & 2.6 &  2.31 & 1.97 &  0.77 &     \\
   J1146-6610  &  J1147.7-6618 &     3.72 &       6.09 & 1.8 &  0.73 & 0.56 &  1.00 &     \\
   J1154-6250 * & J1155.6-6245c &   282.01 &      0.984 & 1.4 &  7.57 & 0.60 &  0.75 &  SNR G296.8-00.3   \\
   J1302-6350 V &  J1302.9-6349 &    47.76 &       826. & 2.6 &  0.01 & 55.53 &  0.32 &  PSR B1259-63   \\
   J1306-4035 * &  J1306.8-4035 &     2.20 &   \nodata  & 4.7 &   & 0.10 &  0.45 &  sic    \\
     J1332-03 *V &  J1331.7-0343 &  1106.40 &   \nodata  & 3.5 &   & 0.69 &  1.06 &  PKS 1328-034   \\
   J1430-6623  &  J1431.5-6627 &   785.44 &      0.226 & 1.3 &  8.44 & 0.47 &  0.78 &     \\
   J1439-5501  &  J1440.2-5505 &    28.64 &      0.235 & 0.7 &  3.01 & 0.31 &  0.35 &  sic    \\
     J1455-59  & J1456.4-5923c &   176.20 &       58.5 & 6.7 &  3.87 &   &  0.81 &  1RXS J145540.4-591320   \\
   J1535-5848 * &  J1534.7-5842 &   307.18 &       3.70 & 3.0 &  2.61 & 3.90 &  1.08 &     \\
   J1545-4550  &  J1545.2-4553 &     3.58 &       45.3 & 2.2 &  0.15 & 0.56 &  1.19 &     \\
   J1550-5418 * & J1550.8-5424c &  2069.83 &       103. & 4.0 &  0.46 & 1.38 &  0.89 &  SNR G327.2-00.1   \\
     J1604-44 *V &  J1604.5-4441 &  1389.20 &   \nodata  & 7.8 &   &   &  0.96 &  PMN J1604-4441   \\
   J1616-5017  &  J1616.6-5009 &   491.38 &       15.5 & 3.5 &  6.89 & 0.78 &  0.82 &     \\
   J1632-4818  & J1631.7-4826c &   813.68 &       47.6 & 5.3 &  4.22 &   &  1.18 &     \\
   J1731-1847  &  J1731.7-1850 &     2.34 &       77.8 & 4.8 &  0.57 & 0.65 &  0.81 &  sic    \\
     J1741-34 * &  J1740.6-3430 &   875.14 &   \nodata  & 4.6 &   & 0.01 &  1.08 &     \\
   J1743-3153  &  J1743.0-3201 &   193.11 &       57.9 & 8.8 &  4.37 & 0.81 &  1.05 &     \\
     J1743-35 * &  J1743.9-3539 &   569.98 &   \nodata  & 4.0 &   & 0.71 &  0.37 &     \\
   J1801-1417  &  J1801.6-1418 &     3.62 &       4.29 & 1.1 &  0.75 & 0.51 &  0.27 &  sic    \\
   J1806+2819  &  J1807.1+2822 &    15.08 &      0.431 & 1.3 &  3.59 &   &  0.83 &  sic    \\
   J1811-1925  &  J1811.5-1925 &    64.71 &      6334. & 5.0 &  0.01 & 8.94 &  0.26 &  sic    \\
     J1813-08 * &  J1812.2-0856 &     4.23 &   \nodata  & 3.3 &   & 0.47 &  0.68 &     \\
   J1838-0549  &  J1838.4-0545 &   235.31 &       101. & 4.1 &  1.40 & 1.33 &  1.02 &     \\
   J1850-0026  &  J1850.3-0031 &   166.64 &       333. & 6.7 &  0.91 & 0.87 &  0.57 &  SNR G032.4+00.1   \\
  J1852+0158g * &  J1852.6+0203 &   185.73 &   \nodata  & 7.6 &   & 0.42 &  1.13 &     \\
  J1852-0002g * & J1851.8-0007c &   245.10 &   \nodata  & 5.6 &   & 0.80 &  0.49 &  SNR G032.8-00.1   \\
  J1855+0455g * &  J1855.2+0456 &   101.01 &   \nodata  & 10.2 &   & 0.00 &  0.25 &     \\
  J1858+0310g * & J1857.9+0313c &   372.75 &   \nodata  & 6.7 &   & 0.66 &  0.77 &  LQAC 284+003   \\
  J1859+0126g * &  J1900.8+0118 &   957.70 &   \nodata  & 9.6 &   & 0.60 &  1.20 &  NVSS J190146+011301   \\
  J1904+0603g * &  J1904.7+0615 &  1974.93 &   \nodata  & 6.1 &   & 0.66 &  0.72 &     \\
  J1906+0646g * &  J1906.2+0631 &   355.52 &   \nodata  & 5.3 &   & 0.65 &  0.80 &  SNR G040.5-00.5   \\
   J1907+0631  &  J1906.2+0631 &   323.65 &       526. & 3.4 &  0.11 & 0.65 &  0.66 &  SNR G040.5-00.5   \\
  J1908+0811g * &  J1908.7+0812 &   181.64 &   \nodata  & 5.6 &   & 0.58 &  0.36 &     \\
   J1911+1051  &  J1911.3+1055 &   190.87 &       69.1 & 10.1 &  5.02 & 0.57 &  0.43 &     \\
   J1915+1150  &  J1915.3+1149 &   100.04 &       539. & 14.0 &  3.15 & 0.63 &  0.25 &  TXS 1913+115   \\
  J1917+1121g * &  J1916.3+1108 &   510.31 &   \nodata  & 7.1 &   & 0.54 &  1.19 &  SNR G045.7-00.4   \\
   J1928+1725 V &  J1929.0+1729 &   289.84 &   \nodata  & 3.7 &   & 0.50 &  0.81 &     \\
  J1929+1731g *V &  J1929.0+1729 &  3995.40 &   \nodata  & 9.2 &   & 0.50 &  0.35 &     \\
   J1930+1852  &  J1930.5+1853 &   137.04 &     11528. & 7.0 &  0.01 & 10.46 &  0.57 &  PWN G54.1+0.3   \\
   J1950+2414  &  J1950.6+2416 &     4.30 &       9.29 & 7.3 & 17.09 & 0.70 &  0.25 &  sic    \\
   J1957+2516 * &  J1957.3+2517 &     3.96 &       17.4 & 2.7 &  0.61 & 1.03 &  0.46 &  sic    \\
   J2015+0756 * &  J2015.3+0758 &     4.33 &   \nodata  & 2.1 &   &   &  0.91 &     \\
  J2051+4434g * &  J2052.3+4437 &  1303.16 &   \nodata  & 13.5 &   & 0.80 &  0.42 &     \\
  J2052+4421g * &  J2052.3+4437 &   375.31 &   \nodata  & 13.5 &   & 0.80 &  0.77 &     \\
     J2055+1545  &  J2055.8+1545 &     2.16 &       79.2 & 3.6 &  0.15 & 0.57 &  0.06 &  sic    \\
     J2327+62 * & J2325.9+6206c &   266.00 &   \nodata  & 4.3 &   &   &  0.91 &  NVSS J232543+620829   \\
\enddata
\end{deluxetable}
\end{singlespace}
\normalsize

\tabletypesize{\scriptsize}
\begin{deluxetable}{lccccll}
\tablecaption{39 MSPs discovered in radio searches of LAT unidentified sources, yet to show gamma-ray pulsations. Six pulsars with X in the 4FGL column are unrelated to the gamma-ray source that guided the radio search. Localization of the two pulsars with no 4FGL name is as yet inadequate for source matching. Distances use DM and the YMW16 model \citep{ymw17}.
Three unpublished pulsars were found using methods as in \citet{ckr+15}.
\label{PSCnotGammaPulse}}
\tablehead{
\colhead{PSR} & \colhead{4FGL }  & \colhead{$P$ } 	  & \colhead{Dist} & \colhead{$10^{-33}L_\gamma$}	& \colhead{Discovery} & \colhead{Reference}  \\
     &         & \colhead{(ms)}  & \colhead{(kpc)}    & \colhead{(erg s$^{-1}$)}    &  &}
\startdata
J0329+50  &  J0330.1+5038 &     3.06 &  0.3  &  0.052 &  GBT 2019 &   \citet{trr+21} \\ 
J0506+50  &  J0506.1+5028 &     3.39 &  1.4  &  0.510 &  FAST 2020 & Wang et al (in prep)  \\ 
J0646$-$54  &  J0646.4$-$5455 &     2.51 &  0.4  &  0.039 &  PKS 2017 &  as in \citet{ckr+15} \\ 
J0657$-$4657  &  J0657.4$-$4658 &     3.95 &  0.5  &  0.085 &  MKT 2022 & TRAPUM collab. (in prep.)   \\ 
J0838$-$2827  &  J0838.7$-$2827 &     3.62 &  0.4  &  0.147 &  MKT 2021 &  TRAPUM collab. (in prep.)  \\ 
J0843+67  &  J0843.3+6712 &     2.84 &  1.6  &   1.13 &  GBT 2017 &  \citet{trr+21}  \\ 
J1008$-$46  &  X\tablenotemark{a}  &     2.72 &  0.4  & \nodata &  GMRT 2021 & Roy et al (in prep)  \\ 
J1036$-$4353  &  J1036.6$-$4349 &     1.68 &  0.4  &  0.069 &  MKT 2021 &  \citet{TRAPUM_FermiSurvey}  \\ 
J1102+02  &  J1102.4+0246 &     4.05 &  3.7  &   2.53 &  GBT 2019 &  \citet{trr+21}  \\ 
J1103$-$5403  &  X &     3.39 &  1.7  & \nodata &  PKS 2009 &  \citet{kjr+11} \\ 
J1120$-$3618  &  X &     5.55 &  1.0  & \nodata &  GMRT 2011 &  \citet{rb13} \\ 
J1304+12  &  J1304.4+1203 &     4.18 &  1.4  &  0.442 &  AO 2017 &  \citet{ThankfulThesis}  \\ 
J1346$-$2610  &  J1345.9$-$2612 &     2.77 &  1.2  &  0.452 &  MKT 2022 &  TRAPUM collab. (in prep.)  \\ 
J1356+0230  &  J1356.6+0234 &     2.83 &  1.8  &  0.757 &  MKT 2021 &  \citet{TRAPUM_FermiSurvey}  \\ 
J1417$-$4402  &  J1417.6$-$4403 &     2.66 &  4.4  &   18.4 &  PKS 2015 &  ``Huntsman'', \citet{Huntsman} \\ 
J1551$-$0658  &  X &     7.09 &  1.3  & \nodata &  GBT 2010 &  \citet{bbc+22} in prep \\ 
J1624$-$39  &  J1624.3$-$3952 &     2.96 &  2.6  &   6.42 &  GBT 2020 &   \citet{trr+21} \\ 
J1634+02  &  J1634.7+0235 &     2.12 &  0.9  &  0.094 &  AO 2020 & Tabassum et al (in prep)  \\ 
J1646$-$2142  &  X &     5.85 &  1.0  & \nodata &  GMRT 2011 &  \citet{rb13}  \\ 
J1709$-$0333  &  J1709.4$-$0328 &     3.52 &  1.0  &  0.902 &  MKT 2021 &   \citet{TRAPUM_FermiSurvey}  \\ 
J1716+34  &  J1716.3+3434 &     2.11 &  2.5  &  0.520 &  AO 2021 & Tabassum et al (in prep)  \\ 
J1727$-$1609  &  J1728.1$-$1610 &     2.45 &  3.7  &   6.07 &  GBT 2017 &  \citet{trr+21} \\ 
J1802$-$4718  &  J1802.8$-$4719 &     3.67 &  1.2  &  0.575 &  PKS 2016 &  as in \citet{ckr+15} \\ 
J1803+1358  &  J1803.2+1402 &     1.52 &  4.6  &   8.44 &  AO 2017 &  \citet{ThankfulThesis}  \\ 
J1814+31  &  J1814.4+3132 &     2.09 &  3.0  &   4.87 &  AO 2017 &  \citet{ThankfulThesis}  \\ 
J1823+1208  &  J1823.2+1209 &     5.21 &  1.7  &  0.646 &  MKT 2021 &  \citet{TRAPUM_FermiSurvey}  \\ 
J1823$-$3543  &  J1823.8$-$3544 &     2.37 &  3.7  &   3.22 &  MKT 2020 &  TRAPUM collab. (in prep.)   \\ 
J1828+0625  &  X &     3.63 &  1.0  & \nodata &  GMRT 2011 &  \citet{rb13} \\ 
J1831$-$6503  &  J1831.1$-$6503 &     1.85 &  1.0  &  0.451 &  MKT 2022 &  TRAPUM collab. (in prep.)   \\ 
J1845+0200  &  J1845.0+0159 &     4.31 &  1.7  &  0.977 &  AO 2016 &  \citet{ThankfulThesis}  \\ 
J1904$-$11  &  J1904.4$-$1129 &     2.62 &  4.7  &   4.80 &  GBT 2020 & \citet{trr+21}  \\ 
J1906$-$1754  &  J1906.4$-$1757 &     2.88 &  6.8  &   18.4 &  MKT 2020 &   \citet{TRAPUM_FermiSurvey}  \\ 
J1910$-$5320  &  J1910.7$-$5320 &     2.33 &  1.0  &  0.360 &  MKT 2022 &   \citet{ProbableJ1910} submitted  \\ 
J1919+23  &  J1919.1+2354 &     4.63 &  2.0  &   1.36 &  AO 2017 &  \citet{ThankfulThesis}  \\ 
J1947$-$11  &  J1947.6$-$1121 &     2.24 &  3.1  &   3.76 &  GBT 2022 &  Strader et al (in prep)   \\ 
J2036$-$02  &  J2036.2$-$0207 &     1.91 &  7.7  &   10.9 &  GBT 2020 &   \citet{trr+21} \\ 
J2045$-$6837  &  J2045.9$-$6835 &     2.96 &  1.3  &  0.421 &  PKS 2017 &  as in \citet{ckr+15} \\ 
J2051+50  &    &     1.68 &  1.4  & \nodata &  GBT 2020 & \citet{trr+21} \\ 
J2333$-$5526  &  J2333.1$-$5527 &     2.10 &  2.5  &   3.13 &  MKT 2021 & \citet{TRAPUM_FermiSurvey}  \\ 
\enddata
\tablenotetext{a}{Search targeted uncatalogued LAT source 605N-0275 (T. Burnett, priv. comm.)}
\tablecomments{\ The Discovery column takes ``{\tt Fermi}'' entries from   \url{http://astro.phys.wvu.edu/GalacticMSPs/GalacticMSPs.txt}. See also \url{http://www.trapum.org/discoveries/} }
\end{deluxetable}

\section{The Gamma-ray Pulsars}\label{cat}
The search methods yielded gamma-ray pulsations for \Nall\ pulsars.  
Table \ref{tab:tallies} breaks down the numbers by different types.
Here we compile their radio fluxes, distances, and, for the MSPs, their proper motions, used to Doppler-correct their spindown properties \citep{Shklovskii}.  
Figure \ref{Aitoff} shows the sky distribution of the \textit{Fermi} pulsars. The LAT's uniform sky coverage is important for detecting MSPs, as these relatively nearby sources are approximately uniformly distributed in Galactic latitude.



\begin{figure}[!ht]
\centering
\includegraphics[width=1.0\textwidth]{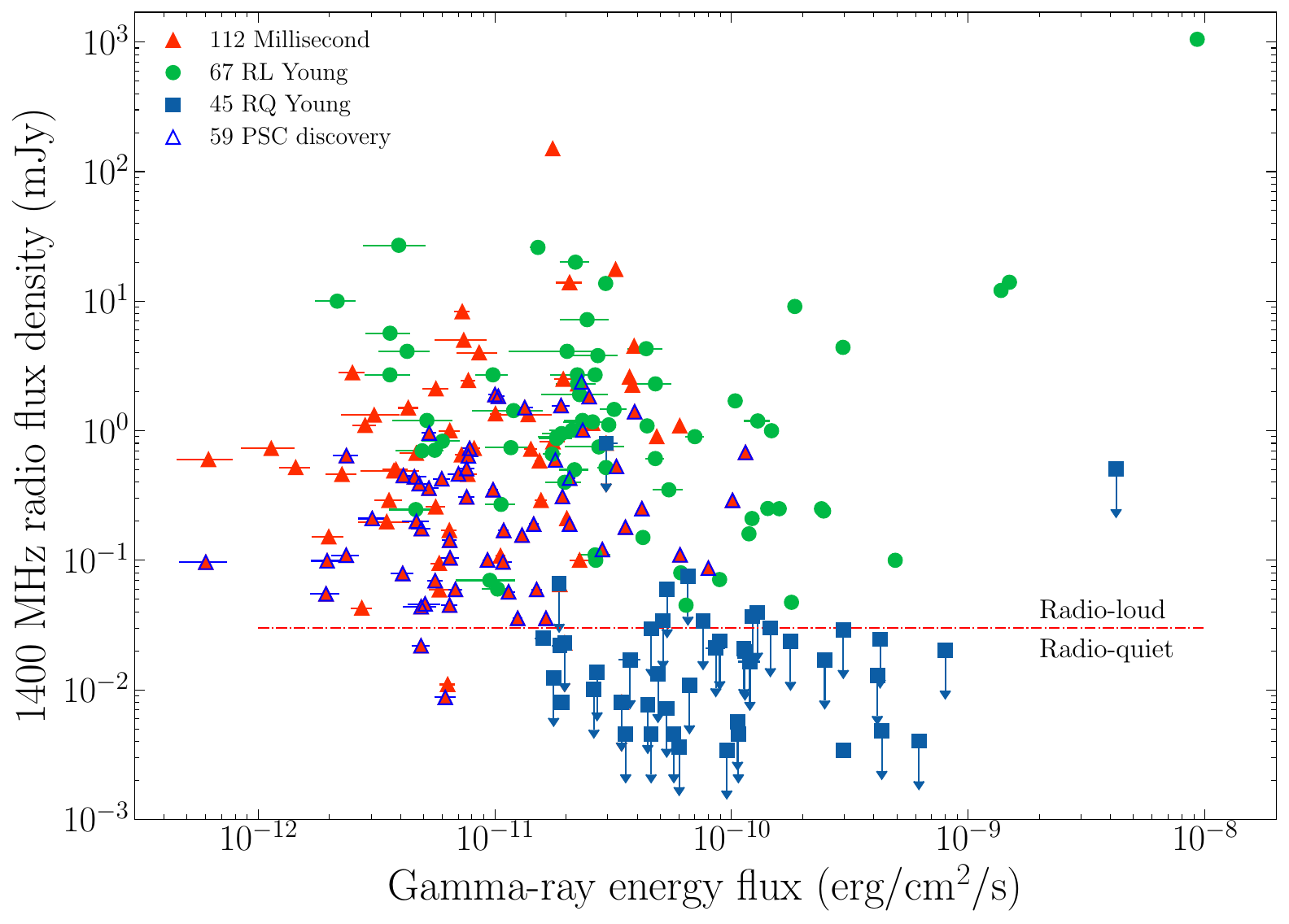}
\caption{The radio flux density at 1400 MHz versus the integral energy flux above 100 MeV. Gamma-ray and radio fluxes are essentially uncorrelated. Dark borders on the triangle symbols indicate MSPs discovered in deep radio searches of previously unidentified LAT sources by the PSC (see Section \ref{psc}). The horizontal line at 30 $\mu$Jy shows our convention to distinguish radio ``loud'' from radio ``quiet'' pulsars.
\label{G100_S1400}}
\end{figure}

\subsection{Radio Intensities}
\label{S1400}

We tabulate measured or derived 1400~MHz flux density values ($S_{1400}$) in Tables~\ref{tbl-charPSR} and \ref{tbl-charMSP}. ATNF \textit{psrcat} provides flux densities $S_{\nu}$ from among 27 different radio frequencies $\nu$; thus as a starting point, we used  {\tt psrqpy} \citep{psrqpy} to extract all $S_{\nu}$ values from {\it psrcat}, as well as the radio spectral index, {\tt SPINDX}. When available, we use $S_{1400}$. Otherwise, we calculate $S_{1400}  =  S_{\nu}(\frac{1400}{\nu})^{\alpha}$ with $\alpha = $ {\tt SPINDX} when present, and $\alpha = -1.7$ otherwise. We use the $S_{\nu}$ value corresponding to the frequency $\nu$ closest to 1400 MHz.
We used  $\alpha = -1.7$ in 2PC, a compromise among published values: $-1.6$ from \citet{Lorimer_1995},  $-1.8 \pm 0.2$ from \citet{Maron_2000},  and $-1.60 \pm 0.03$ from \citet{Jankowski2018}. For MSPs, \citet{sbm+22} found $-1.92\pm 0.06$ with a standard deviation of $0.6$, and \citet{dhm+15} found $-1.81\pm 0.01$. The small formal uncertainties are potentially misleading: for example, the value from \citet{Jankowski2018} is for a sample of 400 pulsars with a log-normal distribution of standard deviation $0.54$, the distribution being sensitive to e.g.\ the frequency range used, yielding the spread in average indices.
Using multiple $S_{\nu}$ values to fit a power law gives $S_{1400}$ values generally compatible with those found above, and possibly more accurate in some cases. However, discrepant $S_{\nu}$ values lead to unrealistic results in some cases, hence
we use the simple prescription described above.

The \textit{psrcat} $S_{\nu}$ values have accumulated over decades, using a variety of instruments and  analysis conventions. Frequency extrapolation introduces additional biases. Fortunately, since v1.68 \textit{psrcat} includes $S_{1400}$ values from two recent studies. \citet{sbm+22} used MeerKAT to obtain $S_{1400}$ for 189 MSPs\footnote{We thank R. Spiewak, M. Bailes et al for sharing their results before publication.}, and \citet{psf+22} analyzed Arecibo data from the PALFA survey for 93 MSPs. We also replace the \citet{hrm+11} values with the updated results reported by \citet{bbc+22}. For a small number of pulsars, we obtained as-yet-unpublished $S_{\nu}$ measurements, noted in the Tables.

In addition to pulsars which are well characterized from radio surveys, those discovered in blind searches are typically searched deeply for radio pulsations \citep{Camilo2009,aaa+10d,Saz_Parkinson_2010,Ray2011,rap+12,pga+12,nck+20}. The 81 blind discoveries were followed by 11 radio detections, 6 of which qualify as `radio-loud', with $S_{1400} > 30\, \mu$Jy (see below). Radio measurements are not yet available for recently discovered LAT pulsars.

Figure \ref{G100_S1400} shows that radio flux densities are essentially uncorrelated with the integrated gamma-ray energy fluxes, $G_{100}$. Population studies aim to constrain a power-law model of radio luminosity in which $L_r \propto \dot E^s$.
\citet{SimonArisBraking} favor $s=1/4$ but recognize that a range of $s$ values match observations, because of the large dispersion. \citet{MeerKAT_IX} find an even smaller index.
Since $L_\gamma \propto \sqrt{\dot E}$ (see Figure \ref{EDotLumG}), flux correlation between the two wavelengths is naturally expected. However, the fraction of neutron star spindown power carried off by radio waves is thousands of times lower, or more, than that of gamma rays, and radio emission is so weakly connected to the braking mechanism that the correlation is invisible.

\begin{figure}[bp]
\centering
\includegraphics[width=0.9\textwidth]{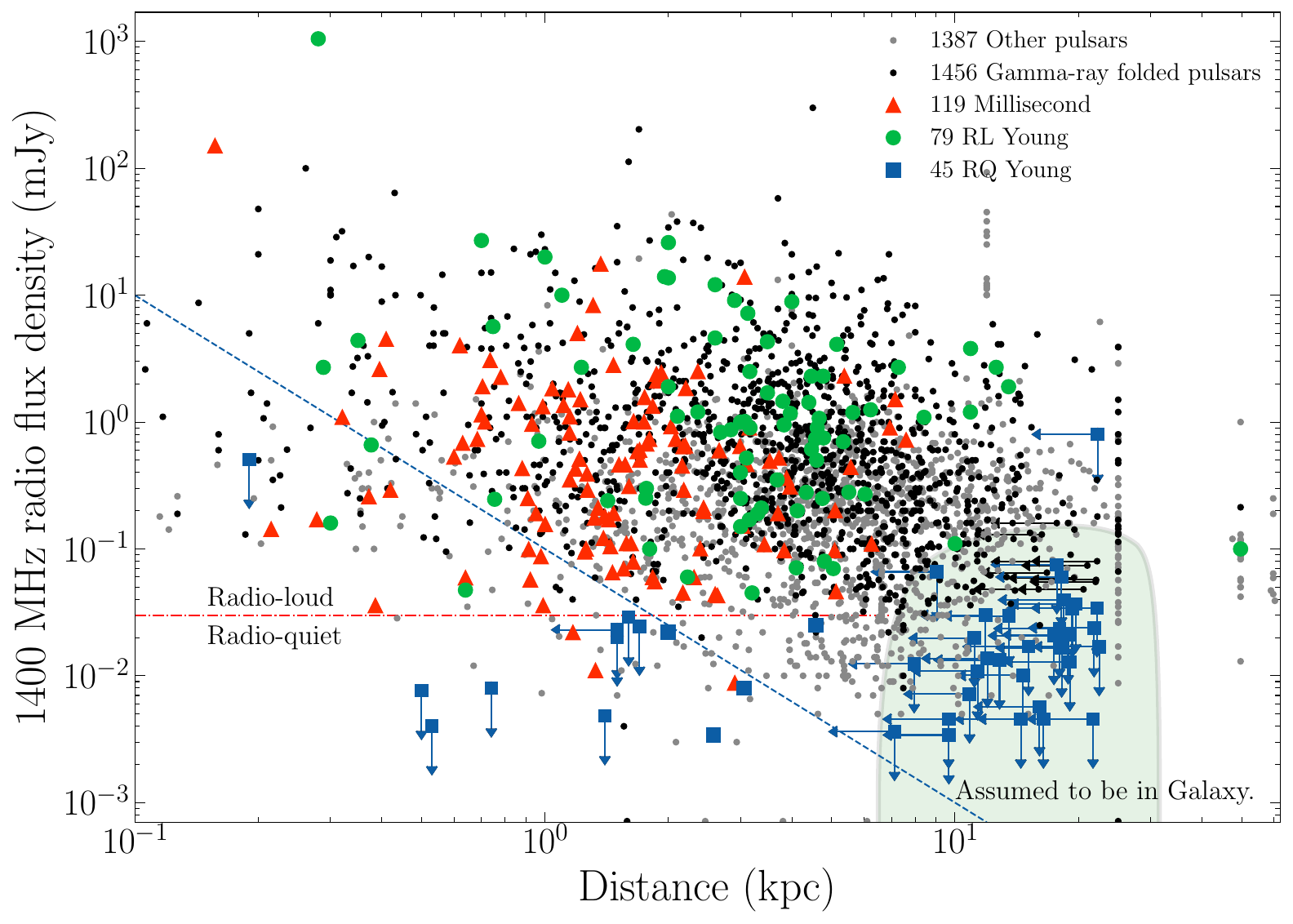}
\caption{Radio flux density at $1400$ MHz versus pulsar distance. 
The horizontal line at 30 $\mu$Jy shows our convention to distinguish radio ``loud'' from radio ``quiet'' pulsars.
The diagonal line shows a threshold in pseudo-luminosity of 100 $\mu$Jy - kpc$^2$.
The pulsars at lower-right are assigned distance limits along the Milky Way's rim in Figure \ref{galaxie}.
DM for the pulsars on the line at 25 kpc saturate the YMW16 model.
\label{S1400Dist}}
\end{figure}

Figure \ref{S1400Dist} shows how the radio flux densities of the gamma-ray pulsars compare with the overall pulsar population, as well as their lack of distance dependence. As in 2PC, we define a pulsar as `radio-loud' if $S_{1400} > 30\, \mu$Jy, and `radio-quiet' if the measured flux density is lower.
The horizontal lines in Figures  \ref{G100_S1400} and \ref{S1400Dist} show the threshold.
Also as in 2PC, the diagonal line in Figure \ref{S1400Dist} shows an alternate threshold at pseudo-luminosity 100 $\mu$Jy-kpc$^2$.

Whether a radio or a gamma-ray beam (or both) sweeps through our line of sight from Earth depends on the beam shapes and orientations. \citet{RadioBeamBroadening} already showed that radio beam size often increases with decreasing frequency. \citet{GriessmeierFR606} searched 27 radio-quiet northern gamma-ray pulsars at 150 MHz but obtained no new detections. This result is compatible with the current understanding of these pulsars' orientations based on their emission geometries, which are derived from LAT light curves and plotted in \citet{JohnstonSmith2020, GriessmeierFR606}, and which are consistent with the radio beam not crossing the Earth's line of sight. 

\subsection{Distances}
\label{dist}

\begin{figure}[!ht]
\centering
\includegraphics[width=0.7\textwidth]{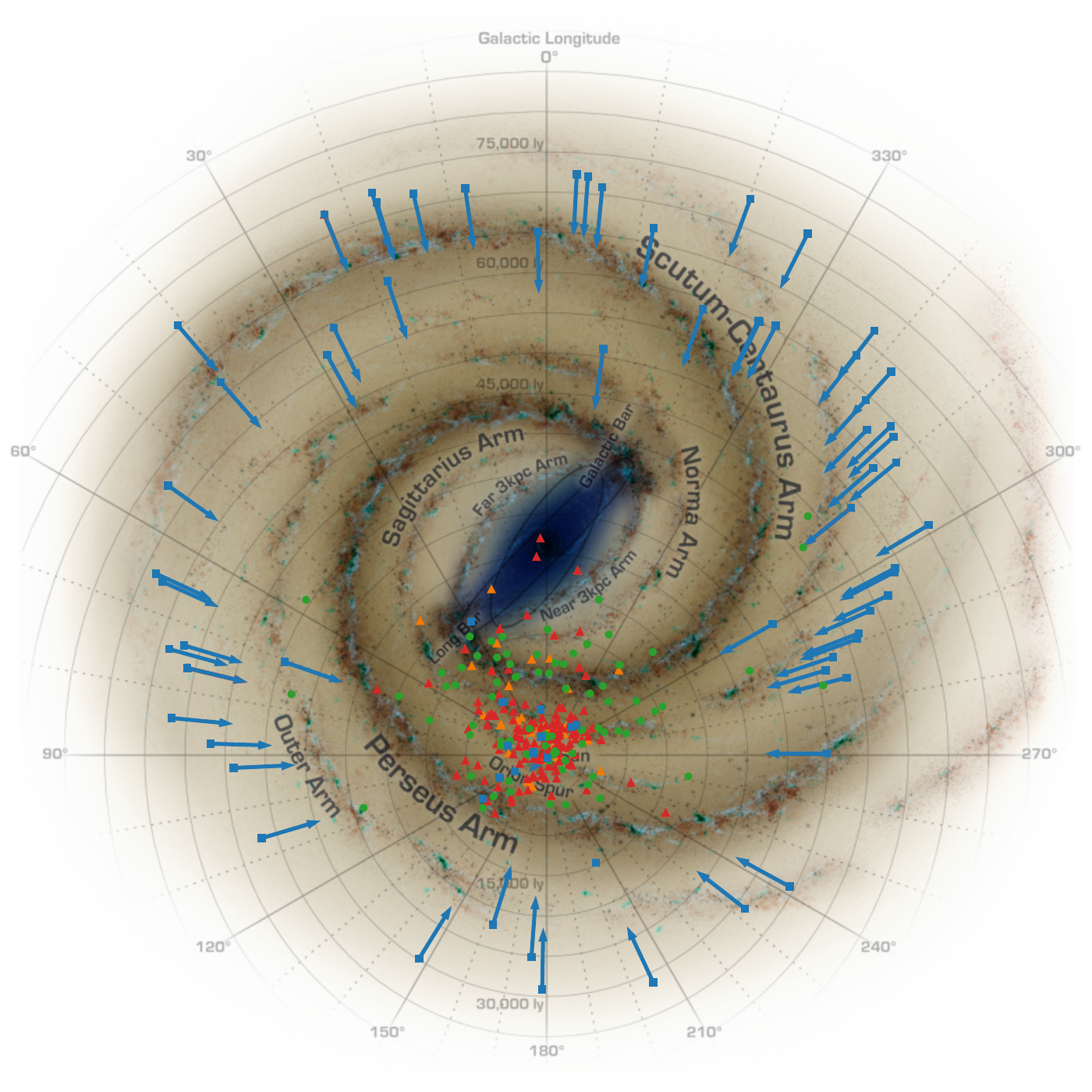}
\caption{ Gamma-ray pulsar positions projected onto the Milky Way model of \citet{AnatomicallyCorrect}.
The MSPs that appear coincident with the Galactic center, PSRs J1823$-$3021A and J1835$-$3259B in the globular
clusters NGC 6624 and 6652, lie well above the Galactic plane. Distance uncertainties are not shown, but can be large for the more distant objects. The squares with arrows indicate the lines of sight toward pulsars for which no
distance estimates exist, placed at the distances where 95\% of the electron column density has been integrated in the YMW16 model. 
Other symbols are as in Figure \ref{PPdotplot}.
\label{galaxie}}
\end{figure}

Converting integrated energy fluxes $G_{100}$ to emitted luminosities $L_\gamma = 4\pi d^2 f_\Omega G_{100}$ requires the source distances $d$. (We set the beaming fraction $f_\Omega=1$, see Section \ref{lumin}.)
Knowing pulsar distances also allows mapping neutron star distributions relative to the Galaxy's spiral arms, as in Figure \ref{galaxie}, and evaluating their scale height above the plane. The various methods to estimate pulsar distances offer widely differing accuracies. Tables \ref{tbl-PSRdist} and \ref{tbl-doppler} list the distances that we use, the estimation methods, and the references. 

Dispersion measure (DM) is the column density of free electrons along the path from Earth to the pulsar, in units of pc cm$^{-3}$. The electrons delay the radio pulse arrival by $\Delta t = {\rm DM} (p\nu^2)^{-1}$ where $\nu$ is the observation frequency in MHz and $p = 2.410 \times 10^{-4}$ MHz$^{-2}$ pc cm$^{-3}$ s$^{-1}$. Given a model for the electron density $n_e$ in the various structures of our Galaxy, integrating DM = $\int_0^d n_e dl$ along the line of sight $dl$ yields the distance $d$ for which DM matches the radio measurement. 
We use mainly the YMW16 model \citep{ymw17}, but often consult the NE2001 model when studying particular lines of sight \citep{Cordes2002}. 
Significant discrepancies between model and true distances occur along some lines of sight, more than three times the probable true distance in the worst cases. 
Examples are PSR J0614$-$3329 \citep{bac+16}; PSR J2021+3651 \citep{PSR2021LAT}; PSR J0248+6021 \citep{tpc+11}; and PSR J0908$-$4913 \citep{jl21}. Figure 4 of \citet{SixWeak} illustrates one way to identify pulsar lines of sight that have DM model peculiarities.

Radio spectra can show Doppler-shifted neutral hydrogen (\textsc{Hi}) absorption or emission lines due to clouds along the line of sight. A Galactic rotation model converts the Doppler-inferred radial velocity to a cloud distance estimate, and the pulsar distance is constrained if it appears to be in or between clouds. This ``kinematic'' method is tagged with a K in the tables. The X tag indicates cases where X-ray absorption below 1 keV yields a column density that can be compared with one obtained from 21 cm radio surveys.

To obtain the distances we first query the ATNF \textit{psrcat} DIST parameter, which is most often DIST\_DM, the  dispersion measure distance using YMW16.
We adopt a 40\% uncertainty on DIST\_DM  as per Figure 12 and Table 5 of \citet{ymw17}.
When \textit{psrcat} has a  DIST\_A value, obtained by another method, we use that instead and propagate the method and reference to our tables. The exception is that we exclude the distances deduced by \citet{wan11} using the LAT $G_{100}$ values and the correlation between $L_\gamma$ and $\dot E$ shown in Section \ref{lumin}, since they would bias our attempts to better determine that correlation.
If the \textit{psrcat} parallax parameter PX is present, DIST\_A = 1/PX and we use the reference for PX. Parallax is measured only for relatively nearby pulsars,  using X-ray or optical images, radio interferometric imaging, or accurate timing\footnote{\url{http://hosting.astro.cornell.edu/research/parallax/} lists known pulsar parallaxes \citep{cbv+09}.}. When converting parallax measurements to distances, the Lutz-Kelker  (LK) effect is an overestimate of parallax values, and hence an underestimate of distances, due to the larger volume of space traced by smaller parallax values \citep{lk73}. 
We use the LK-corrected distance estimates determined by \citet{vwc+12}. 
Good estimates of the distance to an optical companion in a binary system are more frequent than in 2PC because we have more spider MSPs than before, and because of the wealth of information provided by Gaia \citep{gaia18}. For two of our MSPs, PSRs J1227$-$4853 and J1431$-$4715, \citet{jkc+18} find distances using \textit{Gaia} companion parallaxes. They use a spatial distribution prior specific to Galactic pulsars, i.e., not isotropic, which means that the LK-like correction brings some pulsars closer to Earth when they are towards the anti-center, as opposed to farther for standard LK. For these two MSPs, we simply use 1/PX, the same within uncertainties as the \citet{jkc+18} distances.

For several pulsars, measured DM exceeds DM$_{\rm max}$, the maximum modeled by \citet{ymw17} for that line of sight, by between 1 and 50\%. The YMW16 model therefore returns the distance to the Galaxy's edge. Modeled DM saturates much closer to Earth, increasing slowly thereafter. For example, towards PSR J1327$-$0755, 95\% of DM$_{\rm max}$ occurs at $3.7$ kpc, compared to 25 kpc returned by the code, and in better agreement with the $1.7$ kpc NE2001 distance. Therefore, for pulsars with DM $>$ DM$_{\rm max}$, we use the YMW16 distance for
95\% of DM$_{\rm max}$. Other pulsars cited in this catalog for which this occurs are PSR J1102+02 (Table \ref{PSCnotGammaPulse}), and PSRs J1302$-$03, J2051+4434g and J2052+4421G (Table \ref{ColocatedNotPulse}).

For 63 pulsars we have no DM measurement or other distance estimator. We again calculate the YMW16 distance for
95\% of DM$_{\rm max}$, but use it as an upper limit. The method is ``DMM'' in Tables \ref{tbl-PSRdist} and \ref{tbl-doppler}, and these radio-quiet pulsars  are illustrated in Figure \ref{galaxie}.

Both \citet{egc+13} and 2PC discussed the problem of PSR J0610$-$2100 having gamma-ray efficency $\eta = L_\gamma / \dot E > 1$, using the DM distances of $3.3$ and 4 kpc. 
\citet{vbc+22} used LOFAR timing and optical modelling for this system to find $d = 2.2 \pm 0.7$ kpc. \citet{egc+13} point out that the line of sight to the pulsar grazes the electron overdensities bordering the local bubble, in both electron models. A slight extension of the modeled bubble yields the increase in modeled DM needed to reconcile the distances. 
The closer distance for PSR J0610$-$2100 leads to $\eta \approx 50$\%. It also reduces the pulsar's transverse velocity to $v_T \approx  220$ km s$^{-1}$, so that this MSP is no longer among the fastest known. 

\newpage
\startlongtable
\begin{deluxetable}{lccc}
\tabletypesize{\scriptsize}
\tablecaption{Distance estimates for young LAT-detected pulsars. 
\label{tbl-PSRdist}}
\tablewidth{0pt}

\tablehead{
\colhead{Pulsar Name}      & 
\colhead{Distance (kpc)}   & 
\colhead{Method }      &
\colhead{Reference} }
\startdata
 J0002+6216 &   $2.0\pm {0.4}$  &  O & \citet{J0002_VLA}\\ 
 J0007+7303 &   $1.4\pm {0.3}$  &  K & \citet{Pineault1993}\\ 
 J0106+4855 &   $3.1\pm {1.2}$  &  DM & \citet{pga+12}\\ 
 J0117+5914 &   $1.8\pm {0.7}$  &  DM & \citet{bkk+16}\\ 
 J0139+5814 &   $2.6^{+0.3}_{-0.2}$  &  P & \citet{cbv+09}\\ 
 J0205+6449 &   $3.2\pm {0.2}$  &  K & \citet{Xu_2006}\\ 
 J0248+6021 &   $2.0\pm {0.2}$  &  0 & \citet{tpc+11}\\ 
 J0357+3205 &   $<7.1$ & DMM &   \citet{ymw17}\\
\enddata


\tablecomments{The best known distances of the \Nyoung\ young pulsars detected by the {\it Fermi} LAT. 
For the DM method, the reference is for the Dispersion Measurement and the distance was obtained with the YMW16 model for the Galactic electron density  \citep{ymw17}.  DMM is the distance at which the YMW16 DM reaches 95\% of its maximum value for that line of sight. Method P is parallax, K is kinetic (HI absorption), KPLK combines the two and corrects for Lutz-Kelker bias \citep{vwc+12}, X indicates an X-ray measurement of the hydrogen column density, and O englobes optical and other methods.}

\end{deluxetable}
\normalsize


\begin{longrotatetable}
\begin{deluxetable}{lccccccccc}
\tabletypesize{\scriptsize}
\tablecaption{Millisecond pulsar distances and spindown Doppler corrections
\label{tbl-doppler}}
\tablewidth{0pt}
\tablehead{
\colhead{PSR}                               & 
\colhead{$ d $}     & \colhead{Method}   & \colhead{Ref.}   &
\colhead{$ \mu$} & \colhead{ Ref.}   & 
\colhead{$\dot P^{\rm \, shk}$}  & \colhead{$\dot P^{\rm \, gal}$}  &
\colhead{$\dot E^{\rm \, int}$}  & \colhead{$(1-\xi)$} \\ 
\colhead{}  & \colhead{(pc)} & \colhead{} & \colhead{} & \colhead{($\rm mas\; yr^{-1}$)} & \colhead{} &  
\colhead{($\rm 10^{-21}$)} & \colhead{($\rm 10^{-21}$)} & \colhead{($\rm 10^{33} \,erg \, s^{-1}$)} &  }
\startdata
 J0023+0923 &  $1790^{+510}_{-320}$ &    P &  \citet{aab+21a}  & $13.9 \pm 0.1$ & Same  &    $2.57$        & $-0.59$ &  $12.4^{+1.0}_{-0.7}$    & $0.82$ \\ 
 J0030+0451 &  $320\pm 10$ &    P &  \citet{aab+21a}  & $6.4 \pm 0.1$ & Same  &   $0.15$        & $-0.52$ &  $3.602\pm 0.003$ & $1.0$ \\ 
 J0034$-$0534 &  $1340\pm 530$ &    DM &  \citet{aaa+10b}  & $12.1 \pm 0.5$ & \citet{dcl+16}  &    $0.90$        & $-0.40$ &  $26.6\pm 2.2$ & $0.90$ \\ 
 J0101$-$6422 &  $1000\pm 400$ &    DM &  \citet{sbm+22}  & $17.3 \pm 0.5$ & LAT  &    $1.87$        & $-0.51$ &  $8.0\pm 1.8$ & $0.72$ \\ 
 J0102+4839 &  $2310\pm 920$ &    DM &  \citet{hrm+11}  & $4.1 \pm 0.3$ & NRT  &    $0.28$        & $-0.14$ &  $17.6\pm 0.2$ & $0.99$ \\ 
 J0154+1833 &  $1620\pm 640$ &    DM &  \citet{mgf+19}  & $13.6 \pm 1.5$ & Same  &    $1.72$        & $-0.17$ &  $3.8\pm 2.4$ & $0.45$ \\ 
 J0218+4232 &  $3150^{+850}_{-600}$ &    PLK &  \citet{vl14}  & $6.5 \pm 0.1$ & \citet{dyc+14}  &    $0.76$        & $0.11$ &  $241.9^{+0.7}_{-0.5}$    & $0.99$ \\ 
 J0248+4230 &  $1850\pm 740$ &    DM &  \citet{brj+21}  & ( $6.0 \pm 3.3$ ) & Same  &  $<1.8$  &   0.09     &  $>33$   &  $>0.89$    \\
\enddata

\tablecomments{Columns 2 and 3 give the distances for the \Nmsps\ MSPs detected with the {\it Fermi} LAT, and the method used to find them:
P -- parallax; 
LK -- corrected for the Lutz-Kelker effect;
DM -- from dispersion measure using the YMW16 model \citep{ymw17} ; 
O -- optical and other methods.
For DM, the references in Column 4 give the DM measurement.
DMM is a limit on the pulsar distance, corresponding to 95\% of the maximum 
YMW16 DM for that line of sight. Proper motion measurements $\mu$ are listed in Column 5, taken from the references in Column 6,
often the ``Same'' as in Column 4. 
``LAT'' refers to gamma-ray timing, described in Section \ref{LATtiming}. ``NRT'', ``JBO'', ``PKS'', and ``GBT'' indicate that the measurements
were made for this work using data from the Nançay, Lovell, Parkes and Green Bank radio telescopes.
Columns 7 and 8 list the contributions of the Shklovskii effect $\dot P^{\rm \, shk}$ and of acceleration due to the Galactic rotation, $\dot P^{\rm \, gal}$ to the intrinsic spindown rate $\dot P^{\rm \, int}$. The last columns list 
the intrinsic spindown power $\dot E^{\rm \, int} = \dot E (1-\xi)$. See text for how cases with a missing or insignificant $\mu$ measurement, or a distance limit, are handled. }


\end{deluxetable}
\end{longrotatetable}


\subsection{Doppler corrections}
\label{doppler}


Many pulsar characteristics depend on the \textit{intrinsic} spin period $P^{\rm \,int}$ and spindown rate $\dot P^{\rm \,int}$. The Doppler shift of the \textit{observed} period is $P = (1 + v_{\rm \,R}/c)P^{\rm \,int}$,
where $v_{\rm \,R}$  is the pulsar's radial velocity along the unit vector $\textbf{n}$ from the solar system. The Doppler correction to $\dot P$ is obtained by differentiating the equation and separating the effects of the system's proper motion \citep{Shklovskii} from the acceleration due to Galactic rotation (the treatment by \citet{DamourTaylor91} is particularly clear). The result is
\begin{equation}
\dot P^{\rm \,int} = \dot P -\dot P^{\rm \, shk}-\dot P^{\rm \, gal}
\label{DoppEq}
\end{equation}
with
\begin{equation}
\dot P^{\rm \,shk} = \frac{1}{c}\,\mu^2 \,d\,P = k\,\left({\frac{\mu}{\rm mas\, yr^{-1}}}\right)^2 \,\left({\frac{d}{\rm kpc}}\right)\,\left({\frac{P}{\rm s}}\right)
\end{equation}
and
\begin{equation}
\dot P^{\rm \,gal} = \frac{1}{c}\, \textbf{n} \cdot \left(\textbf{a}_{1}-\textbf{a}_{0}\right)P
\end{equation}
where $k=2.43\times 10^{-21}$ for pulsar distance $d$ and proper motion transverse to the line of sight $\mu$. In 2PC we obtained the accelerations $\textbf{a}_{1}$ of the pulsar and $\textbf{a}_{0}$ of the Sun using the Galactic potential model of \cite{Carlberg_Innanen1987} and \cite{Kuijken_Gilmore1989} . Here, we instead use GalPot\footnote{\url{https://github.com/PaulMcMillan-Astro/GalPot}} with the \texttt{PJM17\_best.Tpot} parameter set \citep{GalPot}.

The corrections are negligible for the young gamma-ray pulsars, which all have $\dot P > 10^{-17}$. However, for MSPs $\dot P^{\rm \,int}$ can differ noticeably from observed $\dot P$. Expressing the correction magnitude as 
\begin{equation}
\xi = (\dot P^{\rm \, shk} + \dot P^{\rm \, gal})/\dot P = (\dot P - \dot P^{\rm \, int} ) /\dot P
\end{equation}
leads to $\dot E^{\rm \, int} = \dot E \,(1-\xi)$.  For $\mu \lesssim 10 \, \rm mas\, yr^{-1}$ and $d \lesssim 1$ kpc,   $\dot P^{\rm \, shk}$ and $\dot P^{\rm \, gal}$ have comparable magnitudes and opposite signs. For larger $\mu^2 \,d$, $\dot P^{\rm \, shk}$ dominates.

Table \ref{tbl-doppler} lists $d,\, \mu$, their uncertainties $\delta d,\, \delta\mu$, and the resulting corrections.
When we have no $\mu$ measurement and only a limit for $d$, we leave the last columns empty.
Parentheses around the $\mu \pm \delta\mu$ value indicate that the measurement is insignificant,
$\delta\mu/\mu > 3$, and we express the remaining columns as limits calculated using $\mu_{\rm max} < \mu + 2\delta\mu$.
In parentheses in the $\dot P^{\rm \, gal}$ column are the minimum and maximum values for that direction
for distances from 500 pc to the limit. 
Given a distance but no proper motion value, we calculate $\dot P^{\rm \, gal}$ and limits on $\xi$, using $\dot P^{\rm \, shk} \geq 0$. There are 8 such MSPs, of which PSR J0605+3757 is listed first.
In the same spirit, e.g. PSR J1335$-$5656 has measured proper motion but unknown distance and
we combine the limit for $\dot P^{\rm \, shk}$ with the extremum from the $\dot P^{\rm \, gal}$ range.

Table \ref{tbl-doppler} lists a single MSP with well-measured $\mu$ but $(1-\xi) <0$, implying implausible spin-up, $\dot E^{\rm \, int} < 0$. It is PSR J1024$-$0719, with a binary companion in an atypical multi-century orbit that leads to mis-measured $\dot P^{\rm \, int}$ \citep{bjs+16,kkn+16, gsl+16}. 
Six other MSPs have $\dot E^{\rm \, int} \lesssim 2 \times 10^{33}$ erg s$^{-1}$, some with nominally small uncertainties. The lowest is $\dot E^{\rm \, int} = 9^{+21}_{-8} \times 10^{32}$ erg s$^{-1}$ for PSR J1946+3417. Improved precision and a larger sample will clarify the gamma-ray ``deathline'' for MSPs, below which emission in the LAT range ceases. 
A useful tool to understand the uncertainty on $\dot E^{\rm \, int}$ are plots of $\mu$ vs $d$ 
showing the constraints suggested by the neutron star's transverse velocity $v_{\rm \,T}= \mu d$, minimum 
$\dot E^{\rm \, int}$, and so on. A powerful constraint is to require $\dot E^{\rm \, int} > L_\gamma$ (Section \ref{lumin} addresses luminosity $L_\gamma$).
Examples of $\mu$ vs $d$ plots, or similar, are Figure 11 in 2PC (which includes PSR J1024$-$0719), Figure 1 in \citet{gsl+16}, and Figure 4 in \citet{cpw+18}.

\section{Light Curves and Profile Characterization}
\label{profiles}

\subsection{Gamma-ray and Radio Light Curves}
\label{profiles:lcs}

\begin{figure}[!ht]
\centering
\includegraphics[scale=0.37]{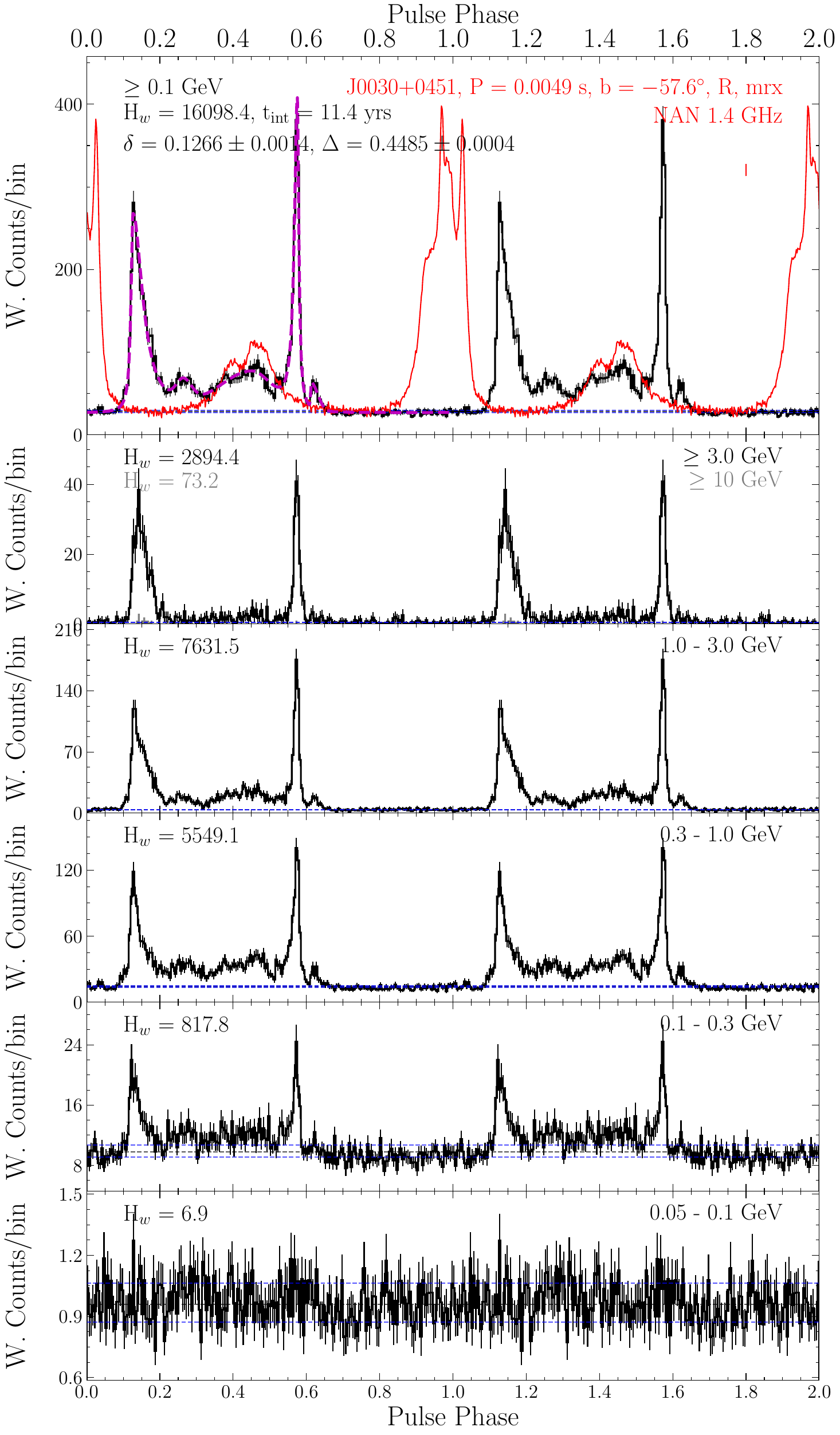}
\caption{Pulse profiles for the MSP PSR~J0030+0451. Weighted LAT pulse profiles in different energy bands are shown in black. The pulse profiles above 10~GeV are in light grey, more apparent for other pulsars shown in Appendix \ref{App-Samples}. The radio pulse profile for this pulsar at 1.4~GHz in shown in red, and a fit of the integrated gamma-ray pulse profile over 0.1 GeV is displayed in magenta. Horizontal blue lines correspond to the expected background levels in the gamma-ray pulse profiles, plus or minus one standard deviation. The quantities at top left correspond to the weighted $H$-test TS for the LAT pulse profile shown in the top panel, and the total integration time. At top right are listed: the pulsar name, its rotational period and Galactic latitude, and the `PSR' and `CHAR' codes defined in the captions for Tables \ref{tbl-charPSR} and \ref{tbl-charMSP} (top row), the radio observatory used to acquire the radio profile (if available), and the corresponding radio frequency (second row). For pulsars with radio profiles, a marker at  top right indicates the phase alignment uncertainty stemming from the dispersion measure. See Section~\ref{profiles} for more details about light curve construction and the fitting procedure.}
\label{profiles:example_lightcurve_0030}
\end{figure}

Figure~\ref{profiles:example_lightcurve_0030} shows an example of gamma-ray light curves in different energy bands for the MSP PSR~J0030+0451. Figure \ref{SampleLC} and Appendix~B show further examples, overlaid with radio profiles when available. The supplementary online material includes all the pulse profiles constructed for this work.


%

After preparation according to  Section \ref{obs}, the photon arrival times were converted to rotational phase using timing models obtained from pulsar timing observations conducted in radio or X-rays, or by direct timing of the LAT photons, as described in Section~\ref{pulsedisc}. In a number of cases, the timing models' validity interval only partially overlapped with the time range of the LAT dataset.  To avoid contaminating off-pulse phase regions by extrapolating inaccurate models, we restricted the LAT datasets to the validity intervals in the cases of (1) timing solutions containing high-order ($\geq 3$) frequency derivatives suggestive of strong timing noise; (2) ``IFUNC'' or ``WAVE'' terms (also used to mitigate timing noise); or (3) ``ORBIFUNC'' or high-order ($\geq 2$) orbital frequency derivatives for pulsars in binary systems, or black widow pulsars that are often subject to timing instabilities. For other objects, generally stable MSPs, all LAT photons were used and we checked that the pulse phases do not drift outside the ephemeris validity intervals (see Figure \ref{SampleLC}). Additionally, we conservatively excluded 10 days of data before and after pulsar glitch epochs, to avoid contamination of off-pulse phase intervals caused by imperfect glitch modeling. At the top of all the plots like Figure~\ref{profiles:example_lightcurve_0030}, $t_{int}$ gives the number of years of LAT data retained.

Pulse profiles as in Figures \ref{SampleLC}  and \ref{profiles:example_lightcurve_0030} were generated by computing histograms of the rotational phases $\phi_j$, weighted by the photon weights $w_j$. The weighted phase histograms repeat over two pulsar rotations to clarify structures near phase 0. The uncertainty for the $i$th histogram bin containing $N_i$ photons was calculated as $\sigma_i^2 = \sum_{j=1}^{N_i} w_j^2 + \left( \max_{j \in \left[1, N_i\right]} w_j \right)^2$.  The first term is the sample variance, while the second term adds in quadrature the maximum possible variance for a bin with one photon.  This is an ad hoc correction for bins with few counts, where by chance the observed sample variance may be far below the typical value: this prescription provides more realistic error bars than that used in 2PC. The number of bins in the weighted histograms was chosen based on the $H$-test TS above 0.1 GeV: 25 bins for pulsars with $H < 100$, 50 bins when $100 \leq H < 1000$, 100 bins when $1000 \leq H < 10000$, and so on. 

Dashed horizontal lines in the figures show the background contribution from neighboring point sources and from diffuse emission. As in 2PC Section 5.1, it was estimated as $b = (\sum_{j=1}^{N} w_j - \sum_{j=1}^N w_j^2)/N_{\mathrm bin}$ where $N$ is the number of sources, excluding the pulsar, and $N_{\mathrm bin}$ is the number of histogram bins. The background uncertainty is dominated by systematic errors in the diffuse background normalization, estimated by changing the background normalization by $\pm 6$\%, also shown as dashed horizontal lines. A pulsar wind nebula (PWN) may surround the pulsar, but with a power-law spectrum extending to high energies. If omitted from the source model, it appears as an excess above background, especially at higher energies. Section 6.2 lists the pulsars for which we added a PWN component to our source model, in addition to the Crab and Vela. 2PC Section 7 describes a search for off-pulse emission, which measured off-pulse spectra for pulsars with an excess above the background level.

For gamma-ray pulsars with no radio emission, we rotated the pulse profiles such that the phase of the first emission component (as determined from the light curve fit above 0.1 GeV, described in the next Section) is at phase 0.1 (see Figure~\ref{App-Samples:example_lightcurve_0007} in the Appendix for an example). When radio profiles are available, a radio fiducial phase at infinite frequency was determined (see below) and placed at phase 0. The gamma-ray light curves were rotated accordingly, to preserve the relative radio/gamma-ray phase offset, including DM delays (see Section \ref{dist}).
We determined the uncertainty on the radio/gamma-ray phase offset stemming from the DM uncertainty neglecting DM time-derivative terms, and show it in the figures with an horizontal error bar at top-right. Figure~\ref{App-Samples:example_lightcurve_1311} in the Appendix shows a case where a large DM uncertainty leads to a large relative phase uncertainty. For some pulsars the DM and its uncertainty were not determined during the radio timing analysis; for these objects we used external DM information, generally from the ATNF pulsar catalog or separate analyses. Those cases are highlighted with a marker underneath the phasing error bar, as illustrated in Figure~\ref{App-Samples:example_lightcurve_2229}. In a very few other cases, the radio data analysis did not yield the information necessary to determine the relative radio/gamma-ray phase offset. For these light curves we indicated the arbitrary relative phase offset with a $\star$ symbol. 

%

Identifying the ``fiducial phase'' at which the magnetic pole of the neutron star crosses the line of sight is of interest.  For some pulsars, this can be inferred from radio polarimetry via the rotating vector model \citep{RVM}, but we do not consider such models here.  As noted in 2PC, for pulsars with a single, symmetric radio peak, the peak maximum can be chosen as an estimator for the fiducial phase. However, many radio profiles in our sample do not consist simply of single, symmetric peaks, so that the choice of the fiducial phase is not always trivial. We adopted the following ``Gaussian smoothing'' approach which reduces complex emission structures by replacing them with the best-fitting Gaussian equivalent.  Each radio profile was inspected visually and one peak or peak complex was chosen as the phase reference. This reference peak or set of peaks was then fitted with a single asymmetric Gaussian curve, or with two asymmetric Gaussian curves, depending on the morphology. Single peaks or ``patchy'' profiles with more than two components were fitted with single asymmetric Gaussian functions. Peaks with two components at their edges were fitted with two asymmetric Gaussians. In the first case, the fiducial phase was simply defined as the best-fit peak phase, and in the second case the fiducial phase was defined as the average of the two best-fit peak phase values.

While the above procedure for defining the radio fiducial phase works well in most cases, a number of radio profiles in the sample are too complex for the procedure to produce an estimate. We inspected the radio profiles and classified them into four categories:

\begin{enumerate}
    \item radio profiles for which one can readily define one peak as the main peak, which has a simple, single or double structure. The procedure for defining the radio fiducial phase should have produced robust results.
    \item radio profiles for which one can readily define one peak as the main peak, whose structure is complex (i.e., patchy profiles). The procedure is believed to be robust.
    \item one can define a radio peak as the main one, but cannot readily determine whether it is isolated, or part of a broader structure with other peaks. The radio fiducial phase estimate may not be robust for these objects.
    \item the profile morphology is too complex for the fiducial phase to be unambiguously chosen, based on the profile alone.
\end{enumerate}

The categories, or ``radio classes'', assigned to the different pulsars are listed in Tables~\ref{tbl-pulsePSR} and \ref{tbl-pulseMSP}. The fiducial phases were finally placed at phase 0 in the radio and gamma-ray pulse profiles, and the gamma-ray phases were rotated accordingly.

%

\subsection{Gamma-ray Light Curve Fitting Above 0.1 GeV}
\label{profiles:lcfits}

Fits of the LAT weighted histograms above 0.1 GeV are displayed in the Figures as magenta lines (see e.g. Figure~\ref{profiles:example_lightcurve_0030}). We followed a similar approach as in 2PC for fitting the LAT pulse profiles, by representing the gamma-ray light curves as wrapped probability density functions (pdfs) of $\phi \in \left[0, 1\right)$, and constructing the pdfs for the individual pulsars as linear combinations of $N_\mathrm{peaks}$ unimodal distributions: 

\begin{equation}
f \left( \phi \right) = \sum_{i=1}^{N_\mathrm{peaks}} n_i\ g_i \left(\phi \right) + \left( 1 - \sum_{i=1}^{N_\mathrm{peaks}} n_i \right).
\end{equation}

In the above expression, $n_i$ terms represent normalization factors, and the $g_i$ terms represent individually normalized distributions and the term in parentheses is a uniform distribution representing an unpulsed component. We chose to fit gamma-ray light curve peaks using asymmetric Gaussian or Lorentzian distributions, wrapped onto a circle: 

\begin{equation}
g \left( \phi \right) = \sum_{k=-\infty}^{+\infty} g^\prime \left(\phi + k \right),
\end{equation}
with $g^\prime$ defined on the real line and $k$ an integer number of rotations. In practice the sum needed to be truncated, and we summed phases over $\left[-N_\mathrm{rot},\ N_\mathrm{rot}+1\right]$ with $N_\mathrm{rot} = 2$. The pdfs were defined as follows: 

\begin{equation}
g^\prime \left( \phi \right) = \dfrac{2}{\pi \left(\sigma_1 + \sigma_2\right) \left(1 + z^2\right)}\ (\mathrm{Asymmetric\  Lorentzian}), 
\end{equation}

\begin{equation}
g^\prime \left( \phi \right) = \sqrt{ \dfrac{2}{\pi} } \dfrac{\exp(-z^2/2)}{\sigma_1+\sigma_2}\ (\mathrm{Asymmetric\ Gaussian}),
\end{equation}
with $z$ defined as $\left(\phi - \phi_0\right) / \sigma$, with $\phi_0$ the phase location of the peak and $\sigma = \sigma_1$ if $\phi < \phi_0$ and $\sigma = \sigma_2$ otherwise. Best-fit normalization, phase and width parameters for each component were determined using maximum likelihood. The logarithm of the likelihood is defined as: 

\begin{equation}
\log L = \sum_{j=1}^N \log \left[ w_j f\left( \phi_j \right) + \left(1 - w_j \right) \right].
\end{equation}

The fit was carried out using the MultiNest Bayesian inference tool \citep{Multinest}. We fitted binned weighted histograms above 0.1 GeV with twice as many bins as displayed in the figures. In some cases the width parameters converged towards the lower edge of the prior; in these cases we doubled the number of phase bins for the fit. For each gamma-ray component in the pulse profiles we tested the different pdfs and chose the one maximizing the log likelihood value. The number of gamma-ray components was determined by adding components at possible peak locations (based on visual inspection) and choosing the model that minimized the Bayesian information criterion (``BIC''). Pulse profile fits were carried out and plotted in the figures for the \npsrfitted{} pulsars with $H$-test values above 100.

In most pulsars with more than two gamma-ray components, one can readily distinguish two ``main'' gamma-ray peaks at phases $\phi_1$ and $\phi_2$ whose separation is denoted as $\Delta = \phi_2 - \phi_1$. In an archetypal gamma-ray pulse profile such as for the Vela pulsar (PSR~J0835$-$4510), a first main gamma-ray peak lags the radio emission peak by $\delta = \phi_1 - \phi_r \simeq 0.1 - 0.2$ (where $\phi_r$ denotes the radio fiducial phase, as discussed previously, if radio emission is also detected), and a second main gamma-ray peak lags the first one by $\lesssim 0.5$ with emission between the peaks. However, in some cases one cannot straightforwardly determine which of the gamma-ray peaks should be labelled as peak 1 and peak 2. Therefore, for each profile with more than one statistically significant emission component, we chose peak 1 and peak 2 based on similarity with the above-described archetype. Then, for each pulsar we considered the following criteria: \\

\begin{itemize}
    \item $\delta < 0.5$,
    \item $\Delta < 0.5$,
    \item $\sigma_2 > \sigma_1$ for the first peak (i.e., is the first peak oriented toward the second one),
    \item $\sigma_1 > \sigma_2$ for the second peak (i.e., is the second peak oriented toward the first one), 
    \item presence of statistically significant emission between the two main peaks.
\end{itemize}
and inspected those meeting fewer than four of the above criteria (i.e., pulse profiles that do not correspond to the ``typical'' Vela-like profile). In some of these cases, peak 1 and peak 2 were inverted based on visual inspection, and a number of objects were flagged as ``uncertain''. 

In addition to those cases where the ordering of the two main gamma-ray peaks is ambiguous, some other pulsars have ``complex'' pulse profiles in which the choice of the two main peaks is itself ambiguous, for various reasons.  In an attempt to list these objects we searched the best-fit pulse profiles for examples that contain at least one peak not overlapping with the two main peaks chosen from initial visual inspection, and whose maximum amplitude is at least half that of the smaller of the two main peaks. The handful of objects that matched this criterion were flagged as ``complex''.

For all these pulsars for which the choice of peak 1 and peak 2 is not obvious, the values for $\delta$ and $\Delta$ are also uncertain. 

\begin{figure}[!ht]
\centering
\includegraphics[width=1.0\textwidth]{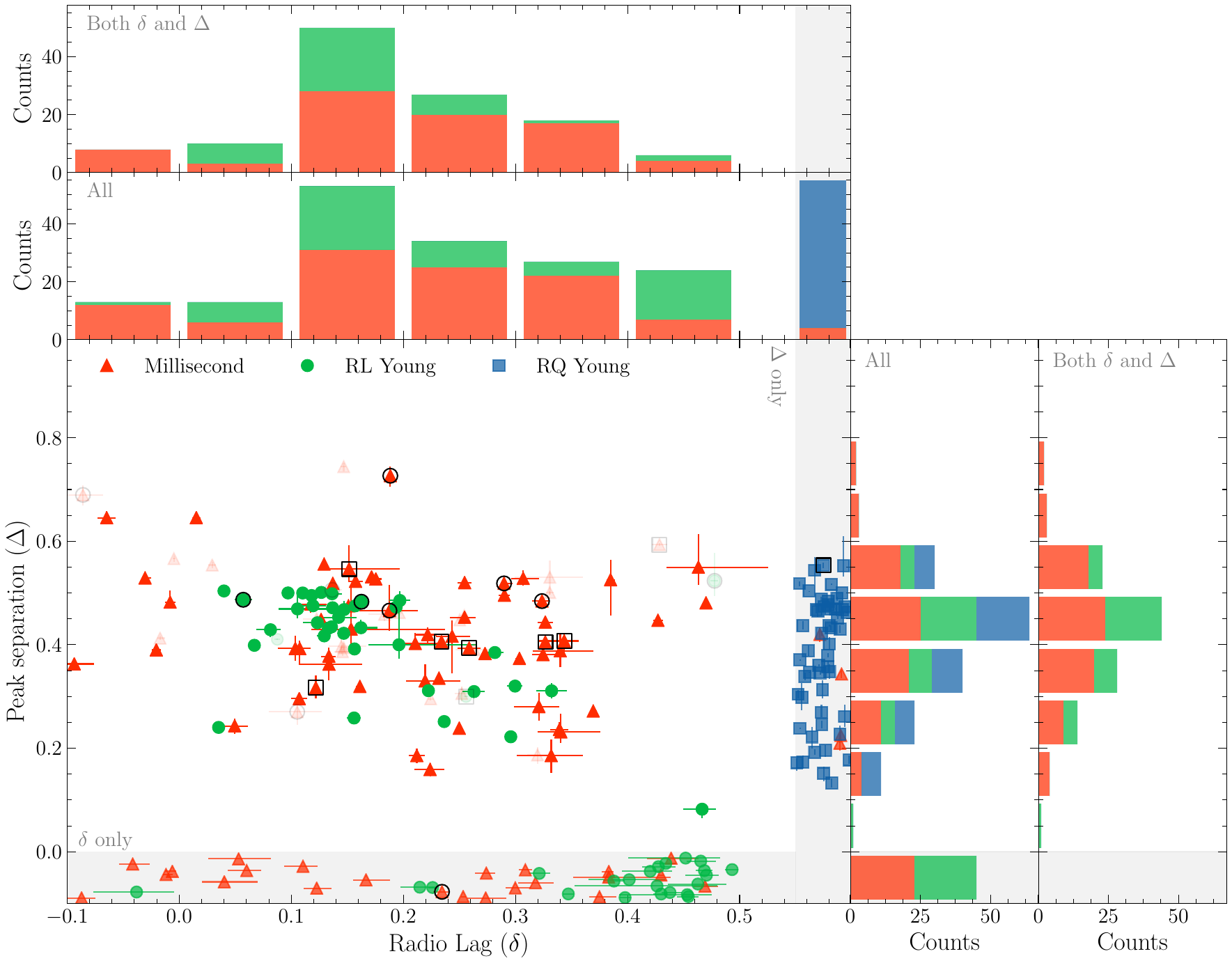}
\caption{Phase separation $\Delta$ between main gamma-ray peaks, versus phase lag $\delta$ of the first gamma-ray peak relative to the radio emission. Values of $\delta$ between 0.9 and 1 are plotted between $-0.1$ and 0 in this graph. Values between 0.5 and 0.9 have been ``folded'' as $\delta^\prime = 1 - \delta$. The markers are the same as in Figure~2. Circled symbols are pulsars for which the ordering of the first and second gamma-ray peaks was considered uncertain. Squares correspond to pulsars with ``complex'' profiles, in which choosing two main peaks is not straightforward. Opaque symbols are pulsars for which the definition of the radio fiducial point is considered robust, other pulsars are shown as transparent symbols (see text for details on the definition of radio fiducial points). Pulsars with $\delta$ and no $\Delta$ measurements (\textit{i.e}, single-peaked gamma-ray pulsars with radio emission) are shown in the gray-shaded region at the bottom, while those with $\Delta$ and no $\delta$ values (\textit{i.e.}, pulsars with at least two gamma-ray components and no radio emission) are plotted in the gray-shaded area at the right. The artificial staggering of the points in these two regions is to improve clarity. At the top and at the right are plotted histograms of the values of $\delta$ and $\Delta$ projected on the axes, for all pulsars shown in the central figure (the `All' panels), and restricting the sample of pulsars to those with $\delta$ and $\Delta$ measurements (the `Both $\delta$ and $\Delta$' panels, \textit{i.e.}, excluding pulsars in the gray-shaded areas).}
\label{profiles:delta_Delta}
\end{figure}

\begin{figure}[!ht]
\centering
\includegraphics[width=1.0\textwidth]{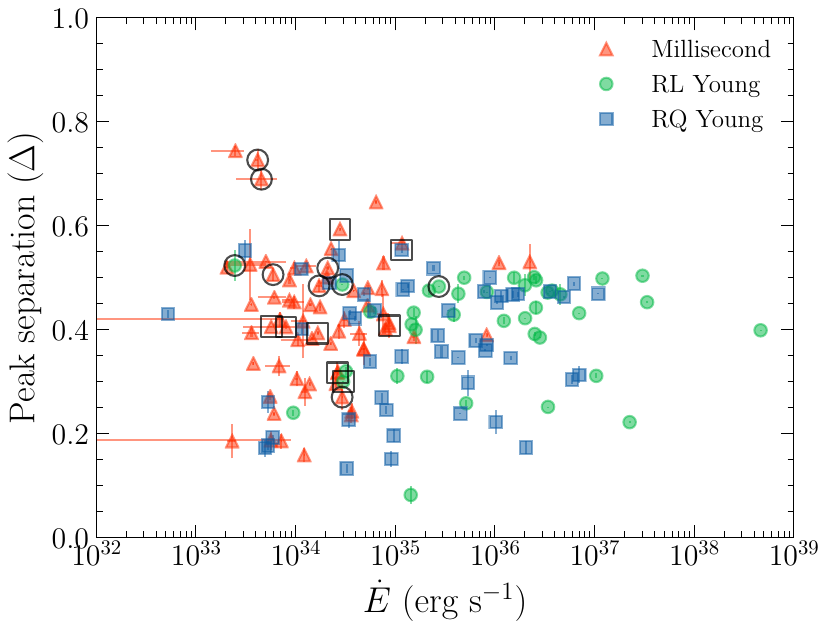}
\caption{Phase separation $\Delta$ between main gamma-ray peaks versus Shklovskii-corrected spin-down power $\edot$. The markers are the same as in Figure~2. See Figure~\ref{profiles:delta_Delta} for the definition of circle and square symbols.}
\label{profiles:Edot_Delta}
\end{figure}

Tables~\ref{tbl-pulsePSR} and \ref{tbl-pulseMSP} list the best-fit parameters. Figure~\ref{profiles:delta_Delta} shows a plot of the separation of the two main gamma-ray peaks, $\Delta$, versus the radio/gamma-ray phase separation, $\delta$. Figure~\ref{profiles:Edot_Delta} shows the $\Delta$ values plotted against the spin-down power values, $\edot$. For the pulsars with radio emission and at least two gamma-ray components (and therefore, with both $\delta$ and $\Delta$ values in Tables~\ref{tbl-pulsePSR} and \ref{tbl-pulseMSP} and in Figure~\ref{profiles:delta_Delta}), and excluding pulsars with profiles considered as complex or with uncertain peak ordering, we find a Spearman rank coefficient of $\sim -0.30$, indicating of a negative correlation between $\delta$ and $\Delta$ with a probability of chance correlation of $\sim 2 \times 10^{-3}$. The anti-correlation of $\delta$ and $\Delta$, already noted in e.g. 2PC, is a well-known prediction of outer magnetospheric pulsar emission models. No strong relation is seen in Figure \ref{profiles:Edot_Delta}.  These results and other trends visible in these figures are further discussed in Section~\ref{disc}. 

We compared $\delta$ and $\Delta$ in Tables~\ref{tbl-pulsePSR} and \ref{tbl-pulseMSP} with the 2PC values, for pulsars in both catalogs. Table~\ref{tbl-comp_2PC_3PC} lists pulsars for which $\delta$ and/or $\Delta$ differ by more than three times the statistical uncertainty, and with an absolute difference $>0.03$. For $\Delta$, no strong inconsistencies appeared: inverted gamma-ray profile fit components, or differing numbers of components or  component phases explain most discrepancies. Similarly, the $\delta$ parameters between 2PC and 3PC are qualitatively similar in most cases: differences typically arise from choices of radio pulse reference phase, or choices of the gamma-ray component used to find $\delta$. The only pulsars with significantly discrepant radio/gamma-ray alignment properties in 2PC and 3PC are PSRs~J2047+1053 and J2215+5135.

We further compared our radio/gamma-ray alignment results to those in the literature but not in 2PC, finding discrepant radio and gamma-ray alignments for PSRs J1341$-$6220, J1431$-$4715, J1646$-$4346, J1731$-$4744, J1816+4510, J1921+1929, J1935+2025, and J2039$-$3616. We stress that the $\delta$ values listed in Tables~\ref{tbl-pulsePSR} and \ref{tbl-pulseMSP}, and the relative radio/gamma-ray alignment results in general, should be taken with a grain of salt due to the numerous systematic effects. In the case of PSR~J0318+0253, whose $\delta$ parameter is not listed in Table~\ref{tbl-pulseMSP}, varying rotational phase references during the commissioning of the FAST telescope prevent us from determining the alignment properly. Different choices for the radio fiducial phase, different central frequencies or frequency bandwidths used for the radio observations, or ill-determined (or varying) DM values can also lead to inconsistent $\delta$ estimates between different analyses, to name but a few examples. Nevertheless, we have established that for almost all pulsars the relative radio and gamma-ray alignments are qualitatively consistent with published results.

%
\clearpage
\startlongtable

\begin{deluxetable}{lccccc}
\tabletypesize{\scriptsize}
\tablewidth{0pt}
\tablecaption{Pulse shape parameters of young LAT-detected pulsars\label{tbl-pulsePSR}}

\tablehead{
\colhead{PSR} & \colhead{Peaks} & \colhead{Radio/gamma phase lag, $\delta$} & \colhead{Peak separation, $\Delta$} & \colhead{Comment} & \colhead{Radio class}} \

\startdata
J0002+6216 & 2 & $0.162_{-0.009}^{+0.015}$ & $0.433_{-0.009}^{+0.015}$ &  & 1 \\
J0007+7303 & 3 & \dots & $0.2383 \pm 0.0014$ &  & \dots \\
J0106+4855 & 2 & $0.057 \pm 0.008$ & $0.4869_{-0.0012}^{+0.0010}$ & u & 1 \\
J0205+6449 & 2 & $0.0400 \pm 0.0024$ & $0.5039 \pm 0.0019$ &  & 1 \\
J0248+6021 & 1 & $0.347 \pm 0.005$ & \dots &  & 1 \\
J0357+3205 & 2 & \dots & $0.192 \pm 0.005$ &  & \dots \\
J0359+5414 & 1 & \dots & \dots &  & \dots \\
J0514$-$4408 & 2 & $0.4776 \pm 0.0034$ & $0.52_{-0.03}^{+0.05}$ & u & 3 \\
J0534+2200 & 2 & $0.067 \pm 0.004$ & $0.39889 \pm 0.00017$ &  & 2 \\
J0540$-$6919 & 1 & $0.96 \pm 0.04$ & \dots &  & 1 \\
J0554+3107 & 2 & \dots & $0.339 \pm 0.008$ &  & \dots \\
J0622+3749 & 2 & \dots & $0.544 \pm 0.027$ &  & \dots \\
J0631+0646 & 2 & $0.332 \pm 0.014$ & $0.311 \pm 0.015$ &  & 1 \\
J0631+1036 & 1 & $0.507 \pm 0.007$ & \dots &  & 2 \\
J0633+0632 & 2 & \dots & $0.4776 \pm 0.0009$ &  & \dots \\
J0633+1746 & 3 & \dots & $0.50398 \pm 0.00015$ &  & \dots \\
J0659+1414 & 1 & $0.226 \pm 0.004$ & \dots &  & 1 \\
J0729$-$1448 & 1 & $0.572 \pm 0.015$ & \dots &  & 2 \\
J0734$-$1559 & 1 & \dots & \dots &  & \dots \\
J0742$-$2822 & 2 & $0.533_{-0.017}^{+0.012}$ & $0.082_{-0.017}^{+0.012}$ &  & 2 \\
J0744$-$2525 & 3 & \dots & $0.468 \pm 0.004$ &  & \dots \\
J0802$-$5613 & 2 & \dots & $0.261 \pm 0.023$ &  & \dots \\
J0835$-$4510 & 4 & $0.133 \pm 0.004$ & $0.431865 \pm 0.000032$ &  & 1 \\
J0908$-$4913 & 2 & $0.0971 \pm 0.0034$ & $0.500 \pm 0.004$ &  & 2 \\
J0940$-$5428 & 1 & $0.454 \pm 0.010$ & \dots &  & 1 \\
J1016$-$5857 & 3 & $0.123_{-0.013}^{+0.008}$ & $0.442_{-0.013}^{+0.009}$ &  & 1 \\
J1019$-$5749 & 1 & $0.566 \pm 0.010$ & \dots &  & 1 \\
J1023$-$5746 & 3 & \dots & $0.4692 \pm 0.0005$ &  & \dots \\
J1028$-$5819 & 2 & $0.1935 \pm 0.0010$ & $0.4727 \pm 0.0004$ &  & 1 \\
J1044$-$5737 & 3 & \dots & $0.3604 \pm 0.0018$ &  & \dots \\
J1048$-$5832 & 3 & $0.1468 \pm 0.0012$ & $0.4221 \pm 0.0009$ &  & 2 \\
J1055$-$6028 & 1 & $0.530_{-0.019}^{+0.011}$ & \dots &  & 1 \\
J1057$-$5226 & 4 & $0.7441 \pm 0.0017$ & $0.3000 \pm 0.0010$ & c & 4 \\
J1057$-$5851 & 1 & \dots & \dots &  & \dots \\
J1105$-$6037 & 2 & \dots & $0.348 \pm 0.014$ &  & \dots \\
J1105$-$6107 & 2 & $0.1268_{-0.0035}^{+0.0027}$ & $0.501 \pm 0.006$ &  & 1 \\
J1111$-$6039 & 2 & \dots & $0.313 \pm 0.018$ &  & \dots \\
J1112$-$6103 & 2 & $0.147 \pm 0.018$ & $0.468 \pm 0.020$ &  & 1 \\
J1119$-$6127 & 1 & $0.469 \pm 0.008$ & \dots &  & 1 \\
J1124$-$5916 & 2 & $0.136 \pm 0.009$ & $0.4985 \pm 0.0011$ &  & 1 \\
J1135$-$6055 & 2 & \dots & $0.173 \pm 0.013$ &  & \dots \\
J1139$-$6247 & 2 & \dots & $0.151 \pm 0.016$ &  & \dots \\
J1151$-$6108 & 2 & $0.081_{-0.013}^{+0.009}$ & $0.429_{-0.014}^{+0.011}$ &  & 2 \\
J1203$-$6242 & 3 & \dots & $0.468 \pm 0.005$ &  & \dots \\
J1208$-$6238 & 2 & \dots & $0.468 \pm 0.007$ &  & \dots \\
J1224$-$6407 & 1 & $0.215 \pm 0.018$ & \dots &  & 1 \\
J1231$-$5113 & 2 & \dots & $0.431 \pm 0.008$ &  & \dots \\
J1231$-$6511 & 1 & \dots & \dots &  & \dots \\
J1350$-$6225 & 3 & \dots & $0.485 \pm 0.005$ &  & \dots \\
J1357$-$6429 & 1 & $0.388 \pm 0.009$ & \dots &  & 1 \\
J1358$-$6025 & 2 & \dots & $0.298 \pm 0.028$ &  & \dots \\
J1410$-$6132 & 1 & $0.537 \pm 0.027$ & \dots &  & 1 \\
J1413$-$6205 & 3 & \dots & $0.3707 \pm 0.0014$ &  & \dots \\
J1418$-$6058 & 3 & \dots & $0.4626 \pm 0.0016$ &  & \dots \\
J1420$-$6048 & 2 & $0.222_{-0.004}^{+0.003}$ & $0.311 \pm 0.005$ &  & 1 \\
J1422$-$6138 & 2 & \dots & $0.196 \pm 0.012$ &  & \dots \\
J1429$-$5911 & 3 & \dots & $0.4738 \pm 0.0030$ &  & \dots \\
J1447$-$5757 & 1 & \dots & \dots &  & \dots \\
J1459$-$6053 & 1 & \dots & \dots &  & \dots \\
J1509$-$5850 & 2 & $0.156 \pm 0.006$ & $0.258 \pm 0.006$ &  & 1 \\
J1513$-$5908 & 1 & $0.321 \pm 0.010$ & \dots &  & 1 \\
J1522$-$5735 & 2 & \dots & $0.465 \pm 0.011$ &  & \dots \\
J1528$-$5838 & 1 & \dots & \dots &  & \dots \\
J1531$-$5610 & 1 & $0.40 \pm 0.04$ & \dots &  & 1 \\
J1615$-$5137 & 2 & \dots & $0.270_{-0.010}^{+0.014}$ &  & \dots \\
J1620$-$4927 & 2 & \dots & $0.246 \pm 0.007$ &  & \dots \\
J1623$-$5005 & 2 & \dots & $0.389 \pm 0.010$ &  & \dots \\
J1624$-$4041 & 2 & \dots & $0.422 \pm 0.005$ &  & \dots \\
J1641$-$5317 & 1 & \dots & \dots &  & \dots \\
J1646$-$4346 & 1 & $0.535 \pm 0.014$ & \dots &  & 1 \\
J1648$-$4611 & 2 & $0.263 \pm 0.010$ & $0.310 \pm 0.011$ &  & 1 \\
J1650$-$4601 & 2 & \dots & $0.358 \pm 0.011$ &  & \dots \\
J1702$-$4128 & 1 & $0.45_{-0.05}^{+0.03}$ & \dots &  & 1 \\
J1705$-$1906 & 1 & $0.562_{-0.007}^{+0.010}$ & \dots &  & 2 \\
J1709$-$4429 & 3 & $0.2364 \pm 0.0004$ & $0.2515 \pm 0.0004$ &  & 1 \\
J1714$-$3830 & 1 & \dots & \dots &  & \dots \\
J1718$-$3825 & 1 & $0.398 \pm 0.005$ & \dots &  & 2 \\
J1730$-$3350 & 2 & $0.129_{-0.006}^{+0.003}$ & $0.417_{-0.007}^{+0.005}$ &  & 1 \\
J1732$-$3131 & 3 & $0.0876 \pm 0.0026$ & $0.4104_{-0.0007}^{+0.0010}$ &  & 4 \\
J1736$-$3422 & 2 & \dots & $0.437 \pm 0.009$ &  & \dots \\
J1741$-$2054 & 2 & $0.035 \pm 0.005$ & $0.240_{-0.006}^{+0.003}$ &  & 2 \\
J1742$-$3321 & 1 & \dots & \dots &  & \dots \\
J1746$-$3239 & 2 & \dots & $0.133 \pm 0.009$ &  & \dots \\
J1747$-$2958 & 3 & $0.156 \pm 0.005$ & $0.392 \pm 0.004$ &  & 1 \\
J1748$-$2815 & 1 & \dots & \dots &  & \dots \\
J1801$-$2451 & 2 & $0.118 \pm 0.007$ & $0.495 \pm 0.007$ &  & 1 \\
J1803$-$2149 & 3 & \dots & $0.379 \pm 0.005$ &  & \dots \\
J1809$-$2332 & 3 & \dots & $0.3465 \pm 0.0007$ &  & \dots \\
J1813$-$1246 & 3 & \dots & $0.489 \pm 0.004$ &  & \dots \\
J1817$-$1742 & 2 & \dots & $0.518 \pm 0.005$ &  & \dots \\
J1826$-$1256 & 2 & \dots & $0.4737 \pm 0.0004$ &  & \dots \\
J1827$-$1446 & 1 & \dots & \dots &  & \dots \\
J1828$-$1101 & 2 & $0.110 \pm 0.005$ & $0.500_{-0.010}^{+0.008}$ &  & 1 \\
J1833$-$1034 & 2 & $0.142 \pm 0.016$ & $0.4530 \pm 0.0023$ &  & 1 \\
J1835$-$1106 & 1 & $0.55_{-0.06}^{+0.02}$ & \dots &  & 1 \\
J1836+5925 & 3 & \dots & $0.5167 \pm 0.0016$ &  & \dots \\
J1837$-$0604 & 2 & $0.197 \pm 0.009$ & $0.486 \pm 0.021$ &  & 1 \\
J1838$-$0537 & 2 & \dots & $0.304 \pm 0.009$ &  & \dots \\
J1844$-$0346 & 1 & \dots & \dots &  & \dots \\
J1846+0919 & 2 & \dots & $0.227 \pm 0.017$ &  & \dots \\
J1856+0113 & 2 & $0.106_{-0.017}^{+0.027}$ & $0.469_{-0.018}^{+0.028}$ &  & 1 \\
J1857+0143 & 1 & $0.43_{-0.07}^{+0.05}$ & \dots &  & 1 \\
J1906+0722 & 2 & \dots & $0.222 \pm 0.023$ &  & \dots \\
J1907+0602 & 3 & $0.282 \pm 0.008$ & $0.3850 \pm 0.0013$ &  & 1 \\
J1913+0904 & 2 & $0.20_{-0.03}^{+0.07}$ & $0.40_{-0.03}^{+0.07}$ &  & 1 \\
J1932+1916 & 1 & \dots & \dots &  & \dots \\
J1952+3252 & 3 & $0.1572 \pm 0.0007$ & $0.4752 \pm 0.0006$ &  & 1 \\
J1954+2836 & 3 & \dots & $0.4521 \pm 0.0025$ &  & \dots \\
J1957+5033 & 2 & \dots & $0.178 \pm 0.009$ &  & \dots \\
J1958+2846 & 3 & \dots & $0.4371 \pm 0.0016$ &  & \dots \\
J2006+3102 & 1 & $0.570 \pm 0.007$ & \dots &  & 1 \\
J2017+3625 & 2 & \dots & $0.401_{-0.010}^{+0.014}$ &  & \dots \\
J2021+3651 & 3 & $0.1369 \pm 0.0019$ & $0.47147 \pm 0.00011$ &  & 1 \\
J2021+4026 & 3 & \dots & $0.553 \pm 0.005$ & c & \dots \\
J2022+3842 & 1 & $0.420_{-0.025}^{+0.020}$ & \dots &  & 4 \\
J2028+3332 & 3 & \dots & $0.432 \pm 0.005$ &  & \dots \\
J2030+3641 & 2 & $0.300 \pm 0.007$ & $0.320 \pm 0.008$ &  & 1 \\
J2030+4415 & 3 & \dots & $0.490 \pm 0.004$ &  & \dots \\
J2032+4127 & 2 & $0.1626 \pm 0.0009$ & $0.4831 \pm 0.0008$ & u & 1 \\
J2043+2740 & 2 & $0.136_{-0.008}^{+0.005}$ & $0.435_{-0.008}^{+0.006}$ &  & 1 \\
J2055+2539 & 2 & \dots & $0.172_{-0.016}^{+0.009}$ &  & \dots \\
J2111+4606 & 3 & \dots & $0.3459 \pm 0.0034$ &  & \dots \\
J2139+4716 & 2 & \dots & $0.55_{-0.02}^{+0.06}$ &  & \dots \\
J2229+6114 & 2 & $0.2959_{-0.0021}^{+0.0016}$ & $0.2221_{-0.0018}^{+0.0012}$ &  & 1 \\
J2238+5903 & 2 & \dots & $0.5002 \pm 0.0005$ &  & \dots \\
J2240+5832 & 2 & $0.119_{-0.009}^{+0.012}$ & $0.475_{-0.009}^{+0.012}$ &  & 1 \\
\enddata

\tablecomments{
Column 2 gives the gamma-ray peak multiplicity.  
Columns 3 and 4 give the radio/gamma phase lag $\delta$ and separation $\Delta$ between the two main gamma-ray peaks. 
In column 5, 'u' indicates uncertain ordering of the first and second gamma-ray peaks, and 'c' means the profile is complex.
For pulsars with radio emission, column 6 lists flags quantifying the robustness of the fiducial phase definition (see text for details).}
\end{deluxetable}

\clearpage
\startlongtable

\begin{deluxetable}{lccccc}
\tabletypesize{\scriptsize}
\tablewidth{0pt}
\tablecaption{Pulse shape parameters of LAT-detected MSPs\label{tbl-pulseMSP}}

\tablehead{
\colhead{PSR} & \colhead{Peaks} & \colhead{Radio/gamma phase lag, $\delta$} & \colhead{Peak separation, $\Delta$} & \colhead{Comment} & \colhead{Radio class}} \

\startdata
J0023+0923 & 2 & $0.321 \pm 0.022$ & $0.280 \pm 0.027$ &  & 1 \\
J0030+0451 & 5 & $0.1266 \pm 0.0014$ & $0.4485 \pm 0.0004$ &  & 2 \\
J0034$-$0534 & 2 & $0.839 \pm 0.004$ & $0.319 \pm 0.006$ &  & 1 \\
J0101$-$6422 & 4 & $0.6733 \pm 0.0032$ & $0.4053 \pm 0.0031$ & c & 1 \\
J0102+4839 & 2 & $0.327 \pm 0.007$ & $0.444 \pm 0.007$ &  & 1 \\
J0218+4232 & 1 & $0.691 \pm 0.006$ & \dots &  & 1 \\
J0251+2606 & 2 & $0.157 \pm 0.007$ & $0.522 \pm 0.007$ &  & 1 \\
J0307+7443 & 2 & $0.935 \pm 0.008$ & $0.645 \pm 0.013$ &  & 1 \\
J0312$-$0921 & 2 & $0.255_{-0.013}^{+0.010}$ & $0.453_{-0.012}^{+0.008}$ &  & 1 \\
J0340+4130 & 2 & $0.7501 \pm 0.0027$ & $0.239 \pm 0.004$ &  & 1 \\
J0418+6635 & 3 & $0.737 \pm 0.017$ & $0.391 \pm 0.005$ &  & 3 \\
J0437$-$4715 & 1 & $0.4299 \pm 0.0023$ & \dots &  & 2 \\
J0533+6759 & 2 & $0.175 \pm 0.005$ & $0.5270_{-0.0028}^{+0.0018}$ &  & 1 \\
J0605+3757 & 2 & $0.812 \pm 0.027$ & $0.47_{-0.04}^{+0.05}$ & u & 2 \\
J0610$-$2100 & 3 & $0.615 \pm 0.004$ & $0.53_{-0.07}^{+0.04}$ &  & 1 \\
J0613$-$0200 & 2 & $0.224 \pm 0.014$ & $0.159 \pm 0.014$ &  & 2 \\
J0614$-$3329 & 2 & $0.1293 \pm 0.0005$ & $0.5558 \pm 0.0004$ &  & 2 \\
J0621+2514 & 2 & $0.340 \pm 0.030$ & $0.388 \pm 0.032$ &  & 2 \\
J0737$-$3039A & 2 & $0.137_{-0.011}^{+0.009}$ & $0.506 \pm 0.012$ & u & 3 \\
J0740+6620 & 2 & $0.1964_{-0.0020}^{+0.0028}$ & $0.4622 \pm 0.0029$ &  & 3 \\
J0751+1807 & 2 & $0.3694 \pm 0.0033$ & $0.272 \pm 0.007$ &  & 1 \\
J0931$-$1902 & 1 & $0.040_{-0.020}^{+0.030}$ & \dots &  & 4 \\
J0952$-$0607 & 1 & $0.053 \pm 0.029$ & \dots &  & 1 \\
J0955$-$6150 & 2 & $0.883 \pm 0.014$ & $0.479 \pm 0.016$ &  & 1 \\
J1012$-$4235 & 2 & $0.290 \pm 0.005$ & $0.496 \pm 0.008$ &  & 1 \\
J1024$-$0719 & 1 & $0.561 \pm 0.024$ & \dots &  & 2 \\
J1035$-$6720 & 2 & $0.307_{-0.011}^{+0.014}$ & $0.528 \pm 0.016$ &  & 1 \\
J1036$-$8317 & 1 & $0.273 \pm 0.019$ & \dots &  & 1 \\
J1048+2339 & 2 & $0.1371 \pm 0.0032$ & $0.518_{-0.011}^{+0.005}$ &  & 1 \\
J1124$-$3653 & 3 & $0.259 \pm 0.010$ & $0.394 \pm 0.005$ & c & 1 \\
J1125$-$5825 & 2 & $0.211_{-0.011}^{+0.021}$ & $0.402_{-0.010}^{+0.020}$ &  & 2 \\
J1142+0119 & 1 & $0.994_{-0.006}^{+0.003}$ & \dots &  & 1 \\
J1207$-$5050 & 2 & $0.255 \pm 0.006$ & $0.520_{-0.013}^{+0.006}$ &  & 1 \\
J1221$-$0633 & 1 & $0.625_{-0.009}^{+0.015}$ & \dots &  & 3 \\
J1227$-$4853 & 3 & $0.656_{-0.022}^{+0.013}$ & $0.407_{-0.024}^{+0.015}$ & c & 1 \\
J1231$-$1411 & 4 & $0.2341 \pm 0.0011$ & $0.4056 \pm 0.0010$ & c & 2 \\
J1301+0833 & 2 & $0.104 \pm 0.016$ & $0.393 \pm 0.024$ &  & 1 \\
J1302$-$3258 & 2 & $0.147 \pm 0.004$ & $0.394 \pm 0.008$ &  & 3 \\
J1311$-$3430 & 3 & $0.906 \pm 0.018$ & $0.3629 \pm 0.0017$ &  & 1 \\
J1312+0051 & 2 & $0.914 \pm 0.019$ & $0.690 \pm 0.020$ & u & 3 \\
J1335$-$5656 & 2 & \dots & $0.209 \pm 0.016$ &  & \dots \\
J1400$-$1431 & 1 & $0.913 \pm 0.013$ & \dots &  & 1 \\
J1446$-$4701 & 2 & $0.339 \pm 0.008$ & $0.236 \pm 0.015$ &  & 1 \\
J1455$-$3330 & 1 & $0.167_{-0.011}^{+0.021}$ & \dots &  & 2 \\
J1513$-$2550 & 2 & $0.983_{-0.002}^{+0.005}$ & $0.413_{-0.003}^{+0.005}$ &  & 3 \\
J1514$-$4946 & 3 & $0.273 \pm 0.004$ & $0.3829_{-0.0022}^{+0.0013}$ &  & 1 \\
J1536$-$4948 & 6 & $0.429 \pm 0.007$ & $0.5934 \pm 0.0030$ & c & 3 \\
J1543$-$5149 & 1 & $0.959 \pm 0.015$ & \dots &  & 1 \\
J1552+5437 & 2 & $0.46_{-0.03}^{+0.06}$ & $0.55_{-0.04}^{+0.06}$ &  & 1 \\
J1555$-$2908 & 2 & $0.029 \pm 0.004$ & $0.554 \pm 0.007$ &  & 3 \\
J1600$-$3053 & 2 & $0.212 \pm 0.007$ & $0.186 \pm 0.015$ &  & 1 \\
J1614$-$2230 & 3 & $0.1716_{-0.0013}^{+0.0017}$ & $0.5312_{-0.0014}^{+0.0017}$ &  & 1 \\
J1625$-$0021 & 2 & $0.992 \pm 0.005$ & $0.482_{-0.012}^{+0.022}$ &  & 1 \\
J1628$-$3205 & 2 & $0.244 \pm 0.016$ & $0.42_{-0.07}^{+0.03}$ &  & 1 \\
J1630+3734 & 2 & $0.2544 \pm 0.0013$ & $0.4567 \pm 0.0023$ &  & 3 \\
J1640+2224 & 2 & $0.332 \pm 0.031$ & $0.185 \pm 0.033$ &  & 2 \\
J1641+8049 & 2 & $0.108_{-0.005}^{+0.003}$ & $0.392_{-0.029}^{+0.015}$ &  & 1 \\
J1649$-$3012 & 1 & \dots & \dots &  & \dots \\
J1653$-$0158 & 2 & \dots & $0.420 \pm 0.014$ &  & \dots \\
J1658$-$5324 & 2 & $0.105 \pm 0.026$ & $0.270 \pm 0.025$ & u & 4 \\
J1713+0747 & 1 & $0.318 \pm 0.016$ & \dots &  & 2 \\
J1730$-$2304 & 1 & $0.300_{-0.009}^{+0.014}$ & \dots &  & 2 \\
J1732$-$5049 & 1 & $0.617_{-0.006}^{+0.004}$ & \dots &  & 2 \\
J1741+1351 & 2 & $0.710_{-0.003}^{+0.007}$ & $0.519_{-0.003}^{+0.007}$ & u & 1 \\
J1744$-$1134 & 2 & $0.8116 \pm 0.0029$ & $0.727 \pm 0.022$ & u & 1 \\
J1744$-$7619 & 2 & \dots & $0.343 \pm 0.008$ &  & \dots \\
J1745+1017 & 1 & $0.988_{-0.003}^{+0.006}$ & \dots &  & 1 \\
J1747$-$4036 & 1 & $0.877 \pm 0.013$ & \dots &  & 1 \\
J1805+0615 & 2 & $0.85_{-0.04}^{+0.08}$ & $0.43_{-0.04}^{+0.08}$ &  & 1 \\
J1810+1744 & 2 & $0.050 \pm 0.013$ & $0.243 \pm 0.015$ &  & 2 \\
J1811$-$2405 & 2 & $0.145_{-0.012}^{+0.022}$ & $0.397_{-0.014}^{+0.023}$ &  & 3 \\
J1816+4510 & 3 & $0.5300 \pm 0.0029$ & $0.481 \pm 0.005$ &  & 1 \\
J1823$-$3021A & 2 & $0.980 \pm 0.004$ & $0.390 \pm 0.005$ &  & 1 \\
J1824$-$2452A & 2 & $0.669 \pm 0.031$ & $0.530 \pm 0.034$ &  & 3 \\
J1827$-$0849 & 1 & \dots & \dots &  & \dots \\
J1832$-$0836 & 1 & $0.7466_{-0.0016}^{+0.0023}$ & \dots &  & 4 \\
J1843$-$1113 & 1 & $0.060_{-0.017}^{+0.013}$ & \dots &  & 1 \\
J1855$-$1436 & 1 & $0.890_{-0.017}^{+0.013}$ & \dots &  & 1 \\
J1858$-$2216 & 2 & $0.748 \pm 0.006$ & $0.306 \pm 0.014$ &  & 3 \\
J1901$-$0125 & 3 & $0.866 \pm 0.007$ & $0.378 \pm 0.008$ &  & 1 \\
J1902$-$5105 & 3 & $0.015 \pm 0.004$ & $0.645 \pm 0.004$ &  & 1 \\
J1903$-$7051 & 1 & $0.531 \pm 0.012$ & \dots &  & 1 \\
J1908+2105 & 2 & $0.778_{-0.012}^{+0.008}$ & $0.419 \pm 0.016$ &  & 1 \\
J1921+0137 & 2 & $0.134 \pm 0.030$ & $0.362 \pm 0.032$ &  & 1 \\
J1939+2134 & 2 & $0.970_{-0.001}^{+0.005}$ & $0.529_{-0.006}^{+0.009}$ &  & 2 \\
J1946$-$5403 & 3 & $0.15_{-0.02}^{+0.05}$ & $0.55_{-0.02}^{+0.05}$ & c & 1 \\
J1959+2048 & 2 & $0.9953 \pm 0.0014$ & $0.566 \pm 0.007$ &  & 4 \\
J2006+0148 & 2 & $0.659_{-0.021}^{+0.035}$ & $0.231_{-0.021}^{+0.035}$ &  & 1 \\
J2017+0603 & 2 & $0.2244 \pm 0.0027$ & $0.2955 \pm 0.0035$ &  & 3 \\
J2017$-$1614 & 2 & $0.183 \pm 0.014$ & $0.459_{-0.013}^{+0.017}$ &  & 3 \\
J2034+3632 & 3 & \dots & $0.225_{-0.011}^{+0.007}$ &  & \dots \\
J2039$-$3616 & 1 & $0.274_{-0.010}^{+0.008}$ & \dots &  & 2 \\
J2039$-$5617 & 3 & $0.107 \pm 0.007$ & $0.296 \pm 0.009$ &  & 1 \\
J2042+0246 & 1 & $0.766 \pm 0.008$ & \dots & u & 1 \\
J2043+1711 & 3 & $0.2503 \pm 0.0028$ & $0.4479 \pm 0.0027$ &  & 3 \\
J2047+1053 & 2 & $0.219_{-0.017}^{+0.032}$ & $0.330_{-0.020}^{+0.033}$ &  & 1 \\
J2051$-$0827 & 1 & $0.38_{-0.03}^{+0.06}$ & \dots &  & 2 \\
J2052+1219 & 2 & $0.675 \pm 0.011$ & $0.381 \pm 0.012$ &  & 1 \\
J2115+5448 & 2 & $0.146_{-0.007}^{+0.005}$ & $0.386_{-0.008}^{+0.006}$ &  & 3 \\
J2124$-$3358 & 3 & $0.8532 \pm 0.0026$ & $0.744 \pm 0.009$ &  & 3 \\
J2129$-$0429 & 3 & $0.6694 \pm 0.0032$ & $0.501_{-0.004}^{+0.003}$ &  & 3 \\
J2214+3000 & 2 & $0.324_{-0.009}^{+0.007}$ & $0.484 \pm 0.012$ & u & 1 \\
J2215+5135 & 3 & $0.573 \pm 0.004$ & $0.447 \pm 0.004$ &  & 1 \\
J2234+0944 & 2 & $0.680 \pm 0.010$ & $0.187 \pm 0.018$ &  & 3 \\
J2241$-$5236 & 5 & $0.1219_{-0.0015}^{+0.0030}$ & $0.317 \pm 0.024$ & c & 1 \\
J2256$-$1024 & 2 & $0.1510 \pm 0.0010$ & $0.4750_{-0.0010}^{+0.0015}$ &  & 2 \\
J2302+4442 & 3 & $0.2318 \pm 0.0020$ & $0.3350 \pm 0.0021$ &  & 2 \\
J2310$-$0555 & 2 & $0.877_{-0.012}^{+0.007}$ & $0.450_{-0.013}^{+0.008}$ &  & 3 \\
J2339$-$0533 & 3 & $0.3036 \pm 0.0034$ & $0.3734 \pm 0.0027$ &  & 2 \\
\enddata

\tablecomments{Same as Table~\ref{tbl-pulsePSR}, for the gamma-ray MSPs.}
\end{deluxetable}


%


\begin{deluxetable}{lccc}
\tabletypesize{\scriptsize}
\tablewidth{0pt}
\tablecaption{Comparisons of radio/gamma phase lags $\delta$ and peak separations $\Delta$ between 2PC and 3PC, for pulsars with apparently discrepant results.\label{tbl-comp_2PC_3PC}}

\tablehead{\colhead{PSR} & \colhead{2PC value} & \colhead{3PC value} & \colhead{Comment}}

\startdata
\multicolumn{4}{c}{Radio/gamma phase lags, $\delta$} \\
\hline
J0030+0451 & $0.1600\pm0.0010$ & $0.1266\pm0.0014$ & Different choice of radio reference phase \\
J0101$-$6422 & $0.145\pm0.005$ & $0.6733\pm0.0032$ & Different choice of gamma-ray reference peak \\
J0102+4839 & $0.259\pm0.004$ & $0.327\pm0.007$ & Qualitatively similar results \\
J0205+6449 & $0.075\pm0.004$ & $0.0400\pm0.0024$ & Qualitatively similar results \\
J0218+4232 & $0.35\pm0.08$ & $0.691\pm0.006$ & Different choice of gamma-ray reference peak \\
J0534+2200 & $0.10900\pm0.00020$ & $0.067\pm0.004$ & Different choice of radio reference phase \\
J0610$-$2100 & $0.236\pm0.006$ & $0.615\pm0.004$ & Different choice of radio reference phase \\
J0742$-$2822 & $0.627\pm0.005$ & $0.533_{-0.017}^{+0.012}$ & Different choice of gamma-ray reference peak \\
J1019$-$5749 & $0.482\pm0.010$ & $0.566\pm0.010$ & Qualitatively similar results \\
J1057$-$5226 & $0.3040\pm0.0029$ & $0.7441\pm0.0017$ & Different choice of radio reference phase \\
J1119$-$6127 & $0.285\pm0.015$ & $0.469\pm0.008$ & Different choice of gamma-ray reference peak \\
J1125$-$5825 & $0.6450\pm0.0017$ & $0.211_{-0.011}^{+0.021}$ & Different choice of gamma-ray reference peak \\
J1410$-$6132 & $0.959\pm0.023$ & $0.537\pm0.027$ & Different choice of gamma-ray reference peak \\
J1509$-$5850 & $0.271\pm0.011$ & $0.156\pm0.006$ & Qualitatively similar results \\
J1514$-$4946 & $0.214\pm0.009$ & $0.273\pm0.004$ & Different choice of radio reference phase \\
J1600$-$3053 & $0.147\pm0.011$ & $0.212\pm0.007$ & Qualitatively similar results \\
J1658$-$5324 & $0.359\pm0.014$ & $0.105\pm0.026$ & Different choice of gamma-ray reference peak \\
J1741$-$2054 & $0.074\pm0.006$ & $0.035\pm0.005$ & Qualitatively similar results \\
J1744$-$1134 & $0.189\pm0.007$ & $0.8116\pm0.0029$ & Different choice of radio reference phase \\
J1747$-$4036 & $0.031\pm0.021$ & $0.877\pm0.013$ & Qualitatively similar results \\
J1801$-$2451 & $0.060\pm0.005$ & $0.118\pm0.007$ & Qualitatively similar results \\
J1810+1744 & $0.894\pm0.018$ & $0.050\pm0.013$ & Qualitatively similar results \\
J1835$-$1106 & $0.139\pm0.006$ & $0.55_{-0.06}^{+0.02}$ & Different choice of gamma-ray reference peak \\
J1907+0602 & $0.2090\pm0.0026$ & $0.282\pm0.008$ & Different choice of radio reference phase \\
J2032+4127 & $0.0990\pm0.0007$ & $0.1626\pm0.0009$ & Qualitatively similar results \\
J2047+1053 & $0.567\pm0.010$ & $0.219_{-0.017}^{+0.032}$ & Inconsistent results \\
J2215+5135 & $0.257\pm0.004$ & $0.573\pm0.004$ & Inconsistent results \\
J2229+6114 & $0.187\pm0.007$ & $0.2959_{-0.0021}^{+0.0016}$ & Qualitatively similar results \\
\hline
\multicolumn{4}{c}{Peak separations, $\Delta$} \\
\hline
J0218+4232 & $0.39\pm0.08$ & \dots & Only one component in 3PC \\
J1119$-$6127 & $0.204\pm0.020$ & \dots & Only one component in 3PC \\
J1124$-$3653 & $0.21\pm0.04$ & $0.394\pm0.005$ & Different component phases \\
J1410$-$6132 & $0.46\pm0.04$ & \dots & Only one component in 3PC \\
J1459$-$6053 & $0.063\pm0.034$ & \dots & Only one component in 3PC \\
J1747$-$4036 & $0.681\pm0.033$ & \dots & Only one component in 3PC \\
J1823$-$3021A & $0.627\pm0.010$ & $0.390\pm0.005$ & Components inverted \\
J1835$-$1106 & $0.421\pm0.011$ & \dots & Only one component in 3PC \\
J2021+4026 & $0.687\pm0.009$ & $0.553\pm0.005$ & Different component phases \\
J2032+4127 & $0.5160\pm0.0010$ & $0.4831\pm0.0008$ & Components inverted \\
J2229+6114 & $0.299\pm0.008$ & $0.2221_{-0.0018}^{+0.0012}$ & Qualitatively similar results \\
J2241$-$5236 & $0.638\pm0.031$ & $0.317\pm0.024$ & Different component phases \\
\enddata

\tablecomments{The columns list the pulsar names, 2PC and 3PC values, and a comment on the apparent discrepancy. Pulsars listed here are those for which $\delta$ and/or $\Delta$ differ between 2PC and 3PC by more than three times the statistical uncertainty, but an absolute difference $>0.03$.}

\end{deluxetable}


\subsection{Energy-resolved Light Curve Fitting }
\label{sec:profs_energy}

The best-fit profiles above 0.1 GeV were then used to fit gamma-ray pulse profiles in narrower energy ranges. To this end, we divided the LAT datasets into logarithmically-spaced energy intervals, and fitted the pulsar pulse profiles in each energy interval independently. The number of energy intervals per decade was determined by the $H$-test TS above 0.1 GeV: one per decade for pulsars with $H$-test TS values between 100 and 1000, two per decade when $1000 \leq H \leq 10000$, and so on.
Profile fits are again done only if $H$-test $>100$ in an energy band: \nwsubbandprofile \, pulsars have at least two such intervals.
Profile parameters were allowed to vary within the same ranges as for the fits over the entire energy range. By default, the number of components for a given pulsar was the same as for the fit over the entire energy range. However,  profile fits in the narrow energy ranges often require fewer components than  for the entire range. We fitted the profiles with $N_\mathrm{peaks} > 2$ components in the individual energy ranges with all possible component combinations, and selected the combination favored by the BIC in most energy intervals. For most objects the number of peaks remained the same, for the others the third component (typically, bridge emission) was discarded. 
 PSR J0007+7303, shown in Appendix \ref{App-Samples}, is an example. 
The peak shapes, Gaussian or Lorentzian, are the same in the sub-bands as in the energy-integrated profile.

\begin{figure}[!ht]
\centering
\includegraphics[width=1.0\textwidth]{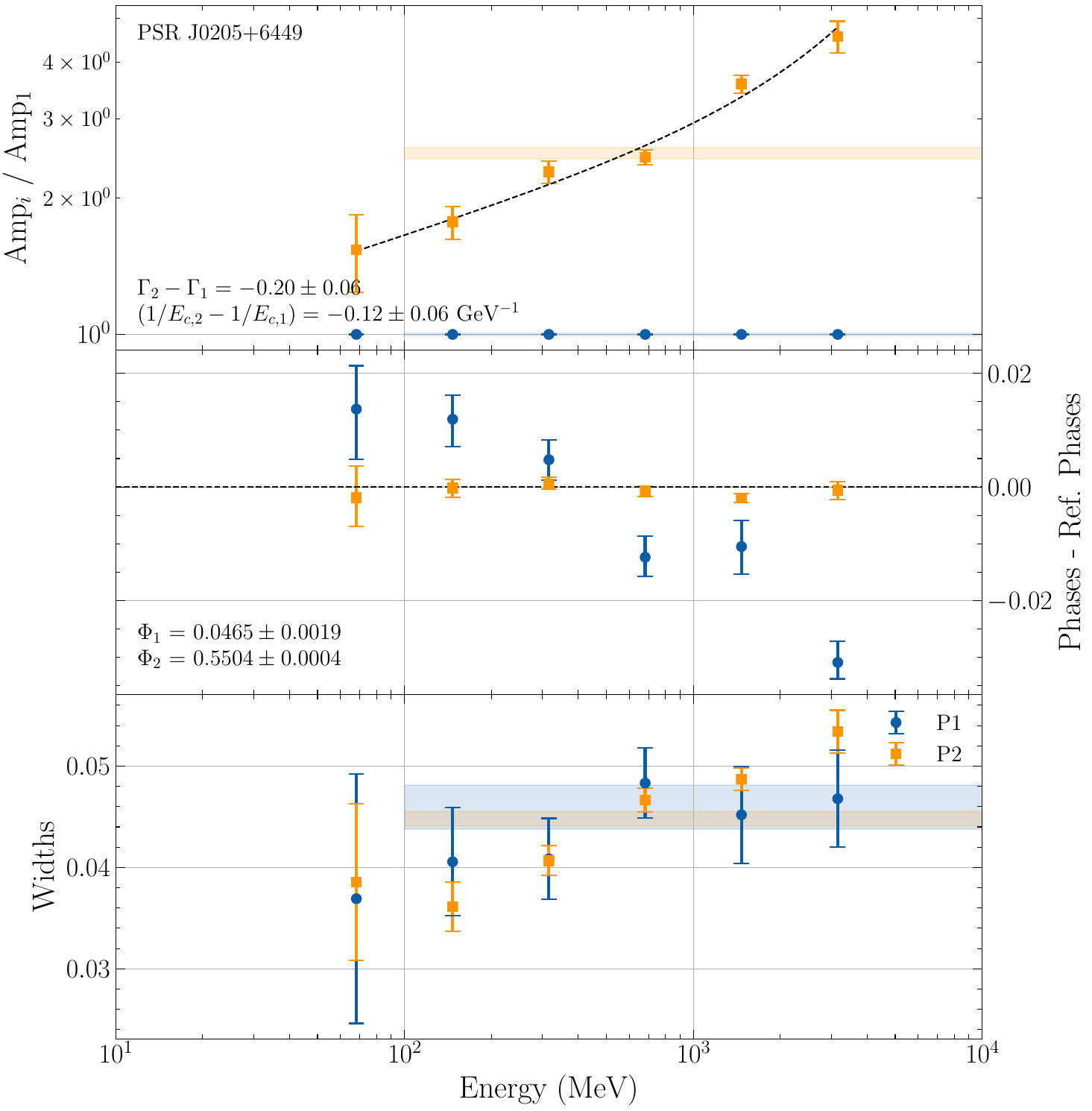}
\caption{Summary of the energy-resolved light curve fitting analysis, for PSR~J0205+6449. \textit{From top to bottom:} ratio of the amplitude of the $i^{\mathrm th}$ pulse profile component to that of the first component as a function of energy; differences between the component phases in the individual energy ranges and those from the global light curve fit (for energies above 0.1 GeV); component widths as a function of energy. PSR~J0205+6449 has two pulse profile components, therefore each panel has two sets of points, with different colors and symbols. Results are shown in energy bins in which the pulsar is detected with an $H$-test TS greater than 300. Fit results for components with peak values smaller than 1/100 of the maximum peak value in the energy bin are not shown. Likewise, peaks with areas smaller than $5 \times 10^{-3}$ of the area of the total pulsed emission and those with widths larger than 20 times the median value are also not plotted. Horizontal bands in the top and bottom panels indicate the reference results from the global light curve fit above 0.1~GeV. The middle panel lists the reference results. The dashed curve in the top panel represents a fit of $\left(\mathrm{Amp_2}/\mathrm{Amp_1}\right)$ versus energy, with the parameters shown in the bottom left corner. }
\label{profiles:Eres_example}
\end{figure}

Figure~\ref{profiles:Eres_example} summarizes the results for one pulsar. The component amplitudes, phases and widths vary with energy, as sometimes seen in other pulsars: see for example the phase-resolved analyses in \citet{FermiCrab, LAT3EGRETpulsars, LATVela2}. In this Figure, results are shown in energy bins in which the pulsar has an $H$-test TS greater than 300. Peaks with amplitudes smaller than 1/100 of the maximum peak value in that energy bin, with areas smaller than $5 \times 10^{-3}$ of the total area of the reconstructed pulsed emission and with widths larger than 20 times the median value are not plotted and not included in the fit described in the following paragraph. As for the pulse profiles described in Section~\ref{profiles:lcs}, all summary plots from the energy-resolved light curve fitting are in the supplementary online material. 

We searched for trends in the amplitude ratios between the first and second gamma-ray peaks as a function of energy, for pulsars with at least two identifiable components. Assuming that the first two components have spectra that follow exponentially cut off power laws, the amplitude ratio can be written as follows:

\begin{equation}
\log_{10} \left(\dfrac{\mathrm{Amp}_2}{\mathrm{Amp}_1} \right) = C - \left(\Gamma_2 - \Gamma_1\right) \times \log_{10} \left(E\right) - \left( 1 / E_{c,\ 2} - 1 / E_{c,\ 1}\right) \times E / \ln{10}, 
\label{P2P1_equation}
\end{equation}

\noindent
where $C$ is an energy-independent constant, and $\Gamma_i$ and $E_{c,\ i}$ respectively denote the spectral index and cutoff energy of the $i^{th}$ component. The top panel of Figure~\ref{profiles:Eres_example} shows the fit using the above expression, and the best-fit parameters. 

\begin{figure}[!ht]
\centering
\includegraphics[width=1.0\textwidth]{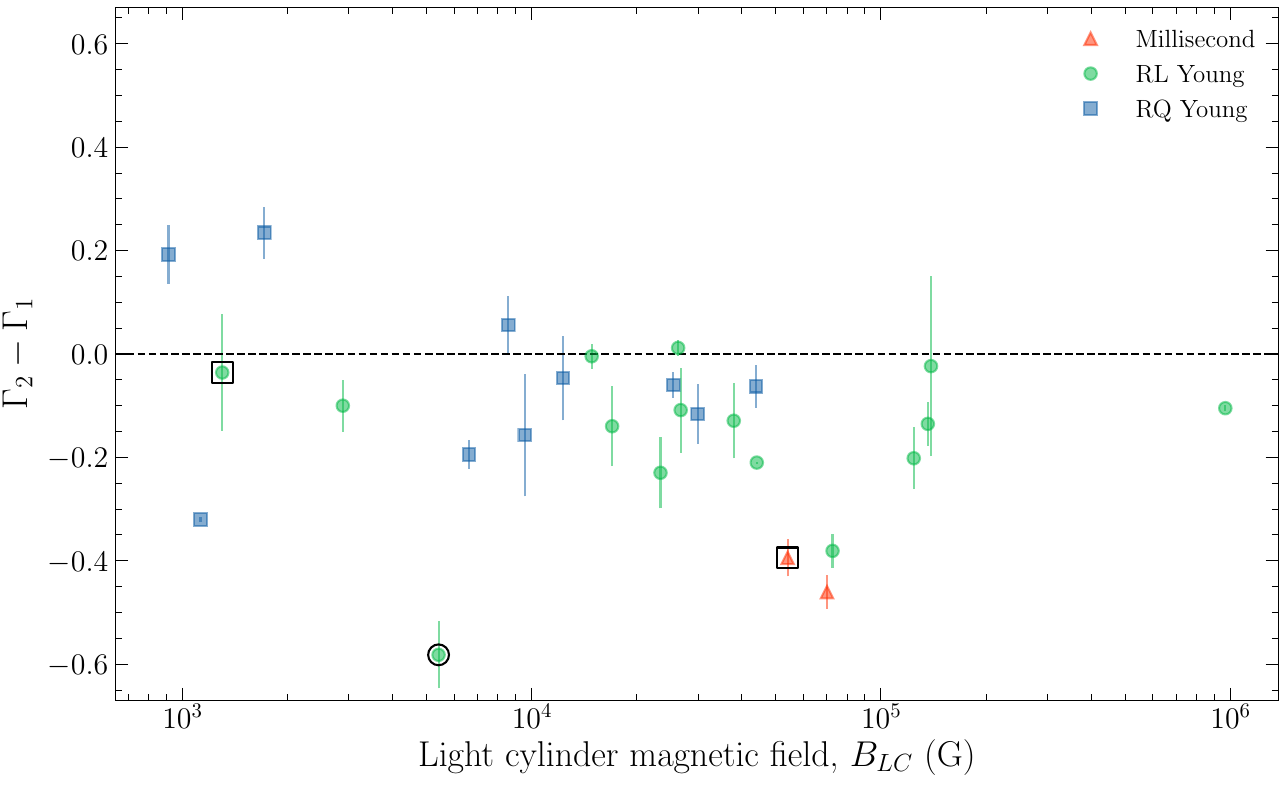}
\caption{$\Gamma_2 - \Gamma_1$ from the energy-resolved light curve fitting, plotted against light cylinder magnetic field, $B_\mathrm{LC}$. See Section~\ref{sec:profs_energy} for details regarding the measurement of this parameter. See Figure~\ref{profiles:delta_Delta} for the definition of circle and square symbols.}
\label{profiles:Eres_P2overP1}
\end{figure}

\begin{figure}[!ht]
\centering
\includegraphics[width=1.0\textwidth]{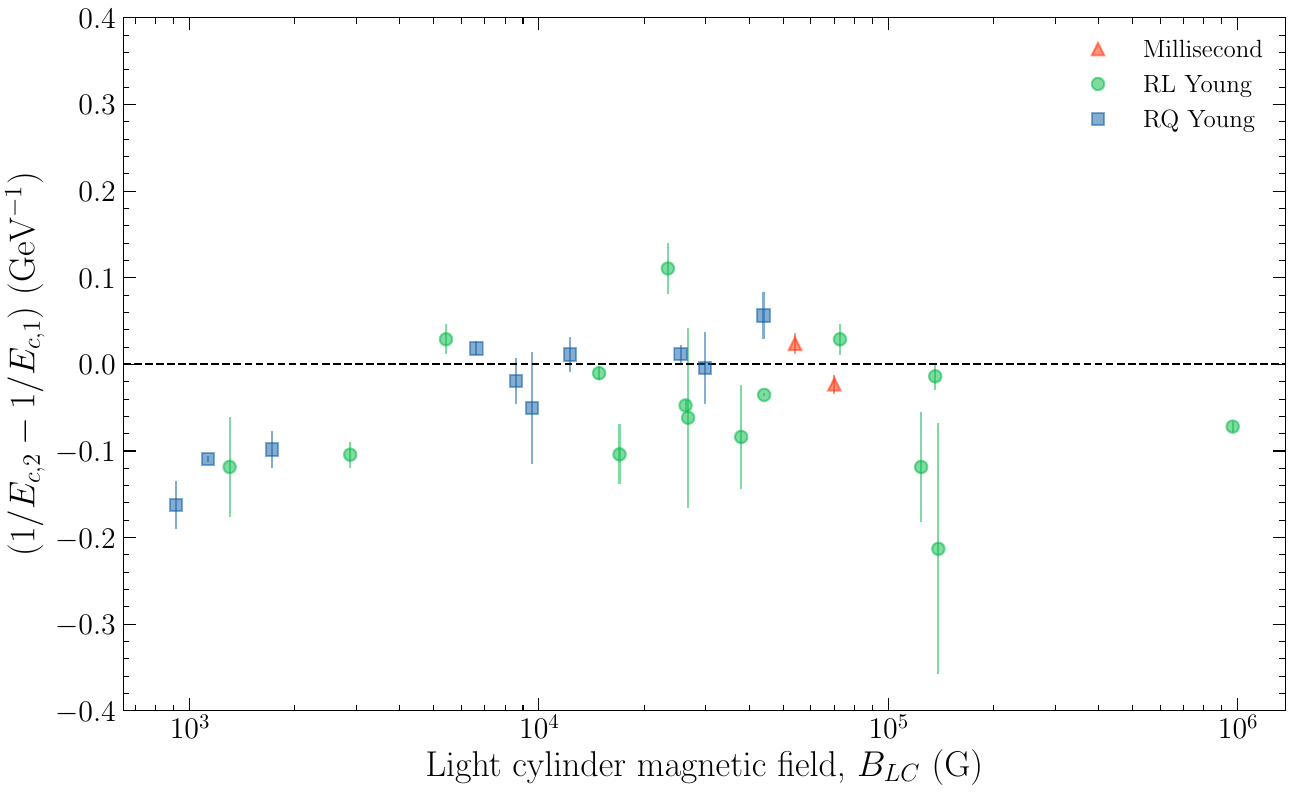}
\caption{$\left( 1 / E_{c,\ 2} - 1 / E_{c,\ 1}\right)$ from the energy-resolved light curve fitting, plotted against light cylinder magnetic field, $B_\mathrm{LC}$. See Section~\ref{sec:profs_energy} for details regarding the measurement of this parameter. See Figure~\ref{profiles:delta_Delta} for the definition of circle and square symbols.}
\label{profiles:Eres_P2overP1_2}
\end{figure}

Figures~\ref{profiles:Eres_P2overP1} and \ref{profiles:Eres_P2overP1_2} show the values of the $\Gamma_2 - \Gamma_1$ and $\left( 1 / E_{c,\ 2} - 1 / E_{c,\ 1}\right)$ parameters, for pulsars with at least six energy intervals with individual $H$-test TS $>300$. Requiring $\geq 6$ data points gives the fit of Equation~\ref{P2P1_equation} enough data points and a wide energy range. The minimum detection significance in the energy bins restricts the study to high-quality profile fits. These criteria select \npsrerestab{} pulsars. Figure~\ref{App-Samples:example_eres_0007} shows that the measured amplitude ratios between the first and second peaks for PSR~J0007+7303 are poorly fit; we hence discarded this pulsar, reducing the sample to \npsrerestabfinal{} objects. The choice of $B_{LC}$ in Figures~\ref{profiles:Eres_P2overP1} and \ref{profiles:Eres_P2overP1_2} is arbitrary; we saw little or no correlation with various abscissa parameters, as discussed further in Section~\ref{disc}. Table~\ref{tbl-P2overP1} summarizes the results for the \npsrerestabfinal{} pulsars. The weighted averages of $\Gamma_2 - \Gamma_1$ and $\left( 1 / E_{c,\ 2} - 1 / E_{c,\ 1}\right)$ are respectively $-0.209(2)$ and $-0.044(1)$ GeV$^{-1}$; clearly incompatible with 0 but dominated by a few measurements with very small statistical uncertainties. The respective unweighted averages are $-0.14(18)$ and $-0.04(7)$ GeV$^{-1}$, where the uncertainties here correspond to the standard deviations on the measured values.



%

%

%

\begin{deluxetable}{lcccc}
\tabletypesize{\scriptsize}
\tablewidth{0pt}
\tablecaption{Summary of energy-resolved fits of the first and second gamma-ray peak amplitude ratios\label{tbl-P2overP1}}

\tablehead{
\colhead{PSR} & \colhead{Code} & \colhead{Comment} & \colhead{$\Gamma_2 - \Gamma_1$} & \colhead{$\left(1/E_{c,2}-1/E_{c,1}\right)$ (GeV$^{-1}$)}
}

\startdata
J0205+6449 & rx &  & $-0.20 \pm 0.06$ & $-0.12 \pm 0.06$ \\ 
J0534+2200 & rx &  & $-0.105 \pm 0.006$ & $-0.072 \pm 0.006$ \\ 
J0614$-$3329 & mbrx &  & $-0.461 \pm 0.032$ & $-0.023 \pm 0.011$ \\ 
J0633+0632 & q &  & $0.23 \pm 0.05$ & $-0.098 \pm 0.022$ \\ 
J0633+1746 & xq &  & $-0.320 \pm 0.005$ & $-0.1093 \pm 0.0034$ \\ 
J0835$-$4510 & rx &  & $-0.2100 \pm 0.0021$ & $-0.0352 \pm 0.0017$ \\ 
J1023$-$5746 & q &  & $-0.06 \pm 0.04$ & $0.057 \pm 0.027$ \\ 
J1028$-$5819 & r &  & $-0.004 \pm 0.024$ & $-0.010 \pm 0.008$ \\ 
J1044$-$5737 & q &  & $-0.16 \pm 0.12$ & $-0.05 \pm 0.06$ \\ 
J1048$-$5832 & r &  & $-0.14 \pm 0.08$ & $-0.104 \pm 0.034$ \\ 
J1057$-$5226 & rx & c & $-0.04 \pm 0.11$ & $-0.12 \pm 0.06$ \\ 
J1124$-$5916 & rx &  & $-0.13 \pm 0.07$ & $-0.08 \pm 0.06$ \\ 
J1231$-$1411 & mbrx & c & $-0.39 \pm 0.04$ & $0.024 \pm 0.012$ \\ 
J1413$-$6205 & q &  & $-0.05 \pm 0.08$ & $0.011 \pm 0.020$ \\ 
J1418$-$6058 & q &  & $-0.12 \pm 0.06$ & $-0.00 \pm 0.04$ \\ 
J1709$-$4429 & rx &  & $-0.11 \pm 0.08$ & $-0.06 \pm 0.10$ \\ 
J1732$-$3131 & r &  & $-0.10 \pm 0.05$ & $-0.104 \pm 0.015$ \\ 
J1809$-$2332 & q &  & $-0.194 \pm 0.028$ & $0.018 \pm 0.008$ \\ 
J1826$-$1256 & xq &  & $-0.060 \pm 0.025$ & $0.012 \pm 0.011$ \\ 
J1833$-$1034 & r &  & $-0.02 \pm 0.17$ & $-0.21 \pm 0.14$ \\ 
J1836+5925 & q &  & $0.19 \pm 0.06$ & $-0.162 \pm 0.028$ \\ 
J1907+0602 & r &  & $-0.23 \pm 0.07$ & $0.111 \pm 0.030$ \\ 
J1952+3252 & rx &  & $-0.381 \pm 0.033$ & $0.029 \pm 0.018$ \\ 
J2021+3651 & r &  & $0.012 \pm 0.016$ & $-0.047 \pm 0.007$ \\ 
J2032+4127 & br & u & $-0.58 \pm 0.07$ & $0.029 \pm 0.017$ \\ 
J2229+6114 & rx &  & $-0.14 \pm 0.04$ & $-0.014 \pm 0.016$ \\ 
J2238+5903 & q &  & $0.06 \pm 0.06$ & $-0.019 \pm 0.027$ \\ 
\enddata

\tablecomments{
In column 3, 'u' indicates that the ordering of the first and second gamma-ray peaks is uncertain, and 'c' indicates that the profile is complex. See Section~\ref{sec:profs_energy} for details on the measurement of the parameters listed in columns 4 and 5.
}

\end{deluxetable}

\clearpage
\section{Phase-averaged Energy Spectra}
\label{spectralSection}

Two significant improvements enhance our pulsar spectral results relative to 2PC.
The first is simply that we have more pulsars, and more photons per pulsar.
The second is a change in our expression for the exponentially cutoff power law spectrum,
a characteristic of pulsars in the GeV domain. 
We replace the classic sub-exponentially cutoff power-law model (`PLEC') with the new `PLEC4' expression introduced in 4FGL-DR3.
Degeneracies between parameters in PLEC4 are reduced and the likelihood
fits converge better, yielding smaller uncertainties. However, some subtleties of the new expression merit the detailed discussion below.


In this Section we first review the motivations behind the exponentially cutoff power law spectrum, highlighting the differences between PLEC used in 2PC and PLEC4 used here. We then describe our results, using the 4FGL-DR3 phase-integrated spectral analysis available for 255 pulsars, and explain some refinements of the spectral analyses that help characterize the pulsars' emission properties, finally yielding \nPLEC{} pulsars with spectral analyses in this catalog. Tables \ref{tbl-psrspec} and \ref{tbl-mspspec}
list parameter values, also available in the electronic version of the catalog. Section \ref{sec:spec_single_pulsars} highlights some features of individual pulsar results.
Section \ref{variability} addresses pulsar flux stability, and in 
Section \ref{lumin} we combine integrated energy fluxes and distances to obtain gamma-ray luminosities and efficiencies.

\subsection{Pulsar Spectral Shapes}
\label{spectra:shapes}
As we will show, the general GeV pulsar spectrum is peaked around $\epeak=1.5$\,GeV (see also Figure \ref{fig-J1227spec} in Appendix \ref{App-Samples}).  It is natural to relate this to the synchrotron spectrum radiated by an ultrarelativistic primary particle \citep{Pacholczyk70,DermerMenon2009}
\begin{equation}
  E^2 dN/dE\equiv S(E,E_c)\propto E^{\frac{4}{3}} \exp[-E/E_c]\equiv E^{2-\Gamma_{\mathrm{sr}}} \exp[-E/E_c].
\end{equation}
Here, the asymptotic spectral index is $\Gamma_{\mathrm{sr}}=2/3$ and the cutoff energy $E_c$ depends on the specific radiation mechanism, e.g. the field line curvature radius $\rho$ with $E_c\propto \gamma_e^3/\rho$ for curvature radiation within the light cylinder \citep{Romani1996}. Farther out, for example in current sheets, the curvature of the particle trajectory may be quite different. 
Other emission mechanisms, such as inverse Compton, can also produce peaked spectra.  In general, pulsars accelerate particles to a range of energies and the observed spectrum is a broader superposition of such basic SED shapes.  A main goal of gamma-ray spectroscopy is to invert this superposition and reveal properties of the underlying particle energy distribution.  In a curvature radiation scenario, e.g., the observed $\epeak$ will roughly track the bulk maximum values of $E_c$, while the width of the peak and the low-energy spectrum ($E\ll\epeak$) encode information about the range of $E_c$.  A nearly monoenergetic distribution of $E_c$ will produce a narrow SED peak and a low-energy $S(E)\propto E^{4/3}$, while a broad range of $E_c$ will produce a broad gamma-ray peak and $\Gamma_0$ of $\sim$1--2 \citep{Romani1996}.

Most previous analyses of LAT pulsar spectra have used the PLEC model:
\begin{equation}
\frac{dN}{dE} \equiv N(E) = N_0 
\left(\frac{E}{\epiv}\right)^{-\Gamma_0}\exp\left[-\left(\frac{E}{E_c}\right)^b\right],
\end{equation}
which is closely related to a monoenergetic curvature emission or synchrotron radiation spectrum, but with the cutoff strength modulated by the parameter $b$. Blends of particle energies and averaging over pulse phases both tend to broaden the peak, yielding ``sub-exponential'' shapes with $b<1$ \citep{LATVela2}.  One innovation since 2PC, adopted beginning with the 8-year 4FGL catalog and justified therein, is to fix $b = 2/3$ instead of $b=1$ for pulsars with lower TS, as this choice better matches the observed spectra.

A good spectral model should have the flexibility to capture the features evoked above, but in most cases measuring them all is not possible.  The LAT sensitivity peaks at $\sim$1\,GeV and the SED slope at 100\,MeV, 
 $2-\Gamma_{100}$, can only be measured for the brightest pulsars.  Furthermore, it is only in the cases of sharply-peaked spectra which quickly transition to a low-energy power law that $\Gamma_{100}$ may be a good proxy for the asymptotic low-energy slope $\Gamma_0$.  Above 1\,GeV, pulsar spectra cut off quickly, making it difficult to measure the cutoff shape and strength.  We show below that for most pulsars, the only reliable features which can be measured are the position and width of the SED peak.  Fortunately, it is these features which strongly correlate with underlying pulsar properties like $\edot$, so they are likely to be the most important in capturing information about the pulsar emission mechanism.

The PLEC formula suffers from large covariance between the model parameters \citep{4FGL-DR3}, 
magnifying the effect of statistical uncertainty.  To reduce these correlations, the spectral analysis 
of 4FGL-DR3 adopted the PLEC4 model (\texttt{PLSuperExpCutoff4} in the 
\textit{Fermitools}\footnote{\url{https://fermi.gsfc.nasa.gov/ssc/data/analysis/software/}. The catalog FITS files tabulate negative $-\Gamma < 0$ values.}, 
specifically designed to minimize the covariance
between the model parameters at a reference energy, $\epiv$:
\begin{align}
  \label{eq:plec4}
  \frac{dN}{dE}= &
  N_0
  \left(\frac{E}{\epiv}\right)^{-\Gamma+\frac{d}{b}}\exp\left[\frac{d}{b^2}\left(1-\left(\frac{E}{\epiv}\right)^b\right)\right]\\
  \equiv & N_0\,
  x^{-\Gamma+\frac{d}{b}}\exp\left[d/b^2\left(1-x^b\right)\right]
   , \quad x \; =\; \frac{E}{\epiv} \quad .
\end{align} 
For PLEC4, $N_0$ is the flux density at $E_0$.  The PLEC4 spectral shape is exactly the same as for the PLEC model, as can be seen by identifying 
\begin{align}
     PLEC & \Longleftrightarrow  PLEC4 \nonumber \\
  E_c & = \ E_0\left(\frac{b^2}{d}\right)^{\frac{1}{b}} \\
  \Gamma_0 & = \ \Gamma-\frac{d}{b} \\
  N_0 & = \ N_0\,\exp\left(\frac{d}{b^2}\right).
\end{align}
In the fits presented below, the reference energies $E_0$ are confined
to a relatively narrow energy range with a typical value of 1.6\,GeV.

\begin{figure} 
\centering 
  \includegraphics[angle=0,width=0.98\linewidth]{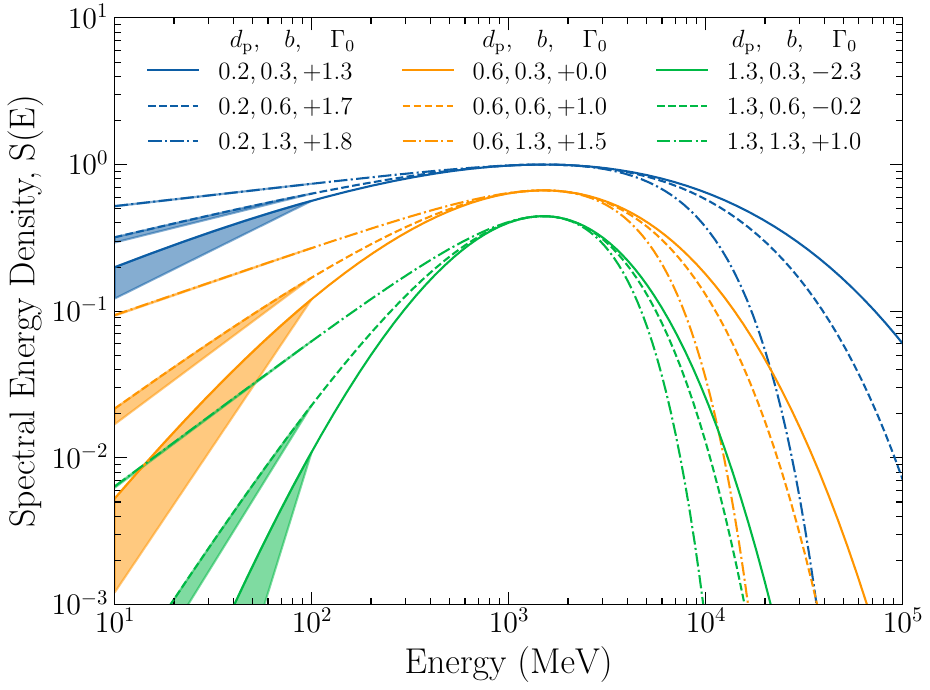} 
  \caption{\label{fig:plec4_demo}Realizations of the PLEC4 model for three values
  of the peak curvature, $\dpeak=0.2,\,0.6,\,1.3$, and the asymmetry parameter
  $b=0.3,\,0.6,\,1.3$.  We set $\epeak=\epiv=1\,$GeV, which uniquely determines $\Gamma$
  and $\Gamma_0$.  The SEDs are normalized to 1 at $\epeak$ but offset by 50\% for clarity.
  Each curve is labeled with the tuple ($d$, $b$, $\Gamma_0$).  The shaded regions
  show the difference between the SED and the asymptotic power law, normalized to 0.1\,GeV.
  } 
\end{figure}

The PLEC4 model parameters $\Gamma$ and $d$ are related to the logarithmic derivatives of the model. The gamma-ray convention defines the photon spectral index as
\begin{equation}
  \Gamma_{E}=-\frac{\partial\log N}{\partial\log E} = 
  \Gamma-\frac{d}{b}\left(1-x^b\right)=\Gamma_0+\frac{d}{b}x^b,
\end{equation}
from which it clear that $\Gamma$ is the photon index at the reference energy ($x=1$) and $\Gamma_0$ is the asymptotic spectral index ($x\rightarrow 0$).
The spectral curvature can be defined as $-\frac{\partial^2\log
N}{\partial\log E^2} = dx^b$, so $d$ is the SED curvature at $\epiv$.

Figure \ref{fig:plec4_demo} shows some example realizations of the PLEC4 model and in particular highlights the role of $d$ in producing narrow (large $d$) or broad (small $d$) peaks.  $b$ controls the overall shape of the spectrum, with larger values of $b$ producing greater asymmetry.  For $b\ll1$ the spectrum becomes more symmetric, with both a slower cutoff and a slower convergence to a power law at low energies.  $b>1$ produces asymmetric spectra with sharp, ``super-exponential'' cutoffs and power-law behavior at low energy.  When $b<0$, the cutoff and power-law regimes are reflected.  For many of these examples, the transition to power-law behavior has not yet happened at 100\,MeV.  In this example, the local spectral index at 100\,MeV agrees with the asymptotic index to within $0.1$ for $b=0.6$, but for $b=0.3$ it differs by 0.3 ($d=0.2$), 0.9 ($d=0.6$), and 1.9 ($d=1.3$).

The energy at which the SED peaks (if the peak exists, is nonzero, and is
finite) is
\begin{equation}
  \label{eq:epeak}
  \xpeak = \frac{\epeak}{\epiv} =
  \left[1+\frac{b}{d}\left(2-\Gamma\right)\right]^{\frac{1}{b}} =
  \left[\frac{b}{d}\left(2-\Gamma_0\right)\right]^{\frac{1}{b}}.
\end{equation}
Note that $b<0$ can yield a valid $\epeak$ if $\Gamma_0\geq2$.  $\epeak$ is a key \textbf{physical} property of
interest, and we can obtain a second key property, the width of the SED
at its peak, by evaluating the curvature at $x_p$:
\begin{equation}
  \label{eq:disaster}
  d \xpeak^b \equiv \dpeak = d + b\left(2-\Gamma\right) =
  b\left(2-\Gamma_0\right) .
\end{equation}
For reference, the maximum expected spectral
curvature, obtained for monoenergetic synchrotron radiation, is $\dpeak=4/3$.

We expect these quantities associated with the SED peak to be well-determined and
relatively independent because $\epeak$ is generally close to $\epiv$.
However, Equation \ref{eq:disaster} indicates that we
\textbf{cannot} measure the peak curvature and the asymptotic index
independently when $b$ is fixed, and since we expect $\dpeak$ to be the
better-determined quantity, we cannot use the PLEC4 model to extrapolate the asymptotic spectral index, $\Gamma_0$.  $\Gamma_{100}$ could be used as a proxy for $\Gamma_0$ (when $\epeak\gg100$\,MeV and the peak is not too broad), but it too suffers from bias: for a typical $\epiv=1.6$\,GeV and with $b=2/3$,
\begin{equation}
  \label{eq:g100}
  \Gamma_{100}= \Gamma_0+0.23d =\Gamma
  -1.26d.
\end{equation} 
This formulation all but maximizes the correlation between the model parameters $\Gamma$ and $d$, undoing the benefits of PLEC4, and like $\Gamma_0$, $\Gamma_{100}$ will potentially be biased if $b=2/3$ does not describe the true SED.

Thus, to use the PLEC4 model to estimate reliable physical properties, we need to answer two primary questions:
\begin{enumerate}
\item For bright pulsars where $b$ can be measured, does the PLEC4 model allow robust
  estimation of other physical parameters of interest, like $\Gamma_{100}$?  That is, can PLEC4 reproduce the true shapes of pulsar SEDs?
\item For fainter pulsars, for which $b$ must be fixed to a canonical value, does the reduced
  parameter space introduce a bias on physical parameters?
\end{enumerate}

\subsection{Spectral Model Fits}

We now address these questions using phase-averaged spectral fits.  
All spectral results in this section rely on the 4FGL-DR3 spectral fitting methodology.  Some results are incorporated directly from that catalog, and we have extended these results using the DR3 model as a baseline.  In particular, we have
\begin{itemize}
    \item Performed spectral fits in which $b$ was free to vary for 119 pulsars with TS $>$ 1000 (c.f. 28 pulsars with TS $>$ 10,000 in 4FGL-DR3).
    \item Fit the 28 $b$-free pulsars of 4FGL with a $b=2/3$ model for comparison.
    \item Re-fit PSRs~J0922$+$0638, J1757$-$2421, J1913$+$1011, and J2205$+$6012 with a PLEC4 model, as opposed to the power law used in 4FGL-DR3.  
    \item Added a power-law component to the spectral models for pulsars with TeV counterparts (PSRs~J0007$+$7303, J0205$+$6449, J1016$-$5857, J1119$-$6127, J1714$-$3830, J1833$-$1034, J1907$+$0602, and J2032$+$4127) to encapsulate potential emission from a pulsar wind nebula and re-optimized the PLEC4 model.
    Crab, Vela and a half-dozen other pulsars were already colocated with TeV power-law sources in 4FGL.
    \item Directly measured the low-energy spectral indices with a power-law  fit of the 50--300\,MeV band, for the 42 pulsars with TS $>$25 in that band and TS$>$1000 overall.
    
\end{itemize}

Of the 255 pulsars, 251 produce a spectral fit with a well-defined $\epeak$.

Here we quickly note a possible point of confusion.  The reference
energy $\epiv$ in Eq. \ref{eq:plec4} is used in fitting models and
evaluating covariance between parameters.  From these results a ``pivot energy'' can be determined at which the uncertainty
on the flux density normalization $N_0$ is minimized.  The results presented here require the covariance matrices from the likelihood fits and so use the reference energy, and thus parameter values will generally differ from those in the FITS files accompanying 4FGL-DR3, which have been scaled to the pivot energy. {\ bf A script provided with the Online Material shows how to correctly unpack the parameters and covariance from the electronic catalog.}

\subsubsection{Distribution of $b$}

\begin{figure}
\centering
  \includegraphics[angle=0,width=0.98\linewidth]{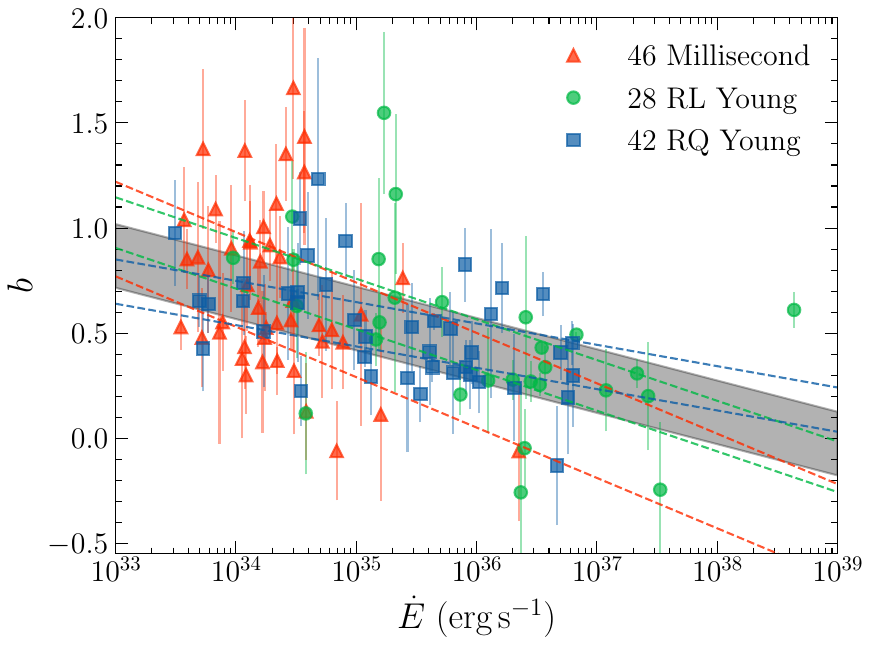}
  \caption{\label{fig:b_vs_edot}The best-fit values of $b$, with
  1$\sigma$ uncertainties, shown as a function of $\edot$.  Maximum likelihood linear models are shown for each class and for the combined population.  The centroid indicates the model fit, while the distance between the dashed lines indicates the intrinsic scatter ($\pm\sigma_i$, see main text).  The dark shaded region indicates the same quantities for the full population with measured values of $b$.}
\end{figure}

First, we consider the distribution of best-fit $b$ values, plotted
against $\edot$ in Figure \ref{fig:b_vs_edot}.  It is clear that $b$ does not take on a constant value, so we used maximum likelihood to fit a linear model $b=b_0 + m_b\log_{10}(\edot/10^{36}\,\ergs)$ to each class and to the full population.  We assumed that each measured value follows a Gaussian distribution with the measurement uncertainty, and we also included an intrinsic scatter, $\sigma_i$, which is added in quadrature to the measurement error.  

\begin{figure}
\centering
  \includegraphics[angle=0,width=0.98\linewidth]{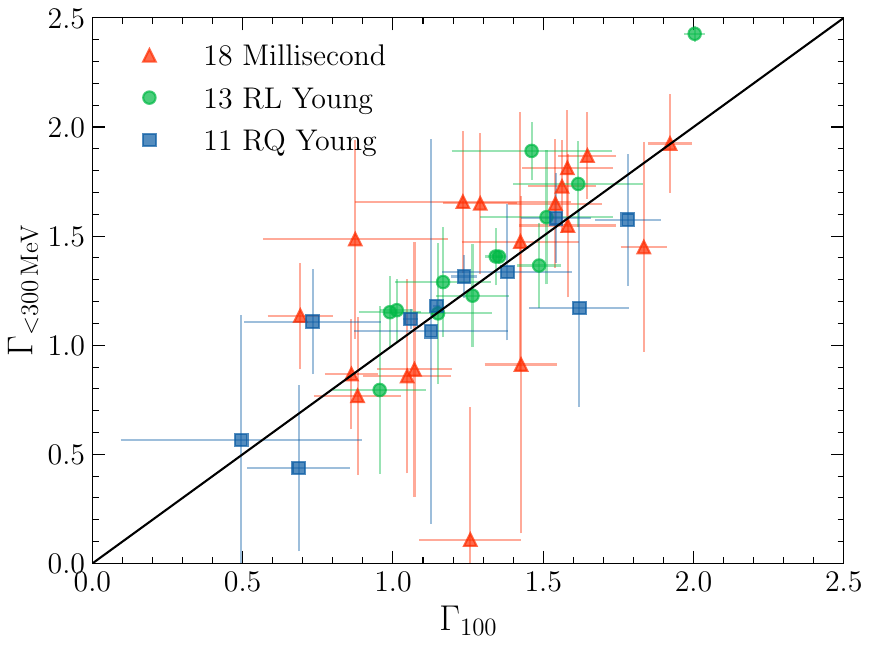}
\caption{\label{fig:lowe_indx_comp}The comparison between spectral
  indices measured directly with a low-energy fit (y-axis) and the
  predicted local index at 100\,MeV from $b$-free models.}
\end{figure}

We excluded the Crab pulsar from the fits. In Figures \ref{fig:lowe_indx_comp} through \ref{fig:g100_vs_edot} its well-measured parameters generally deviate from those of most other pulsars. With its high $\dot E$, including it changes the fit results substantially, such that they no longer describe most other gamma-ray pulsars well. If the Crab is atypical only in having both high $\dot E$ and good statistics, then perhaps our simple linear fit breaks down for high $\dot E$.
But the Crab is also atypical in being a rare pulsar with much stronger flux density in the MeV range than in the GeV range, as we further discuss in Section \ref{disc:lumANDspec}.

The results, depicted in Figure \ref{fig:b_vs_edot}, are (1) there is strong evidence (log likelihood improves by $>$40) of intrinsic scatter, i.e. $\sigma_i>0$, and (2) there is no evidence (log likelihood improves only by a few) for a different linear relation amongst the classes.  Therefore, we report a single model for all pulsars, finding $b_0=0.42$, $m_B=-0.15$, and $\sigma_i=0.15$.  

These values are intended as a guide only: the simple model does not fully capture the observed variations, and the fitting approximates the measurement uncertainty on $b$ as a Gaussian rather than using the full likelihood surface.  Nonetheless, it is clear that many pulsars fit with a $b=2/3$
model will suffer appreciable biases.  In particular, estimates of $\Gamma_0$ and $\Gamma_{100}$ for MSPs (with $\edot$\,$\sim$\,$10^{34}$\,erg\,s$^{-1}$)
will tend to be too soft, and those of RL pulsars (with $\edot>10^{36}$\,erg\,s$^{-1}$) too
hard.  We analyze these biases more quantitatively below.  With reference to Figure \ref{fig:plec4_demo}, we also see that the
distributions of $b$ indicate that MSP spectra hew more closely to the power-law with exponential cutoff model, while young pulsars, particularly
high-$\edot$ RL pulsars, have substantial curvature over the full LAT
band and more gradual high-energy cutoffs.
This is suggestive of MSPs possessing an intrinsically narrower radiating particle distribution than do their RL and RQ counterparts; how this might connect to the smaller magnetospheres of MSPs remains to be understood.

In Appendix \ref{sec:spectral_notes}, we study the change in model parameters between the 119 pulsars which have
$b$-free and $b=2/3$ fits available.  In particular, we find that $d$ and $\Gamma$ agree well between the models, while $\Gamma_{100}$ is irrecoverably biased.  The physical properties $\dpeak$ and
$\epeak$ are robustly measured in both $b$-free and $b=2/3$ fits and we conclude that 
we can make inferences on their distribution using the entire
population of pulsars.  On the other hand, the low-energy spectrum is
unreliable, and values inferred from the $b=2/3$ model could be very
far from the true values.

Figure \ref{fig:lowe_indx_comp} establishes the converse, that the
$b$-free PLEC4 model yields a good estimate of $\Gamma_{100}$, at
least for the small sample for which model
comparison is possible.  The figure shows the indices resulting from
the 119 low-energy fits, of which 42 have TS$>$25 in the
50--300\,MeV band.  The geometric mean energy of this band is
122\,MeV, but the improvement of LAT sensitivity with energy, along
with the use of likelihood weights, means the effective energy is
greater.  Nevertheless, the agreement is good for all 3 pulsar
classes\footnote{The one clear outlier is the Crab pulsar, which is indistinguishable from the pulsar wind nebula in the narrowband analysis, yielding a softer index.}, and consequently we conclude that the $b$-free $\Gamma_{100}$
measurements are reliable.  If additionally $\epeak\gg100$\,MeV and $\dpeak$ is not too small, $\Gamma_{100}$ may approximate $\Gamma_0$ well.  We discuss this further below. 

\subsection{Physical Properties}

We can thus make inferences about three physical properties: $\dpeak$
and $\epeak$ for the full population of 251 pulsars, and
$\Gamma_{100}$ for the 119 $b$-free pulsars.  Unless otherwise noted, results from $b$-free fits are preferred and presented whenever available.

\begin{figure*}
\centering
  \includegraphics[angle=0,width=0.98\linewidth]{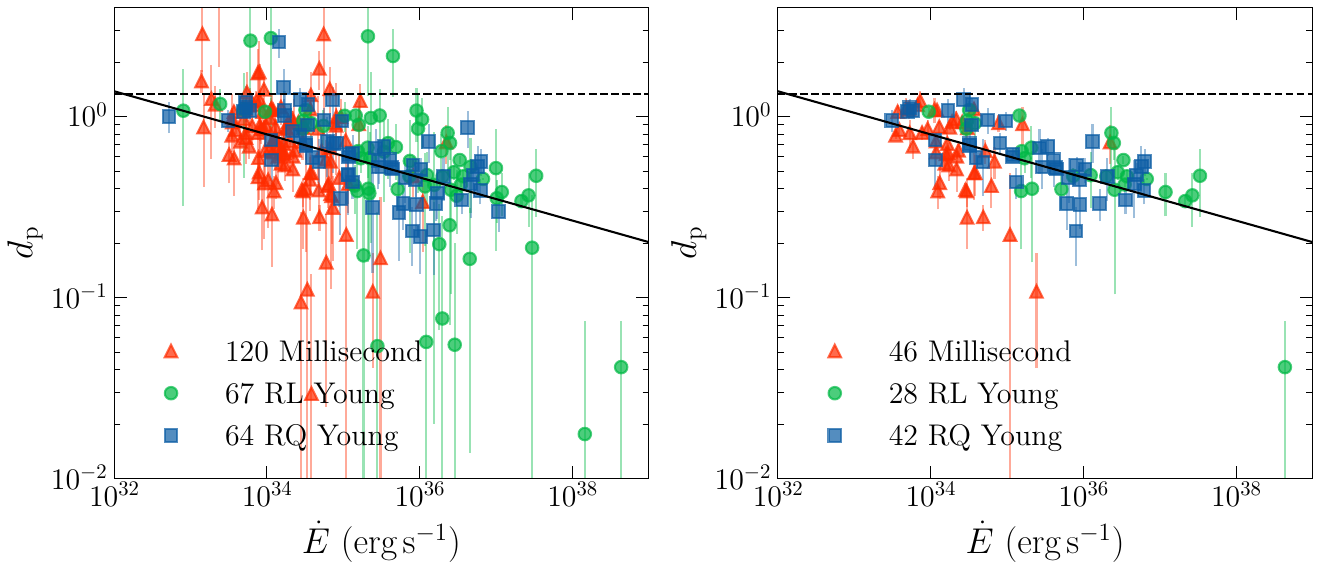}
  \caption{\label{fig:dpeak_vs_edot}Dependence of the spectral
  peak curvature $\dpeak$ on $\edot$, showing a
  well-defined trend.  The horizontal dashed line at 4/3 indicates the
  expected peak curvature for a monoenergetic curvature radiation
  spectrum.  The lefthand panel shows results for all 251 pulsars
  (i.e. mixes $b$-free and $b=2/3$ fits), while the righthand panel
  shows only the 116 pulsars with a $b$-free fit (and a well-defined
  peak energy).  The solid line indicates a best-fit relation to the merged sample (lefthand panel). }
\end{figure*}

\begin{figure}
\centering
  \includegraphics[angle=0,width=0.98\linewidth]{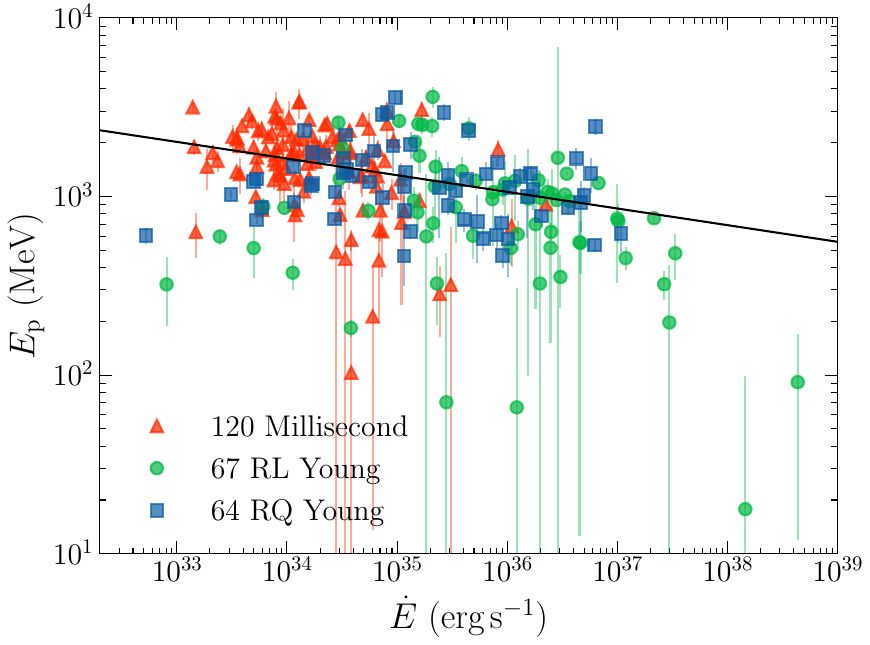}
  \caption{\label{fig:epeak_vs_edot}Dependence of the spectral
  peak energy on $\dot{\mathrm{E}}$, with low-$\edot$ pulsars having a peak energy roughly twice that of high-$\edot$ pulsars.  The samples and best-fit model are as in the left panel of Figure \ref{fig:dpeak_vs_edot}.}
\end{figure}

First, we consider the distribution of spectral curvature, $\dpeak$, or
equivalently, (inverse) SED peak widths, with results shown in Figure
\ref{fig:dpeak_vs_edot}.  $\dpeak$ depends strongly on $\edot$,  varying
in width by a factor of about 3 over the main population.  The narrowest
peaks---primarily of MSPs, but with members of each class---saturate
at about $4/3$, corresponding to a monoenergetic synchrotron spectrum. 
Only PSR~J1827$-$1446, a radio-quiet pulsar, lies out of family
with $\dpeak=2.6\pm0.5$ and one of the largest $\epeak$ values (see below).
To quantify the relation, we fit a simple power-law model to the full population (left panel, excluding Crab), finding $\dpeak = 
0.46(\edot/10^{36}\,\ergs)^{-0.12}$ with an additional scatter of 17\% added in quadrature.  

Next, we consider the distribution of peak energies, $\epeak$, shown in Figure \ref{fig:epeak_vs_edot}. 
 As with $\dpeak$, there is a trend with $\edot$, which we characterized as $\epeak=1.1\,\mathrm{GeV}(\edot/10^{36}\,\ergs)^{-0.09}$ with an additional scatter of 30\%.  This $\edot$ evolution leads to a distinct difference between the pulsar classes, with median
values of 1.7\,GeV (MSP), 1.2\,GeV (RQ), and 0.9\,GeV (RL).
Other population variables, such as the magnetic field at the surface or light cylinder, do not reduce the larger intrinsic scatter.

\begin{figure}
\centering
  \includegraphics[angle=0,width=0.98\linewidth]{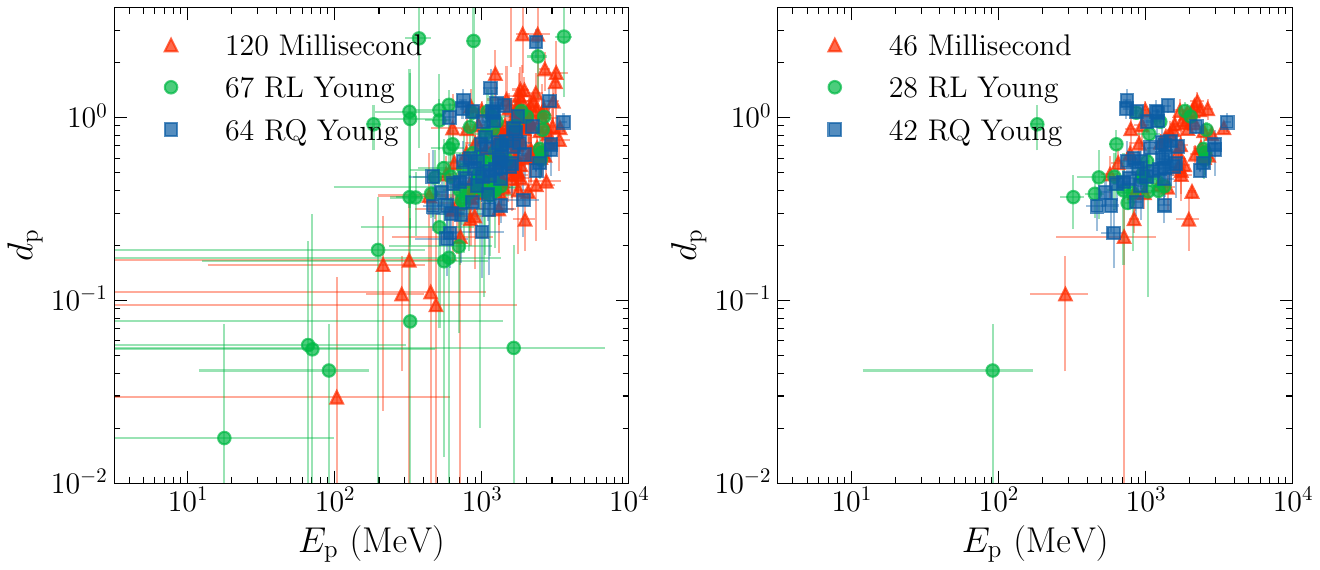}
  \caption{\label{fig:dpeak_vs_epeak}Dependence of the spectral
  curvature $\dpeak$ on the peak energy, $\epeak$. Panel selections are as in Figure
  \ref{fig:dpeak_vs_edot}.}
\end{figure}

$\dpeak$ and $\epeak$ are also correlated (Figure \ref{fig:dpeak_vs_epeak}), but this relation appears to be entirely captured by $\edot$ evolution.  Using the single-variable models determined above removes the correlation to the level of the intrinsic scatter in $\epeak$.
In general, we arrive at a picture of larger peak widths at lower peak energies, a paradigm clearly illustrated by Crab-like pulsars with very broad peaks with, presumably, substantial contributions from secondary particles.
The large widths are such that some pulsars, like the Crab and Vela, emit into the TeV range \citep{HESSVela2023}, implying very high particle energies, which emission models need to accommodate.

\begin{figure}
\centering
  \includegraphics[angle=0,width=0.48\linewidth]{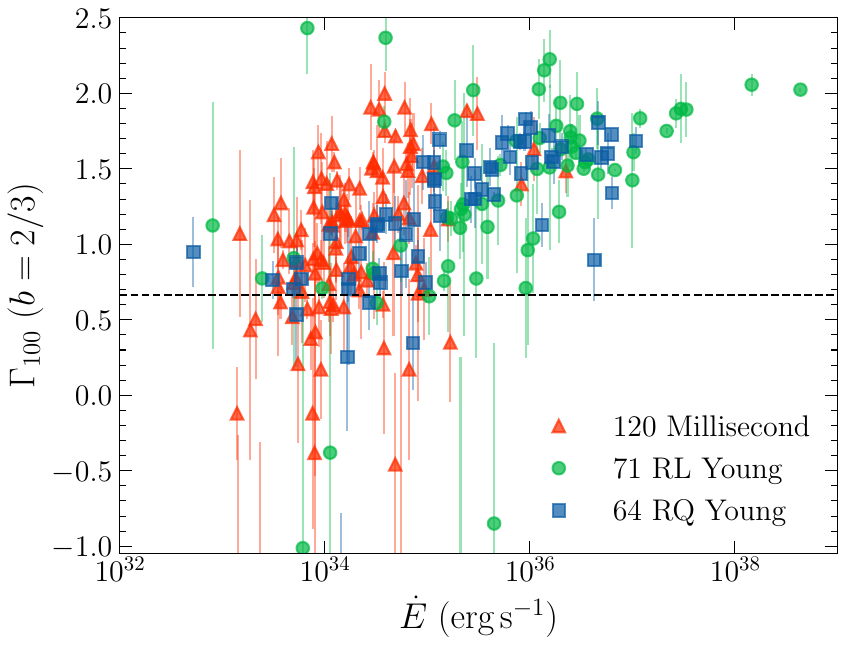}
  \includegraphics[angle=0,width=0.48\linewidth]{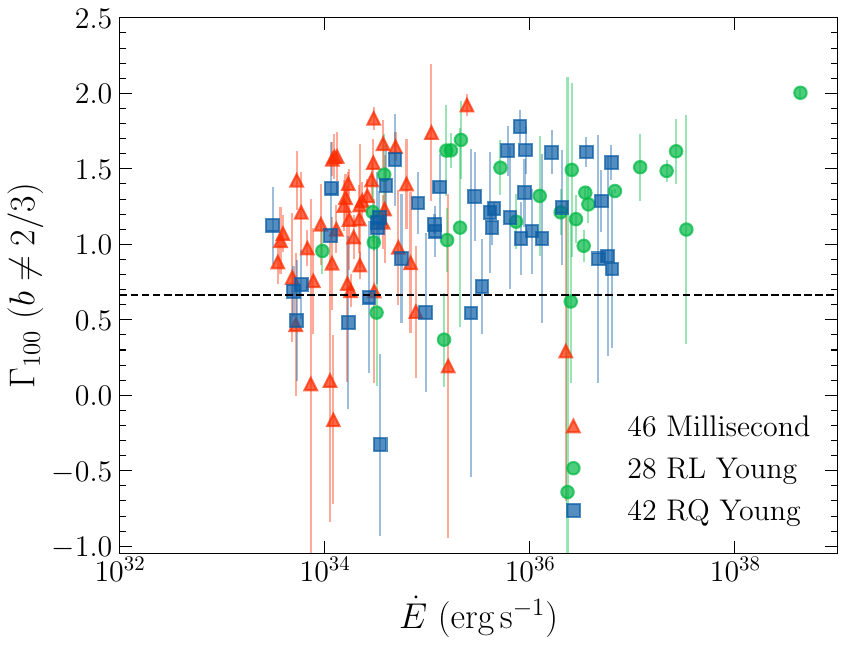}
  \caption{\label{fig:g100_vs_edot}Dependence of the low-energy
  spectral index $\Gamma_{100}$ on $\edot$.  The left panel shows fits
  in which every pulsar has $b=2/3$, revealing a strong but erroneous
  trend.  The right panel shows only the 119 unbiased $b$-free fits.
  The horizontal dashed line indicates the monoenergetic limit, 2/3.
  }
\end{figure}

Finally, we consider $\Gamma_{100}$, a proxy for the asymptotic spectral index.  As an
illustration of the possibility of making an incorrect inference about
the physical properties of the pulsar population, in the first panel
of Figure \ref{fig:g100_vs_edot} we show the results of 251 $b=2/3$
fits, revealing an apparently significant trend with hardening spectral
indices for increasing $\edot$.  However, this result is largely due
to the bias relating $\Gamma_{100}$ to the spectral curvature and thus
reproduces the trend shown in Figure
\ref{fig:dpeak_vs_edot}.  In the right panel, we show the $b$-free
fits, which we demonstrated provide relatively unbiased estimators for
$\Gamma_{100}$.  The trend is reduced or absent: the bulk of the
pulsars with $\edot>10^{35}$erg\,s$^{-1}$ have a spectral index near $1.2$, and
somewhat lower ($\approx 1$) for lower $\edot$.
Below this (possible) threshold, harder spectra become apparent, with
values approaching the monoenergetic limit of 2/3.  These pulsars have
the highest values of $\epeak$, and so for them $\Gamma_{100}$ may be 
closer to $\Gamma_0$.

\begin{figure}[!ht]
\centering
\includegraphics[width=0.48\textwidth]{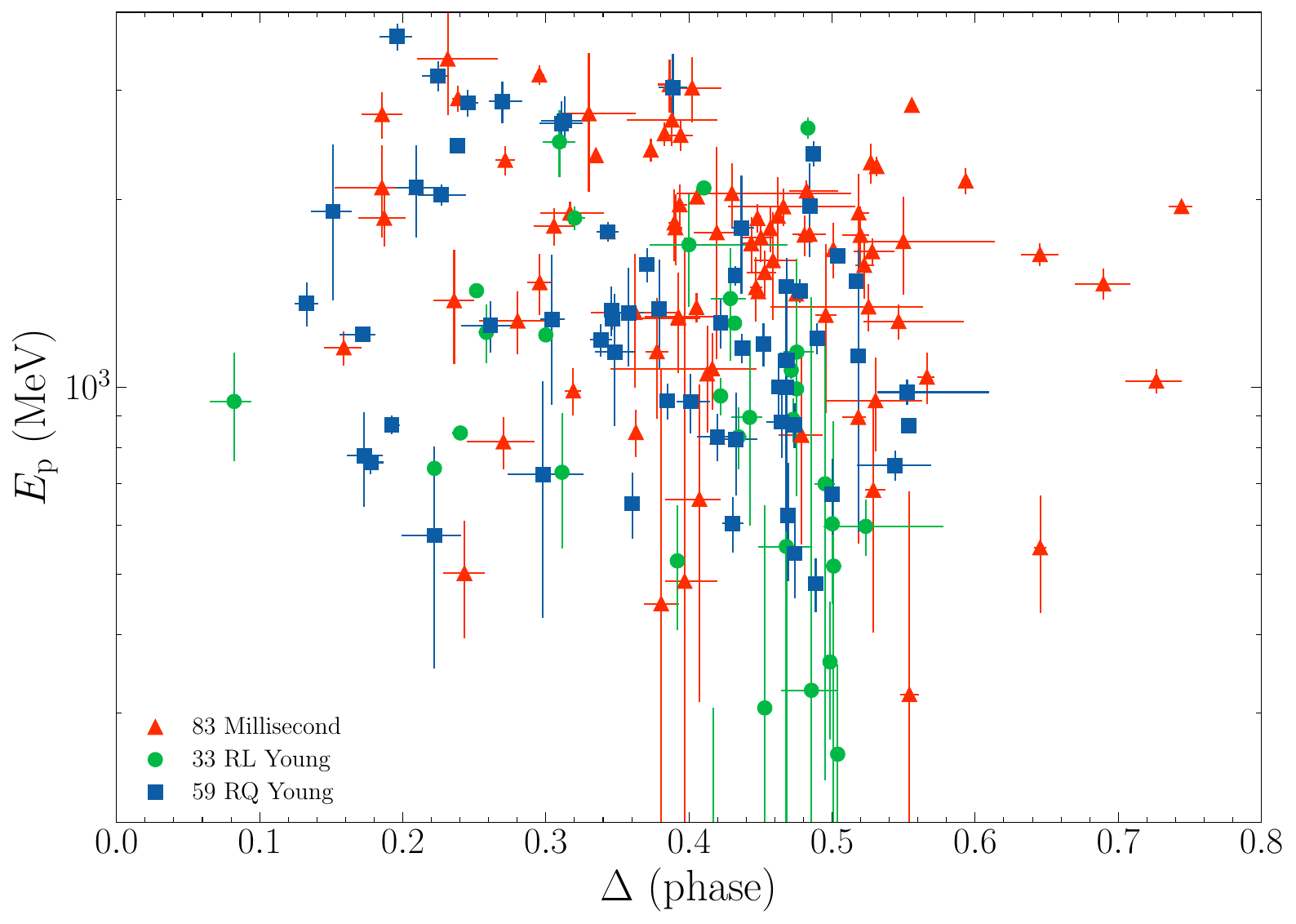}
\includegraphics[width=0.48\textwidth]{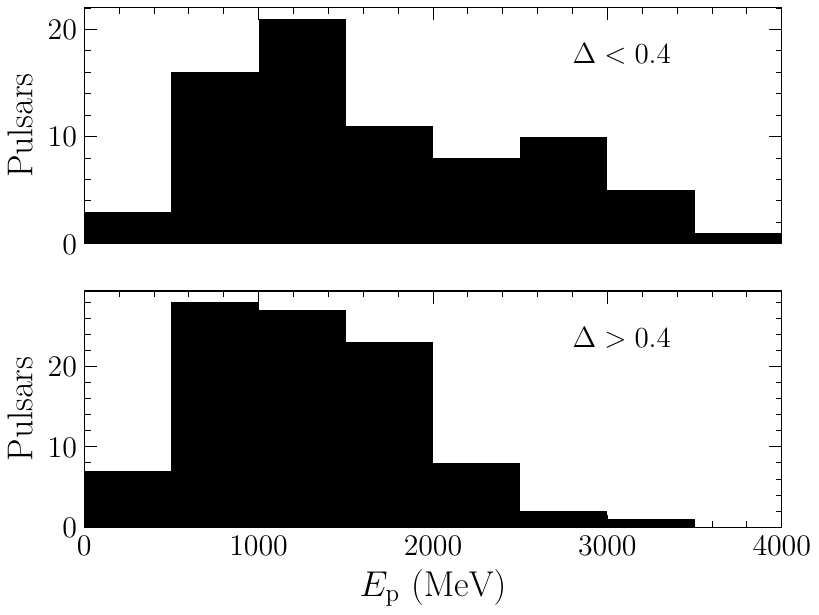}
\caption{Left: The SED peak energy $\epeak$ versus the phase separation $\Delta$ between the two principal gamma-ray peaks. The Crab and PSR J1730-3350 are off-scale with $\epeak \approx 50$ MeV.  Right: $\epeak$ distributions for pulsars with closely- (top) and widely- (bottom) spaced peaks. Pulsars with closely-spaced peaks perhaps have the highest $\epeak$ values.
\label{fig:epeak_Delta}}
\end{figure}

Both observed SED shapes and pulse profile shapes should depend on the magnetic field strength and configuration, as well as on ``geometry'', meaning for example the inclination $\alpha$ of the magnetic axis relative to the rotation axis; the inclination $\zeta$ of the rotation axis relative to the line of sight; and the radius of the light cylinder $R_{LC} = cP/2\pi$. 
We thus searched for correlations between measured profile and spectral parameters.
Figure \ref{fig:epeak_Delta} suggests that pulsars with classic Vela- or Crab-like gamma-ray pulse profiles,
with $\Delta \lesssim 0.5$ rotations, have typical SED peak energies. The highest $\epeak$ values seem to
occur mainly for pulsars where the second gamma-ray peak follows the first more closely. 
This is perhaps related to the evidence that the gamma pulse farther from the magnetic pole generally has the harder spectrum (Figure \ref{profiles:Eres_P2overP1}). 

\subsection{Summary of Results}
We summarize our main results here:
\begin{itemize}
    \item The broadband shape, governed by the parameter $b$, evolves with $\edot$, yielding  slowly-decaying spectra for high-$\edot$ pulsars and sharply-cutoff spectra for low-$\edot$ pulsars.
    \item Pulsar spectral energy densities (SEDs; $E^2 dN/dE$) are most effectively characterized by the curvature at the SED peak, $\dpeak$, which is inversely proportional to the SED peak width.  $\dpeak$ anti-correlates with $\edot$ with low (17\%) scatter: see Figure \ref{fig:dpeak_vs_edot}.
    \item The energy at which the SED peaks, $\epeak$, also varies inversely with $\edot$ with somewhat larger (30\%) scatter: see Figure \ref{fig:epeak_vs_edot}.
    \item Consequently, the spectra of MSPs are more sharply peaked than those of young pulsars, and reach the highest peak energies.
    \item The asymptotic ($E\rightarrow0$) spectral index, $\Gamma_0$ lies out of the LAT band.  The spectral index near 100\,MeV, $\Gamma_{100}$, is close to $\Gamma_0$ when $\epeak\gg100\,$MeV, and it can be estimated for 116 of the brightest pulsars (Figure \ref{fig:g100_vs_edot}).  These indices appear to take on an almost universal value of about 1.2, though low-$\edot$ pulsars, particularly MSPs, may have harder spectra that approach the `monoenergetic limit' of 2/3.
    \item Thus, for our LAT pulsars, nearly all of the rotational energy that is converted to electromagnetic radiation is concentrated in a single decade of the spectrum.
    \item The SED peak energy $\epeak$ seems to be highest for pulsars with pulse phase separations $\Delta<0.4$ (Figure \ref{fig:epeak_Delta}). This, coupled with the observation that the second pulse tends to have a harder spectrum than the first (Figure \ref{profiles:Eres_P2overP1}), is new empirical
    evidence that the same underlying magnetosphere configurations affect both profile and spectral shapes.
\end{itemize}

\subsection{Notes on Individual Pulsars}
\label{sec:spec_single_pulsars}

Twenty gamma-ray pulsars in Tables \ref{tbl-psrspec} and \ref{tbl-mspspec} are flagged with $\dag$ symbols, meaning that no 4FGL-DR3 source with Test Statistic $>25$ is colocated with the pulsar. 
The pulsed signature is detectable in these cases, in spite of a low gamma-ray flux or large local background.

\citet{GuillemotJ0737} provide spectral parameters for the double pulsar PSR J0737$-$3039A via a dedicated analysis. PSR J1513$-$5908 (B1509$-$58) is brightest in the MeV domain, with a spectral cutoff below 100 MeV, measured by \citet{Kuiper_J1846-0258}. The colocated extended source 4FGL J1514.2$-$5909e corresponds to the pulsar wind nebula MSH 15$-$52 and its spectral parameters do not apply to the pulsar.
Similarly, PSR J1023+0038 is a transitional pulsar, and the 4FGL-DR3 spectral results are dominated by the data obtained after it transitioned to its accreting state, where the (unpulsed) gamma-ray flux is greatly enhanced.

PSR J1748-2815 was co-located with a TS $\sim$300 source in the LAT 8-year catalog. 
Prior to 4FGL-DR1 the LAT team adopted a new model for the diffuse Galactic gamma-ray emission that is
brighter around the Galactic center, and the source disappeared, although its pulsations remain significant with the model weights used here. 

For PSR J1531$-$5610, 2PC mistakenly reported a luminosity $L_\gamma$ over ten times smaller than the value in Table \ref{tbl-psrspec}. An outlier in 2PC, with corrected $L_\gamma$ its efficiency $\epsilon = L_\gamma/\edot$ becomes quite typical.

\clearpage
\startlongtable
\begin{deluxetable}{lrrrrrcc}
\tabletypesize{\scriptsize}
\tablecaption{Spectral fitting results for young LAT-detected pulsars.
\label{tbl-psrspec}}
\tablewidth{0pt}
\tablehead{
\colhead{PSR\tablenotemark{a}}       & \colhead{Energy Flux}  & \colhead{$\Gamma_{100}$} & \colhead{$\epeak$}  & \colhead{$\dpeak$} & \colhead{$b_{\rm free}$}  & \colhead{Luminosity}	        & \colhead{Efficiency\tablenotemark{b}}\\
\colhead{}                     & \colhead{($\rm erg \, cm^{-2}\, s^{-1}$)} & \colhead{}         & \colhead{($\rm GeV$)} & \colhead{}  	    & \colhead{}  	     & \colhead{($\rm 10^{33}\, erg \,s^{-1}$)} & \colhead{(\%)} \\
\colhead{}                              & \colhead{($\times  10^{-11}$)}            & \colhead{}         & \colhead{}         &  \colhead{}  & \colhead{}  	      & \colhead{}				& \colhead{}
}
\startdata
 J0002+6216  &    $1.9 \pm 0.14 $   &   $1.6 \pm 0.30 $   & $ 0.82 \pm 0.16$ &  $ 0.65 \pm 0.09$  & $ 0.85 \pm 0.39$  & $9.01 \pm 0.69 \pm 3.2$ &   $5.9 \pm 0.4\pm 3$ \\ 
   J0007+7303  &    $43.0 \pm 0.33 $   &   $1.2 \pm 0.04 $   & $ 2.44 \pm 0.03$ &  $ 0.36 \pm 0.01$  & $ 0.56 \pm 0.04$  & $101.23 \pm 0.76 \pm 39$ &   $22.7 \pm 0.2\pm 10$ \\ 
   J0106+4855  &    $1.9 \pm 0.07 $   &   $1.2 \pm 0.20 $   & $ 2.37 \pm 0.11$ &  $ 0.67 \pm 0.06$  & $ 1.05 \pm 0.26$  & $21.57 \pm 0.81^{{+21}}_{{-14}} $ &   $73 \pm 3\pm 70$ \\ 
   J0117+5914 \dag &   \nodata &  \nodata & \nodata &  \nodata  & \nodata  &  \nodata &   \nodata \\ 
  J0139+5814 \dag &   \nodata &  \nodata & \nodata &  \nodata  & \nodata  &  \nodata &   \nodata \\ 
  J0205+6449  &    $6.4 \pm 0.20 $   &   $1.6 \pm 0.22 $   & $ 0.26 \pm 0.10$ &  $ 0.45 \pm 0.08$  & $ 0.20 \pm 0.26$  & $78.8 \pm 2.5 \pm 9.5$ &   $0.295 \pm 0.009\pm 0.04$ \\ 
   J0248+6021  &    $2.9 \pm 0.21 $   &   $1.7 \pm 0.26 $   & $ 0.77 \pm 0.10$ &  $ 1.02 \pm 0.12$  & $ 1.16 \pm 0.38$  & $14.1 \pm 1.0 \pm 2.7$ &   $6.6 \pm 0.5\pm 1$ \\ 
   J0357+3205  &    $6.0 \pm 0.18 $   &   $0.73 \pm 0.23 $   & $ 0.87 \pm 0.03$ &  $ 0.95 \pm 0.05$  & $ 0.64 \pm 0.14$  &  \nodata &   \nodata \\
\enddata
\end{deluxetable}


\clearpage
\startlongtable
\begin{deluxetable}{lrrrrrrrrcc}
\tabletypesize{\scriptsize}
\tablecaption{Spectral fitting results for LAT-detected millisecond pulsars.
\label{tbl-mspspec}}
\tablewidth{0pt}

\tablehead{
\colhead{PSR\tablenotemark{a}}       & \colhead{Energy Flux}  & \colhead{$\Gamma_{100}$} & \colhead{$\epeak$}  & \colhead{$\dpeak$} & \colhead{$b_{\rm free}$}  & \colhead{Luminosity}	        & \colhead{Efficiency\tablenotemark{b}}\\
\colhead{}                     & \colhead{($\rm erg \, cm^{-2}\, s^{-1}$)} & \colhead{}         & \colhead{($\rm GeV$)} & \colhead{}  	    & \colhead{}  	     & \colhead{($\rm 10^{33}\, erg \,s^{-1}$)} & \colhead{(\%)} \\
\colhead{}                              & \colhead{($\times  10^{-12}$)}            & \colhead{}         & \colhead{}         &  \colhead{}  & \colhead{}  	      & \colhead{}				& \colhead{}
}

\startdata
 J0023+0923  &    $7.8 \pm 0.6 $   &  \nodata & $ 1.28 \pm 0.15$ &  $ 0.54 \pm 0.11$  & \nodata  & $3.03 \pm 0.21^{{+2.0}}_{{-1.00}} $ &   $19 \pm 1^{{+10}}_{{-6}} $  \\ 
  J0030+0451  &    $60.0 \pm 1.1 $   &   $0.88 \pm 0.15 $   & $ 1.42 \pm 0.03$ &  $ 0.58 \pm 0.03$  & $ 0.53 \pm 0.11$  & $0.743 \pm 0.013 \pm 0.026$ &   $21.3 \pm 0.4\pm 0.8$ \\ 
   J0034$-$0534  &    $20.0 \pm 0.6 $   &   $1.5 \pm 0.16 $   & $ 0.99 \pm 0.09$ &  $ 0.32 \pm 0.03$  & $ 0.65 \pm 0.23$  & $4.38 \pm 0.14^{{+4.2}}_{{-2.8}} $ &   $14.7 \pm 0.5\pm 10$ \\ 
   J0101$-$6422  &    $13.0 \pm 0.4 $   &   $0.87 \pm 0.31 $   & $ 1.34 \pm 0.07$ &  $ 0.53 \pm 0.05$  & $ 0.43 \pm 0.20$  & $1.562 \pm 0.053^{{+1.5}}_{{-1.00}} $ &   $13.3 \pm 0.4\pm 10$ \\ 
   J0102+4839  &    $15.0 \pm 0.7 $   &   $1.2 \pm 0.33 $   & $ 1.70 \pm 0.17$ &  $ 0.41 \pm 0.05$  & $ 0.48 \pm 0.25$  & $9.59 \pm 0.48^{{+9.2}}_{{-6.1}} $ &   $55 \pm 3\pm 50$ \\ 
   J0154+1833 \dag &   \nodata &  \nodata & \nodata &  \nodata  & \nodata  &  \nodata &   \nodata \\ 
  J0218+4232  &    $48.0 \pm 1.2 $   &   $1.9 \pm 0.07 $   & $ 0.33 \pm 0.08$ &  $ 0.26 \pm 0.02$  & $ 0.76 \pm 0.17$  & $57.4 \pm 1.4^{{+35}}_{{-20}} $ &   $23.5 \pm 0.6^{{+10}}_{{-8}} $  \\ 
  J0248+4230  &    $1.9 \pm 0.3 $   &  \nodata & $ 1.80 \pm 0.26$ &  $ 1.39 \pm 0.41$  & \nodata  & $0.79 \pm 0.11^{{+0.76}}_{{-0.51}} $ &   $2.1 \pm 0.3\pm 2$ \\
\enddata

\tablenotetext{a}{A dagger (\dag) means the point source has TS$<25$, hence we show no spectral fit results.}
\tablenotetext{b}{Overestimated distances or the assumed beaming factor, $f_\Omega = 1$, can result in an efficiency $>$ 100\%.}

\tablecomments{Unbinned maximum likelihood spectral fit results  
for the LAT millisecond (MSP) gamma-ray pulsars, using the PLEC4 model (Eq. \ref{eq:plec4} in Section \ref{spectralSection}). 
Column 2 lists the phase-averaged integral energy flux in the 0.1 to 300 GeV energy band, $G_{\rm 100}$. 
Column 3 lists $\Gamma_{100}$, the photon index at 100 MeV (Eq. \ref{eq:g100}), reliable only for fits where the $b$ parameter is left free, and not shown when fixed to $b=2/3$.  
Columns 4 and 5 list the energy $\epeak$ of the maximum of the SED (Eq. \ref{eq:epeak}), and the SED curvature d$_p$ at the maximum (Eq. \ref{eq:disaster}). Column 6 lists the fitted value of $b$ for pulsars with TS $>1000$ (fixed to $b=2/3$ otherwise).
Tabulated values are from the $b$-free fit when available, and the $b=2/3$ fit otherwise.
Columns 7 and 8 give the total gamma-ray luminosity $L_\gamma$ in the 0.1 to 100 GeV energy band, and the gamma-ray conversion efficiency $\eta \equiv L_\gamma/\dot{E}$, assuming
$f_\Omega=1$ as described in Section \ref{lumin}. $\dot E$ is not Doppler corrected. 
The first uncertainties in $L_\gamma$ and $\eta$ come from the statistical uncertainties in the spectral fit, whereas the second are due to the distance uncertainty.
The strong dependence of these quantities on distance (see Table \ref{tbl-PSRdist}) and beaming factor means that these values should be considered with care.}
\end{deluxetable}


\subsection{Searches for Pulsar Flux Variability}
\label{variability}
In 4FGL-DR3, fluxes for sources with {\tt Variability\_Index} $<24.7$ are considered stable. 
Three pulsars exceed this threshold: PSR~J2021$+$4026, cited in Section \ref{pulsedisc} with {\tt Variability\_Index} $= 285$, and the two transitional MSPs J1023$+$0038 and J1227$-$4853 with {\tt Variability\_Index} $= 58$ and $1470$, respectively, which exhibit large flux changes when switching between accretion- and rotation-powered states, but appear to be stable emitters like all other gamma-ray MSPs when in the rotation-powered state. 
In contrast, $\dot E$ of PSR J0540$-$6910 in the Large Magellanic Cloud increased by a third, with no corresponding change in the gamma-ray flux \citep{B0540_spindown_change},
contrary to what the correlation between $L_\gamma$ and $\dot E$ seen in Figure \ref{EDotLumG} suggests ({\tt Variability\_Index} $=11.1$).
A search for variability in our pulsar sample yielded no new cases.

The linear increase in pulsed significance (weighted Htest) with time of PSR J2256$-$1024 changes abruptly near MJD 56600. It has {\tt Variability\_Index} $= 18.3$.
A dedicated analysis confirms that the pulsar flux is constant, but that nearby 4FGL J2250.0+3825, associated with a blazar, has a very bright flare peaking at the same time.
The softer significance slope is thus due to increased background rather than decreased signal.
Exposure changes due to pointing changes also cause Htest vs MJD slope changes. 

PSR J2116+1345 is an MSP discovered using the Arecibo radio telescope in a search of a pulsar-like LAT source.
Gamma-ray pulsations were found using Arecibo and GMRT timing after we froze the list of pulsars reported in this catalog. Its corresponding source, 4FGL J2117.0+1344, has {\tt Variability\_Index} $= 16.2$. Here too, pulsed significance is not linear with time. Analysis showed that the decreased rate of significance growth between MJD 56900 and 57900 is due to the flaring activity of 4FGL J2035.4+1056 (PKS 2032+107) which is 10.5° away from the pulsar.

\citet{SpiderGammaEclipses} discovered dips in the gamma-ray flux of ``spider'' MSPs at orbital phases when the companion star lies between Earth and the pulsar. 
The 4FGL-DR3 fluxes are integrated in both rotational and orbital phase, these eclipses hence do not affect {\tt Variability\_Index}. Orbital modulations \citep[see for example][or the references cited in Table \ref{spiders}]{An2020+J2339} similarly do not influence {\tt Variability\_Index}.

\subsection{Luminosity}
\label{lumin}
Most of the low-frequency Poynting flux generated by the dipole radiation of the rotating neutron star converts to a wind of electrons and ions towards and/or beyond the light cylinder,
carrying away angular momentum and energy, braking the rotation.
About 10\% of the known pulsar population---those in this catalog---convert much of their spindown power into intense gamma-ray beams.
Pulsar energetics contribute a piece to the broader puzzle of Galactic ecology, as recently illustrated by the PeV energy detection by LHAASO of the Crab nebula \citep{LhaasoCrabPevatron} and by the \Fermi\ observation of GeV-band halos around 3 bright HAWC sources \citep{HAWC_LAThalos}, and gamma-ray pulsars allow the most direct look at the acceleration mechanisms.

Integrating the LAT spectral energy flux $E \frac{dN}{dE}$ above 100 MeV yields $G_{100}$, which is tabulated in 4FGL. The range of $G_{100}$ values for the gamma-ray pulsars is apparent in Figure \ref{G100_S1400}. We can then calculate the luminosity
\begin{equation}
 L_\gamma = 4\pi d^2 f_\Omega G_{100},
\label{LumEq}
\end{equation}
using the distance $d$ to the pulsar, and the beaming fraction $f_\Omega$.

\begin{figure}[!ht]
\centering
\includegraphics[width=1.0\textwidth]{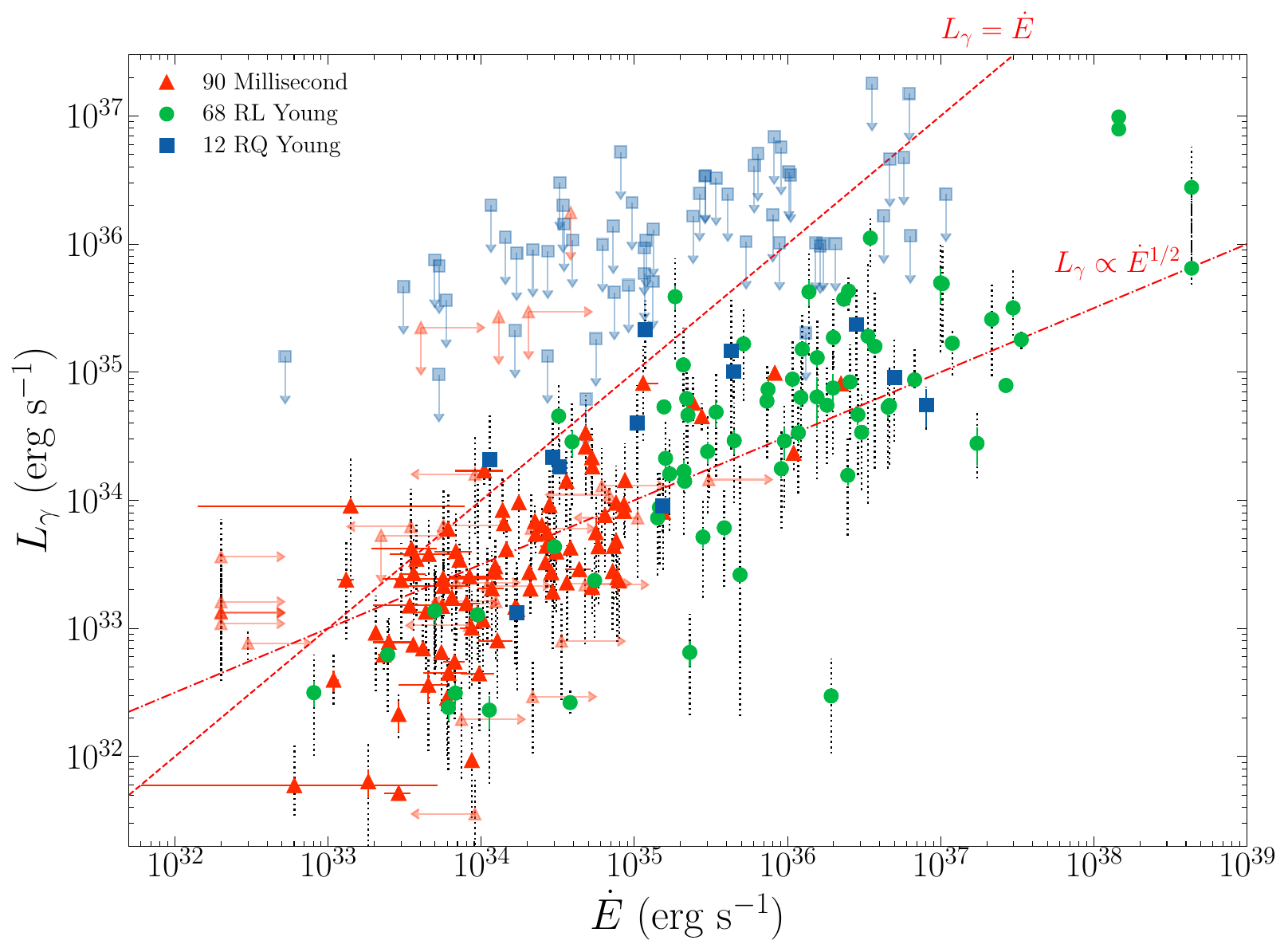}
\caption{
Gamma-ray luminosity $L_\gamma = 4\pi f_\Omega d^2 G_{100}$ in the 0.1 to 100 GeV energy band versus spindown power $\dot E$.
The vertical error bars from the statistical uncertainty on the energy flux $G_{100}$ are colored in the electronic journal version.  
The vertical error bars due to the distance uncertainties are black-dashed, and generally larger. 
Doppler corrections (Section~\ref{doppler}) have been applied to MSPs with known proper motions, leading to visible horizontal error bars in many cases, and when the corrections yield negative $\dot E$, we place the limit at $\dot E> 2\times 10^{32}$ erg s$^{-1}$ for clarity.
Luminosity upper limits use distance upper limits, while $\dot E$ limits result from insignificant proper motion measurements or undetermined distances. We use a light color shade to plot these limits in order to de-emphasize them, and we do not include them in the tallies in the legend.  
The upper diagonal line indicates 100\% conversion of spindown power into gamma-ray flux: pulsars above this line, may have smaller distance $d$, and/or the assumed beam correction $f_\Omega \equiv 1$ is wrong. 
The lower diagonal line indicates the heuristic luminosity $L_\gamma^h\propto\sqrt{\edot}$, to guide the eye.
For the Crab and PSR J0540$-$6919 in the Large Magellanic Cloud, at far right, the upper points include the X-ray energy flux.
\label{EDotLumG}}
\end{figure}

\begin{figure}[!ht]
\centering
\includegraphics[width=1.0\textwidth]{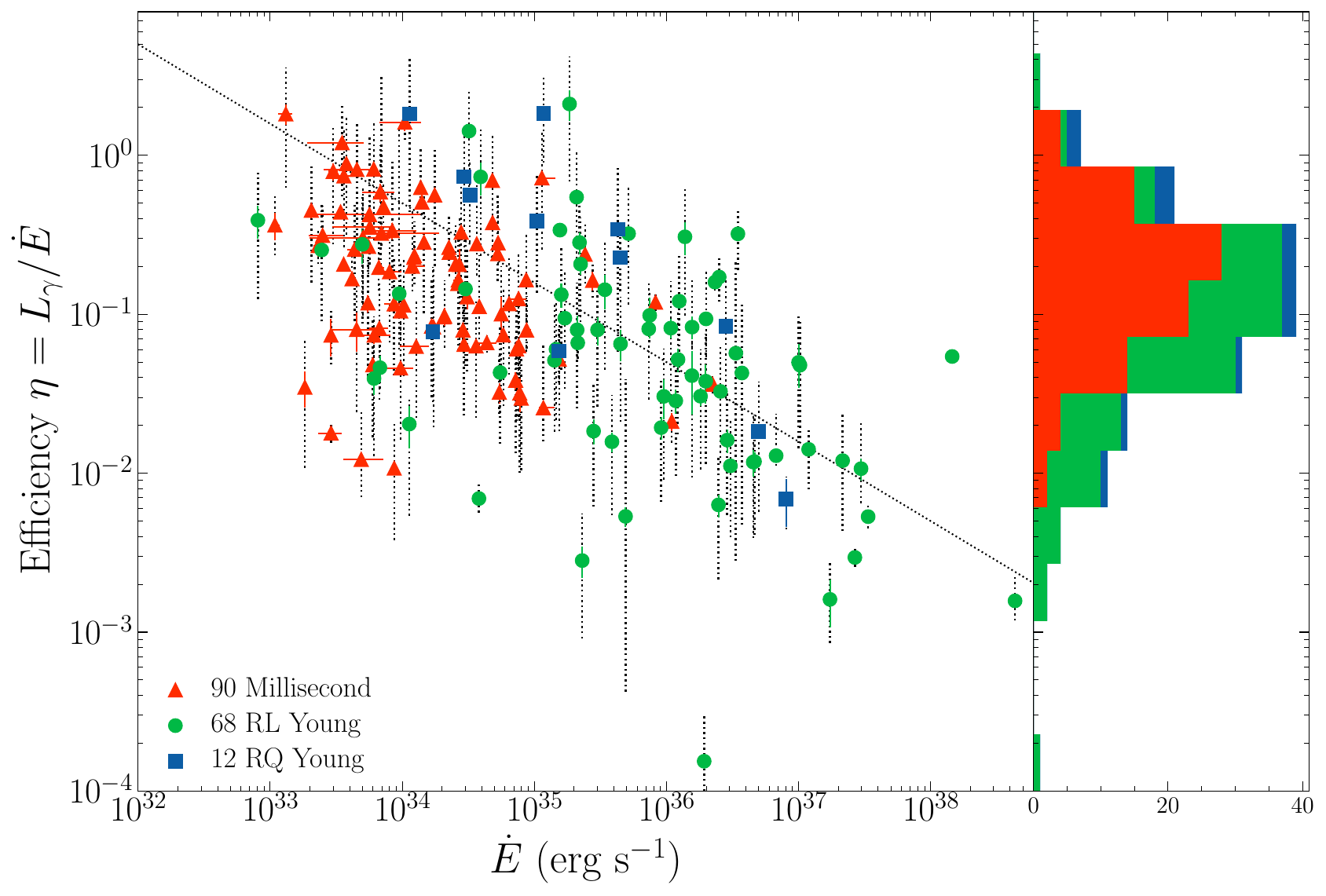}
\caption{
Gamma-ray efficiency $\eta= L_\gamma/\dot E$ versus spindown power $\dot E$. The error bars are as in Figure \ref{EDotLumG}. 
The dashed diagonal shows $\eta \propto \dot E^{-1/2}$, corresponding to the heuristic luminosity $L_\gamma^h$, to guide the eye.
The correlation is more striking than in 2PC.
\label{Eff_Edot}}
\end{figure}

Figure \ref{EDotLumG} shows that the luminosity (Eq. \ref{LumEq}) is highly correlated with the available braking power.
However, at a given $\dot E$ there are two decades of dispersion in the observed efficiency $\eta = L_{\gamma} / \dot E$,
shown in Figure \ref{Eff_Edot}. Distance errors cause some of the dispersion. 
The total radiated power likely depends on the inclination $\alpha$ between the neutron star's magnetic and 
rotation axes, adding more pulsar-to-pulsar variation.
The factor $f_\Omega$ in Eq.~(\ref{LumEq}) is the ratio of total beam power averaged over the whole sky
to the power in the beam slice illuminating the Earth averaged in phase. It is defined in 2PC Eq. 16, and by \citet{AtlasII}, and depends on both $\alpha$ and on the
inclination $\zeta$ between the rotation axis and the line of sight. 
Different models predict different pulse profile shapes and thus $f_\Omega$,
and in any case observational difficulties cause large uncertainties in the $\alpha, \, \zeta$ model inputs. 
Lacking $f_\Omega(\alpha, \, \zeta)$, our choice of $f_\Omega = 1$ likely contributes to the observed  $\eta$ dispersion. A key test of a successful theory of pulsar radiation is its predictions of $L_\gamma$ for given pulsars. {\bf \citet{Kalapotharakos2023}  generally find $f_\Omega < 1$ for the LAT pulsar sample, which would make our $L_\gamma$ values over-estimates.}

The upper diagonal line in Figure \ref{EDotLumG} shows 100\% efficiency, $ L_\gamma = \dot E$, 
for converting spindown power into gamma rays.
Progress since 2PC in determining pulsar distances and MSP proper motions has eliminated pulsars more than $1\sigma$
above the line. 
Furthermore, $R_{\rm NS} = 10$ km is an outdated convention, and $\dot E \propto M_{\rm NS} R_{\rm NS}^2$ is $\approx 2$ times larger than what we use, reducing all $\eta$ values. 
The neutron star mass values $M_{\rm NS}$ in our sample probably have variance $<20$\% \cite[see for example Figure 5 in][]{MeerKAT7Shapiro}, contributing imperceptibly to the luminosity dispersion.


The lower diagonal shows the ``heuristic'' gamma-ray pulsar luminosity, 
\begin{equation}
 L_\gamma^h = \epsilon_{\gamma} \edot \approx  \edot \sqrt{\frac{10^{33}\,{\rm erg \, s}^{-1}}{\edot}} = 10^{33}\, {\rm erg \, s}^{-1} \sqrt{\frac{\edot}{10^{33}\,{\rm erg \, s}^{-1}}} ,
 \label{heurLumEq}
 \end{equation}
motivated by the idea that above some minimum open field-line voltage 
$V \simeq \sqrt{10^{-5}\dot E }$ volts, electron-positron cascades occur in an acceleration region with a self-governing size that produces a gamma-ray efficiency $\epsilon_{\gamma}\propto 1/V$ \citep{Arons96}. Recent particle-in-cell simulations of pulsar magnetospheres also support the idea of a critical pair production rate below which pulsar magnetospheres become ``dead'' \citep[e.g.][]{Kalapotharakos18, Philippov18, Chen20}.  
The scaling of $L_\gamma$ with $\dot E$ depends on whether gamma rays are radiated primarily via synchrotron or curvature processes \citep{Kalapotharakos19}.  We discuss our results in the context of these simulations further in Section \ref{disc}.


\section{The Pulsars Not Seen}
\label{not_seen}
Here we address known pulsars not seen with the LAT, in spite of characteristics shared with gamma-ray emitters.
PSR J1846$-$0258 (Section \ref{J1846-0258section}) is a unique case of a pulsar seen in LAT data only below 100 MeV \citep{Kuiper_J1846-0258}, not using the methods of this catalog.
Finally, we present the LAT's sensitivity for future pulsar discoveries.

\subsection{Radio and X-ray pulsars}
\label{xraypsrs}
Figure \ref{EdotFractions} shows that about 5\% and 35\% of young pulsars and MSPs, respectively, are seen with the LAT in the $\edot \approx 10^{33}$ erg s$^{-1}$ decade, climbing to over 60\% and 80\% at high $\dot E$.
Removing pulsars discovered using LAT sources as seeds, the low $\edot$ fractions decrease to about 3\% and 25\%.
2PC Table 13 listed 28 pulsars with $\edot > 10^{36}$ erg s$^{-1}$ that were not seen at the time,
of the 64 then known. Since 2PC, half of these 28 have been detected and are in this catalog, while the total number increased to 71. The majority of those not yet seen are young X-ray pulsars, discussed below.

With a distance estimate, we calculate the pulsar's ``heuristic'' flux $G_h = L_{\gamma}^h /4\pi d^2$,
which converts to the units of the spectral measurements using $G_{100}/G_h = 3.33 \times 10^{-27} = 10^{-26.48}$. 
Figure \ref{EdotFractions} shows the fraction of detected pulsars rising steadily with $G_h$ above the LAT detection threshold near $G_{100} \approx 10^{-12}$ erg cm$^{-2}$ s$^{-1}$ (detailed in Section \ref{sensitivity}). 
Here, removing pulsars discovered in searches of unidentified LAT sources moves the MSP points closer to those of the young pulsars.
\citet{ThousandFold} discuss whether the non-detection of lower $\dot E$ pulsars is due to a deathline, meaning that the emission mechanisms cease below a minimum braking power, or comes from the small number of pulsars close to Earth and the low detection fraction for low $G_h$.

For the non-detections at high $\edot$, for the young pulsars, \citet{JohnstonSmith2020} find that beam geometry and pulsar inclination dominate (the beam misses the Earth). \citet{DetectedNondetected} observe that broad radio pulses are a good proxy for the inclinations unfavorable to gamma detection. 
An example is the young pulsar PSR J1302-6350 (B1259-63), with large $\sqrt{\edot}/d^2$ and co-located with a LAT source (see Table \ref{ColocatedNotPulse}), yet showing no gamma-ray pulsations \citep[][]{gammaB1259,ThousandFold}. GeV flares occur during periastron every 3.4 years, when the neutron star moves out of the stellar disk of the companion star \citep[][]{MoreGammaB1259flares}. Here, the absence of gamma-ray pulsations is likely due to the beam orientation.  Some gamma-ray pulsars are detected at much lower efficiency than expected from the general trend (e.g. PSRs J0659$+$1414, J0940$-$5428 and J1740+1000 near $\edot = 4,\, 23,$ and $190 \times 10^{34}$ erg s$^{-1}$, respectively, in Figure \ref{Eff_Edot}). If these objects are typical gamma-ray pulsars, then Earth lies well outside of the standard strong beam, and a weak fringe of this beam or an alternate emission component may be responsible for such ``sub-luminous'' gamma-ray pulsars \citep[][]{subluminous}.

Different factors contribute to the larger fraction of MSP gamma-ray detections.
For gamma-rays emitted in the inner magnetosphere, the small light cylinders broaden the beams. However, current sheet emission models predict similar beams for MSPs and young pulsars. MSPs have lower $\dot E$, and thus lower $L_\gamma$ than young pulsars (see Figure \ref{EDotLumG}) but they are also generally closer to Earth.
Figure \ref{fig:epeak_vs_edot} shows that $E_{\mathrm p}$ is slightly higher for MSPs, facilitating detection, as was highlighted by \citet{Kalapotharakos2017} using 2PC data.
The details of which MSPs are not seen by the LAT remain to be clarified.

Another category of as-yet unseen pulsars is those for which the radio rotation ephemerides do not suffice for multi-year gamma folding. An example is the 33 radio MSPs found in deep radio searches targeting unidentified LAT sources that have not yet been confirmed as gamma-ray pulsars (Tables \ref{tab:tallies} and \ref{PSCnotGammaPulse}).
It also occurs for pulsars found in a survey, but never included in a long-term timing campaign: 
psrcat lists 278 pulsars with $30 < P < 800$ ms but without $\dot P$. Nearly half were recently discovered by the GPPS\footnote{\url{http://zmtt.bao.ac.cn/GPPS/GPPSnewPSR.html}} and CRAFT\footnote{\url{http://groups.bao.ac.cn/ism/CRAFTS/}} surveys, both on the FAST telescope \citep{hww+21}, with improved measurements ongoing.  Some of the others have been added to the timing program for \Fermi{} at the Parkes radio telescope, when their declination and radio brightness allow. Of those found to have large $\edot$, a fraction may be detected in LAT data in the coming years.

X-ray timing with \textit{RXTE} enabled the gamma-ray discovery of PSR J0540$-$6919 in the Large Magellanic Cloud (LMC) \citep{B0540_gamma_discovery}. Subsequent \textit{Swift} timing
revealed a rare 30\% increase in $\edot$ and a near-zero braking index \citep{B0540_braking_index}.
\textit{XMM-Newton} timing assisted the LAT pulsation searches for PSR J2022+3842 \citep[][]{Limyansky_thesis}. 

Several of the high-$\dot E$ pulsars unseen in LAT data mentioned above are radio-quiet X-ray pulsars.
Timing noise for such young pulsars is generally high, and while X-ray observations made
over a few years with \textit{XMM} or \textit{Chandra} yield astrometry and the spindown rate, they cannot give the phase-connected ephemeris necessary for multi-year gamma-folding. 
Thus, for many years, only pulsars which have low timing noise and/or are bright enough for routine observations with 
\textit{RXTE} or \textit{Swift} could be gamma-ray phase-folded with the same level of confidence as we have for our thousand radio ephemerides. 

NICER has provided new capabilities, allowing precision X-ray timing for more pulsars than previously possible.
An example is the study of PSR J1813$-$1749, the third pulsar in 2PC Table 13, by \citet{hgp+20}. The pulsar has no 4FGL counterpart and gamma folding yields a null result. \citet{5NicerMSPs} present NICER timing for 5 other high-$\edot$ X-ray pulsars, unseen with the LAT.

X-ray and optical studies of unidentified 4FGL sources have led to the pulsar candidates listed in Table \ref{spiders}. Many of these have led to MSP discoveries in radio pulsation searches using the methods described in Section \ref{psc} or gamma-ray MSP discoveries in restricted blind searches (Section \ref{blindsearch}), once the orbital parameters are better constrained by further multi-wavelength studies. Transitional MSPs, of which PSR J1023+0038 is the archetype \citep[see][and Appendix A]{J1023transition} , may ``turn on'' and show pulsations at some future date. In a similar vein, \citet{ProbablePSRJ1015} identified a point source in the heart of an X-ray pulsar wind nebula that merits pulsation searches.

\tabletypesize{\scriptsize}
\begin{deluxetable}{llrll}
\tablewidth{0pt}
\tablecaption{Candidate ``spider'' and transitional MSPs (tMSPs) discovered in deep searches of previously unidentified LAT catalog sources. BW means black widow, RB means redback, and DStMSP means that a ``disk state'' is presumed to have obscured pulsed detections to date. ``PL'' in place of the SED peak energy (in GeV) indicates that the best spectral fit uses a power law instead of log parabola. Two tMSPs seen to pulse in gamma rays are PSRs J1023+0038 and J1227$-$4853.
All LAT names are 4FGL except for 2FGL J0846.0+2820, detected only in 2009.
\label{spiders}}
\tablehead{
\colhead{LAT name} & \colhead{Type}  & \colhead{$E_{\rm peak}$ }  & \colhead{Class, Assn name} &\colhead{Reference} }
\startdata
 J0212.1+5321 & RB & 1.20 &  bin  1SXPS J021210.6+532136  & \citet{ProbablePSRJ0212_Li, ProbablePSRJ0212_Linares}  \\
 J0336.0+7502 & BW & 1.71 &  lmb  4FGL J0336.0+7502  & \citet{ProbablePSRJ0336}   \\
 J0407.7$-$5702 & DStMSP & PL & ( --  --) & \citet{ProbablePSRJ0407}   \\
 J0427.8$-$6704 & DStMSP & PL &   LMB  1SXPS J042749.2-670434  & \citet{ProbablePSRJ0427}  \\
 J0523.3$-$2527 & RB & 1.77 &  bin  CRTS J052316.9-252737  & \citet{ProbablePSRJ0523}  \\
 J0540.0$-$7552 & DStMSP & 0.59 & (--  --) & \citet{ProbablePSRJ0540}   \\
 J0846.0+2820 & RB &    &   & \citet{ProbablePSRJ0846}   \\
 J0935.3+0901 & RB & 0.50 & ( --  --) & \citet{ProbablePSRJ0935}  \\
 J0940.3$-$7610 & RB & 1.13 &  lmb  2SXPS J094023.5-761001  & \citet{ProbablePSRJ0940}   \\
 J1120.0$-$2204 & RB+ & 1. &  (--  --)   & \citet{ProbablePSRJ1120}       \\
 J1408.6$-$2917 & BW & 0.18 & bin 4FGL J1408.6$-$2917 & \citet{Swihart23} \\
 J1544.5$-$1126 & DStMSP & 0.36 &  lmb  1RXS J154439.4-112820  & \citet{ProbablePSRJ1544}     \\
 J1702.7$-$5655 & RB     & 0.54 &  (--  --)   & \citet{ProbableJ1702}    \\
\enddata
\end{deluxetable}
\normalsize


%


\subsection{PSR J1846-0258}
\label{J1846-0258section}
PSR J1846-0258 is a young ($\tau \approx 723$ yr), high magnetic field ($5 \times 10^{13}\,  \text{G}$) pulsar located in the Kes 75 SNR \citep{Gotthelf2000}.
Although radio quiet \citep{Archibald2008}, it has been timed regularly in X-rays by RXTE, \textit{Swift}, INTEGRAL, and NICER.
PSR J1846-0258 behaves largely as a rotation-powered pulsar, but has twice exhibited magnetar-like outbursts \citep{Gavriil2008,Blumer2021}.
These outbursts have garnered interest in this source as a ``magnetar-pulsar transitional object'', possibly able to shed light on the evolutionary relationship between rotation-powered pulsars and magnetars.
Additionally, it is one of 18 ``soft" gamma-ray pulsars \citep{KuiperHermsen2015}, with strong emission in the MeV range. The X-ray spectrum suggested that the LAT would be able to detect the pulsed high-energy tail, aiding research into the soft pulsar emission mechanism \citep{Torres2018}.

PSR J1846-0258 is the only pulsar with a published gamma-ray detection (pulsed $4.2\sigma$, via unweighted $Z^2_{m=1}$, see Eq. \ref{Eq:Ztest} in Sec.~\ref{pulsedisc}) that is not detected using the methods of this catalog \citep{Kuiper_J1846-0258}. It has no 4FGL-DR3 counterpart source, and thus no $L_\gamma$ measurement.
Following \citet{Kuiper_J1846-0258}'s prescription, we confirmed this detection. It increases to  $4.3\sigma$ using Pass 8 data.
As discussed in Section ~\ref{pulsedisc}, timing solutions may produce significance vs time curves which peak outside the ephemeris validity range.
This is indeed the case with PSR J1846-0258, with a maximum H-Test ($Z^2_{m=1}$) of $4.4\sigma$ ($5.0\sigma$) occurring $\sim484$ days after the end of the published 3295 day timing solution (including a 153 day loss of coherence).
We also used NICER to construct an independent $\sim$770 day timing solution outside the time range covered by \citet{Kuiper_J1846-0258}, which we applied to similarly filtered LAT data.
This did not yield a LAT detection, which given the 4$\times$ shorter span of the NICER data is consistent with expectations from the previous detection \citep{Limyansky_thesis}.

\subsection{Flux Upper Limits and Sensitivity}
\label{sensitivity}

Discovery of most of the pulsars in this list started with an initial phase-integrated source detection of the type applied to all \Fermi-LAT catalog entries.
We summarize the 4FGL procedure: for an all-sky model defined by \texttt{pointlike} \citep{KerrThesis} which includes all current sources, a new trial source is introduced at the position of each of HEALPix nside=256 ($\sim$1/4 deg) pixels. Its flux is optimized using each of five fixed trial spectral shapes. Three are power laws, but two are peaked, at 1 and 2 GeV. The one with the highest TS, if greater than 10, is used as a seed to add to the model. The model is re-optimized with now free spectral parameters, including the new source or sources. All sources are also relocalized. All that survive with TS$>10$ are used for additional iterations. Finally the list is used as seeds for the full catalog \texttt{gtlike} analysis. After this stage only sources with TS$>25$ are included in the final catalog.
Similarly, 2PC started from the 3FGL catalog, which however used only a single power-law trial spectrum.

\begin{figure}[!ht]
\centering
\includegraphics[width=1.0\textwidth]{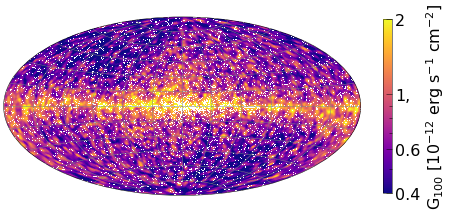}
\caption{Hammer-Aitoff projection of the estimated LAT 12-year sky-survey energy flux pulsar detection sensitivity.
\label{SeattleSkySensitivity}}
\end{figure}

Construction of an estimated sensitivity threshold map, shown in Figure \ref{SeattleSkySensitivity}, followed a similar procedure. For each HEALPix pixel location a trial source was introduced, now with a pulsar-like spectrum, and the flux likelihood function determined. Instead of optimizing it as before, the value corresponding to its 95\% cumulative probability was recorded for that point.  A more precise procedure would have instead inserted simulated trial data with varying intensity, applying the full subsequent analysis procedure, to determine the threshold curve. 

\begin{figure}[!ht]
\centering
\includegraphics[width=0.9\textwidth]{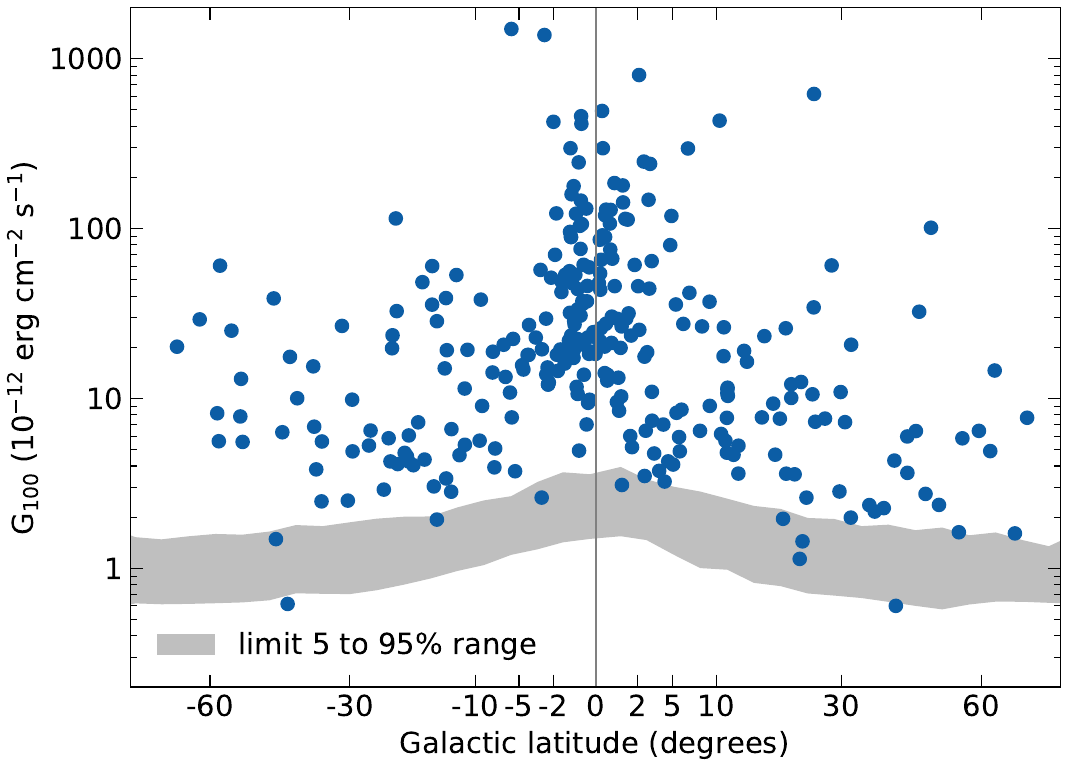}
\caption{ Integral energy flux from 0.1 to 100 GeV, $G_{100}$, versus Galactic latitude $b$ (scaled as $b^{0.65}$ for clarity).
The shaded area shows the 5\% to 95\% percentile range of the 12-year pulsar detection sensitivity in the sky directions at the corresponding latitude. The points correspond to the gamma-ray pulsars in this catalog.
\label{SeattleSkySensitivityProjection}}
\end{figure}

We assume that our procedure yields the correct position dependence. Converting the sensitivity estimates to a 50\% completeness threshold requires an ad hoc scale adjustment, which we obtain by comparing the distribution of sensitivity values to the detected pulsar fluxes, shown in Figure \ref{SeattleSkySensitivityProjection} as a function of latitude.
This detection threshold analysis considers only phase-integrated source detections.

As discussed in 2PC and by \citet{SixWeak}, and further demonstrated by the 15 pulsars in this catalog that were detected via their pulsations but have TS below the 4FGL threshold,
narrow gamma-ray pulses can facilitate source discovery. Such pulsars appear in Figure \ref{SeattleSkySensitivityProjection} near the shaded area.

Finally, a requirement for inclusion in the catalog is that no new candidate source be within $0^\circ.5$ of an existing one. Therefore, in the map in Figure \ref{SeattleSkySensitivity}, we set such pixels around each 4FGL source to {\tt nan}.  The all-sky sensitivity threshold map is provided in FITS format with the online material.

\clearpage

\section{Discussion}
\label{disc}

\subsection{Phase-averaged Spectral Trends}
\label{disc:lumANDspec}

In Section \ref{spectralSection}, we established that the energy where the SED culminates, $\epeak$, and the SED's curvature at that energy, $\dpeak$,
are robust quantities that can be reliably estimated when using the
general $b$-free SED model or the restricted $b=2/3$ case.  $\epeak$,
$\dpeak$, and $b$ all evolve with $\edot$, but the scatter of $b$ and $\epeak$ is much larger
compared to $\dpeak$.  For 90\% of the pulsar sample, 0.5\,GeV$<$$\epeak$$<$2.5\,GeV, with young,
high-$\edot$ pulsars at the bottom of the range
and older, unrecycled pulsars and MSPs at the top.

\begin{figure}
\centering
  \includegraphics[angle=0,width=0.98\linewidth]{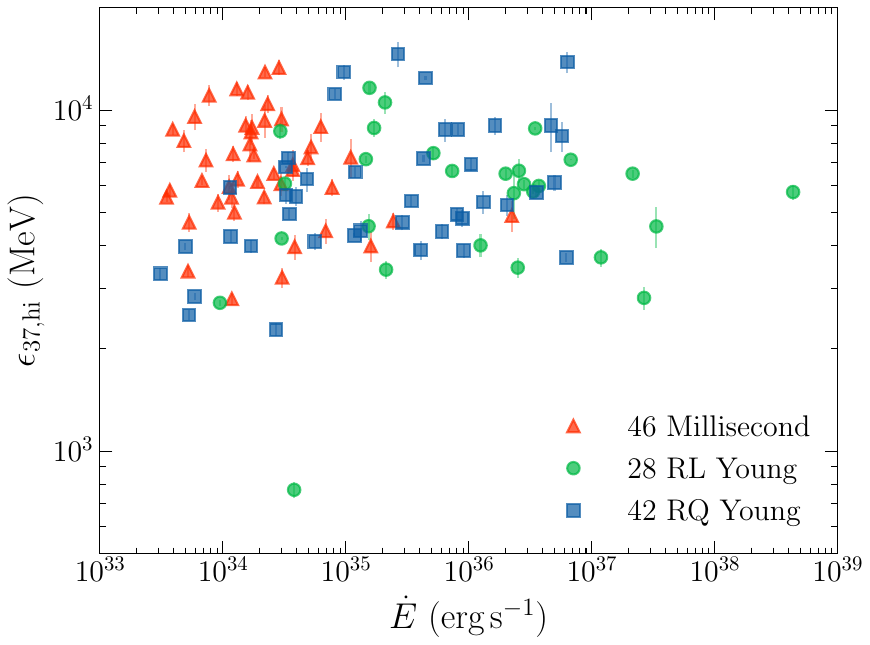}
  \caption{\label{fig:eps37_edot}One realization of the cutoff energy, $\epshi$, takes on a nearly constant value, shown here in the estimates for the $b$-free sample.}
\end{figure}

\begin{figure}
\centering
  \includegraphics[angle=0,width=0.98\linewidth]{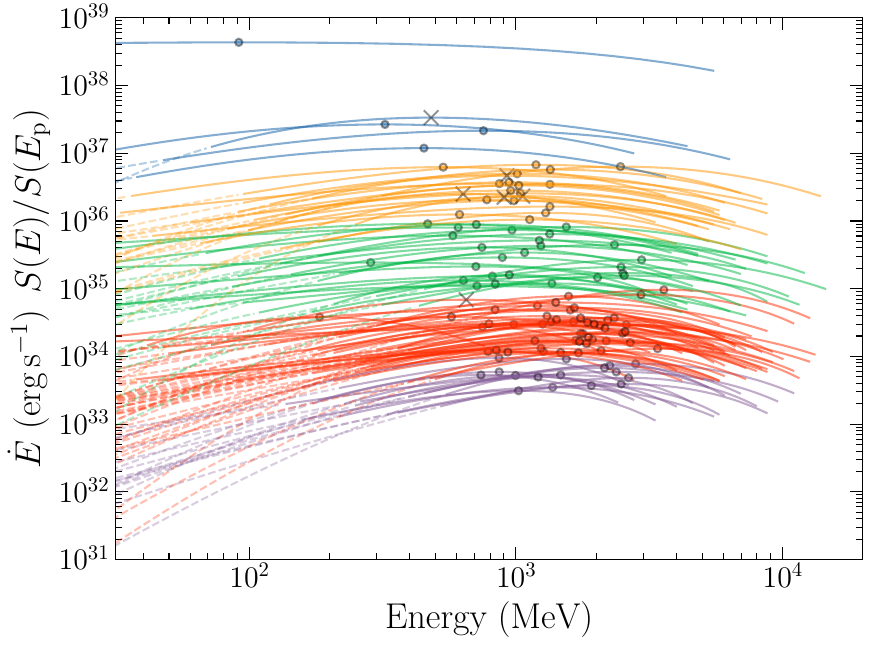}
  \caption{\label{fig:sed_edot}The spectral energy density,
  $S(E)$, for 116 pulsars with an estimate of $b$.  All pulsars have a
  valid $\epeak$, marked with a circle ($b>0$) or a cross (six pulsars
  with $b<0$).  The spectra are normalized such that $S(\epeak)=\edot$.
 Curve colors in $\edot$ decades guide the eye.
  Curves are dashed below $\epslo$ and end at $\epslo$.  }
\end{figure}

A consequence of the opposed $\edot$ evolution of $\epeak$ and $\dpeak$ is a nearly-universal cutoff energy, $\epshi$, at which the SED has fallen $1/e\approx0.37$ from the peak (Figure \ref{fig:eps37_edot}).  Another way of looking at this universality is to consider individual spectral shapes.  Figure \ref{fig:sed_edot} shows the best-fit
SED shapes for the $b$-free pulsars, encoded according to $\edot$.  The region between $\epshi$ and $\epslo$ (the mirror 1/$e$ point below $\epeak$) is highlighted,
allowing a visual estimate of the effective maximum energy
and width.  The upper peak bounds cluster in the 3--10\,GeV range, independently of $\edot$.  Thus, we
have the simple descriptive result: almost all $\gamma$-ray pulsars
emit their power within a two-decade (0.1--10\,GeV) envelope, with the
Crab-like, flat-spectrum, high-$\edot$ pulsars nearly filling the
envelope, while sharply-peaked MSPs concentrate their power into only the
higher decade. In contrast with this absolute energy scale, the spectral curvature as estimated by $\dpeak$ varies strongly with $\edot$.  Selection bias, if any, contributes weakly to this result, as we discuss below.

In  Section \ref{spectralSection}, we also showed that the low-energy
spectral index $\Gamma_{100}$ could not be measured reliably for
$b=2/3$ models, and of the population of 116 pulsars for which it can
be measured, the values tend to cluster around 1.2 with a scatter of a
few tenths and only a very modest evolution with $\edot$.  This result
is clearly visible in Figure \ref{fig:sed_edot}, in which it can
further be seen that many of these SEDs have not transitioned to a
power law at 100\,MeV, making these measurements effectively upper
limits on the asymptotic spectral index.
The low-energy spectral shape is thus another frequent
property of gamma-ray pulsars, taking on values which are well in
excess of the monoenergetic synchrotron radiation spectral index ($\Gamma_0 = \Gamma_{\mathrm{sr}}=2/3$)
for all but few of the lowest-$\edot$ MSPs. 

As stated, the curvature evolves strongly with $\dot E$, and depends on the spectral steepness beyond $\epeak$ and  $\epsilon_{37,\mathrm{hi}}$. Indeed, \citet{Lyutikov2012} and \citet{RichardsLyutikov2018} argue that the shape of the high-energy spectral tail is particularly rich in information about the highest energies reached by the particle acceleration mechanisms, and thus, whether and when the radiation-reaction limit is reached. The tail's shape can reveal, for example, the presence of an inverse Compton ``rebound'', and also informs about the best candidates for $>$10\,GeV emission to be observed by Cherenkov telescopes. 
\citet{BarnardVenter2022} calculate the very high energy Vela spectrum radiated by primary particles in the pulsar magnetosphere and current sheet. 
Measuring the high-energy spectral shape with \textit{Fermi} alone is, however, challenging: the photon rates decrease rapidly, and possible contributions from a surrounding PWN and/or SNR must be handled carefully. We envisage future studies dedicated specifically to these measurements.

The low- and high-energy spectral shapes, and thus the curvature, depend on
the underlying acceleration of leptons; on variations in the local
conditions that govern the balance between acceleration and
radiation, observational convolutions of which depend on pulse phase; and on the extent and re-acceleration of pair cascades.
Disentangling these effects may not be possible with only GeV data,
but the robust evolution of spectral shape demonstrated here is an
important input in constraining models of pulsar emission.
Future successful theories should also reproduce
the approximately constant spectral indices at 100\,MeV, 
the weak decline of $\epeak$ with pulsar $\dot E$, curvatures which track
$\edot$, and the overall shape which evolves from flat,
Crab-like spectra to MSP-like spectra with well-defined power-law
indices and sharp peaks.  These measured trends also provide new
ingredients for population syntheses: by adopting a particular
spectral shape for a given synthesized $\edot$, more reliable
estimates of the LAT sensitivity and selection effects can be
obtained.

The physical implication of these results may be somewhat diluted
since the spectra are phase-averaged, with multiple distinct
acceleration regions (whose observational sampling depends on the pulse phase) contributing to and broadening the measured
spectral features.  On the other hand, we can point to the relatively
tight clustering of $\dpeak$ about the $\edot$ scaling relation, which
suggests that these accelerators must have broadly similar properties.
Indeed, while we established a general trend of harder trailing peaks in Section \ref{sec:profs_energy}, these differences do not seem to be large enough to add appreciable scatter to the observed $\dpeak$ values.  Furthermore, at least some MSPs approach the shape of monoenergetic curvature radiation, 
further limiting the possible variations in spectral shape with phase.  
On the other hand, the preference for $b<1$ for high-$\edot$ pulsars may indicate a richer dependence on phase.

\citet{Kalapotharakos19} examine the mechanisms and configurations at play in the current sheet just outside the light cylinder, and conclude that curvature radiation dominates over synchrotron radiation yielding a ``Fundamental Plane'' (FP) such that $L_\gamma \propto \epsilon_{\mathrm cut}^{4/3}B_S^{1/6}\dot E^{5/12}$. The spectral cutoff energy $\epsilon_{\mathrm cut}$ tracks the pulsar’s highest photon energies, which reflect the highest attained particle energies. \citet{Kalapotharakos2022} developed a proxy for $\epsilon_{\mathrm cut}$ called  $\epsilon_{c1}$, obtained by imposing a PLEC shape with $b=1$ on the 4FGL spectra and then finding the energy $\epsilon_{c1} > \epeak$ where the SED falls to 10\% of its maximum. This is similar to our $\epsilon_{37,{\mathrm hi}}$ shown in Figure \ref{fig:eps37_edot}, and we include \texttt{PLEC\_E10\_b23} and \texttt{PLEC\_E10\_bfr} in the 3PC catalog FITS file, with the important difference that we use the $b= 2/3$ and $b$-free PLEC4 fits that give the best LAT photon data likelihoods.
\citet{Kalapotharakos2022, Kalapotharakos2023} thus improved the FP data fits as compared to their earlier work, finding good agreement with their FP exponent predictions. Further improvement could include selecting pulsars with the most reliable distances, and reviewing the choice of the $\epsilon_{\mathrm cut}$ estimator in light of what we established here regarding measurement bias, and improving the uncertainties used in the FP fits. Applying $f_\Omega$ corrections to $L_\gamma$ for pulsars with reliable estimates of the inclination angles $\alpha, \zeta$ ; including Shklovskii corrections for the MSPs ; and perhaps correcting $\dot E$ for pulsars with neutron star mass measurements may also improve FP tests.

\subsection{Selection Bias}
Are our results representative of all gamma-ray pulsars? 
Here we address whether observation bias might skew our $\epeak$ distribution,
and prospects for discovering pulsars with lower $\epeak$.
In 4FGL-DR3, over 2000 sources have $\epeak < 500$ MeV. 
They were fit with the \texttt{LogParabola} function rather than PLEC4, but that affects the $\epeak$ values only marginally. 
The LAT detection and analysis chain is clearly sensitive to SEDs with maxima below what we see in our sample.

Half of our gamma-ray pulsars were discovered using radio rotation ephemerides of pulsars
discovered independently of the LAT, code ``R'' \footnote{The discovery codes U (LAT unidentified target), P (Pulsar Search Consortium), G (gamma-ray blind search) and R, X (radio or X-ray ephemeris) are defined in the caption for Table \ref{tbl-charPSR} and are included in the electronic catalog.}, generally in radio surveys covering large sky areas, with selection biases a priori unrelated to gamma-ray properties. 
Conceivably, neutron star orientations 
such that both radio and gamma-ray beams sweep the Earth
could correlate with gamma-ray spectral shape. 
The ``X'' pulsars in our sample mitigate
such a putative radio-selection bias somewhat.

On the other hand, the average $\epeak$ of ``U'' pulsars is higher than for ``R'' pulsars, with
``P'' MSPs having higher $\epeak$ than the ``G'' pulsars.
MSPs generally have lower $\dot E$ than young radio-quiet pulsars, so this reflects the $\epeak$ vs $\dot E$ trend we report.
Our target selection procedures may have tended to favor larger $\epeak$, to be addressed in future work.
Because the $\epeak$-$\edot$ anti-correlation is fairly modest with substantial scatter, its slope does not change substantially when excluding LAT-induced discoveries. 

We thus confirm that our highly diverse pulsar sample is striking in its spectral homogeneity: 
almost all gamma-ray pulsars are ``GeV pulsars'', with $0.5 < \epeak < 2.5\,\mathrm{GeV}$ and following the same $\edot$ trends. 

Exceptions exist. Upon its EGRET discovery, \citet{EGRETB0656p14} highlighted the low gamma-ray efficiency of PSR J0659+1414. \citet{subluminous} cast it as the archetype of a hypothesized category of pulsars sub-luminous in gamma rays. It stands out here as having very low $\epeak$ and $\dpeak$ (Figure \ref{fig:dpeak_vs_epeak}) and the lowest $\epshi$
(the isolated dot at lower left in Figure \ref{fig:eps37_edot}).

Another clear exception is the Crab: although its measured $\epeak = 90 \pm 80$ MeV and spectral shape might be construed as a high-$\edot$ continuation of the trend in Figure \ref{fig:sed_edot}, its true SED peak occurs in the soft gamma-ray band \citep{KuiperHermsen2015}.  A further two pulsars, PSRs B1509$-$58 and J1846$-$0258, have spectral shapes similar to GeV pulsars but with values of $\epeak$ well outside the continuum of GeV pulsars.
Both pulsars have $\edot \approx 10^{37}$\,erg\,s$^{-1}$ and little or no LAT signal above 100 MeV.
\citet{KuiperHermsen2015} highlight them as archetypes of a pulsar population with SED peaks in the MeV range,
and show 18 high-$\dot E$ pulsars with SEDs rising in the keV range. Half of the 18 are in this catalog, as are 2 of 5 similar pulsars they discuss.
PSR B1509$-$58 has the best-defined MeV peak. They argue that some of the others may not be detected by the LAT because their SEDs also peak below the LAT's range, as for PSR~J1846$-$0258, at the very limit of our instrument's sensitivity (see Section \ref{J1846-0258section}). However, the SEDs shown by \citet{KuiperHermsen2015} leave open an alternate explanation: 
PSRs J0205+6449 and J2229+6114 indeed have keV slopes similar to PSR~J1846$-$0258, yet are bright up to a GeV.
The as-yet undetected pulsars may simply be too faint, located as they are in very high background regions.
Phase-connected rotation ephemerides for a few of the radio-quiet, noisy pulsars are unavailable.
Others may be GeV pulsars, but with gamma-ray beams that simply miss the Earth.
\citet{HardingKalapo2017MeVpsrs} further discuss the MeV pulsars.

To see whether the paucity of $\epeak < 500\,\mathrm{MeV}$ pulsars reflects magnetospheric emission processes or stems from observational limits, we start with a tally of LAT candidate sources. 
There are 238 unidentified non-variable LAT sources with $\epeak < 500\,\mathrm{MeV}$, undetected beyond 10 GeV,
and significantly curved (well fit with a \texttt{LogParabola} function). Of these, 122 have localisation ellipse semi-major axes with 95\% C.L. $<10$ arcmin, suitable for radio searches (see Table \ref{tab:radio-search}). Restricting the search to the 87 sources with $|b| < 10^\circ$ would favor young, energetic pulsars.
Blind searches for radio-quiet pulsars require a large gamma-ray count rate, even more so if timing noise, glitch rates, or background are high. Unfortunately, only 4 (25) of the 238 targets have TS$>200$ (100). The 4 sources are 4FGL J0340.4+5302, J0426.5+5434, J1749.8-0303 and J1805.7+3401. All were searched several times for radio pulsations but none have been reported to date.

Many high-$\edot$ pulsars have already been found in radio and X-ray searches, effective at finding such sources because high-$\edot$ pulsars tend to have broad radio beams, and because searches can target the supernova remnants near to which such young sources likely remain. 
But high-$\edot$ pulsars are young, and therefore rare, limited by the Galactic supernova rate. 
What is rare is far, and thus faint. The subset of high-$\edot$ pulsars for which only the spectral tail reaches the LAT's sensitivity range is thus even rarer. \citet{Kuiper_J1846-0258}'s detection of PSR J1846$-$0258 importantly demonstrates the LAT's ability to see such objects. 
Learning how many more $\epeak < 500\,\mathrm{MeV}$ pulsars exist will likely have to wait for a
sensitive, large field-of-view MeV instrument such as AMEGO-X \citep{Caputo-2022-JATIS} and ASTROMeV \citep{DeAngelis-2021-ExA}. It would allow the missing SED segments for the known and candidate low-$\epeak$ pulsars to be filled in, and to search for pulsations from new ones. 
But again, the ultimate {low $\epeak$ population is likely to be small compared to that for GeV pulsars, thus
we propose that the trends shown here do capture the spectral character of most gamma-ray pulsars.



\subsection{Light Curve Trends}
\label{disc:lcTrends}

The $237$ energy-integrated profile fits in this catalog generally
confirm the broad categories of light curve shapes known since EGRET,
consolidated in 1PC and 2PC: a majority of gamma-ray pulsars have two principal peaks separated
by $\Delta \approx 0.4 \pm 0.15$ rotations (see Figure \ref{profiles:delta_Delta}). 
For radio-detected pulsars, the first peak generally trails the radio pulse by $\delta \approx 0.2 \pm 0.2$ in phase. 
Gamma-ray emission from near or beyond the light cylinder can produce the observed anti-correlation between $\Delta$ and $\delta$, 
as discussed in 2PC and explicitly reproduced by the models of \citet{Kalapotharakos2014, Kalapotharakos2023}. About 40 pulsars have a single broad gamma-ray peak. 
Very few pulsars deviate from these trends: only two young pulsars in Table \ref{tbl-pulsePSR} are flagged as ``complex'', appearing to have a third peak half-way between the ``standard'' two peaks (one is EGRET pulsar B1055-52). The number increases to 7 for the MSPs in Table \ref{tbl-pulseMSP}. Overall, the MSP and young pulsar profiles are more similar than they differ.

On the other hand, \nwsubbandprofile \, pulsars have enough accumulated photons to allow profile fits in multiple energy sub-bands, fits not done for 2PC.
For the first time, we quantify how profiles evolve with energy for a large sample.
Figures \ref{profiles:Eres_P2overP1} and \ref{profiles:Eres_P2overP1_2} highlight the 28 brightest profiles, fit in at least 6 energy sub-bands.
Reminiscent of the spectral results, profile morphology and evolution are remarkably homogeneous given the large
diversity of pulsar types and environments. Here follow two examples.

In the second panel from the top\footnote{Summary plot examples are Figures \ref{profiles:Eres_example} and \ref{App-Samples:example_eres_0007} to \ref{App-Samples:example_eres_2229}. 
The full sample is in the online supplement.}, 
the energy-resolved fit summaries show the phase evolution for P1 and P2, and P3 for the few cases where it is a strong component.
Changes in the peak separation $\Delta$ are nearly always small, $\lesssim 0.1$ at most.  
For $<5$ pulsars, such as PSR~J2229$+$6114 shown in Figure \ref{App-Samples:example_lightcurve_2229}, $\Delta$ appears to change with energy.
However, inspection of the profiles in Figure \ref{App-Samples:example_lightcurve_2229} shows that individual profile components vary to accommodate slight pulse shape changes, but $\Delta$ in fact varies little.


The summary plots also illustrate the other example. Whereas P3 for the Vela pulsar increases
by $0.1$ in phase over the LAT energy range, visibly detaching from P1 and moving across the bridge region, this turns out to be the only known pulsar where a gamma-ray peak phase varies substantially with $E_\gamma$.
Three pulsars, PSRs J0633+1736 (Geminga), J1747-2958, and J1836+5925 (``the next Geminga''), 
show P3 evolution in the summary plots. Again, inspection of the profiles shows it comes from weak broadening of P2 rather than a peak drift as seen for Vela.

Such weak dependence of pulse phase on energy does not necessarily imply that the region illuminating an observer is uniform. Relativistic aberrations and time delays create caustics such that photons emitted at different distances from the neutron star remain in narrow phase ranges. Thus, whether different electron energies occur at different magnetospheric depths or not, we could see little light curve evolution with photon energy.
Nevertheless, since the spectral energy distributions suggest populations of gamma-ray emitting electrons with a narrow range of maximum energies for the lowest $\dot E$ pulsars, broadening gradually for systems with more available power,
the light curves also encourage vision of emission zones along caustic viewing lines that change little with electron energy. 
Figure \ref{fig:epeak_Delta} attempts to relate spectral and profile properties: it suggests that when the line of sight cuts across the emission region such that the viewer sees two closely-spaced peaks (small $\Delta$), then the line of sight sometimes samples a region with the highest electron energies (large $\epeak$).

Indeed, how the line of sight intersects the emission region is key. 
Beginning with \citet{Watters09}, and continuing as mastery of magnetospheric processes evolves \citep[e.g.][]{Venter2009,ContopKalapo2010,BaiSpitko2010,ArkaDubus2013,Kalapotharakos2014,HardingAtlas-like,Cerutti2016,Kalapotharakos18,Philippov18,PetriAtlas-like,Cerutti2020,Kalapotharakos2023},
modelers have provided ``atlases'' allowing comparison between observed and calculated pulse profiles.
Sky maps show beam intensity versus neutron star latitude and longitude, that is, as a function of the inclination $\zeta$ between the rotation axis and the line of sight, 
and the rotational phase $\phi$, with other parameters fixed. 
A cut across a sky map yields the profile sampled by a given line of sight.
The dominant ``other'' parameter is the magnetic inclination $\alpha$, but e.g. period or magnetic field strength have more or less importance for different models. 
Some theories favor emission from plasmoids in current sheets beyond the light cylinder, where magnetic reconnection accelerates electrons to high energies to then radiate synchrotron gamma rays \citep{PetriStripedWind,Cerutti2020}. Other models favor electric field acceleration near the current sheet and curvature radiation.  In either case the picture is quite different from the various ``gaps'' that previously dominated the field \citep[][]{PhilippovKramer2022},
and the resulting sky maps differ in consequence. 
The particle-in-cell (PIC) calculations of the relativistic magneto-hydrodynamic (MHD) processes consume substantial computer resources. Realistic atlases are currently out of reach, with progress in kinetic plasma simulations expected in the coming years.

Figure \ref{fig:dutycycle} histograms the gamma-ray duty cycle $D$, the total phase for which the fitted profile is $>25\%$ of the range from its baseline to the maximum. (The online catalog also includes $D$ with 10\% and 50\% thresholds, and \textit{psrcat} includes radio pulse width.) 
\citet{SixWeak} highlighted some pulsars with large duty cycle, and discussed how detection sensitivity depends on $D$.
With two exceptions, the young radio-loud pulsars have $D < 0.4$. In contrast, 3 times more young radio-quiet pulsars have $0.3 < D < 0.5$ as loud ones. 

Radio detection occurs when $|\zeta - \alpha| \lesssim \rho$, that is, along a broad diagonal
swath of the plane of neutron star inclinations $(\alpha,\, \zeta)$  \citep[see Figure 3 of ][]{JohnstonSmith2020}. The opening angle of the radio cone is $\rho = 3 \sqrt{\pi h/2Pc}$, for emission height $h$. 
Gamma-ray detection favors the $\alpha > 30^\circ$ and $\zeta > 30^\circ$ part of the plane.
Finally, long duty cycles arise when the line of sight slices a skymap in a $\zeta$ range with emission extended in $\phi$, but these ``flat'' beam shapes favor small $\alpha$ \citep[see e.g. Figure 15 of ][]{Kalapotharakos18}. The tallies in Figure \ref{fig:dutycycle} are in good qualitative agreement with this simple geometric beam model. 
Thus, study of the high duty cycle pulsars may help constrain $(\alpha,\, \zeta)$ and/or discriminate between models. The small light cylinders characteristic of MSPs make for broader beams, the likely reason for the MSP $D$ tail in Figure \ref{fig:dutycycle}.

If $D$ could constrain the ``beaming fraction'' $f_\Omega$ used to calculate luminosity $L_\gamma$ (Eq. \ref{LumEq}), the dispersion in the $L_\gamma$ versus $\dot E$ could be reduced.
Large $D$ pulsars, with their distinctive skymaps,
could be particularly useful. But they are rare: the $\alpha$ range producing the right beam shape is limited, and the line of sight must fall into a narrow $\zeta$ range. PSR J1857+0143, the young radio-loud outlier in Figure \ref{fig:dutycycle}, may be an example. \citet{sjk+21} show its polarization position angle versus phase at $1.4$ GHz, but unfortunately scattering excludes a rotating vector model (RVM) estimate of $(\alpha,\, \zeta)$ at this frequency while its narrow radio pulse width poorly constrains the RVM at higher frequencies.  Table 6 points out unidentified LAT sources co-located with known pulsars for which we have not seen gamma-ray pulsations: some may be $D \approx 1$ pulsars and could benefit from a  $(\alpha,\, \zeta)$ study.

The differences in spectral hardness (Figure \ref{profiles:Eres_P2overP1}) between the peaks nearer (P1) and farther (P2) from the magnetic pole, from the
observer's point of view, is another powerful discriminant, mentioned above. 
\citet{Harding2021} and \citet{BarnardVenter2022} address this specifically.
Early LAT studies measured spectra in as many phase bins as the statistics would allow \citep{Geminga1,LAT3EGRETpulsars,LATVela2,MeganThesis}. The observed phase evolution of the spectral parameters is smooth, and the range of parameter values is moderate (the cutoff energy can vary by a factor of 2). A simple analysis, measuring on-peak spectral shape in the peak phase intervals we provide, is easier to perform than a full phase-resolved spectral analysis that samples light curves at many points. 
Furthermore, a larger sample of pulsars is bright enough for on-pulse measurements than for full phase-resolved analysis, yet a simple analysis is well-suited to the current model accuracy. Such future work is encouraged.


\begin{figure}
\centering
  \includegraphics[angle=0,width=0.98\linewidth]{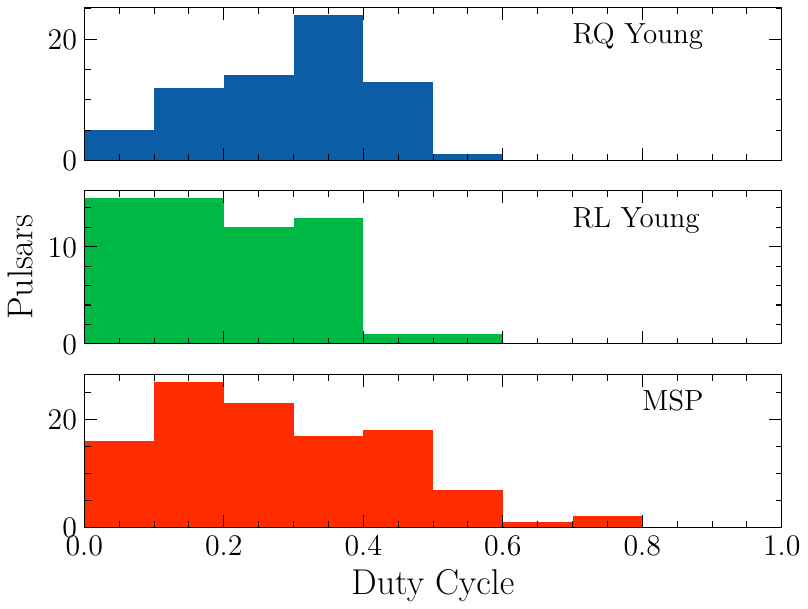}\\
  \caption{\label{fig:dutycycle} Duty cycle $D$ of the gamma-ray pulse profiles (fraction of a rotation that the fitted profile exceeds 25\% of the maximum amplitude, measured from the background level). More young radio quiet pulsars have large duty cycles than do the young radio loud pulsars.}
\end{figure}

\subsection{Luminosity and Pulsar Populations}
\label{disc:pops}
Our large gamma-ray pulsar sample, along with the known radio pulsars, is a starting point for a census of the total Milky Way population. Characterizing the population sheds light on many open questions. These include the supernova rate and neutron star evolution. The large number of ``spider'' binary MSPs caused a renaissance in the study of binary evolution and pulsar recycling \citep{Strader19,SpiderShocks,Swihart23}. The number and luminosity of high-energy pulsars helps determine the relative contributions of pulsars and putative dark matter annihilation to the observed cosmic positron spectrum \citep[][]{Venter2015,MWpsrsPositrons} and, via pulsar wind nebulae, to the leptonic cosmic ray spectrum generally. Applying emission models to a synthetic population and then simply tallying the observed numbers of, for example, radio-loud versus radio-quiet pulsars tests predictions of beam geometry \citep[][]{JohnstonSmith2020}.

The strong dependence of $L_\gamma$ on $\dot E$ (Figure \ref{EDotLumG}) and the LAT's sensitivity (Figure \ref{SeattleSkySensitivity}) are key inputs. Key tests of a synthesis's fidelity are the pulsar latitude distribution (Figure \ref{Aitoff}) and the cumulatative logN-logS distribution of our pulsar sample, shown in Figure \ref{fig:logNlogS}.
The slopes in the logN-logS plot are related to the spatial distributions of the sources:
a flux-limited sample of a population of sources having a narrow luminosity distribution has slope -3/2 if the population is isotropic, and -1 if the distribution is disk-like. 
Indeed, in Figure \ref{fig:logNlogS} the MSPs have a steeper slope, being more isotropic, while the young pulsars have a slope close to -1.  The turnover at low fluxes is the detection threshold, lower for objects away from the intense background in the plane. Distance uncertainties do not affect the logN-logS plot. However, an extrapolation from the observed to the unresolved population needs to take selection biases into account. One example is that most of our young radio-loud pulsars were discovered in radio surveys near the plane -- results sensitive to details of the high-latitude population may suffer from bias. Another is that target sources for gamma-ray blind searches for radio-quiet pulsars have a higher threshold than shown in Figure \ref{SeattleSkySensitivity}, explained in Section \ref{blindsearch}.

\begin{figure}
\centering
  \includegraphics[angle=0,width=0.8\linewidth]{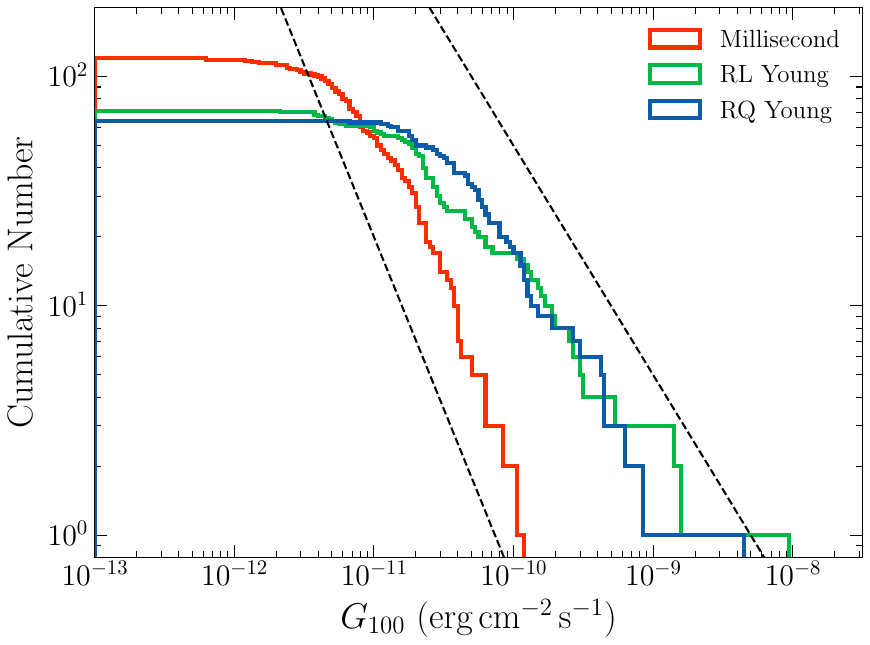}\\
  \caption{\label{fig:logNlogS}The cumulative number of pulsars detected with $>$100\,MeV flux over $G_{100}$, ``logN-logS''.  The lines show slopes of $-1$ and $-3/2$.}
\end{figure}

The scatter around the $L_\gamma$ versus $\dot E$ correlation (or, equivalently, of efficiency versus $\dot E$ shown in Figure \ref{Eff_Edot}) is a combination of physics and uncertainties.
The ``physics'' bias includes beaming, since $f_\Omega = 1$ in our $L_\gamma$ computation. We use a canonical moment of inertia $I = 10^{45}$ g cm$^2$ for all pulsars, but the true average is probably twice that, since neutron star radii are closer to 14 km than to 10 km, shifting the x-axis of all $\dot E$ plots to the right. Furthermore, especially MSPs have a range of masses, of order only 60\%, neglected in our single $I$ value. Luminosity may also depend on parameters other than $\dot E$, such as the inclination angles, which would contribute to the scatter. The correlation is stronger at higher spindown powers, and is less pronounced for the MSPs. It remains to be understood whether this is due to high energy processes ``shrinking'' as $\dot E$ tends toward the ``deathline'' near $10^{33}$ erg s$^{-1}$, or is because e.g. of the compactness of the MSP magnetosphere.

Highly polarized radio beams generally occur only for pulsars with $\edot > 10^{34}$ erg$ s^{-1}$ \citep{PolnEnergeticPSRs, PolnPSRs2018}. 
\citet{ThousandFold} suggested it is an unlikely coincidence that both radio polarization and gamma-ray emission require the same minimum spin parameters for independent reasons -- that these two phenomena share the same deathline may mean that the same electron populations produce both signals. 

\subsection{Towards 4PC and 400 gamma-ray pulsars}
A small number ($\sim 20$) of our gamma-ray pulsars are below the 4FGL catalog detection threshold. Thus we expect that the bulk of future pulsed discoveries will correspond to the thousand or more currently unidentified LAT sources. Requiring non-variability and a pulsar-like spectrum reduces this to a few hundred. It thus appears unlikely that the LAT will ever more than double the size of the current sample.  

However, another hundred gamma-ray pulsars is possible, and dozens are likely.
Machine-learning methods applied to the 4FGL sources typically find $\sim 70$ pulsar-like unassociated objects with $|b|>10^\circ$ \citep{MirabelMachineLearning}.
Future high-sensitivity radio surveys with SKA are predicted to find $27,000$ young pulsars and $3,000$ MSPs \citep{SKApsrCensus}. Radio timing solutions for those co-located with LAT source would allow gamma-ray folding, at least for the years following the radio discovery. A radio position, period, and orbital period for the binaries would improve the sensitivity and decrease the computing cost for gamma-ray blind searches. Without waiting for SKA, TRAPUM at MeerKAT and FAST are already finding new pulsars. Sustained timing of the new discoveries will require thoughtful management of available resources. The insights we have achieved in this work into the shapes of pulsar spectra will help quantify how many of the unidentified LAT sources are likely to be pulsars without pulsed detections.  

We intend to maintain the online catalog (see the footnote at the end of Section \ref{intro}). We will add new gamma-ray pulsars as they accumulate. If new analyses are performed, e.g. on-peak spectral analysis, we will add those results. 

Pulsar observations enable investigations over a spectacular range of topics, from fundamental physics to variations of the interstellar medium, including several aspects of stellar evolution, and more. Our gamma-ray pulsar sample is so large that even sub-samples selected for specific studies have more members than did the entire collection acquired by \textit{Compton Gamma-ray Observatory}, over twenty years ago. Discoveries enabled by exploiting the sample will expand for many years to come, at the same time that \Fermi{} LAT keeps finding more, and better characterizes those already known. 


\clearpage

\appendix
\label{appendices}

\section{Radio Timing for This Work}
\label{TimingAppendix}
%
%
\textbf{PSR J0653+4706: }
PSR J0653+4706 was discovered with LOFAR \citep{LOFAR} at 135\,MHz in a targeted pulsar survey towards unidentified \Fermi\ LAT gamma-ray sources \citep{pbh+17, bph+17}.
The survey used semi-coherent dedispersion to minimize the effects of
dispersive smearing at these low frequencies while retaining sensitivity to
MSPs \citep{LOFARmspSearching}. The pulsar has
a 4.76\,ms spin period and a dispersion measure of
$\mathrm{DM}=25.74 \pm 0.14$\,pc\,cm$^{-3}$, placing it at a distance of $\sim900$\,pc
with both the NE2001 and the YMW16 models. It is in a 5.84\,day binary system with a
$M_\mathrm{c}\ga0.21$\,M$_\odot$ binary companion \citep{bph+17a}. The timing solution used in the catalog is obtained from LOFAR timing observations covering a 2.8\,yr timespan. Subsequently, a $5.2$ year solution using Jodrell Bank Observatory L-band observations confirmed the model.

The LOFAR data indicate a flux density of $5 \pm 2$ mJy at 150 MHz. An 1800 s Green Bank Telescope observation at 820 MHz, with 200 MHz bandwidth, yielded a $37.2 \sigma$ detection. For 2 polarizations, gain $G = 2$ K Jy $^{-1}$, system temperature T$_{sys} = 29$ K, and a 12\% duty cycle seen in both the 150 and 820 MHz profiles, we obtain a flux density of $260\, \mu$Jy, calculated using the PRESTO {\texttt sum\_profiles.py} routine.

Analysis of 101 positive detections at 1400 MHz with the Lovell telescope at Jodrell Bank Observatory yields flux densities of $89 \pm 11$ and $10 \pm 3\, \mu$Jy for the main peak and interpulse, respectively. The uncertainties are statistical, and include variations due to interstellar scintillation. The total is $S_{1400} = 99 \pm 23\,  \mu$Jy, having added the 20\% systematic calibration uncertainty in quadrature.
The spectral index obtained from these three measurements agrees well with the average value of $1.7$ of the pulsar population.

The JBO ROACH timing data show no significant pulsar proper motion. We found {\tt PMDEC} $= -1.5(2)$ and {\tt PMRA} $= 0.5(1.8)$. The Doppler correction is hence $0.98 <  (1-\xi) <1$, with $ (1-\xi) =  \dot E^{int}/\dot E $ as in Section \ref{doppler}.

\textbf{PSR J1023+0038: }
PSR~J1023$+$0038 was the first transitional millisecond pulsar to be discovered \citep{Archibald2009+J1023}, and it underwent a transition from a rotationally-powered MSP state to an accretion-powered state in 2013 \citep{J1023transition}. A faint source coincident with the pulsar was detected by the LAT in the pre-transition data \citep{2FGL}, and the flux from this source increased five-fold when accretion began \citep{J1023transition}.

Using the radio timing ephemeris of \citet{Archibald2013+J1023}, we folded the pre-transition data (from the beginning of the LAT data to 2013 June), discovering significant pulsations, with $H=62$. This ephemeris contains a single orbital period derivative to account for long-term trends in the orbital period, but does not account for shorter-timescale orbital phase variations seen in radio timing observations. We refined this solution using an unbinned timing analysis on the LAT data, also using only one orbital period derivative, resulting in an increased $H=92$.  

We also folded the post-transition data using a timing model built from X-ray timing by \citet{Jaodand2016+J1023} and X-ray and optical timing from \citet{Illiano2023+J1023}. We fit a polynomial function to the orbital phase deviations measured by \citet{Illiano2023+J1023}, and folded the LAT data using this orbital model and the spin-down model of \citet{Jaodand2016+J1023}. No pulsations were detected. We also searched for pulsations with slightly different spin periods and spin-down rates, and with slight shifts in orbital phase, using a sliding window approach, but again did not detect any significant pulsations. 

This non-detection implies that the flux increase due to accretion is not an enhancement of the pulsed flux (seen in other non-accreting redbacks, e.g. \citealt{An2020+J2339,cnv+21}) , but is instead an additional unpulsed component. If the five-fold increase in gamma-ray flux is indeed due to unpulsed emission from the accretion process, and the gamma-ray pulsations remain unaffected by the accretion,  then the pulsed fraction would be reduced to $20\%$ of the pre-transition level, and this would leave the pulsations undetectable to our analysis. Like PSR~J1227$-$4853 \citep{jrr+15}, it therefore remains unclear whether or not this system emits gamma-ray pulsations in the accretion-powered state. 

\textbf{PSR J1827$-$0849: }
Applying the methods of \citet{ClarkEatHomeI} and Section \ref{blindsearch} to LAT data around the position of a steep-spectrum radio source coincident with unidentified LAT source 3FGL J1827.6-0846, found by \citet{FrailUnIds}, we discovered gamma-ray pulsations with spin period $2.24$ ms in 2016. The radio positional uncertainties of $\sim 2^{\prime \prime}$ have greatly reduced the computing cost and allowed us to increase our sliding coherence window length to $2^{23}$ s (or $\sim 97$ days) and still perform the search on the ATLAS cluster\footnote{\url{https://www.aei.mpg.de/43564/atlas-computing-cluster}} in about 1 day per source.

The associated radio source has flux density $229.9 \pm 14.8$ mJy in the GMRT 150 MHz All-Sky Survey, and $777 \pm 126$ mJy in VLSS$_r$ at 74 MHz. NVSS has no detection, giving $S_{1400} < 0.8$ mJy per beam. Table 8 lists an insignificant proper motion estimate from LAT timing as per Section \ref{LATtiming}. C.J. Clark and co-workers are studying other LAT unidentified sources co-located with steep-spectrum radio sources and will detail their methods and results in future work.

\textbf{PSR J1833$-$3840: }
This $1.87$ ms pulsar was discovered in 2015 by F. Camilo, in observations of unidentified LAT sources using the Parkes radio telescope, using the methods reported by \citet{ckr+15}.
Later, \citet{FrailUnIds2} independently highlighted it in a search for steep radio spectrum pulsar candidates in LAT unidentified sources using the GMRT 150 MHz all-sky survey (TGSS ADR1). Fewer than 10\% of known pulsars have such high spindown power, $\dot E = 1.1 \times 10^{35}$ erg s$^{-1}$, in the Milky Way or in globular clusters. The orbital period is $0.9$ days. In 2020, then-undergraduate Camryn Phillips phase-connected GBT timing observations during \texttt{PINT} code development she was conducting. We phase-folded the gamma-ray data using her ephemeris and obtained the pulsed detection reported here. 

\textbf{PSR J1852-1310: }
The $P=4.31$ ms isolated pulsar PSR J1852$-$1310 was discovered in a 1900\,s pointing on MJD 57793 towards a LAT unidentified source 
with a pulsar-like spectrum, at 800\,MHz using the GUPPI backend on the Green Bank radio telescope. 
It was initially named PSR J1852$-$13.
Routine timing observations at 1400\,MHz with the Lovell telescope at Jodrell Bank Observatory began on MJD 57818. 
As of MJD 59927, 124 arrival times have been acquired using ROACH, allowing a phase-connected rotation ephemeris,
provided with the online material.
The spindown power is $\dot E = 4.95 \times 10^{33}$ erg s$^{-1}$. 
For DM $= 44.95 \pm 0.01$\,pc\,cm$^{-3}$ in the direction of this pulsar, the YMW16 model gives distance $1.27 \pm 0.5$ kpc.
The integral energy flux of co-located 4FGL J1852.2$-$1309 thus gives $L_\gamma = (5.8 \pm 1.0) \times 10^{32}$ erg s$^{-1}$,
for a gamma-ray efficiency of 11\%. 

GBT timing observations at 1400 MHz between MJD 58698 and 59050 yielded 49 times of arrival. 
Combining these with the JBO data slightly improves the MSP's proper motion measurement, and we obtain
 {\tt PMDEC} $= -6.6 \pm 1.4 $ and {\tt PMRA} $=  -1.84 \pm 0.25$. 
The Doppler correction is hence $(1-\xi) = 0.91$, as defined in Section \ref{doppler},
such that  $\dot E^{\mathrm int} = (4.6 \pm 0.2) \times 10^{33}$ erg s$^{-1}$.

Gamma-ray phase folding $14.5$ years of data using the JBO timing model revealed pulsations with test statistic $H = 122$ 
using the simple weights parameter $\mu_w = 4.3$. 
Model weights increase this to $H = 133$. 
$H$ increases steadily for the $8.5$ years before the start of the timing model, indicating that the MSPs
spindown rate is stable, well-described using only spin frequency and its first derivative.
The radio source is bright ($250 \sigma$ in the discovery observation) and the profile has sharp features, hence this MSP may rate inclusion in gravitational wave searches.

\textbf{PSR J1857+0943: }
This $5.36$ ms pulsar has long been known for its stable rotation rate \citep{srs+86}.
Gamma-ray phase folding using the rotation ephemeris provided by \citet{rsc+21} yielded H-test $> 30$, using simple weights. Located in a high background region, $(l,b) = (42.2,   3.12)$ degrees, the gamma-ray profile is ill-defined but appears to be dominated by a single narrow peak. The timing parallax distance from Table \ref{tbl-doppler} and the integral energy flux of the co-located LAT source 4FGL J1857.2+0941 give $L_\gamma = (1.27 \pm 0.31 ^{0.6} _{-0.9}) \times 10^{33}$ erg s$^{-1}$, where the first uncertainty comes from the flux measurement and the second from the distance, about 30\% of the spindown power.

\textbf{PSR J1901$-$0125: }
The same procedure as described for PSR J1827$-$0849, above, led to the 2016 discovery of $2.79$ ms gamma-ray pulsations. F. Camilo then re-analyzed GBT radio data from previous observations of the LAT source position using the LAT rotation ephemeris and discovered radio pulsations and measured the DM.  Table 3 lists $S_{1400} = 2.50$ mJy, extrapolated from the \citet{FrailUnIds} GMRT flux density of $362 \pm 15$ mJy at 150 MHz and radio spectral index $-2.23$. 
Table 8 lists a $2.4\sigma$ proper motion measurement using the methods of Section \ref{LATtiming}. Clark and co-workers are studying other LAT unidentified sources co-located with steep-spectrum radio sources and will detail their methods and results in future work.

\textbf{J2256$-$1024: } \citet{csm+20} found $\mu = 9 \pm 3$ mas yr$^{-1}$ using GBT timing data.  NRT has also monitored this pulsar, but with a substantial gap between the end of the GBT data and the beginning of the NRT observations.  This gap and limited frequency coverage cause a strong degeneracy between the spindown parameters, time-varying dispersion measure, and orbital frequency evolution.  We therefore perform a hybrid analysis, using the methods of Section \ref{LATtiming}, in which we use LAT data to determine the ``long-term'' parameters (F0, F1, FB0, FB1, and FB2).  We then fix these parameters and fit for the astrometry using the radio data, finding a proper motion of PMRA $ = 2.80 \pm  0.05$, PMDEC $ = -5.30 \pm  0.11$ from which $\mu = 6.0 \pm 0.1$  mas yr$^{-1}$.  A systematic error comparable to or larger than the statistical uncertainty is likely, due to the inhomogeneous data used in the analysis.


\section{Sample light curves, profile evolutions, and spectra.}\label{App-Samples}
\renewcommand{\thefigure}{B-\arabic{figure}}
\setcounter{figure}{0}

\makeatletter
\@addtoreset{figure}{section}
\makeatother

Figures~\ref{App-Samples:example_lightcurve_0007} to \ref{App-Samples:example_lightcurve_2229} show examples of LAT pulsar light curves in different energy bands, with the radio pulse profiles overlaid when available. See Section~\ref{profiles} for details about the construction of these light curves. All the other pulse profiles are available in the supplementary online material, detailed in Appendix~\ref{online}. \\

Figures~\ref{App-Samples:example_eres_0007} to \ref{App-Samples:example_eres_2229} show examples of energy-resolved light curve fitting analysis summaries, as described in Section~\ref{profiles}. The plots for the remaining pulsars are in the supplementary online material. \\

\begin{figure}[!ht]
\centering
\includegraphics[scale=0.37]{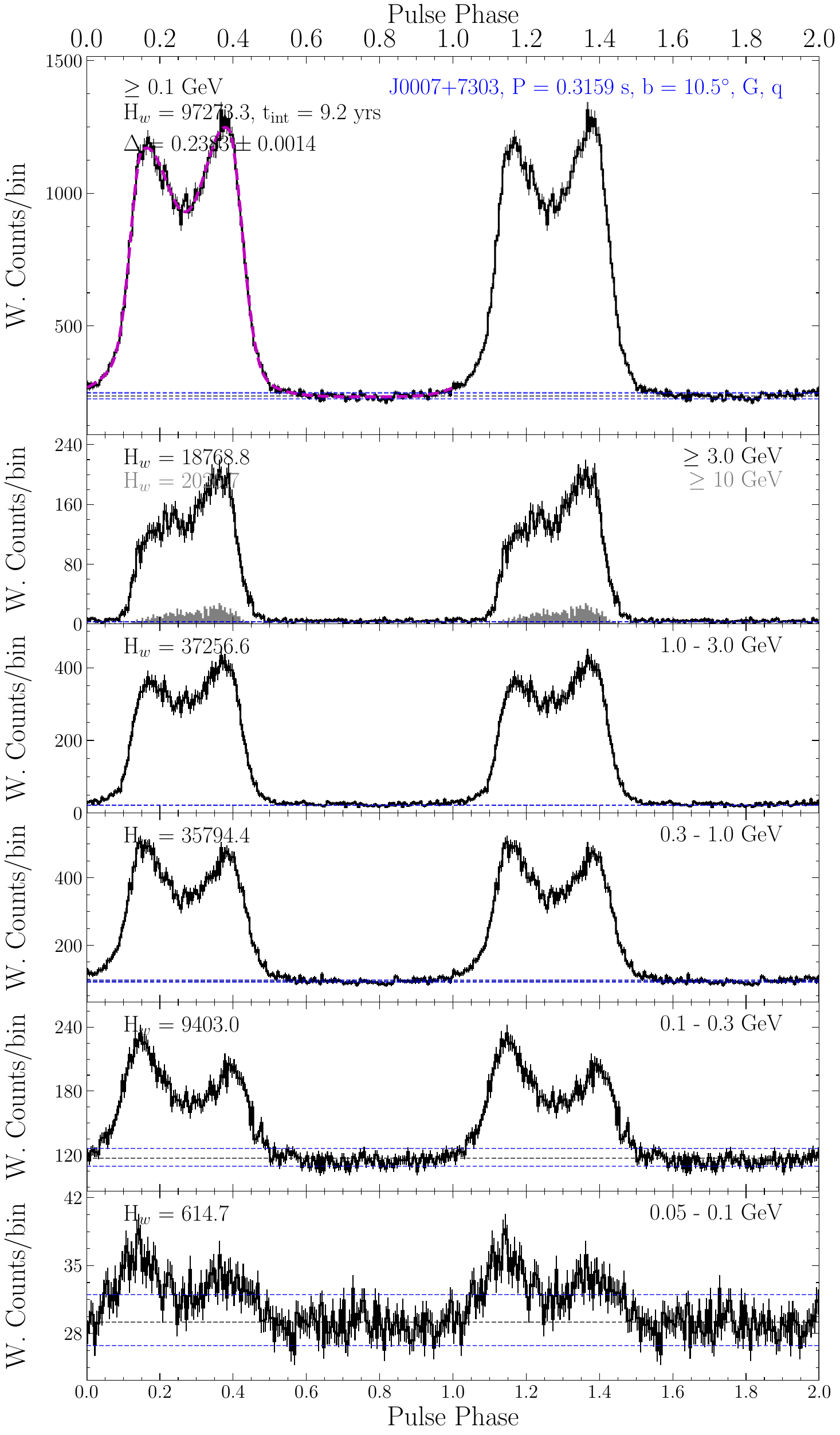}
\caption{Same as Figure~\ref{profiles:example_lightcurve_0030}, for PSR~J0007+7307.}
\label{App-Samples:example_lightcurve_0007}
\end{figure}

\begin{figure}[!ht]
\centering
\includegraphics[scale=0.37]{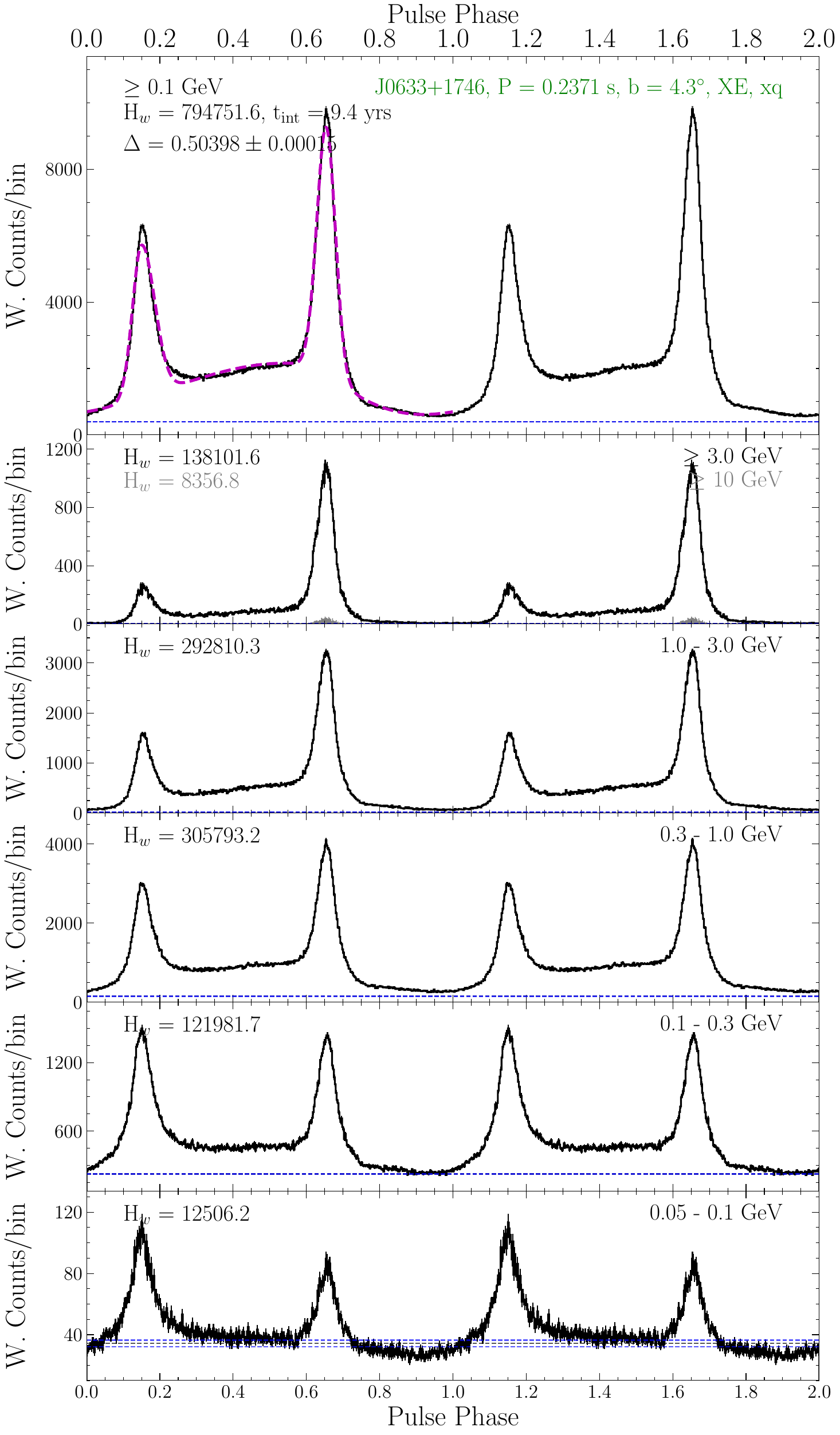}
\caption{Same as Figure~\ref{profiles:example_lightcurve_0030}, for PSR~J0633+1746 (the Geminga pulsar).}
\label{App-Samples:example_lightcurve_0633}
\end{figure}

\begin{figure}[!ht]
\centering
\includegraphics[scale=0.37]{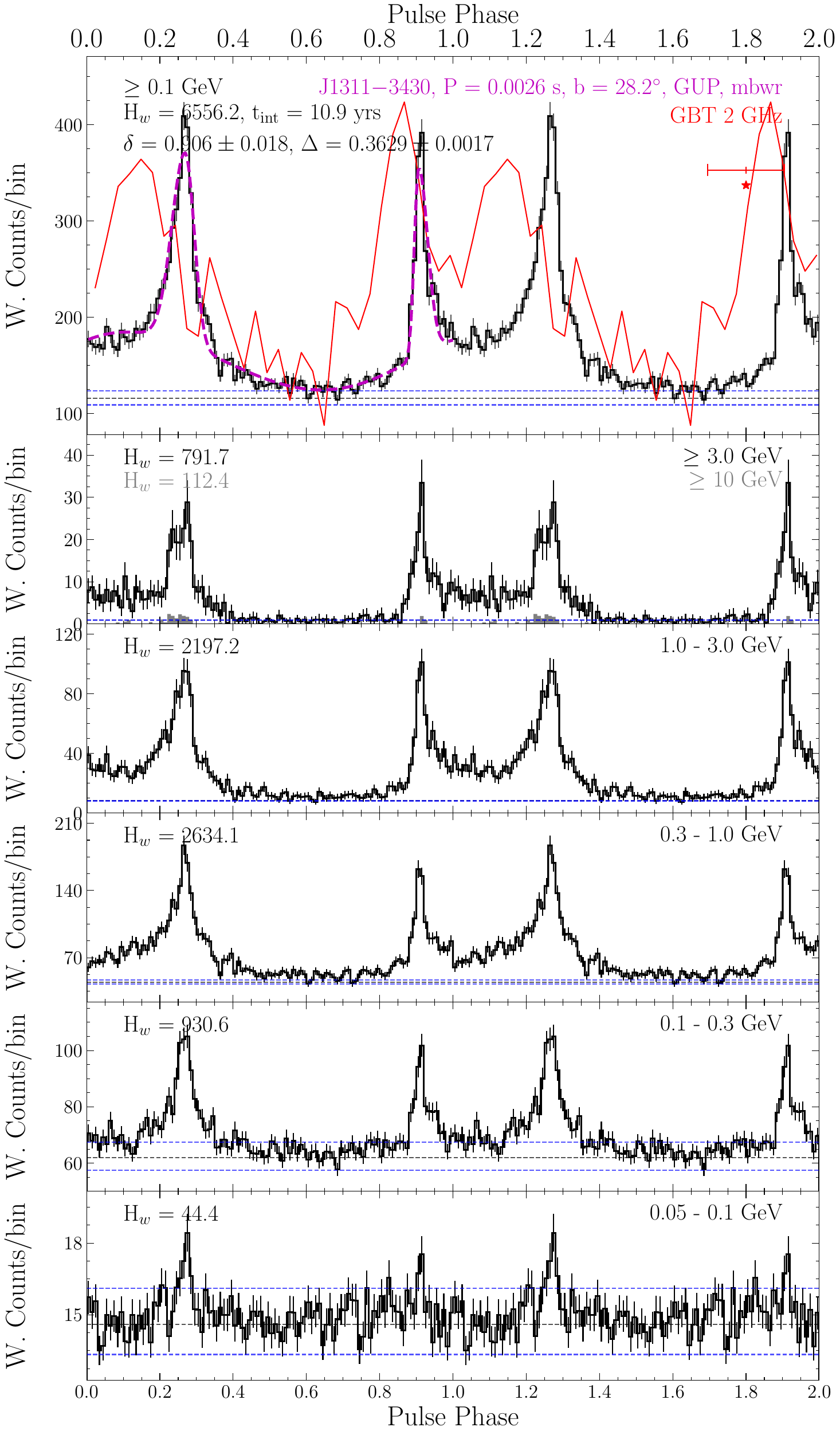}
\caption{Same as Figure~\ref{profiles:example_lightcurve_0030}, for PSR~J1311$-$3430. The $\star$ symbol indicates that we lacked information
needed for absolute phase alignment of the radio and gamma-ray profiles (see Section \ref{profiles:lcs}).}
\label{App-Samples:example_lightcurve_1311}
\end{figure}

\begin{figure}[!ht]
\centering
\includegraphics[scale=0.37]{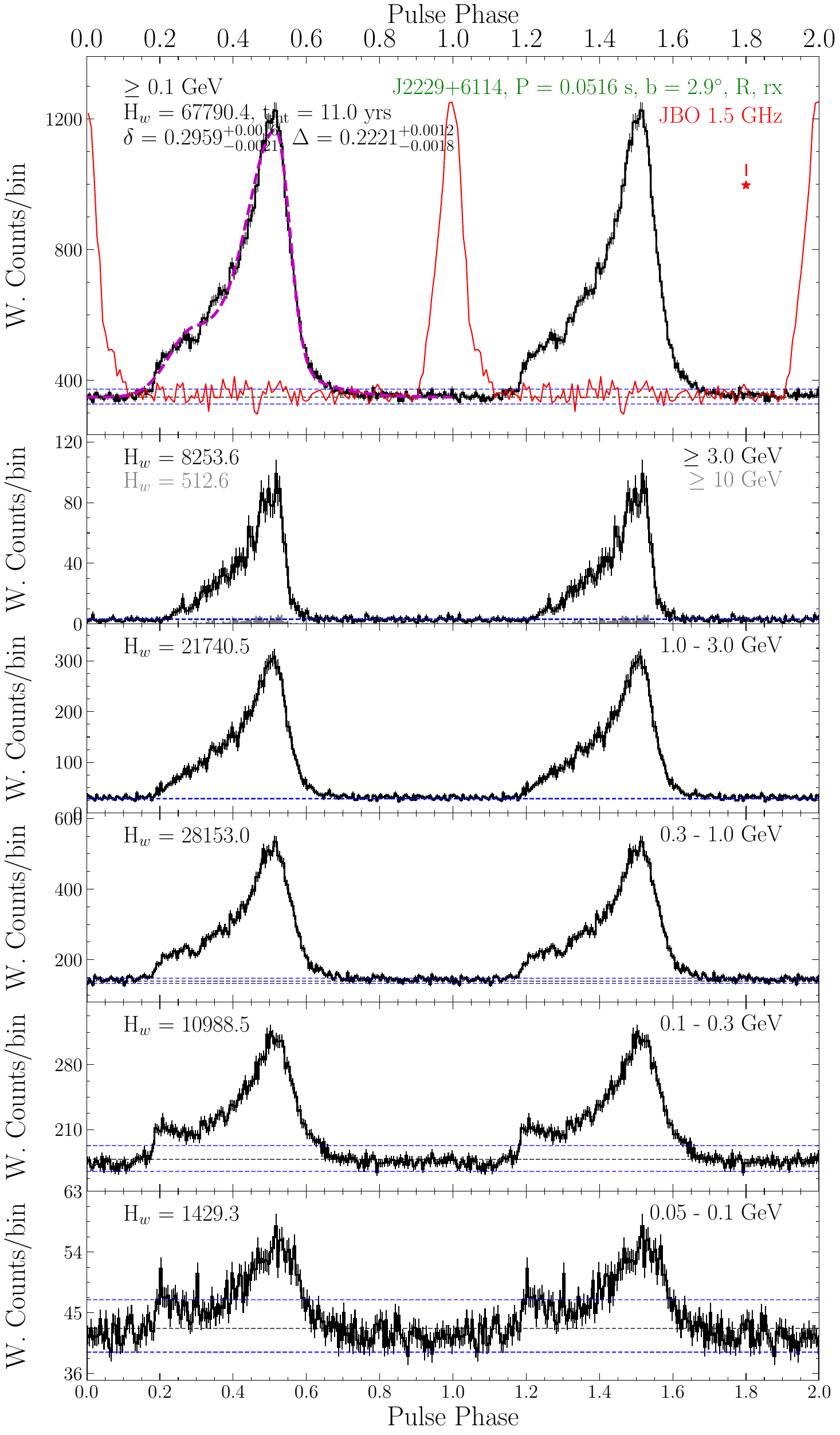}
\caption{Same as Figure~\ref{profiles:example_lightcurve_0030}, for PSR~J2229+6114.
The $\star$ symbol indicates that we lacked information
needed for absolute phase alignment of the radio and gamma-ray profiles (see Section \ref{profiles:lcs}).}
\label{App-Samples:example_lightcurve_2229}
\end{figure}

\begin{figure}[!ht]
\centering
\includegraphics[width=1.0\textwidth]{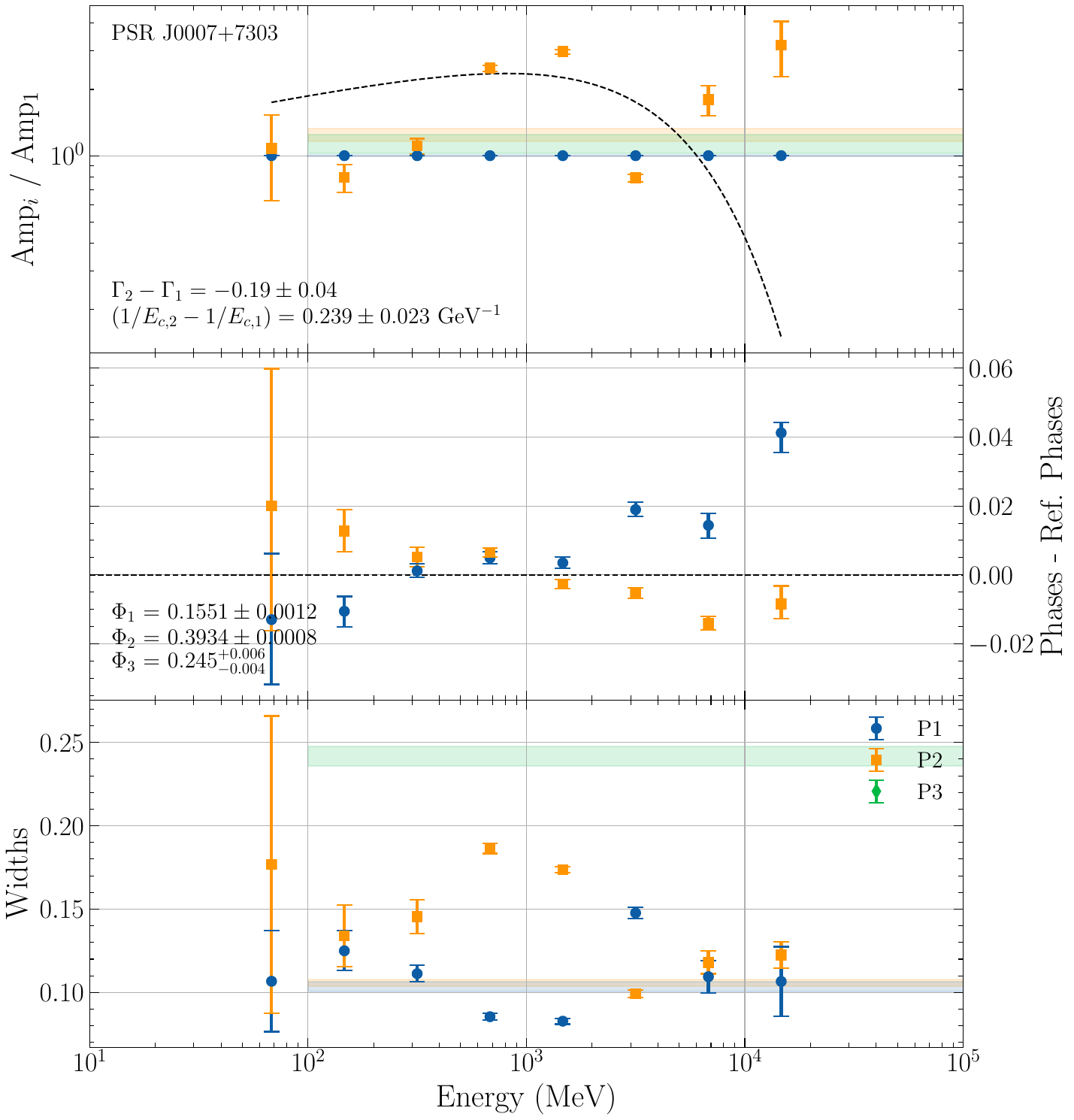}
\caption{Same as Figure~\ref{profiles:Eres_example}, for PSR~J0007+7303.
Here, the P3 (bridge) component was too weak to fit in the different energy intervals, hence the points are not shown (see Section \ref{sec:profs_energy}). }
\label{App-Samples:example_eres_0007}
\end{figure}

\begin{figure}[!ht]
\centering
\includegraphics[width=1.0\textwidth]{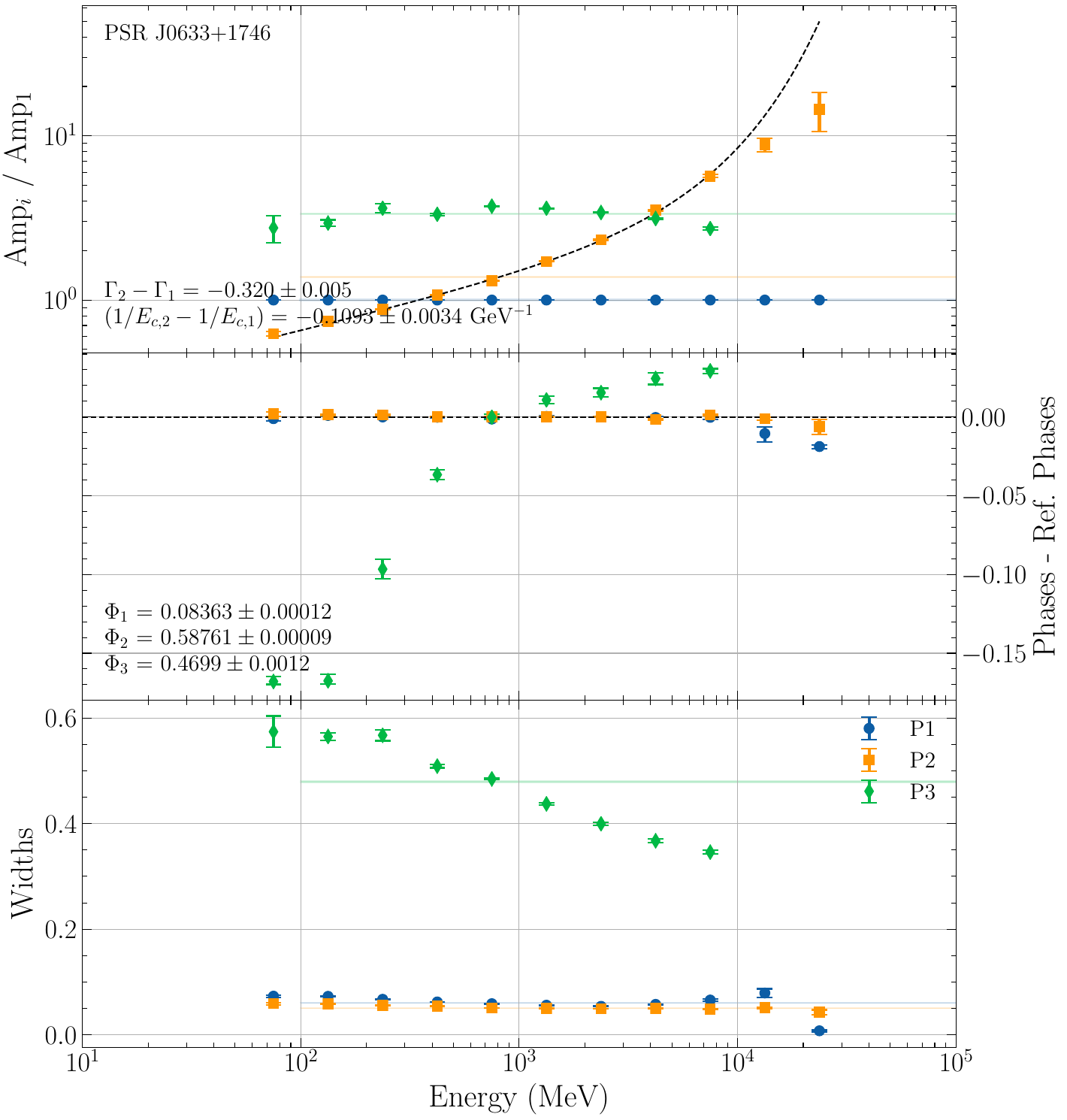}
\caption{Same as Figure~\ref{profiles:Eres_example}, for PSR~J0633+1746 (the Geminga pulsar).}
\label{App-Samples:example_eres_0633}
\end{figure}

\begin{figure}[!ht]
\centering
\includegraphics[width=1.0\textwidth]{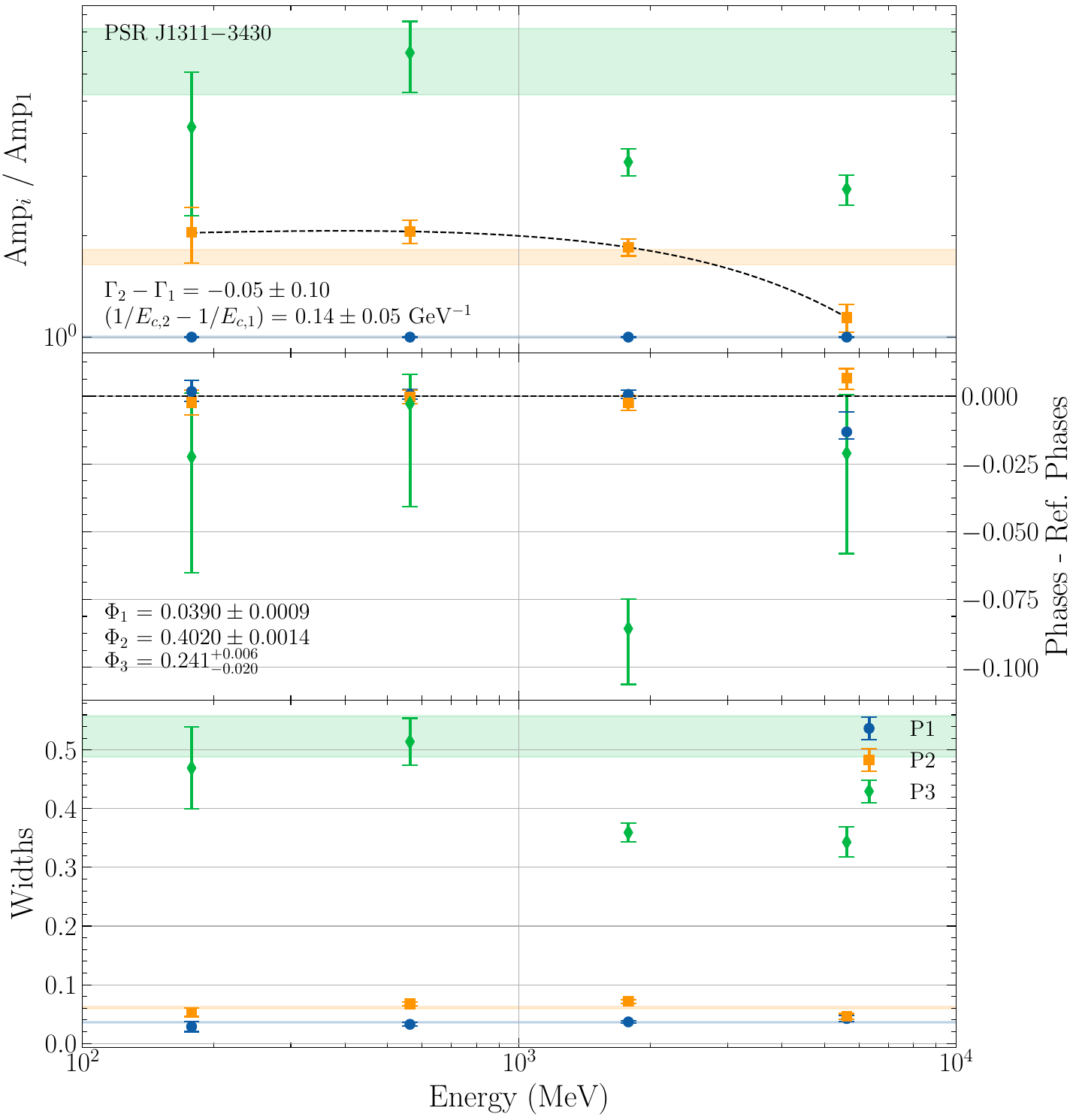}
\caption{Same as Figure~\ref{profiles:Eres_example}, for PSR~J1311$-$3430.}
\label{App-Samples:example_eres_1311}
\end{figure}

\begin{figure}[!ht]
\centering
\includegraphics[width=1.0\textwidth]{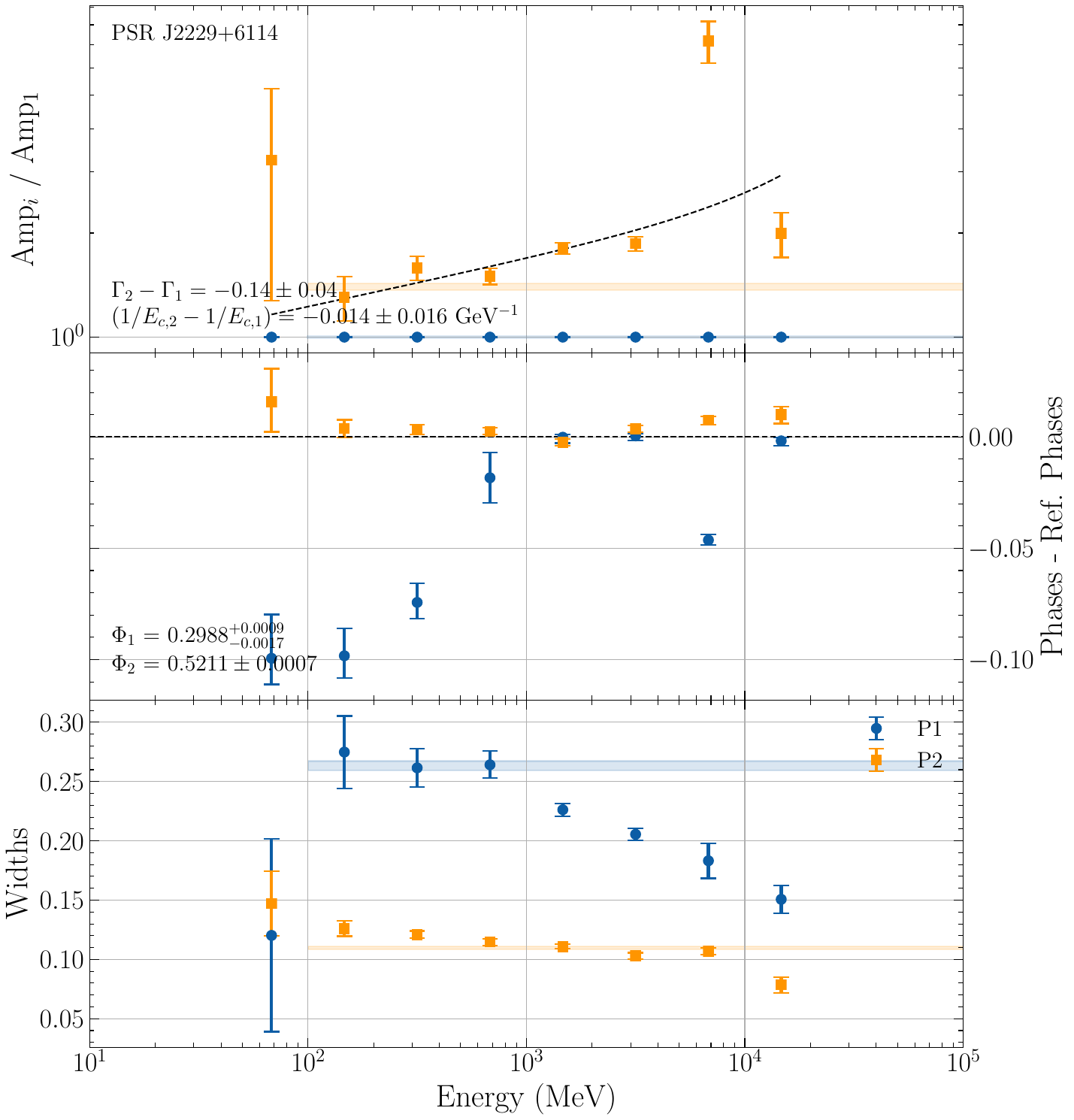}
\caption{Same as Figure~\ref{profiles:Eres_example}, for PSR~J2229+6114.}
\label{App-Samples:example_eres_2229}
\end{figure}


Figure ~\ref{fig-J1227spec} shows sample spectral energy distributions (SEDs) provided as part of 
4FGL-DR3\footnote{\url{https://fermi.gsfc.nasa.gov/ssc/data/access/lat/12yr_catalog/}}. 
The supplementary online material for this pulsar catalog includes a compilation of the SEDs of the
gamma-ray pulsars, in PDF format. It also includes the spectral parameters with their covariance matrices and the data points necessary to reproduce the SEDs, and a python script that performs the task.

\begin{figure}[!ht]
 \centering
\includegraphics[width=0.47\textwidth]{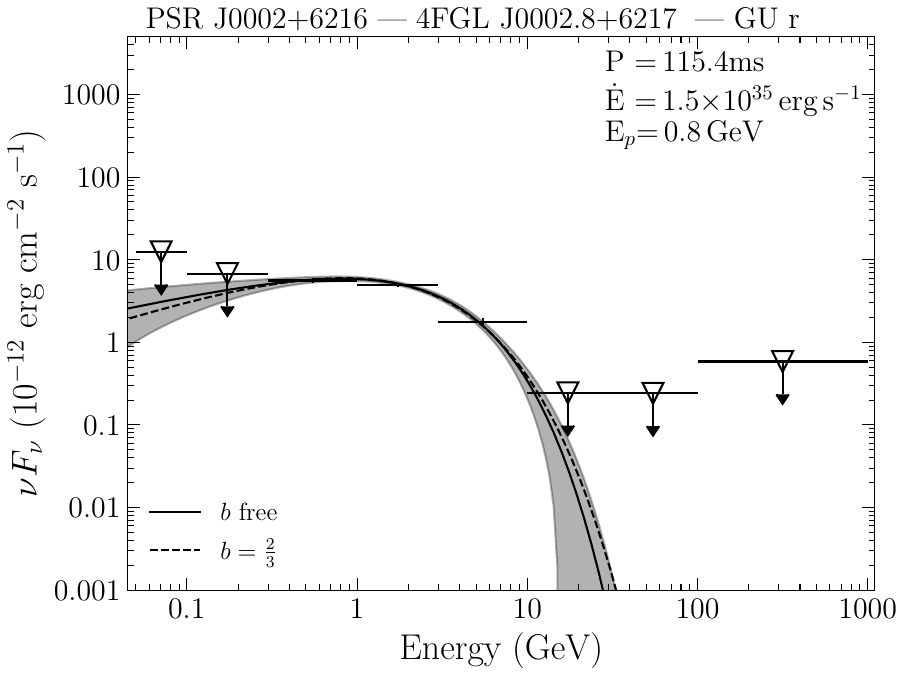}
\includegraphics[width=0.47\textwidth]{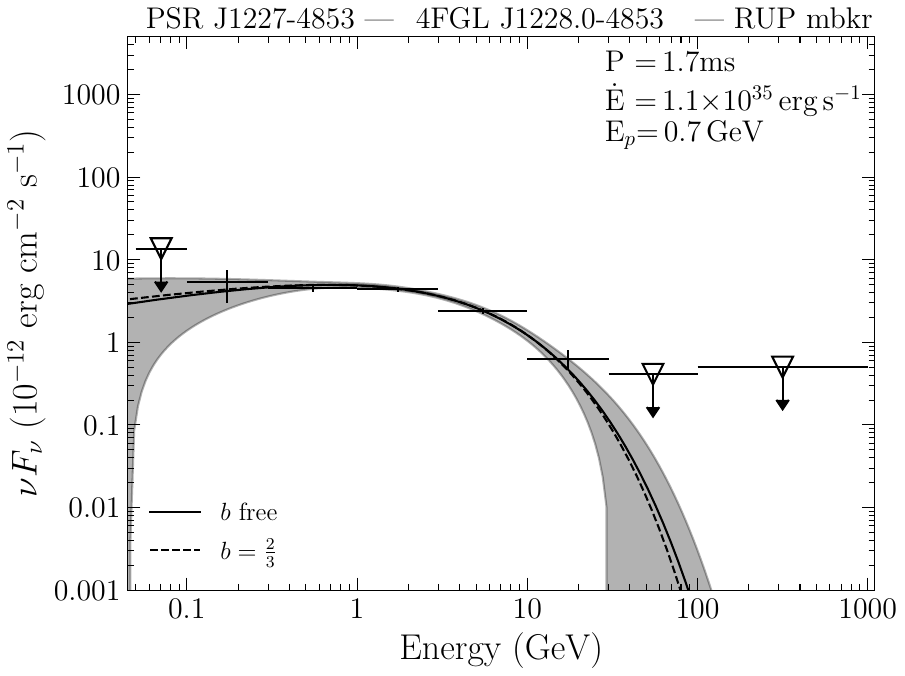}
\includegraphics[width=0.47\textwidth]{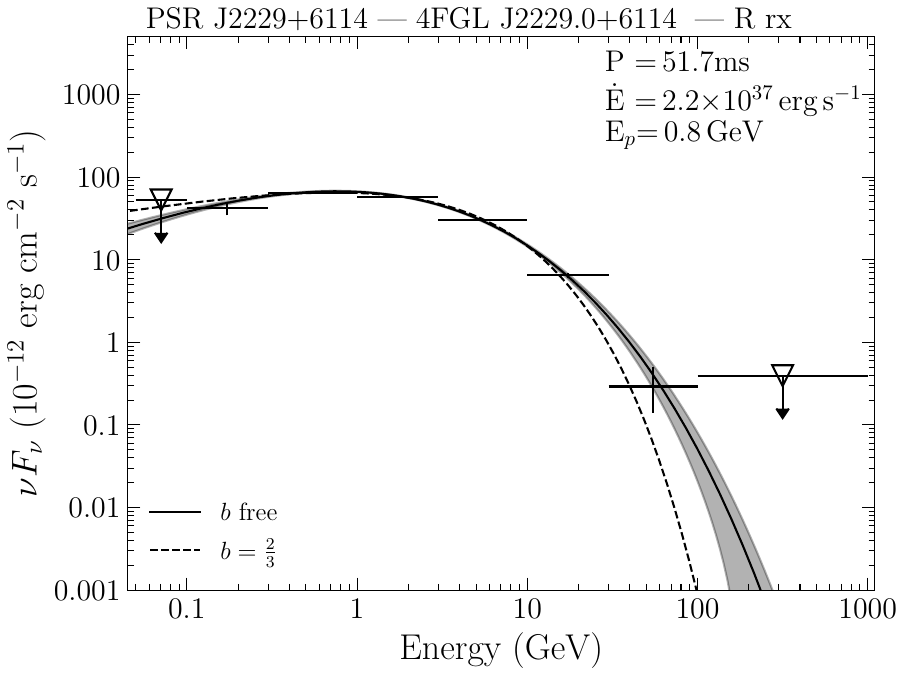}
\includegraphics[width=0.47\textwidth]{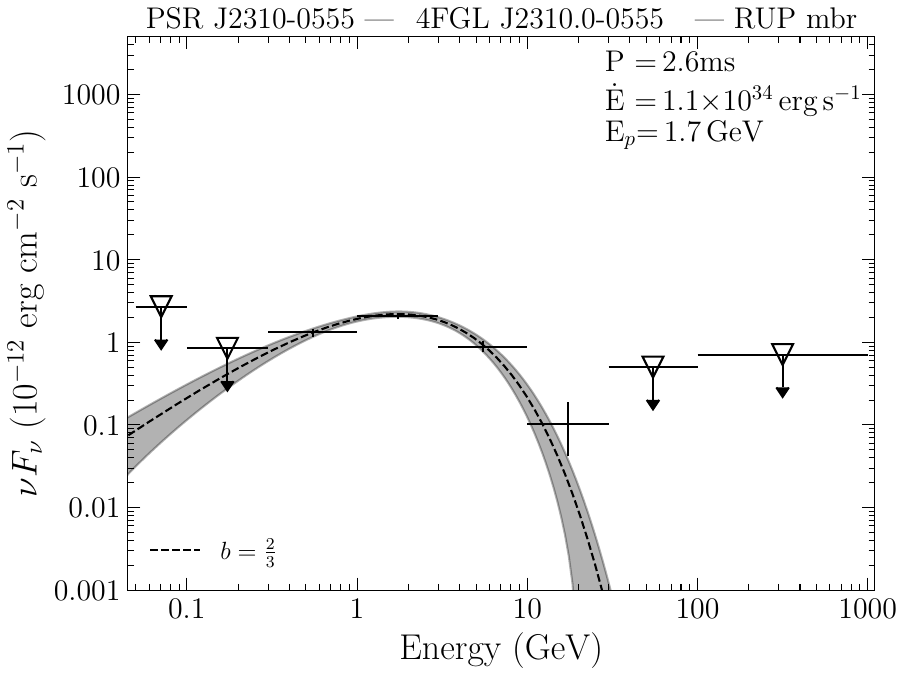}
\caption{Typical gamma-ray spectral energy distributions (SEDs). The plots were made using the python script \textit{sed\_plotter.py} and data files in the Online Material. The SEDs for all the gamma-ray pulsars detected as point sources (TS $>25$) are in the Online Material as well.
Similar to those provided with 4FGL-DR3, they use the refits done for some pulsars, described in Section \ref{spectralSection}.  When  $b_{fr}$ exists, both $b_{23}$ and $b_{fr}$ fits are shown, with the $b_{fr}$ uncertainty envelope. 
\label{fig-J1227spec}}
\end{figure}

\clearpage 

\section{Additional Notes on Spectral Analysis}
\label{sec:spectral_notes}
\renewcommand{\thefigure}{C-\arabic{figure}}

We study here the distribution of the model parameters and derived
physical quantities as they vary among the \nbfree{} pulsars which have both
$b$-free and $b=2/3$ fits available.  These results are presented in Figure
\ref{fig:model_comp}.  In brief, they are
\begin{itemize} 
  \item $\Gamma$, the spectral index at the reference energy $\epiv$, is
    stable, varying by $<$10\% over the
    effective range in $b$.
  \item The curvature at the reference energy, $d$, varies more than
    $\Gamma$, by up to $\sim$50\%.  If considering $b<1$, the total
    range is only about $\pm$20\%.
  \item $\Gamma_{100}$ varies almost linearly with $b$ before
    saturating at $b=1.2$.  The shift is up to $1$ unit harder for
    pulsars with $b$ near 0, and $0.5$ units softer for (mostly) MSPs
    with $b\geq1.2$.  Thus, clearly, freeing $b$ profoundly
    alters the inferred low-energy spectral index.
  \item The curvature at the peak, $\dpeak$, is stable if the
    intrinsic value of $b>0.5$.  For pulsars with spectra more
    consistent with small $b$, the $b$-free pulsars become sharper,
    with a difference in log-width ranging up to $2\times$.  It is
    not (yet) clear if this difference primarily ``fixes'' the true
    width of the peak or delivers a better low-energy index.
  \item The peak energy $\epeak$ varies by $<$10--15\% in the ``core''
    $b$ range and $<25$\% overall.  The largest deviations are for the
    MSPs with fit values $b>1$ which move to higher $\epeak$.
\end{itemize}
In summary, the physical properties measured at the peak, $\dpeak$ and
$\epeak$, are robustly measured in both $b$-free and $b=2/3$ fits and
so we can make inferences on their distribution using the entire pulsar sample.  
On the other hand, the low-energy spectrum is
unreliable, and values inferred from the $b=2/3$ model could be very
far from the true values.

\begin{figure*}
\centering
  \includegraphics[angle=0,width=0.48\linewidth]{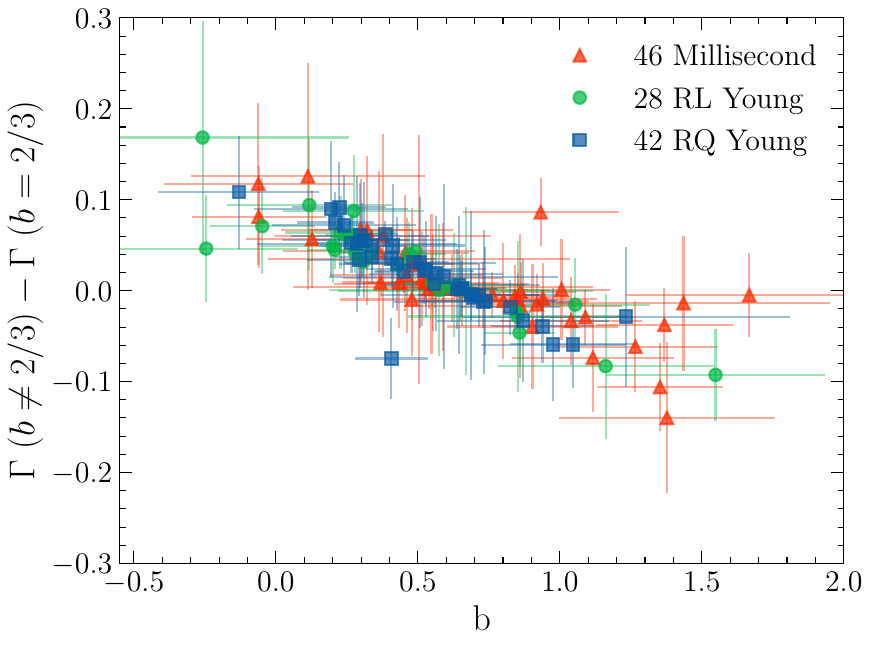}
  \includegraphics[angle=0,width=0.48\linewidth]{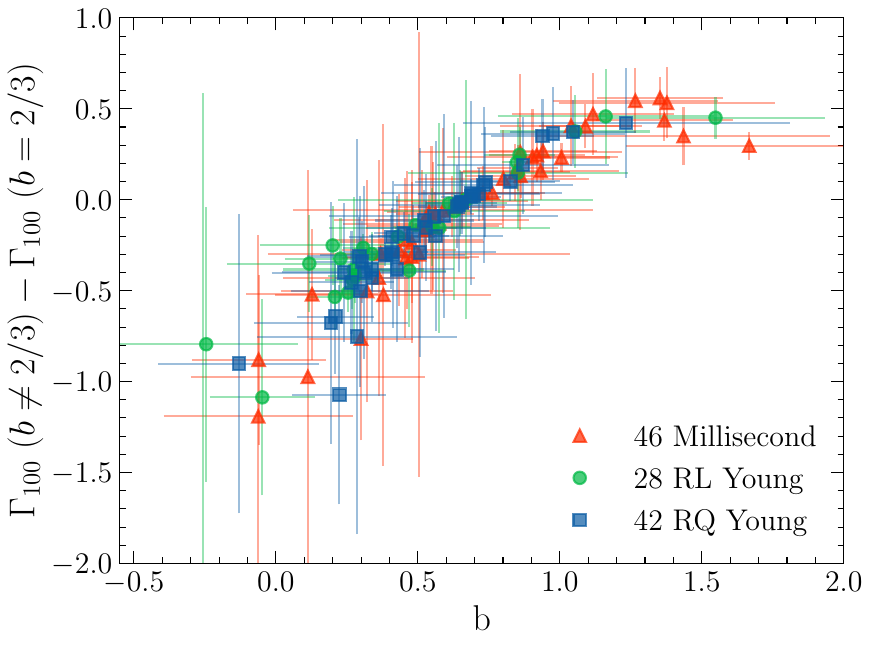}
  \includegraphics[angle=0,width=0.48\linewidth]{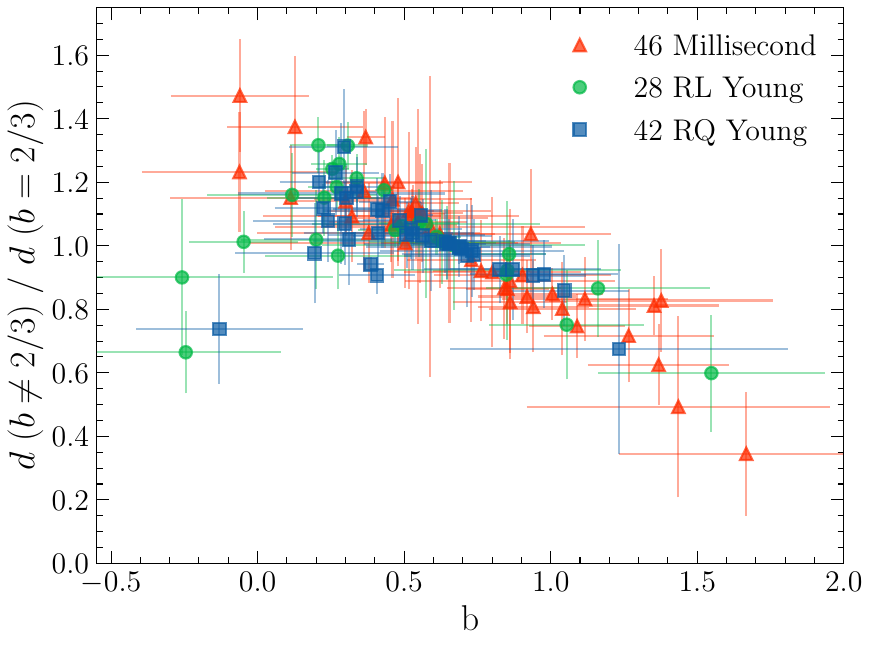}
  \includegraphics[angle=0,width=0.48\linewidth]{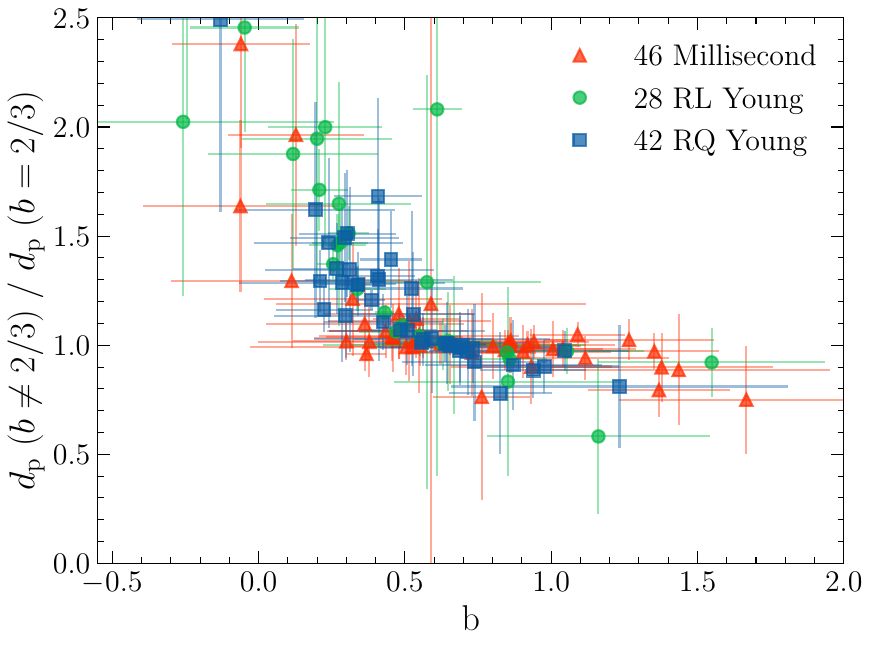}
  \includegraphics[angle=0,width=0.48\linewidth]{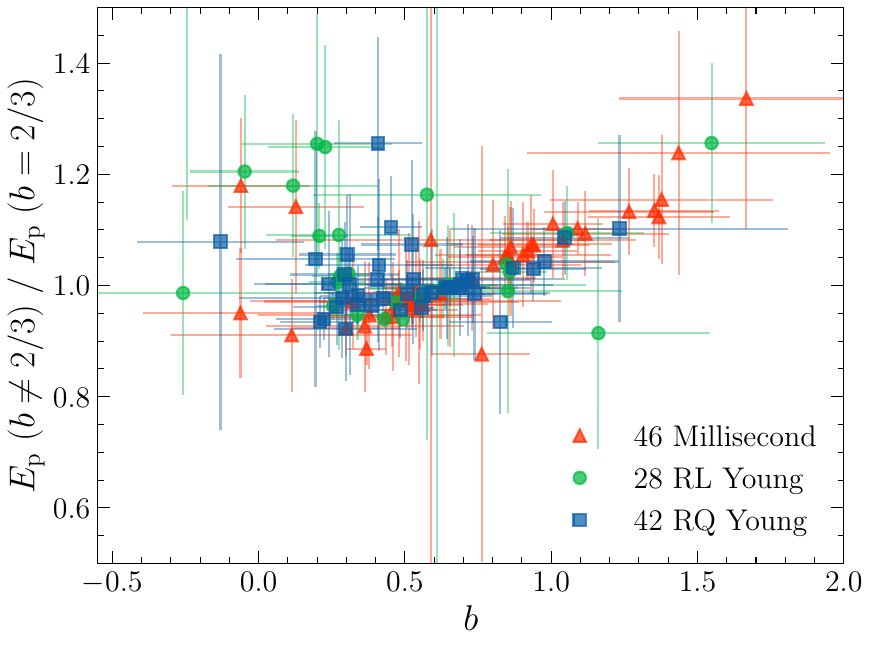}
  \caption{\label{fig:model_comp}Comparisons of base and derived
  parameters between the $b=\frac{2}{3}$ and $b$-free populations.
  See main text for further discussion.}
\end{figure*}
\clearpage

\section{Description of the online catalog files
\label{online}}
Online material is available through this Journal as well as at \url{https://fermi.gsfc.nasa.gov/ssc/data/access/lat/3rd_PSR_catalog/}. Here we describe the files posted as of publication. Additional material may be added in the future. The ``xxx'' in the file names is a multi-digit timestamp of the file creation date. Contact the authors to request missing information.

\textbf{3PC\_Catalog+SEDs\_xxx.fits,  3PC\_Catalog\_xxx.xls}: The pulsar catalog ``flat file'', in FITS and spreadsheet formats. Each has 294 rows, one for each gamma-ray pulsar. The 112 columns contain basic pulsar properties (name, position, spin parameters, distance, radio flux density, et cetera) as well as the spectral and profile parameters described in this article. The FITS version includes a subtable (``HDU'') with the detailed spectral fit parameters and covariance matrix needed to reproduce the Spectral Energy Distributions (SEDs), provided in \texttt{3PC\_SEDs.tgz} (individual plots) and \texttt{3PC\_AllSEDs.pdf} (all plots concatenated into a single file). 
\texttt{3PC\_SEDPlotter.tgz} contains a python script that reads 
\texttt{3PC\_Catalog+SEDs\_xxx.fits} to generate the SED plots.
Most Figures and Tables in this article were created from the FITS and spreadsheet files: the python script \texttt{Example\_3PC\_CatalogReader.py} is a simple example.
\texttt{README\_3PC\_Catalog+SEDs\_xxx.txt provides more detail}.

\textbf{3PC\_ProfileData\_xxx.tgz}: Contains two files for each pulsar, one in FITS format and the other in simple text (ascii), containing the weighted gamma-ray phase histograms and, when they exist, the fit to the histogram and/or the radio profile. The python script 
\texttt{Example\_3PC\_PlotProfileData.py} illustrates how to read the files to reproduce the plots provided in \texttt{3PC\_ProfilePlotsPDF\_xxx.tgz}  and
\texttt{3PC\_ProfilePlotsPNG\_xxx.tgz}. 
The plots of gamma-ray peak evolution with energy, as in Figures \ref{profiles:Eres_example} and \ref{App-Samples:example_eres_0007} to \ref{App-Samples:example_eres_2229}, are provided in
\texttt{3PC\_LC\_Eres\_pdf\_xxx.tgz, 3PC\_LC\_Eres\_png\_xxx.tgz}.
\texttt{README\_ProfileData.txt} provides details. 

\textbf{3PC\_TimingModels\_xxx.tgz}: A compressed archive of the 294 rotation ephemerides (``.par files'') used to calculate the gamma-ray rotational phases using software such as \textsc{Tempo2} or PINT. \texttt{README\_3PC\_TimingModels.txt} provides details.

\textbf{Phased, weighted FT1 FITS files:} available at \url{https://heasarc.gsfc.nasa.gov/FTP/fermi/data/lat/catalogs/3PC/photon/}. Two subdirectories called \texttt{3deg\_50MeV} and \texttt{15deg\_20MeV} each contain one file for each gamma-ray pulsar, in the standard \Fermi{} LAT data format, with additional columns PHASE and MODEL\_WEIGHT.  The 277 files in  \texttt{3deg\_50MeV} include photons with energies $> 50$ MeV, within $3^\circ$ of the pulsar position, and were used to create the gamma-ray phase histograms. They occupy roughly 30 Gb of disk space, with a large variation in single file size depending on the background intensity (generally decreasing away from the Galactic plane) and pulsar intensity.
The files in \texttt{15deg\_20MeV} are on average $15\times$ larger, 
including photons with energies $> 20$ MeV within $15^\circ$ of the pulsar position.
They are suitable for spectral analyses (e.g.  using the \texttt{gtlike} \Fermi{} tool)
or to explore the lowest LAT photon energies.
\texttt{README\_3PC\_PhotonData\_.txt} describes the FT1 files in more detail.

\textbf{3PC\_sensitivity\_xxx.fits :} The all-sky \Fermi{} LAT pulsar sensitivity, in FITS format, useful for e.g. population syntheses. Example\_3PC\_SensitivityMap\_xxx.py is a python script that reads the FITS file and generates a plot as in Figure \ref{SeattleSkySensitivity}. 
\texttt{README\_3PC\_SensitivityMap\_xxx.txt} provides details.

Finally, we provide HTML files for each pulsar, to allow a visual summary of the information for each gamma-ray pulsar, named \texttt{3PC/J0007+7303\_LAT.html} and so on.


\hspace{2.0in}

\textit{Acknowledgments:} {\small 
The {\it Fermi}-LAT Collaboration acknowledges generous ongoing support from a number of agencies and institutes that have supported both the development and the operation of the LAT as well as scientific data analysis. These include the National Aeronautics and Space Administration and the Department of Energy in the United States, the Commissariat \`a l'Energie Atomique and the Centre National de la Recherche Scientifique / Institut National de Physique Nucl\'eaire et de Physique des Particules in France, the Agenzia Spaziale Italiana and the Istituto Nazionale di Fisica Nucleare in Italy, the Ministry of Education, Culture, Sports, Science and Technology (MEXT), High Energy Accelerator Research Organization (KEK) and Japan Aerospace Exploration Agency (JAXA) in Japan, and the K.~A.~Wallenberg Foundation, the Swedish Research Council and the Swedish National Space Board in Sweden.\newline
Additional support for science analysis during the operations phase is gratefully acknowledged from the Istituto Nazionale di Astrofisica in Italy and the Centre National d'\'Etudes Spatiales in France. This work performed in part under DOE Contract DE-AC02-76SF00515. \newline
The Parkes radio telescope is part of the Australia Telescope which is funded by the Commonwealth Government for operation as a National Facility managed by CSIRO.  The Green Bank Telescope is operated by the National Radio Astronomy Observatory, a facility of the National Science Foundation operated under cooperative agreement by Associated Universities, Inc. The Arecibo Observatory is part of the National Astronomy and Ionosphere Center (NAIC), a national research center operated by Cornell University under a cooperative agreement with the National Science Foundation. The Nan\c cay Radio Observatory is operated by the Paris Observatory, associated with the French Centre National de la Recherche Scientifique (CNRS). The Lovell Telescope is owned and operated by the University of Manchester as part of the Jodrell Bank Centre for Astrophysics with support from the Science and Technology Facilities Council of the United Kingdom. 
The International LOFAR Telescope and the Westerbork Synthesis Radio
Telescope are operated by ASTRON, the Netherlands Institute for Radio
Astronomy.
This work made extensive use of the ATNF pulsar catalog \citep{ATNFcatalog}.
IG, AKH, and DAS thank the ISSI (Bern, Switzerland) for financial support of Team 459 meetings that helped to improve the present work.}

\vspace{5mm}
\facilities{Fermi(LAT)}

\software{Tempo2 \citep{Hobbs2006}, 
PINT \citep{Luo21},
astropy \citep{astropy},  
numpy \citep{numpy}
}

\bibliographystyle{aasjournal}
\bibliography{3rdPulsarCatalog,psrcat}



\end{document}